\documentclass[pdflatex,sn-mathphys-ay]{sn-jnl}

\usepackage{graphicx}%
\usepackage{multirow}%
\usepackage{amsmath,amssymb,amsfonts}%
\usepackage{amsthm}%
\usepackage{mathrsfs}%
\usepackage[title]{appendix}%
\usepackage{xcolor}%
\usepackage{textcomp}%
\usepackage{manyfoot}%
\usepackage{booktabs}%
\usepackage{algorithm}%
\usepackage{algorithmicx}%
\usepackage{algpseudocode}%
\usepackage{listings}%
\usepackage{soul, CJK}
\usepackage[dvipsnames]{xcolor}
\usepackage{placeins}
\usepackage{subcaption}
\usepackage{rotating,tabularx}
\usepackage{arydshln}
\usepackage{siunitx}

\usepackage{upgreek}

\theoremstyle{thmstyleone}%
\theoremstyle{thmstyletwo}%

\theoremstyle{thmstylethree}%

\def \bp{$\upbeta$\,Pic}

\def\fei{Fe\,{\sc i}}
\def\feii{Fe\,{\sc ii}}

\def\ovi{O\,{\sc vi}}

\def\mnii{Mn\,{\sc ii}}

\def\mgii{Mg\,{\sc ii}}

\def\nii{Ni\,{\sc i}}
\def\niii{Ni\,{\sc ii}}

\def\nai{Na\,{\sc i}}

\def\cai{Ca\,{\sc i}}
\def\caii{Ca\,{\sc ii}}

\def\siii{Si\,{\sc ii}}

\def\siiv{Si\,{\sc iv}}

\def\aliii{Al\,{\sc iii}}

\def\cii{C\,{\sc ii}}
\def\ciii{C\,{\sc iii}}
\def\civ{C\,{\sc iv}}
\begin{document}

\title[Article Title]{Exocometary physics: material release and tails}



\author[1]{\fnm{Théo} \sur{Vrignaud}}\email{vrignaud@iap.fr}

\author[2]{\fnm{Dennis} \sur{Bodewits}}\email{dennis@auburn.edu}
\equalcont{These authors contributed equally to this work.}

\author[3]{\fnm{Jake} \sur{Hanlon*}}\email{jake.hanlon.23@ucl.ac.uk *}
\equalcont{These authors contributed equally to this work.}


\author[4]{\fnm{Matthew M.} \sur{Knight}}\email{knight@usna.edu}
\equalcont{These authors contributed equally to this work.}

\author[5]{\fnm{Tim D.} \sur{Pearce}}\email{tim.pearce@warwick.ac.uk}
\equalcont{These authors contributed equally to this work.}

\author[6,7]{\fnm{Darryl Z.} \sur{Seligman}}\email{dzs@msu.edu}
\equalcont{These authors contributed equally to this work.}

\author[8,9,10]{\fnm{Dimitri} \sur{Veras}}\email{dimitri.veras@aya.yale.edu}
\equalcont{These authors contributed equally to this work.}

\author[3,11]{\fnm{Geraint H.} \sur{Jones}}\email{geraint.jones@esa.int}
\equalcont{These authors contributed equally to this work.}

\affil[1]{\orgdiv{Institut d'Astrophysique de Paris}, \orgname{CNRS, UMR 7095, Sorbonne Université}, \orgaddress{\street{98bis boulevard Arago}, \city{75014 Paris}, \country{France}}}

\affil[2]{\orgdiv{Department of Physics}, \orgname{Auburn University}, \orgaddress{\street{Edmund C. Leach Science Center}, \city{Auburn}, \postcode{36849}, \state{AL}, \country{USA}}}

\affil[3]{\orgdiv{Mullard Space Science Laboratory}, \orgname{University College London}, \orgaddress{\street{Holmbury St. Mary}, \city{Dorking}, \postcode{RH5 6NT}, \state{Surrey}, \country{United Kingdom}*}}

\affil[4]{\orgdiv{Physics Department}, \orgname{United States Naval Academy}, \orgaddress{\street{572C Holloway Rd}, \city{Annapolis}, \postcode{21402}, \state{MD}, \country{USA}}}

\affil[5]{\orgdiv{Department of Physics}, \orgname{University of Warwick},  \city{Coventry}, \postcode{B5 4TF}, \country{UK}}

\affil[6]{\orgdiv{Department of Physics and Astronomy}, \orgname{Michigan State University}, \orgaddress{\street{ \,}, \city{East Lansing}, \postcode{48824}, \state{MI}, \country{USA}}}

\affil[7]{\orgdiv{NSF Astronomy and Astrophysics Postdoctoral Fellow}}

\affil[8]{\orgdiv{Centre for Exoplanets and Habitability}, \orgname{University of Warwick},  \city{Coventry}, \postcode{B5 4TF}, \country{UK}}

\affil[9]{\orgdiv{Centre for Space Domain Awareness}, \orgname{University of Warwick},  \city{Coventry}, \postcode{B5 4TF}, \country{UK}}

\affil[10]{\orgdiv{Department of Physics}, \orgname{University of Warwick},  \city{Coventry}, \postcode{B5 4TF}, \country{UK}}

\affil[11]{\orgdiv{European Space Technology Centre (ESTEC)}, \orgname{European Space Agency}, \orgaddress{\street{Keplerlaan 1}, \city{2200 AG 
 Noordwijk}, \country{Netherlands}}}

\affil[*]{Corresponding author}

\abstract{Despite decades of observations, the physical processes governing mass loss from small bodies beyond our Solar System remain poorly constrained.  These ``exocomets" are often treated as analogs of Solar System comet, yet the stellar environments they inhabit spans a wide range in terms of luminosity, stellar wind intensity, and evolutionary stage, leading to potentially very diverse physical behaviors. 
Within our Solar System, small bodies lose gas and dust through a range of mechanisms, including sublimation, desorption, impacts, and/or sputtering. Once released, the composition and dynamics of the ejecta are then altered by additional processes, such as dust sublimation, ionization, and radiation pressure. In extrasolar systems, these mechanisms unfold under vastly different radiative and plasma conditions, leading to a rich diversity of mass-loss pathways and observable signatures.

This work reviews our understanding of the mechanisms driving mass loss from small bodies and the subsequent evolution of ejecta in diverse stellar environments. We compare the physical and chemical mechanisms that drive gas and dust production, such as sublimation, thermal and photon desorption, and investigate how they scale with stellar luminosity, temperature, and activity. We then examine the processes that modify the composition of the ejecta (e.g., dust sublimation, dissociation, or ionisation) and its dynamics (e.g., radiation pressure or stellar winds). To illustrate how these processes vary across different stellar environments, we use four well-studied planetary systems as case studies: the Sun, $\beta$\,Pictoris, AU~Microscopii, and WD~1145+017. By exploring how cometary tails behave under such diverse conditions, this work provides a physical framework for interpreting exocometary activity and sheds light on why A-type stars, such as the famous $\beta$\,Pictoris, are over-represented in the population of exocomet-hosting stars.}





\maketitle

\section{Introduction}
\label{Sect. intro}

Small bodies in the Solar System, such as comets and asteroids, are relics from the era of planet formation. When their orbits are altered, they can redistribute water, noble gases, and organic materials across a wide range of heliocentric distances \citep{Rubin2019}. In our Solar System, these objects can be studied directly through telescopic observations and spacecraft missions, providing detailed insights into their composition and evolution. Beyond our Solar System, however, small bodies are typically studied indirectly. Observations around other stars often capture the collective properties of multiple bodies rather than the unique characteristics of single objects \citep{Strom2020}. When small bodies release gas and dust into space, whether by sublimation, thermal desorption, or other processes, this ``activity" offers a window into their chemical composition. Yet even within the Solar System, sublimation represents only one pathway for mass loss. Thermal and photon desorption, plasma sputtering, micrometeroid impacts, and electrostatic lofting also shape the evolution of airless bodies and asteroids. Linking these ejecta back to the physical properties of the parent body requires an understanding of storage and release mechanisms, the physical and chemical processes that alter these materials, and an accurate interpretation of their excitation to determine relative abundances \citep[e.g., the recent review by][]{Bodewits2022}.

Solar System small bodies experience a wide range of physical environments that shape their behavior. For example,  many comets sublimate as they approach the Sun, forming characteristic gas and dust tails. Some objects, like the near-Sun  asteroid (3200) Phaethon, reach surface temperatures exceeding 1000~K \citep[see the review by][]{Jones2018}, leading to the thermal desorption of refractory elements such as alkali metals \citep{Zhang2023}. In contrast, more distant objects remain too cold for significant sublimation; for example, water and CO$_2$, typically sublimate only within  2.5~au and 13~au of the Sun, respectively  \citep[e.g.,][]{Meech2005}. Around other stars, small bodies may encounter even more extreme conditions. For example, exocomets around $\upbeta$ Pictoris (\bp) pass close to a hot A-type star, exposing them to intense radiation and high temperatures, while potential exocomets orbiting AU~Microscopii (hereafter AU~Mic), a highly variable M-dwarf,  endure intense stellar winds. These diverse environments add complexity to exocometary behavior and evolution.

This work reviews the physical processes governing mass loss and tail formation for small bodies and explores how these processes vary across different stellar environments. We compare Solar System comets with exocomets, examining the mechanisms driving gas and dust ejection, their dependence on surface composition, and the physical factors altering the ejected material. We also discuss observational constraints, including differences between absorption and emission features, and how these processes serve as diagnostics of local environments. To illustrate these key concepts, four case-study stars are used throughout this work: the Sun, \bp, AU~Mic, and WD 1145+017, each hosting a unique planetary system (see the stellar properties in Table \ref{table: case study stars}). These stars provide a framework for assessing whether Solar System comets can serve as valid analogs for exocomets.

\begin{table}[h!]
  \renewcommand{\arraystretch}{1.2}
    \begin{tabular}{ c c c c c c c c}   
    \hline
    \noalign{\smallskip}
    \hline
    \noalign{\smallskip}
    \textbf{Star} & Distance &  Mass            & Radius          & T$_{\rm eff}$       & Luminosity          & Age   & Planets \\
                  & (pc)     &  {[}$M_\odot${]} & {[}$R_\odot${]} &    [K]              & {[}$L_\odot${]}     & [Gyr] &         \\

            \noalign{\smallskip}
            \hline
            \noalign{\smallskip}

AU~Mic            & 9.7  & 0.6             & 0.82            & 3\,665              & 0.10                    & 0.023  & b, c, d, e?      \\
Sun               &      & 1.00            & 1.00            & 5\,770              & 1.00                    & 4.57  &    \\
Beta Pic          & 19.3 & 1.75            & 1.51            & 8\,052              & 8.7                    & 0.020  & b, c        \\ 
WD 1145+017       & 142  & 0.63            & 0.012           & 15\,020             & 0.02$^a$                    & 0.22   &  b$^b$       \\

    \noalign{\smallskip}
    \hline
    \end{tabular} 

    \caption{\small Properties of the four stars chosen as case-studies. \\
    \footnotesize 
    $\boldsymbol{[}\textbf{a}\boldsymbol{]}$ The progenitor of WD 1145+017 was an A-type star, which lived $550 \pm 100$ Myr before becoming a white dwarf \citep{izquierdo2021}.\\
    $\boldsymbol{[}\textbf{b}\boldsymbol{]}$ As of 2025, the dust activity around WD1145+017 has completely ceased, suggesting that planet b has likely been destroyed \citep{Aungwerojwit2024_WD_activity}.\\
    \textbf{References}: \\ 
    AU~Mic: \cite{donati_magnetic_2023};\\
    Sun: \cite{Prsa_2016_IAU, Bonanno_2002_Sun}; \\
    \bp:  \cite{Crifo1997, miret-roig_dynamical_2020, Gray2006};\\
    WD1145+017: \cite{gaia_collaboration_gaia_2018, izquierdo2021, Farihi2017}; \\ 
    }

\label{table: case study stars}
\end{table}

\begin{table}[h!]
    \begin{tabular}{ c c c c c c}   
    \hline
    \noalign{\smallskip}
    \hline
    \noalign{\smallskip}
    \textbf{Planet} & Semi-major axis &  Period & Mass &  Radius   & T$_{\rm eq}$    \\

            \noalign{\smallskip}
            \hline
            \noalign{\smallskip}
            
            AU~Mic b & 0.065 au & 8.5\,d & $10.2_{-2.7}^{+3.9} M_\oplus$ & $3.96 \pm 0.15 R_\oplus$ & 600\,K \\
            \noalign{\smallskip}
            AU~Mic d$^a$ & 0.085 au & 12.7\,d & $1.0\pm0.5\, M_\oplus$ & $1.02 \pm 0.14$ $R_\oplus$ & 520\,K \\
            \noalign{\smallskip}
            AU~Mic c & 0.11 au & 18.9\,d & $14.2_{-3.5}^{+4.8} M_\oplus$ & $2.52 \pm 0.25 R_\oplus$ & 450\,K \\
            \noalign{\smallskip}
            AU~Mic e$^b$ & 0.16 au & 33.4\,d & $35.2_{-5.4}^{+6.7} M_\oplus$ &                          & 380\,K \\

            \noalign{\bigskip}

            \bp\ c & 2.7 au & 3.3 yr & $8.2 \pm 0.8 M_J$ & $1.1 \pm 0.1 R_J$ & 1300\,K \\
            \noalign{\smallskip}
            \bp\ b & 10.0 au & 23.6 yr & $11.7 \pm 2.3M_J$ & $1.5 \pm 0.2 R_J$ & 1600\,K \\

            \noalign{\bigskip}
          
            WD 1145+017 b & 0.0054 au & 4.5\,h & $\sim 1.6\times10^{-5}\, M_\oplus$ & $\sim 0.03 R_\oplus$ & 1100\,K \\

    \noalign{\smallskip}
    \hline
    \end{tabular} 

    \caption{\small Known planets in the WD1145+017, AU~Mic and \bp\ systems. \\
    \footnotesize         $\boldsymbol{[}\textbf{a}\boldsymbol{]}$ AU~Mic d was detected through TTV; the parameters provided here correspond to the favored solution in \cite{Wittrock_2023_AU_mic_d}. \newline     $\boldsymbol{[}\textbf{b}\boldsymbol{]}$ AU~Mic e has not been confirmed.\\
    \textbf{References}: \\ 
    AU~Mic: \cite{donati_magnetic_2023, Wittrock_2023_AU_mic_d}; \\
    \bp : \cite{Nowak2020, Feng2022_betapic, Bonnefoy2014_betapic}; \\
    WD1145+017: \cite{vanderburg2015, rappaport_drifting_2016}; 
    }

\label{table: case study stars - planets}
\end{table}

\begin{figure}[h!]
\centering
\includegraphics[scale = 0.28, trim = 50 0 50 45, clip]{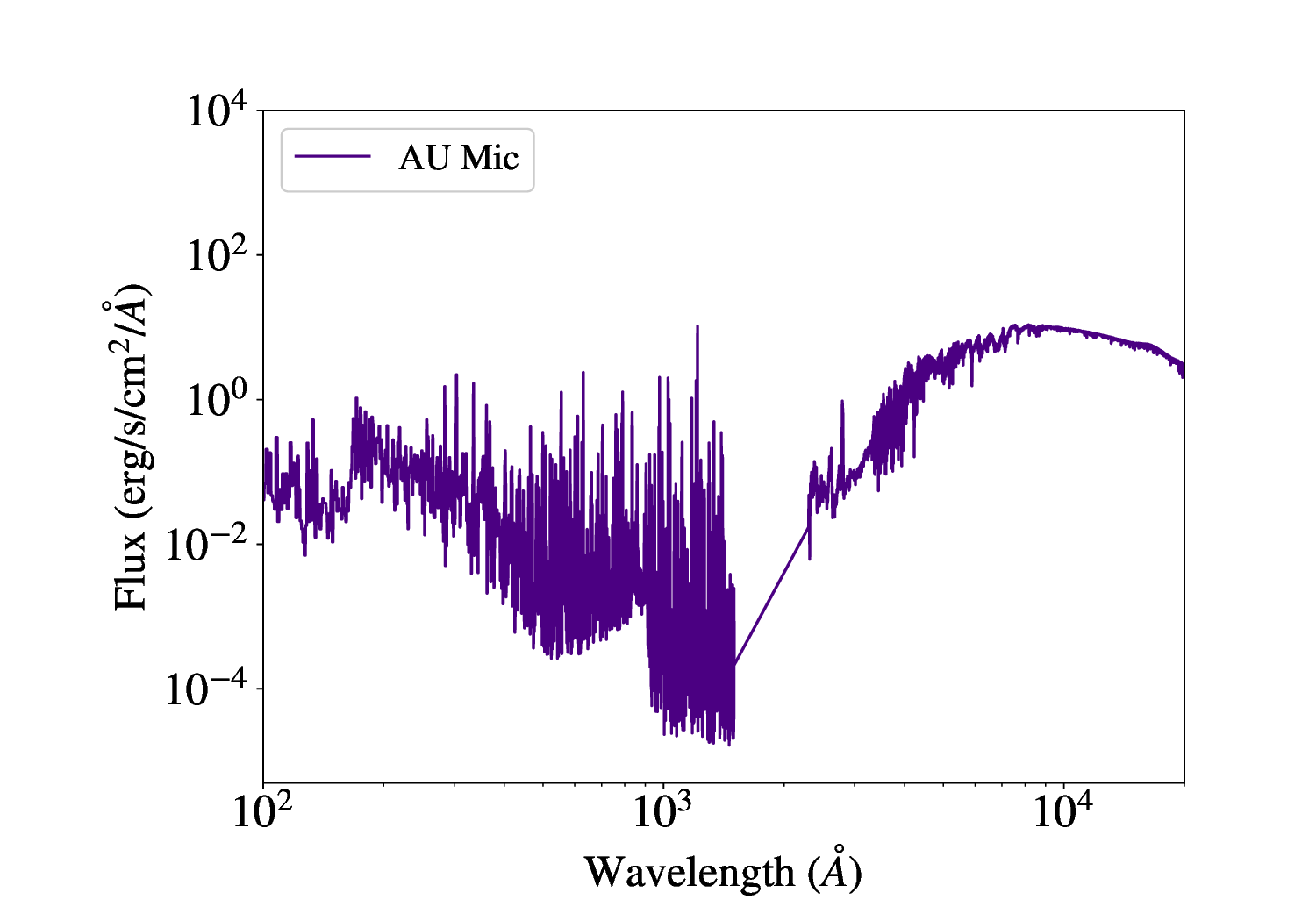}
\includegraphics[scale = 0.28, trim = 50 0 50 45, clip]{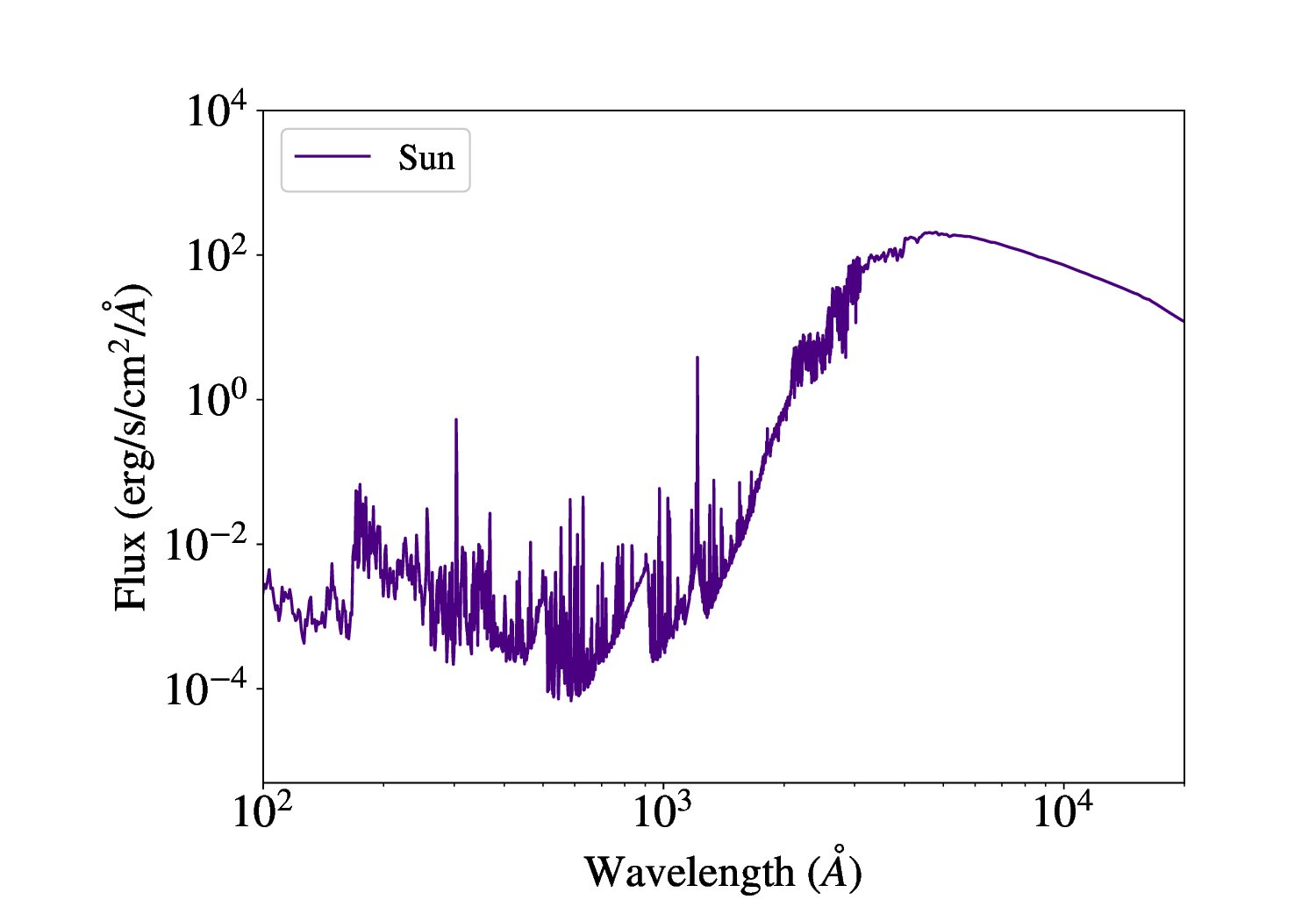}
\includegraphics[scale = 0.28, trim = 50 0 50 45, clip]{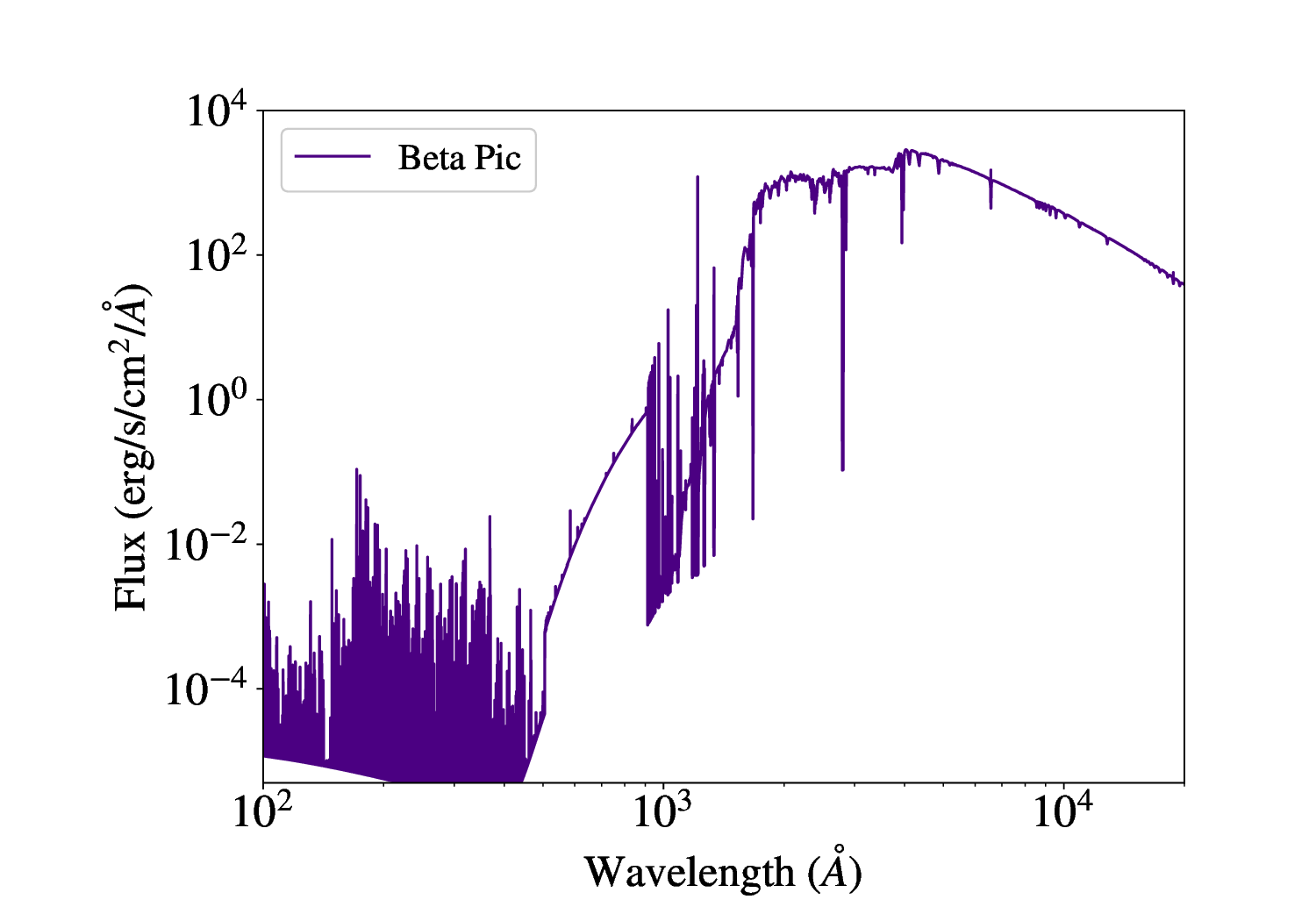}
\includegraphics[scale = 0.28, trim = 50 0 50 45, clip]{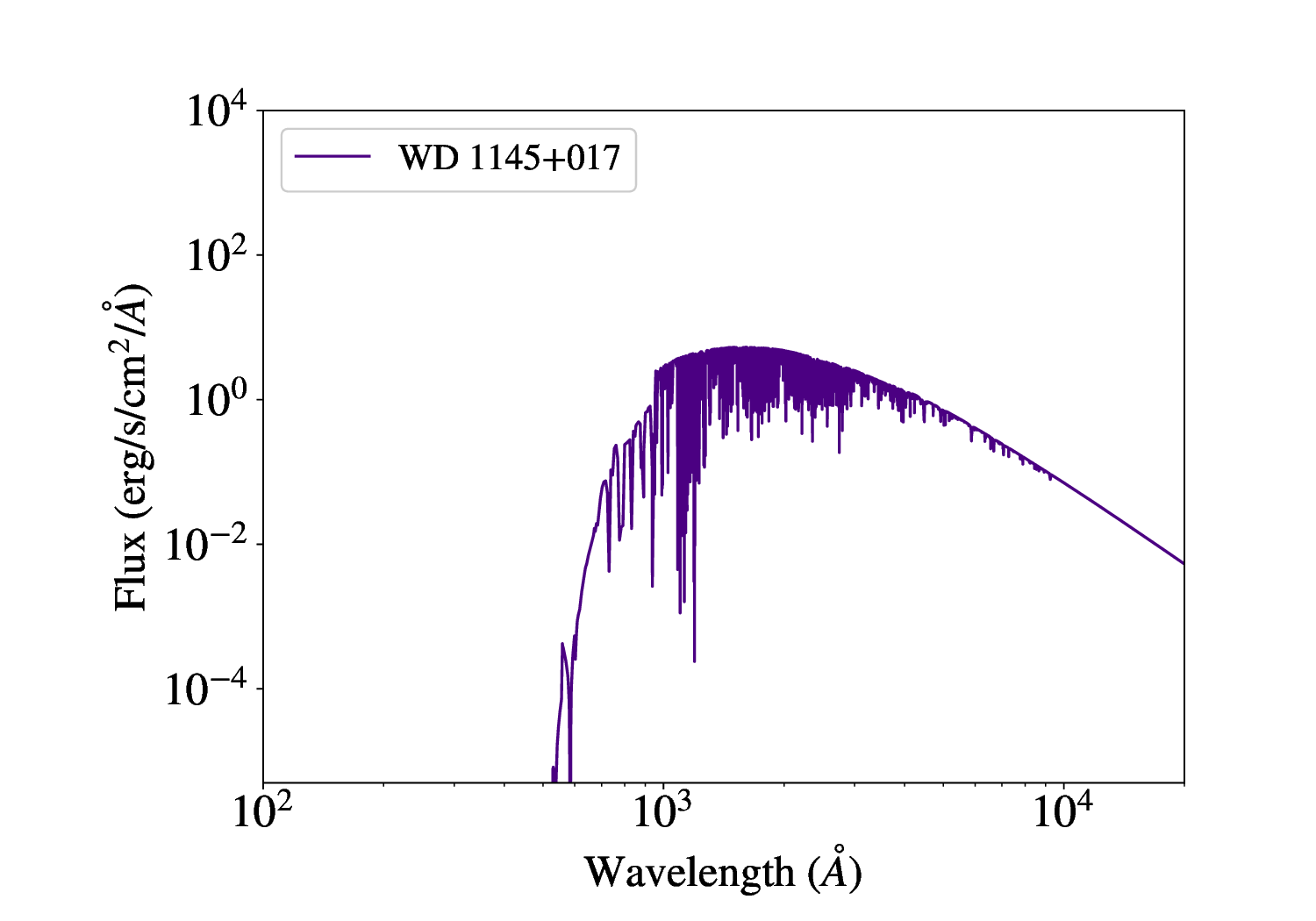}
\caption{\small Spectral energy distribution used for our case study stars. The fluxes are all provided for distances of 1~au.}
\label{fig:spectra case study stars}
\end{figure}

To model the influence of the Sun, \bp, AU~Mic and WD 1145+017 on nearby small bodies, we adopt the following spectral energy distributions (also shown in Fig.~\ref{fig:spectra case study stars}):

\begin{itemize}
    \item {\bf AU~Mic: } We use the spectrum from \cite{Feinstein2022}, reconstructed by combining observations and synthetic models. This spectrum corresponds to a quiescent stellar activity.
    \item {\bf The Sun:} We use the Solar Spectral Irradiance Reference Spectra for Whole Heliosphere Interval \citep{Woods2009}, measured in March 2008 during a period of moderately low solar activity. This spectrum spans the wavelength range from 1 to 24\,000 \AA.
    \item {\bf \bp:} For the photosphere, we use a synthetic spectrum from the PHOENIX library \citep{PHOENIX}, with $T_\text{eff} = 8000$\,K, $\log(g) = 4.5$, $[\text{Fe/H}] = 0$, and $[\alpha\text{/H}] = 0$. This spectrum was convolved with the rotational velocity of the star, 130 km~s$^{-1}$. At wavelength shorter than 1300 \AA, we adopt the synthetic spectrum of \cite{Wu2025_betapic}, which includes chromospheric and coronal emission. This spectrum, calculated with CLOUDY \citep{Ferland2013_cloudy}, This model, computed with CLOUDY \citep{Ferland2013_cloudy}, is calibrated to reproduce the observed flux in the \ciii\ and \ovi\ emission lines.
    \item {\bf WD1145+017: } A spectrum provided by \cite{Bourdais2024} is adopted, incorporating absorption lines from metallic species in the white dwarf’s polluted photosphere. 
\end{itemize}

These spectral models will help evaluate the environmental effects on cometary activity across different stellar systems, guiding our understanding of how small bodies evolve under diverse radiative conditions, how these environments shape detectable spectral features, and how those features can be linked to the physical and chemical properties of the small bodies.

\newpage
\section{Mass loss processes}

\subsection{Diversity in the Solar System}

\begin{figure}[h!]
    \includegraphics[width = 0.491\linewidth]{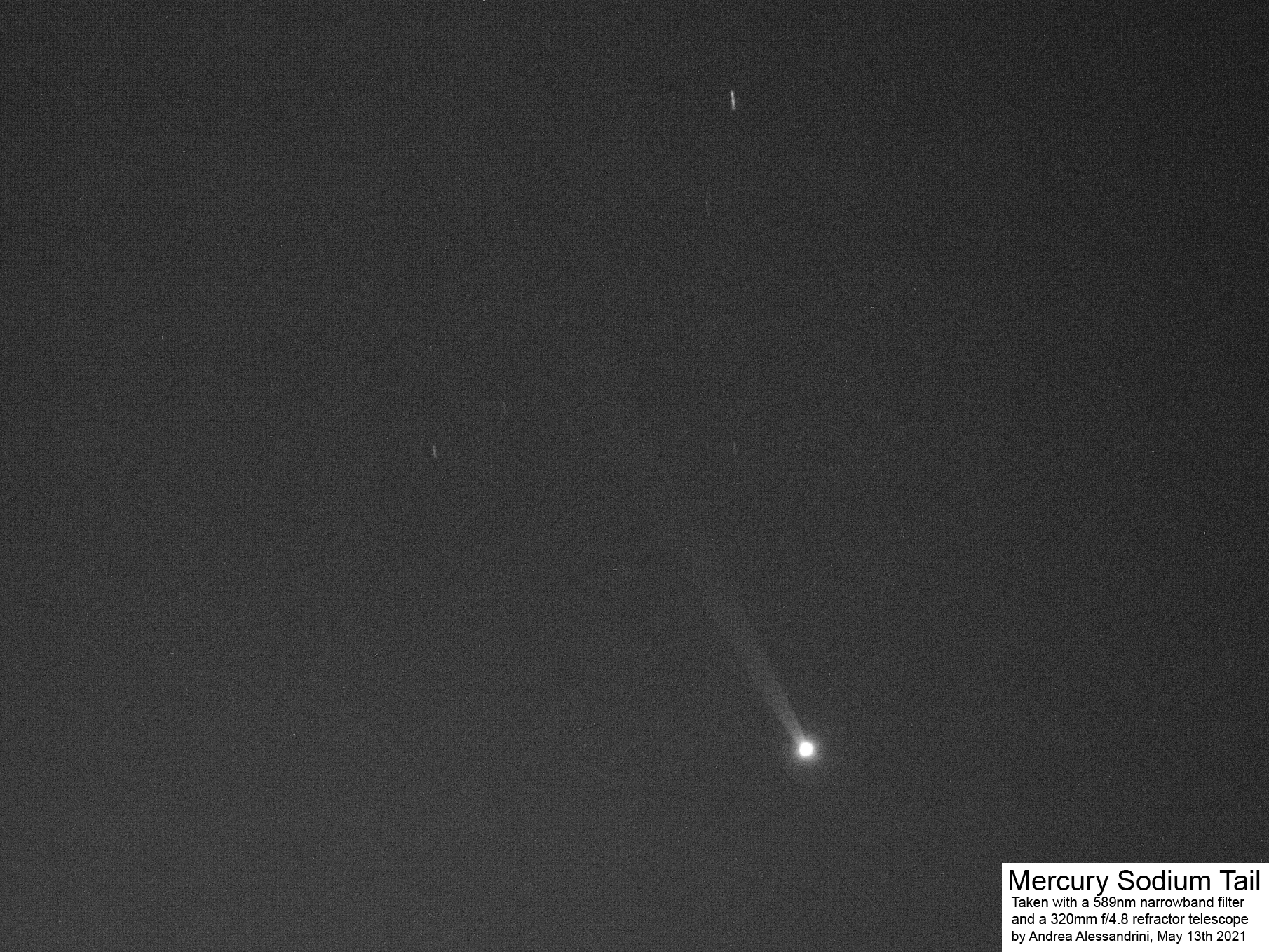}
    \includegraphics[width = 0.509\linewidth]{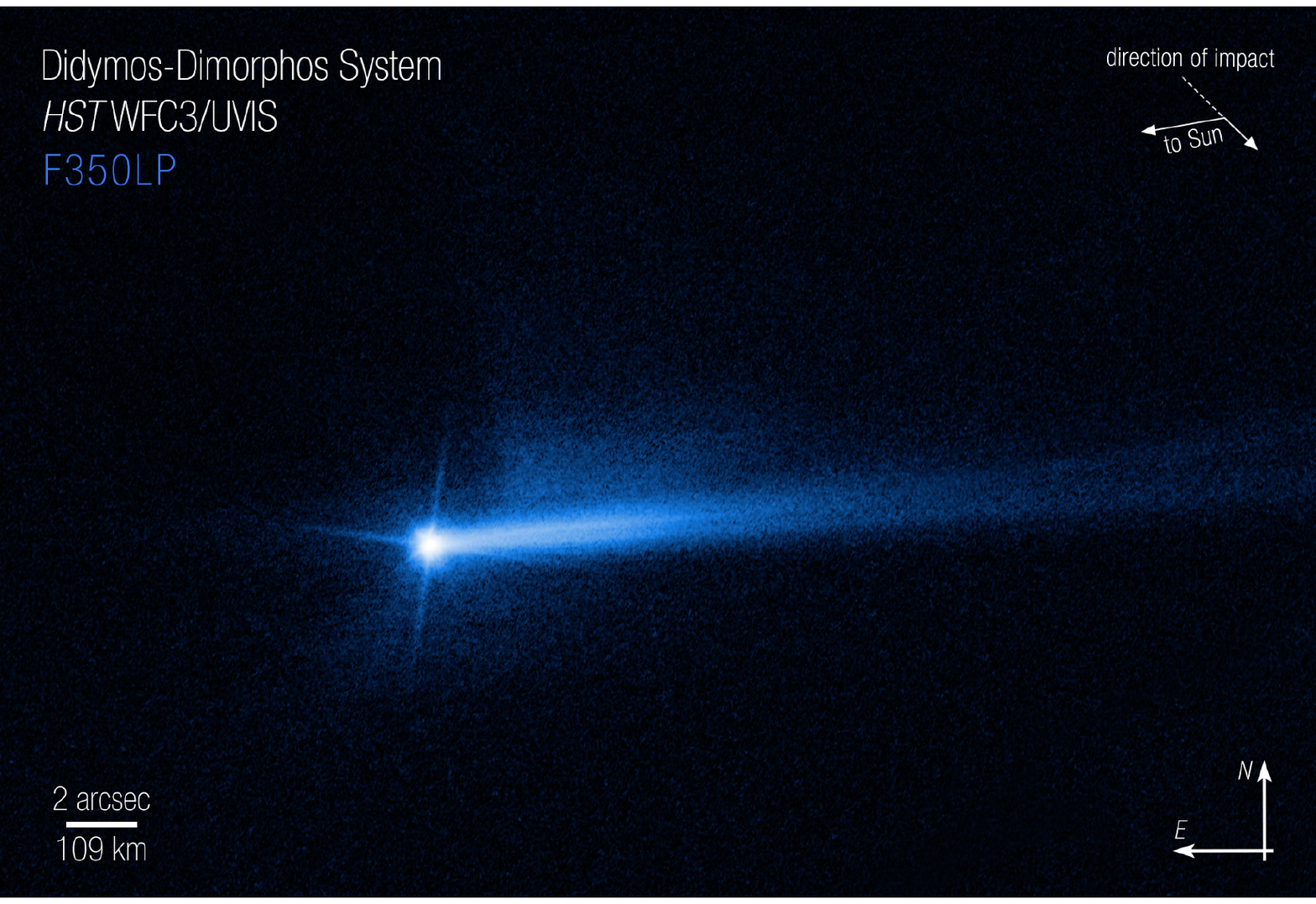}
    \includegraphics[width = 0.58\linewidth]{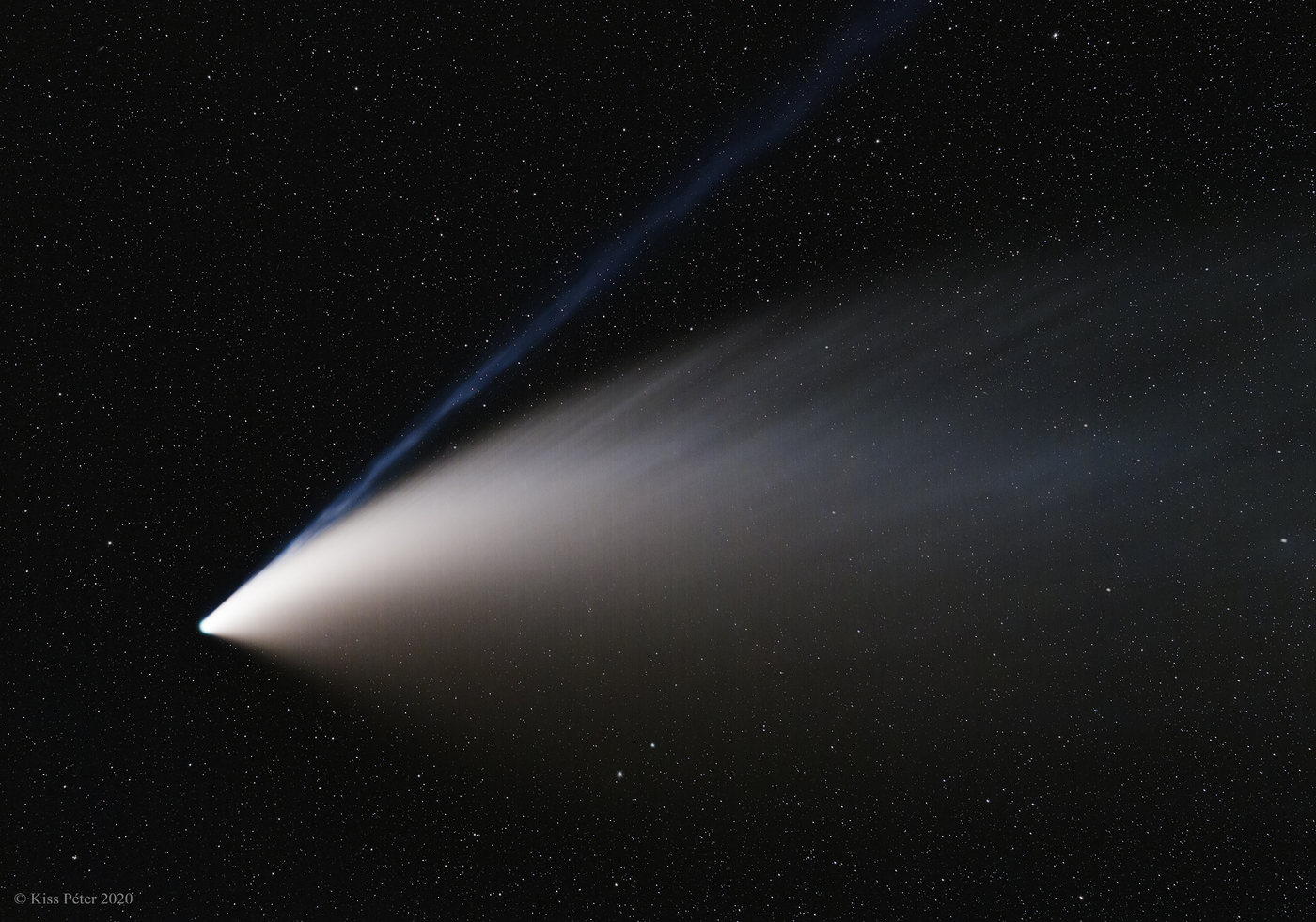}
    \includegraphics[width = 0.42\linewidth]{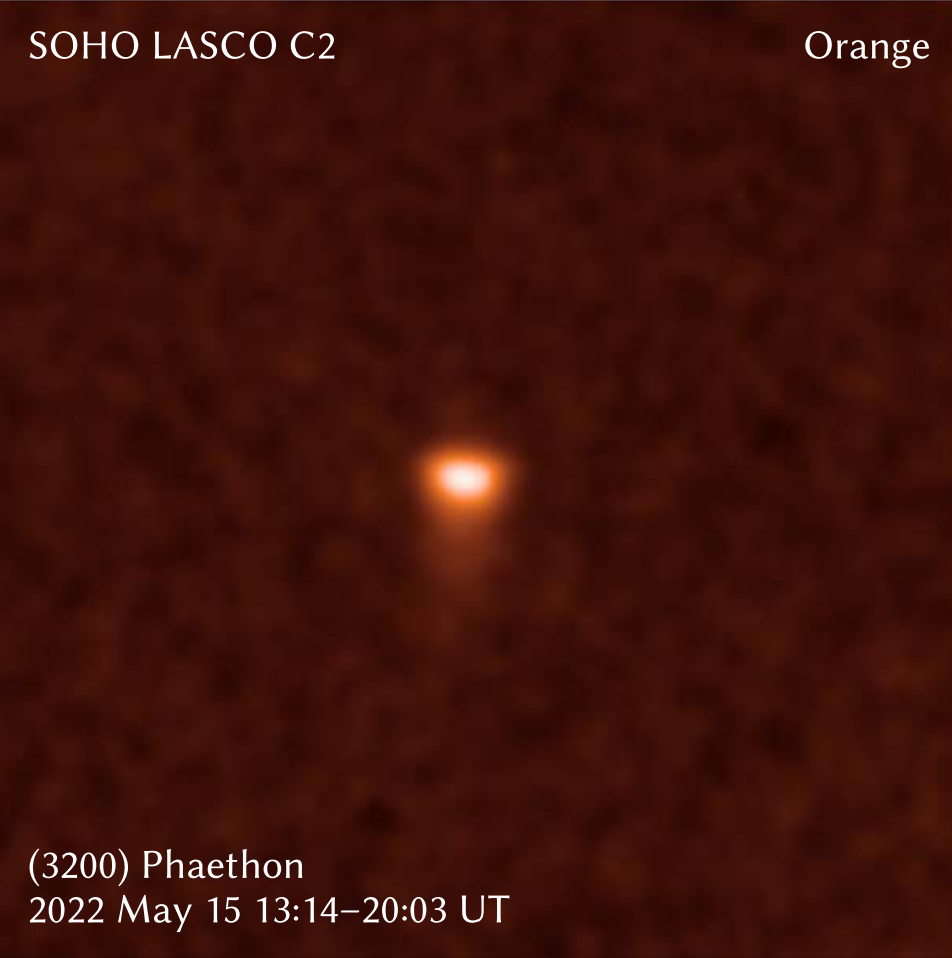}
    \caption{\small Example of mass loss in Solar System objects.  \newline
    \textbf{Upper left}: Sodium tail of Mercury, observed through a narrowband filter centered on the Na D lines around 589~nm. (Credit: Andrea Alessandrini). Mercury's sodium tail is produced from a variety of processes, including photon and thermal desorption, meteroid impact, and ion sputtering. \newline
    \textbf{Upper right}: HST/WFC3 captured multiple tails emerging from the asteroid Dimorphos after impact with the DART spacecraft (Credits: NASA, ESA, STScI, Jian-Yang Li (PSI); Joseph DePasquale). \newline
    \textbf{Lower left}: Gas and dust tails of C/2020 F3 (NEOWISE), driven by the sublimation of volatiles (Credits: Kiss Péter, 2020). \newline
    \textbf{Lower right}: Sodium tail of (3200) Phaethon \citep{Zhang_2023}, produced from the thermal desorption of Na atoms at the surface of the asteroid.}
    \label{fig:images_mass_loss}
\end{figure}

Activity among Solar-System bodies spans a remarkable range of environments and mechanisms, from volatile sublimation in icy comets to impact vaporization and sputtering on airless surfaces (see Fig.~\ref{fig:images_mass_loss}). Table~\ref{table: mass loss examples} summarizes the principal drivers of mass loss and the conditions under which they dominate. The Sun loses about $3{\times}10^9$~kg~s$^{-1}$ through radiation and the solar wind \citep{Pitjeva2021}. Atoms and molecules in Earth's exosphere can reach escape velocity and are lost to space \citep[e.g.,][]{Howe2020}. Owing to its lower gravity, Mars lost most of its water over time \citep[e.g.,][]{Pollack1987} and  continues to release hydrogen produced by the photodissociation of H$_2$O \citep{Mayyasi2023_mars_H_escape}. Mercury, the Moon, and Phaethon exhibit ongoing mass loss, as evidenced by the detection of sodium and potassium in their exospheres, likely driven by photon-induced desorption or sputtering \citep[see also Table \ref{table: mass loss examples}]{Wurz2022}. Comets, volatile-rich remnants from the era of planet formation,  release gases as their ices sublimate when heated by the Sun, dragging refractory particles along to form characteristic tails. Asteroids lose mass through electrostatic lofting, as well as in collisions that eject fragments or, in extreme cases, disrupt the entire body (see Section~\ref{sec: macroscopic processes}).

The mechanisms driving mass loss in small bodies can generally be categorized into macroscopic processes -- such as collisions, tidal breakups, rotational instability, or thermal fracturing -- which shape the long-term evolution of minor  bodies \citep{Jewitt2014}, and microscopic effects -- including  sublimation, plasma sputtering, micrometeoroid impacts, and photon- or thermally-driven desorption -- which act on shorter timescales but are equally important for the gradual alteration of surfaces  \citep[e.g., see the review by][]{Wurz2022}. In the inner Solar System, volatile sublimation is by far the most efficient mechanism for material release \citep[][see also Table \ref{table: mass loss examples}]{Rickman2010_cometary_dynamics, Cochran2015_composition_comets, Bockel2017_cometary_ices}. However, whether this remains true for small bodies orbiting stars of different spectral types or at different evolutionary stages is still uncertain. Since the efficiency of each mass-loss process depends on the local radiation field and plasma environment, the dominant mechanism ultimately varies with both orbital distance and stellar type.

\begin{figure}[h!]
\centering
    \includegraphics[scale = 0.57,     trim = 0 0 0 0,clip]{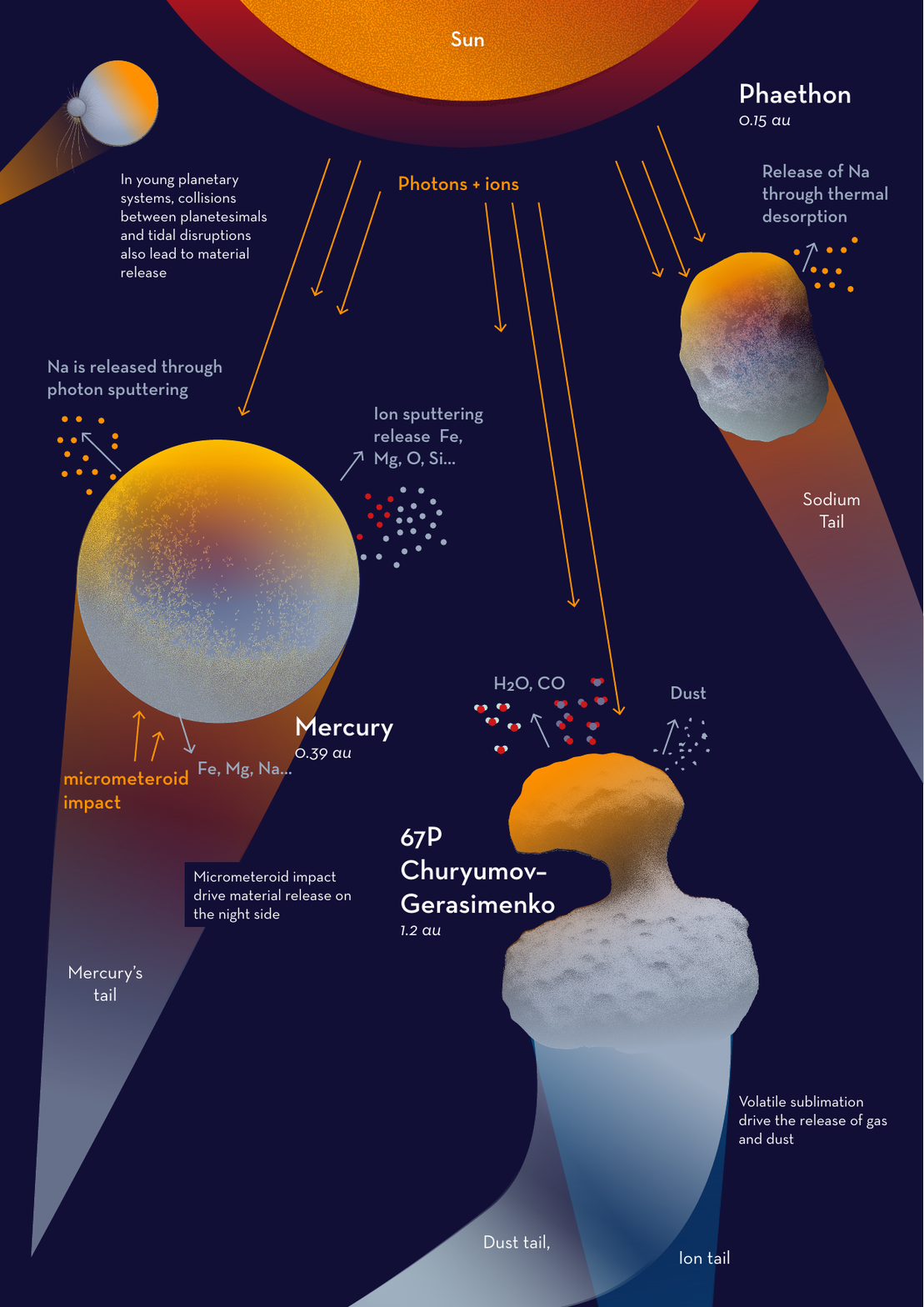}      
    \vspace{0.2 cm}
    \caption{\small Summary of the various physical processes that trigger material release from Solar System objects. Depending on the nature of the object (e.g. rocky asteroid, terrestrial planet, or comet) and its heliocentric distance, different mechanisms dominate the mass-loss process.}
    \label{Fig. infographic}
\end{figure}

In the Solar System,  mass loss processes can often be inferred directly from observational evidence. For example, the recurring activity of main-belt comets implied sublimation driven activity \citep[e.g.,][]{Hsieh2006} well before it was directly detected \citep{Kelley2023}. In contrast, non-recurrent activity coupled with unusual morphology identified a natural impact event on asteroid (596) Scheila \citep[cf.][]{Bodewits2011}. For small bodies beyond the Solar System, however, disentangling these mechanisms is much more challenging. The following subsections therefore review the various processes that may trigger mass loss from minor bodies. This broad perspective is essential for identifying better Solar System analogs for exocomets and understanding how their activity is influenced by both internal properties and external environments. The various processes driving material release for solar system objects are summarized in Fig. \ref{Fig. infographic}.

\begin{table}[h!]
    \begin{tabular}{ c c c c c }   
    \hline
    \noalign{\smallskip}
    \hline
    \noalign{\smallskip}
         \textbf{Object} &  Release mechanism(s)& Material & Rate (kg/s) & References\\

            \noalign{\smallskip}
            \hline
            \noalign{\medskip}
         Phaethon&  Thermal desorption& Na & 0.02 & 1 \\
         \noalign{\bigskip }
                    & Thermal desorption,   &  \multirow{2}{*}{H, He, K, Al,}    &      &    \\
         Mercury    & photon desorption, sputtering,   &  \multirow{2}{*}{Fe, Na, Ca, Mg ...}        & $\>$0.01 &  2 \\
                    &  micrometeroid impact &                                             &      &    \\
        \noalign{\bigskip }
               & photon desorption, sputtering,    &                      &          &       \\
        Moon   & micrometeroid impact,                &  H$_2$O, Na  ...   &  0.001   &  3    \\
               & electrostatic lofting                &                      &          &       \\

        \noalign{\bigskip }
         \multirow{2}{*}{Asteroids} & Impact, rotational spin up, & \multirow{2}{*}{Dust} & \multirow{2}{*}{Variable} &  \multirow{2}{*}{4}\\ 
                                    & electrostatic lofting       &                       &                            &\\ 

         \noalign{\bigskip }
                             & Thermal desorption,   &  \multirow{2}{*}{OH, H$_2$O, H$_2$, O}    &      &    \\
         Ganymede    & sublimation, sputtering    &  \multirow{2}{*}{O$_2$, O$_3$}        & 30 &  5 \\
                    &  micrometeroid impact &                                             &      &    \\
            \noalign{\bigskip }
         \multirow{2}{*}{Comets\,$^{\rm a}$} &  \multirow{2}{*}{Sublimation} & H$_2$O, CO$_2$, CO,            &  Up to $10^6$ & \multirow{2}{*}{6} \\
                                 &                               &  Na, Fe, Ni, dust ...          &  (Hale-Bopp)  &  \\
    
    \noalign{\medskip}
    \hline
    \end{tabular} 

    \caption{\small Examples of mass loss processes  in a variety of Solar System objects. \\ References: [1] \cite{Zhang_2023}, [2] \cite{McClintock2018}, [3] \cite{Stern1999}, [4] \cite{Jewitt2012}, [5] \cite{Galli2022}, [6] \cite{Biver2022}. \\ $\boldsymbol{[}\textbf{a}\boldsymbol{]}$ \footnotesize So far, over 100 species have been observed in comets. See \cite{Biver2022} for a complete review.}

\label{table: mass loss examples}
\end{table}


\subsection{Processes linked to the sublimation of volatiles}

\subsubsection{Volatile release}
\label{Sect. Volatile sublimation}

\begin{figure}[h!]
\centering
\includegraphics[scale = 0.7]{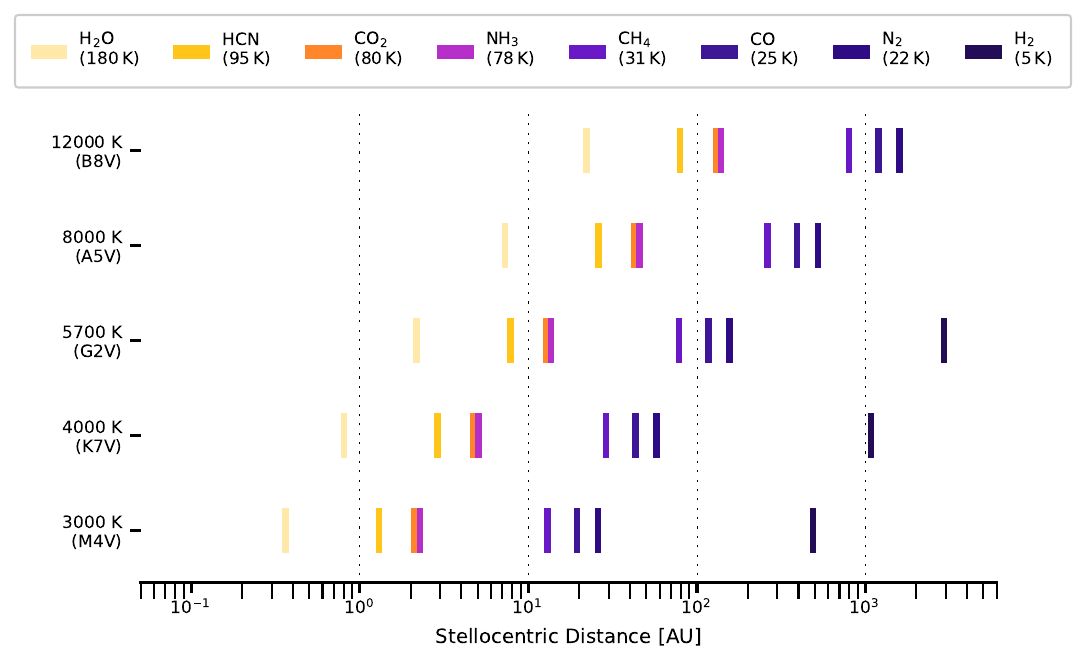}
\caption{\small Location of the sublimation fronts of a variety of volatiles for 5 stars, spanning a wide range of effective temperatures. The values are from Table 1 in \cite{Meech2005}, and scaled to each star assuming the sublimation distance of each molecule scales with $L_\star^{0.5} \propto T_{\rm eff}^2 R_{\star}$, where $L_\star$, $ T_{\rm eff}^2$ and  $R_{\star}$ are the stellar luminosity, effective temperature and radius, respectively.}
\label{fig:sublimationlines MS}
\end{figure}

\begin{figure}[h!]
\centering
\includegraphics[scale = 0.7]{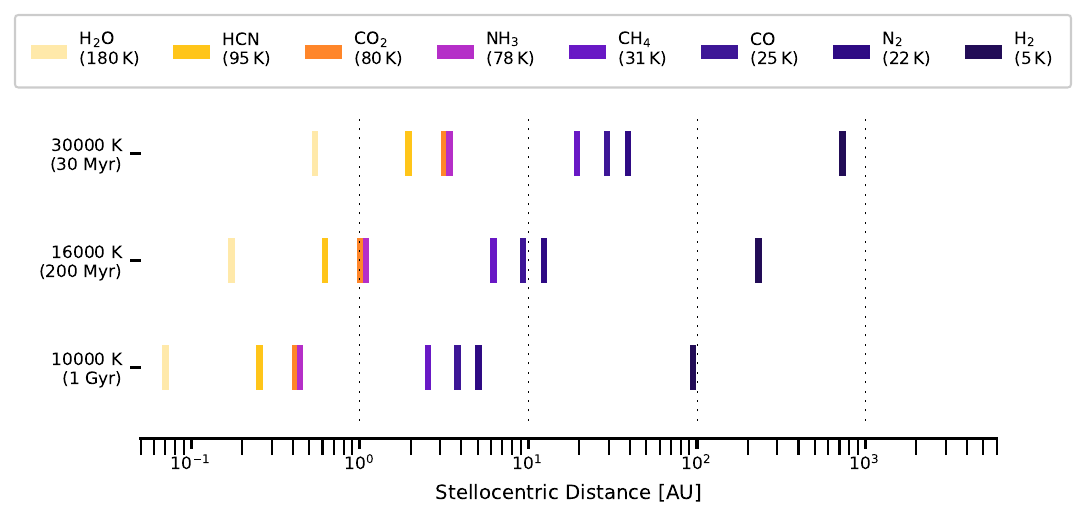}
\caption{\small Same as Fig. \ref{fig:sublimationlines MS} but for three white dwarfs with different ages (known as ``cooling ages" in the white dwarf community). Again, values were derived from \cite{Meech2005}, scaling the sublimation distance with $L_\star^{0.5}$. The age-luminosity relationship of white dwarfs was taken from \cite{veras2022} (Eq. 9). The second row ($T_{\rm eff} = 16\,000$ K) corresponds to WD 1145+017.}
\label{fig:sublimationlines WD}
\end{figure}

Sublimation is the primary mechanism driving activity in most classical comets. As comets approach the sun, rising temperatures trigger the sublimation of various ices. The primary cometary volatiles that contribute to the outgassing are known to be H$_2$O, followed by CO$_2$ and CO \citep[the relative abundances of CO and CO$_2$ can vary by more than an order of magnitude,][]{harringtonpinto2022}. 
The volatility of an ice is set predominantly by the latent heat of sublimation. The latent heats for sublimation of H$_2$O, CO$_2$, and CO are around 51, 26, and 8 ${\rm kJ\,mol^{-1}}$ respectively \citep[see for instance the NIST database,][]{NIST_ASD}. Therefore, these three primary drivers of activity represent high, medium, and low volatility species. In the Solar System, the sublimation fronts of these materials are located at $\sim 2.5$ au (H$_2$O), $13$ au (CO$_2$), and $120$ au \citep[CO; see Table 1 of][]{Meech2005}. However, in other planetary systems, these ice lines shift depending on the luminosity of the host star.  Fig.~\ref{fig:sublimationlines MS} and \ref{fig:sublimationlines WD} show the location of sublimation fronts for various volatiles  around different stars, including main sequence stars (Fig.~\ref{fig:sublimationlines MS}) and white dwarfs (Fig.~\ref{fig:sublimationlines WD}). The position of an ice line varies significantly with stellar type. Around M and K dwarfs, the sublimation line of H$_2$O is located within 1~au, whereas it shifts to $\sim8$~au around an A5V star (such as \bp), and 25~au around an even hotter B8V star. These differences have profound implications for comet activity, as comets orbiting hotter stars would begin outgassing volatiles at much greater distances. In extreme cases, such as of OB stars, the extreme sublimation rate of volatiles could even lead to the complete disruption of the cometary nucleus before it can reach periastron (see the discussion, Sect. \ref{Sect. Discusion}).

The velocity of the gas outflow is typically given by the ambient sound speed,  $c_s=\sqrt{ \gamma k_B T/(\mu m_{\rm u})}$ \citep{probstein1969}, where $T$ is the  temperature of the outgassing material, $\gamma$ is the adiabatic index of the outflow, $\mu m_{\rm u}$ is the mass of the species, and $k_B$ is the Boltzmann constant. In the Solar System, typical temperatures in the inner comae are in the range $10^1 - 10^3$ K \citep{Rodgers2004_cometsII}, and generally rise with increasing distance to the nucleus.

Different from most other volatilization mechanisms, sublimation is not confined to the surface; it can also occur in the interior, depending on the thermal conductivity of the nucleus and the duration or intensity of heat exposure \citep{Huebner2006}. Gases released from the deeper layers of the nucleus may migrate toward the surface and, depending on the thermal environment, refreeze in subsurface layers when the comet moves away from the Sun \citep{Parhi2025, Luspay2022, Filacchione2016}.  This internal sublimation and redeposition process highlights the complexity of volatile storage and release in comets. Consequently, the chemical composition of gases in the coma may not directly reflect  the original abundances of ices present in the nucleus \citep[see][]{Brasser2015_comet_fading, Seligman2022}.

In white dwarf planetary systems (Fig.~\ref{fig:sublimationlines WD}), volatile-rich bodies deliver material to the stellar photosphere, where their presence is detected through atmospheric contamination \citep{farihi2013,raddi2015,gentilefusillo2017,xu2017,hoskin2020}. However, unlike in the Solar System, these deposited volatiles are observed as individual elements rather than molecules, because the bonds holding the latter together are broken apart by the immense gravity of the white dwarf. Hence, the original composition of accreted material has to be reconstructed from the elemental composition of the deposited metals, which mix with the native photospheric elements (a combination of hydrogen and helium). For instance, the ice fraction of the comets can be inferred from the excess oxygen content \citep{farihi2013}, which is likely produced from the sublimation of H$_2$O, CO, and CO$_2$ (the most common oxygen-bearing molecules, Fig. \ref{fig:sublimationlines WD}). In the case of a prominent nitrogen signature \citep{xu2017}, the key molecules sublimated are likely HCN and NH$_3$.

\subsubsection{Dust release}
\label{sec: dust escape}
In many cases, detecting dust in a coma or tail is a far easier way to identify cometary activity than directly observing volatile escape. The sublimation of volatiles (see Section~\ref{Sect. Volatile sublimation}) is the primary mechanism driving dust ejection from cometary nuclei. The rate and onset distance of dust production vary depending on the comet’s composition, thermal properties, and active surface fraction. Generally, the closer a comet is to the Sun, the greater its dust production, as increased insolation leads to the sublimation of a larger range of volatile species. Empirical correlations exist between water production rates and a comet’s visible brightness, as dust reflects sunlight \citep{Jorda2008}. 

Robotic surveys are enabling comet discoveries at increasingly large distances from the Sun discoveries, and pre-discovery observations in their archival data allow us to study the activity evolutions even farther out \citep{Meech2017}.  For example, C/2017 K2 (Pan-STARRS) was detected with sustained activity at 23.75 au pre-perihelion \citep{Hui2018}, well beyond the expected limit for CO$_2$  sublimation, providing strong evidence for CO-driven activity \citep{Meech2017}.  The large  discovery distances of C/2017 K2  and a handful of other long-period comets (e.g. C/2014 UN271, \cite{Farnham2021_C_2014_distant_activity}; C/2019 E3 \cite{Hui2024_distant_activity_C_2019}) suggest that CO sublimation may play a more significant role in cometary activity at large distances than previously assumed. JWST observations of long period comet C/2024 E1 at 7 au pre-perihelion revealed that its activity was driven by CO$_2$, and detected no emission from H$_2$O or CO \citep{Snodgrass2025}. The exact mechanisms governing cometary activity beyond 10~au remain uncertain, and future observations of distant comets and Centaurs will be essential to further constrain these processes \citep[e.g.,][]{Faggi2024, Pinto2023}.         


The Rosetta mission to comet 67P/Churyumov–Gerasimenko provided an unprecedented view of cometary volatile release, revealing complex associations between different gases and shedding light on the mechanisms governing cometary activity. Unlike previous flyby missions, Rosetta remained in close proximity to 67P for over two years, which made possible a temporally resolved chemical analysis of the comet’s coma \citep{Combi2020}. The mission confirmed that H$_2$O and CO$_2$ outgassing originate from distinct regions of the nucleus, supporting the idea that minor species are embedded in separate ice matrices within the nucleus \citep{Luspay2015}. Seasonal variations played a significant role in altering volatile production rates, with different molecules correlating with H$_2$O or CO$_2$ depending on the comet’s orbital position (see Section~\ref{sec: thermal desorption}). These findings were complemented by ground-based observations, which, while unable to directly detect CO$_2$ due to atmospheric absorption, used ethane (C$_2$H$_6$) as a proxy for CO$_2$ sublimation \citep{Saki2024}. 

Since cometary nuclei vary in shape, activity, and rotation state, a range of coma features such as fans, shells, and jets can form \citep[e.g.,][]{Farnham2009}. Modeling dust grain dynamics in a cometary coma is particularly challenging, as it requires accounting for a wide range of gas-dust interactions and particle collisions \citep[e.g.,][]{combi_hale-bopp_2002, tenishev_numerical_2011}.  Once dust is released from the nucleus, the motion of the smallest dust particles is initially governed by drag forces from sublimated volatiles, whereas larger grains are mostly affected by radiation pressure. Once in the coma, dust-dust collisions can alter the properties of the ejected grains. Continuing with 67P as a Solar System example, modeling by \cite{planes_dust-dust_2024} found that grain erosion, fragmentation, and grain density (physical density rather than number density) increases with the number of collisions, whilst larger grain speeds (150\,m\,s$^{-1}$) prevent the coagulation of grains. Consequently, collisions between dust grains in the coma can determine the properties of dust grains later detected in the tail.

As demonstrated by 67P, dust production is a highly dynamic process, influenced by a combination of sporadic events on the nucleus surface, grain-grain interactions in the coma, and varying external forces. As a result, modeling the dust production rate of an exocomet may be challenging, and simplifications have to be made. \cite{lecavelier_des_etangs_library_1999} proposed the production of dust is dependent on the luminosity $L_\star$ of the host star and the stellocentric distance $r$ of an exocomet, such that:
\begin{equation}
\label{eq:exo dust prod}
    P\left(r\right) = P_0 \left(\frac{r_0}{r}\right)^2 \left(\frac{L_*}{L_\odot}\right)
\end{equation}
where $P_0$ is the dust production rate at $r_0 = 1$~au (which depends on the comet size and composition), and $L_\odot$ the solar luminosity. The dependence on luminosity implies comet-like objects orbiting smaller and less luminous stars such as the M-dwarf AU~Microscopii could struggle to form well defined comae and dust tails.


This approach was implemented by \citep{bodman_kic_2016}, which analyzed photometric variations in the KIC 8462852 light curve. The study concluded a family of comets could explain the unusual dips in flux. This approach also assumes dust production ceases when an exocomet is beyond a distance $r_\mathrm{crit} = 3\sqrt{L_\star/L_\odot}$. The ability of a minor body to eject dust is essential to produce asymmetric photometric variations that are identifiable as exocomets. Since dust production varies with stellar luminosity, our ability to photometrically detect exocomets could be confined to stars with suitable luminosities. This is discussed later in Sect. \ref{sec: case study stars}.

Universally modeling the size distribution of ejected cometary dust is also a difficult task. Influential work developed by \cite{lecavelier_des_etangs_library_1999} modeled asymmetric photometric variations by assuming a size distribution which followed:
\begin{equation}
    \label{eq: size distribution LDE}
    \frac{\mathrm{d} n}{\mathrm{d} s} = \frac{\left(1-s_0/s\right)^m}{s^n},
\end{equation}
\begin{equation}
    m = \frac{n\left(s_p - s_0\right)}{s_0},
\end{equation}
where $s$ is the radius of a dust grain and $n$ its number density. This description is based on observations of cometary grains in the solar system \citep{Hanner1983}. The values of $s_0$ (minimum grain size) and $s_p$ (peak size distribution) are found to vary in time, with dust distribution peaking at smaller sizes for smaller heliocentric distances \citep{Newburn1985}. For comet 67/P, the value of $\alpha$ was also found to increase sharply after perihelion \cite{Fulle2004}. These temporal variations are not fully understood; they may result from changes in the nucleus region exposed to stellar irradiation \citep{Fulle2004}, or to the fragmentation of dust grains in the inner coma at short heliocentric distances \citep{Gronkowski2009}.

\begin{figure}[h!]
    \centering
    \includegraphics[width=0.8\linewidth]{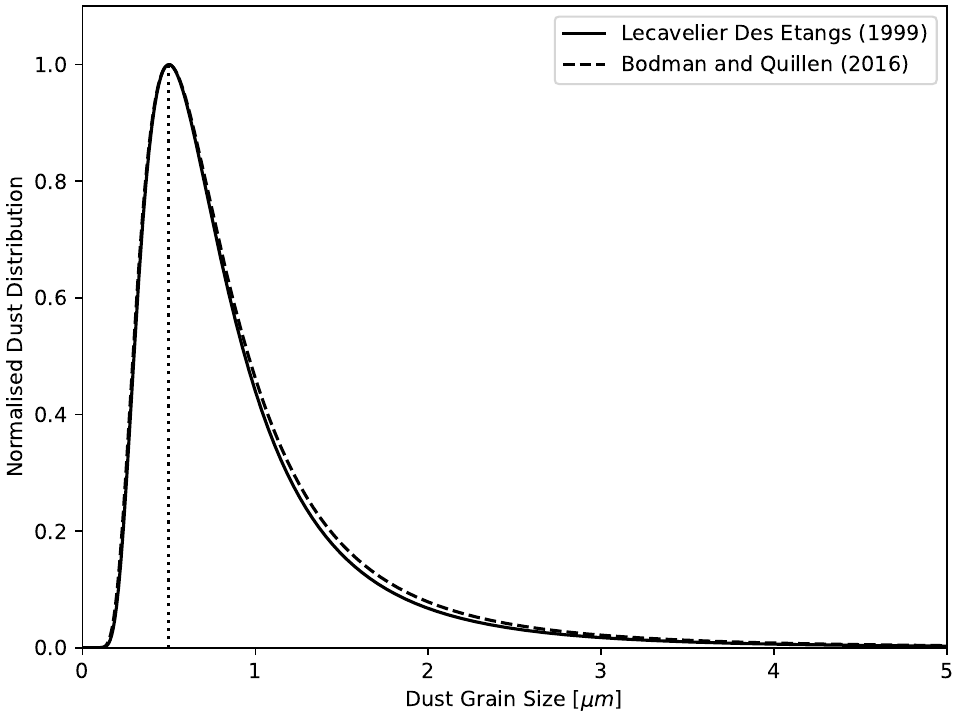}
    \caption{\small Comparison of the distribution models used by \cite{lecavelier_des_etangs_library_1999} and \cite{bodman_kic_2016}. In this example, $s_p = 0.50$ \textmu m, $s_0 = 0.10$, \textmu~m, $n = 4.2$, and $\alpha = 4$.}
    \label{Fig. Dust size distrib}
\end{figure}

Models considered by \cite{lecavelier_des_etangs_library_1999} are summarized in Table \ref{table: dust distribution}. In a slightly different way, \cite{bodman_kic_2016} decided to adopt a power law distribution coupled with an exponential cut-off for small particles outlined by \cite{fink_calculation_2012}. In this case, the size distribution took the form
\begin{equation}
    \label{eq: dust dist exp}
    \frac{\mathrm{d} n}{\mathrm{d} s} \propto \left(\frac{s_p \alpha}{s}\right)^\alpha \exp\left(-\frac{s_p \alpha}{s}\right)
\end{equation}
where $s_p = 0.50$ $\textmu m$ and $\alpha = 4.0$. The two approaches produce similar distributions (see Fig. \ref{Fig. Dust size distrib}), with the main difference being \cite{bodman_kic_2016} using fewer parameters than \cite{lecavelier_des_etangs_library_1999}. Since the absorption cross-section of a dust grain is dependent on its size (see Sect. \ref{Sect. Dust sublimation}), considering realistic grain size distributions is crucial to the interpretation of asymmetric transit events in the light curve of nearby stars, and link the observed events to the properties, e.g. dust production rate, of the transiting exocomets.

\begin{table}[h!]
\begin{tabular}{cccc}
\hline
\textbf{Periastron {[}au{]}} & $S_p$ {[}\textmu m{]} & $S_0$ {[}\textmu m{]} & n \\ \hline
q \textless\ 0.50             & 0.20                      & 0.05                      & 4.2        \\
q \textless\ 0.50             & 0.25                      & 0.10                      & 4.2        \\
q \textgreater\ 0.50          & 0.50                      & 0.10                      & 4.2        \\ \hline
\end{tabular}
\caption{\small Parameters used by \cite{lecavelier_des_etangs_library_1999}.}
\label{table: dust distribution}
\end{table}

Finally, we note that volatile sublimation is not the only mechanism responsible for dust release. For instance, dust can also be ejected through electrostatic lofting. On airless bodies, the day side can become positively charged as incident solar photons remove electrons from the surface, while electrons in the solar wind accumulate preferentially on the night side, creating a localized electric field. This process results in negatively charged dust grains that are repelled by the surface and lofted into space. Electrostatic lofting has been proposed as a contributing factor to dust activity on near-Earth asteroid Bennu \citep{hartzell_evaluation_2022}, Phaethon \citep{kimura_electrostatic_2022}, and 67P itself \citep{nordheim_surface_2015}, demonstrating that dust ejection is not solely reliant on volatile sublimation.


\subsection{Other microscopic processes}

While volatile sublimation is the main mechanism that leads comets to lose gas and dust, other processes play important roles in the mass loss of rocky bodies, such as asteroids, terrestrial planets and moons, whose surfaces are largely devolatilized. Different from sublimation, which can release materials from great depth, the following processes only affect the topmost layers. The impact of these processes on exocomets (or, more generally, exo-small bodies) has never been studied. However, they could represent important drivers of activity for objects exposed to different stellar environments than the Sun.

\subsubsection{Thermal desorption}
\label{sec: thermal desorption}

Thermal desorption refers to the release of species that are adsorbed onto or trapped within another material as a result of heating. Unlike sublimation, thermal desorption does involve breaking chemical bonds but instead overcomes physical adsorption forces or diffusive barriers between the matrix and the molecule or atom being released \citep{Collings2004}. Volatile species commonly released through thermal desorption include H, He, Ar, Ne, H$_2$, O$_2$, N$_2$, H$_2$O, OH, and CO$_2$, as observed on the Moon \citep{Stern1999} and Mercury \citep{Killen2007, Wurz2022}. Some of these species, such as H$_2$O and CO$_2$, can freeze out on the night side of these bodies and are subsequently released at dawn due to solar heating. 

Another example of a body experiencing thermal desorption is (3200) Phaethon, which experiences surface temperatures up to 1000~K near perihelion \citep[at 0.14 au, see][]{Masiero2021}. At such temperatures, the thermal desorption of Na atoms becomes efficient, leading to a peak loss rate of \(5.5 \times 10^{23}\) atoms~s$^{-1}$ \citep{Zhang2023}. The sodium is sequestered beneath a devolatilized surface layer about 0.1~m thick that acts as a diffusive barrier. If this barrier were disrupted or absent, the sodium production rate could increase by six orders of magnitude, potentially generating enough pressure to lift dust from the surface \citep{Masiero2021}. By contrast, on Mercury (perihelion 0.31~au), the release of sodium atoms by thermal desorption is marginal and limited to the subsolar point \citep{Wurz2022}. This illustrates the strong sensitivity of thermal desorption to  stellocentric distance and, more generally, to the incoming radiative flux. As a result, thermal desorption is expected to play an even more prominent role around stars hotter and more luminous than the Sun, leading asteroids and refractory grains to rapidly lose atoms as they approach their host stars. 


Thermal desorption processes are also linked to matrices found on comets (e.g., clathrates), where H$_2$O and CO$_2$ ices serve as storage reservoirs for minor species. Observations have shown correlations between the release of minor volatiles (e.g., C$_2$H$_6$, CH$_3$OH, HCN, \ldots), and major volatile species (H$_2$O and CO$_2$) in cometary comae \citep{DelloRusso2016, Luspay2018, Saki2024}. Such correlations are consistent with laboratory results \citep{Collings2004,Kruczkiewicz2024}. It has also been suggested that amorphous water ice in the nuclei of distant comets may phase transition into crystalline ice, releasing minor species and driving activity \citep[cf.][]{Jewitt2009,GuilbertLepoutre2012}. Although amorphous ice is thermodynamically unstable, at low temperatures its crystallization timescale is longer than the age of the Solar System; only when the ice is heated above $\sim$100~K does crystallization proceed efficiently \citep[][]{Jewitt2009}. This temperature-dependent phase transition is exothermic, and the energy released can drive additional activity at distances beyond which water-ice sublimation is expected. Consequently, there might be multiple `snow lines' for a given species: one where the species itself sublimates, and others where trapped volatiles are released as the matrix holding them either undergoes a phase transition or sublimates.

\subsubsection{Photon and electron desorption}
\label{Sect. Photon and electron desorption}

Photon-stimulated desorption (PSD; also known as photodesorption or photosputtering) is a low-energy process capable of releasing atoms and molecules from a surface when sufficient UV flux is present. In our Solar System, PSD plays a significant role in the release of sodium (Na) and potassium (K) from airless bodies such as Mercury and the Moon \citep{Yakshinskiy1999}, while on icy moons like Europa and Enceladus, it contributes to the ejection of water molecules and their fragments \citep[e.g.,][]{Plainaki_2010_europa,Plainaki2012}.

In PSD,  a photon is absorbed by an atom or molecule adsorbed on a surface, promoting it into an unbound excited state. The photon energies involved \citep[4 -- 10~eV, corresponding to 3000 -- 1200~\AA,][]{Wurz2022}, are typically insufficient to break chemical bonds within minerals,  restricting PSD to species that are already loosely bound to the surface. The required photon energy depends on the bonding strength of the adsorbed species:  water molecules are held to surfaces by weak hydrogen bonds or van der Waals forces \citep{Fillion2022}, while alkalis such as Na and K are often strongly bound to metallic or oxide surfaces, requiring higher energies for desorption \citep{Yakshinskiy2000}. PSD yields vary significantly between species, ranging from $\approx 10^{-3}$~photon$^{-1}$ for H$_2$O to $\approx 10^{-4}$~photon$^{-1}$ for Na and K, and are probably even lower for heavier atoms like Ni and Fe. The yield  also depends on surface composition, temperature, and photon flux intensity \citep{Bahr2012, Fillion2022}.

The efficiency of PSD scales with the incident photon flux, making Mercury's PSD rates higher  than those of the Moon. The same scaling implies stronger PSD  around hotter stars with elevated UV flux, or during stellar flares, which can temporarily enhance PSD rates by up to an order of magnitude \citep{Welsh2006}. However, PSD can be attenuated by the presence  of gas above the solid surface, which absorbs UV photons. For example, water vapor has strong absorption cross-sections between 50 and 175~nm \citep{Phillips1977, Parkinson2003}, and when production rates typically exceed $10^{27}$ molecules~s$^{-1}$ (30 kg~s$^{-1}$) for a $\sim$km-wide nucleus, the resulting coma becomes optically thick to UV photons within several kilometers above the surface \citep[e.g.,][for 67P]{Bodewits2016}. PSD yields for H$_2$O are approximately an order of magnitude higher than for D$_2$O, potentially leading to isotopic fractionation over time \citep{Fillion2022}. This phenomenon may influence the interpretation of isotopic ratios.

Electron-stimulated desorption (ESD) is a process very similar to PSD, but involves impact by low-energy electrons instead of photons. This process has been proposed to explain part of the release of Na and K from Mercury; however, disentangling the impacts of ESD and PSD is difficult, due to the close similarity between the two processes \citep{Wurz2022}. The cross-sections for ESD and PSD are of similar magnitude; for instance, the cross-section of Na is typically $10^{-20}$\,cm$^{-2}$ for PSD and $10^{-19}$\,cm$^{-2}$ for ESD \citep{Wurz2022}. In the Solar System, the UV photon flux is usually much higher than the electron flux, making PSD the dominant desorption process \citep{Killen2007}. As with PSD, ESD can be suppressed by collisions with ambient gas, as inelastic interactions with molecules such as H$_2$O
rapidly degrade electron energies \citep{Cravens1986, Itikawa2005}. 

Thermal, photon, and electron desorption processes thus provide efficient, non-sublimative mechanisms for mass release from refractory surfaces. In the solar system, these processes lead to the formation of tenuous tails around bodies like Mercury, the Moon or Phaethon, which would however be too faint to produce detectable signatures in the solar spectrum if viewed from an extrasolar perspective. Nevertheless, their strong dependence on stellar radiation and particle flux implies that, in exocometary systems orbiting luminous or active stars, these mechanisms may rival or even exceed classical volatile sublimation as sources of transient gas and dust.

\subsubsection{Sputtering}
The surfaces of airless bodies can lose material through sputtering, which is the momentum transfer from a projectile (typically ions) to molecules or atoms of a solid. In our Solar System, sputtering contributes to the formation of tenuous atmospheres around the icy moons of Jupiter and Saturn \citep{Plainaki_2010_europa, Plainaki2012, Galli2022}, along with photon-desorption. Mercury has a rather weak magnetosphere, resulting in localized sputtering at magnetic poles releasing O, Si, Ca and Fe, among other species \citep{Killen2007}. Earth's Moon, which spends most of its time outside Earth's magnetosphere, also loses H, O, Ca, Na, and Fe atoms through sputtering \citep{Wurz2022, Vorburger2012, Poppe2016}. Sputtering occurs through two main mechanisms: nuclear-elastic collisions between the incoming ion and atoms/molecules, known as knock-on sputtering, and electronic sputtering, where a particle is released through the electronic excitation of a dissociative state, similar to photon/electron desorption \citep{Johnson2013, Sigmund1981}.

Sputtering yields depend heavily on the nature of the collision partners and the energy of the incident ions. For water ice, ions with energies below 1 keV/amu (typical of solar wind velocities) predominantly induce nuclear (knock-on) sputtering, resulting in the release of water molecules and their fragments \citep{Shi1995, Fama2008}. At higher ion energies, electronic sputtering becomes more effective as electronic excitations dominate the energy deposition processes \citep{Johnson2013}. Conversely, for refractory materials and metals, knock-on sputtering is generally the primary mechanism due to their denser and more robust atomic structures, which favor momentum transfer during collisions \citep{Johnson1991, Cassidy2005}.

A third type of sputtering is potential sputtering, which occurs when highly charged ions eject material from a surface through the release of Coulomb potential energy. Unlike nuclear and electronic sputtering, this mechanism relies on the ion's high charge state rather than its kinetic energy \citep{Bodewits2010}. As highly charged ions approach or impact a surface, they undergo rapid neutralization as electrons transfer from the surface to the ions. This neutralization process releases significant energy, capable of disrupting the surface and ejecting particles. Consequently, ions with higher charge states in the solar wind can significantly increase the sputtering yield \citep{Tielens2005}.
Overall, the energy of the incident ion plays a critical role in determining sputtering efficiency. At low energies, the projectile may not have sufficient momentum to cause substantial material ejection. At extremely high energies, the ion may penetrate too deeply into the material, depositing most of its energy below the surface and reducing the effective sputtering yield \citep{Johnson2013}. Optimal sputtering occurs at intermediate energy ranges where surface interactions dominate. Solar wind disturbances, such as Coronal Mass Ejections, can temporarily enhance sputtering rates by introducing higher fluxes, or particles with higher charges \citep{Killen2012_moon, Killen2012_mercury}. In fact, comparing sputtering yields under different solar wind conditions, and at different heliocentric distances, could provide insights into the role of sputtering processes around different host stars. In this context, the Moon, Mercury, and Sun-grazing asteroids could serve as excellent analogs to explore the effects of stellar winds on airless bodies \citep{Wurz2022, Poppe2016}.

\subsubsection{Micrometeorite impacts}
\label{Sect. Micrometeorite impacts}

The impacts of micrometeorites on unprotected planet surfaces drive numerous processes, including exospheric generation, impact gardening, surface contamination, and electrostatic effects \citep{Szalay2018}. One key outcome is the formation of impact plumes, which consist of surface materials fragmented into particles ranging from macroscopic debris to atoms and molecules. 
These plumes contribute significantly to the exosphere, particularly through the release of low-volatile and refractory species \citep{Stern1999, Killen2016, Janches2021}. During periods of low solar activity, micrometeorite impacts may dominate particle release processes across an entire planetary surface, especially at night, when other mechanisms, such as solar wind sputtering and PSD, are absent \citep{Cassidy2021, Pokorny2018, Nie2024}. The high impact speeds of micrometeorites result in substantial vaporization of surface materials \citep{Collette2014}.

The energy involved in micrometeorite impacts is substantial, producing vapor clouds at temperature ranging from 2500 to 5000~K \citep{Eichhorn1978}, and perhaps as hot as 15000 K \citep{Cassidy2021}, far exceeding typical surface temperatures. At Mercury, for example, the temperature of the impact plumes can reach up to 10 times higher than the surface temperature on the day side \citep{Killen2016}. The volume of material vaporized during impacts depends on the mass, density, and velocity of the projectile, as well as the composition of the surface material \citep{Collette2014}. 

Recent findings by \citet{Nie2024} further support this by showing that micrometeorite impacts are responsible for approximately 70\% of the Moon's exosphere, as determined from isotopic analyses of lunar soil samples. The remaining 30\% of the exosphere is maintained by solar wind sputtering, emphasizing the crucial role of impact vaporization in sustaining tenuous atmospheres around airless bodies.

\subsection{Macroscopic processes}
\label{sec: macroscopic processes}

In young and compact extrasolar systems, macroscopic processes, such as tidal disruptions or collisions between planetesimals, can also play a major role in shaping the circumstellar environment. These processes can generate substantial amounts of dust and gas, leading to the formation of detectable, comet-like tails. 

Since the Sun is much older than any of our other case-study systems, a direct comparison of the macroscopic processes presently observed as mass-loss mechanisms is not particularly helpful. Collisions are infrequent in the current solar system, but there is evidence from the cratering record of past higher rates. As the planets accreted $\sim$4.5~Gyr ago, the cratering rate declined. This was likely followed by a period of enhanced bombardment between $\sim$3.5 and 4.0-4.2 Gyr, known as the Late Heavy Bombardment \citep{Wetherill1975}. Although there is some debate about the details of the Late Heavy Bombardment, including if there was ever a minimum prior to a later spike \citep[see, e.g., the review by][]{Bottke2017}, the modern cratering rates on the Moon, Mars, and Mercury are unarguably lower than in the early solar system \citep[cf.][]{Strom2005, Strom2015}. 

Despite today's lower cratering rate, collisions are still relatively common in the asteroid belt, with several asteroid families apparently having formed within the last million years \citep[see review by][]{Nesvorny2015}. The importance of collisions in sculpting the asteroid population has been well-established for decades, primarily driven by asteroids' size-frequency distribution \citep[cf.][and references therein]{Bottke2005}.
Regular monitoring of Jupiter, Saturn, and the Moon has occasionally revealed evidence of impacts by asteroids and/or comets, with the impact of D/1993 F2 (Shoemaker-Levy 9) into Jupiter the prime example \citep[cf.][]{Weaver1994, Hammel1995}. Several main-belt asteroids have also been observed to release material as a direct consequence of recent impacts \citep{Jewitt2012}.


Some extrasolar systems show evidence for a much stronger collisional activity.
As discussed in Mustill et al. (submitted) and \cite{Bannister2025SSR}, the widespread detection of dusty debris discs around mature stars necessitates ongoing, destructive collisions in planetesimal belts \citep[e.g.][]{Wyatt2008, Hughes2018, Pearce2024Review}. This is because the collisional lifetimes of detected grains are much shorter than the stellar ages, and these grains are typically located in discs $\sim$100s of au from their stars, so they are thought to be released via planetesimal collisions rather than from sublimating comets. Further evidence for extrasolar collisions comes from observed-dust quantities around stars decreasing with stellar age, which is consistent with the picture that planetesimal belts collisionally erode over time \citep[e.g.][]{Wyatt2007AStars, Lohne2008, Matra2025}. Short-term spikes in dust production are also seen around some stars, which are thought to signify collisions between large asteroids \citep[e.g.][]{Su2019}, and resolved images of some debris discs show tail-like structures thought to arise from a single impact event \citep[e.g.][]{Jackson2014, Jones2023, Rebollido2024}.

Besides collisions, minor bodies can also be disrupted through tidal interactions with stars and massive planets. In the solar system, there is evidence for tidal disruptions of comets due to close approaches to the Sun and Jupiter. The aforementioned comet Shoemaker-Levy~9 was conclusively shown to have disrupted in 1992 during a close approach to Jupiter, 2 years before its eventual impact into the planet \citep{Sekanina1994, Asphaug1996}. The Kreutz group of sungrazing comets (see Section~\ref{Sect. Solar System comets}) are thought to have been produced by ongoing, hierarchical fragmentation driven in part by tidal forces near the Sun \citep[see review by][]{Marsden2005}. 
Extrasolar systems also show tidal disruption; the most-convincing indications come from white-dwarf stars, where there is evidence of disintegrating comet-sized bodies inside the star's tidal-diruption radius \citep[e.g.][]{Jura2003, vanderburg2015}. The disintegration of these small bodies lead to the formation of tail structures, which produce asymmetric transit events in the stellar light curve \citep[e.g.,][]{gansicke2016}.

\FloatBarrier

\section{Production of secondary species}

This section investigates the various processes that alter the composition, excitation and ionisation states of the material released by comets or exocomets, and how these processes scale across different stellar environments.

\subsection{Dust grain sublimation}
\label{Sect. Dust sublimation}

When dust grains are released from cometary nuclei, they become directly exposed to stellar radiation. This can trigger their sublimation, which is the process by which minerals (e.g. Mg$_2$SiO$_4$, forsterite) transition from solid to gas phase. For most minerals, this transition is accompanied by the breaking of chemical bonds \citep[e.g.,][]{Steinmeyer2023_sublimation}; as a result, the sublimation product are generally atoms (e.g. Mg, Si, O) or simpler molecules (e.g. SiO). This process likely occurs in the \bp\ system, leading to the release of extreme amounts of refractory atoms and ions by exocomets \citep{Beust1990}. The goal of this section is to investigate the stellar conditions under which dust sublimation becomes efficient, considering the physical properties of the grains, such as their size and composition.

The sublimation rate of a dust grain is directly linked to its temperature, which in turn is controlled by the energy equilibrium of the grain \citep[see for instance][]{Kimura2002}. Taking into account the flux received from the star, the flux emitted by the grain, and the energy used in sublimation, this energy equilibrium is written:
\begin{equation}
    \Omega \int \sigma_{\rm abs} (s, \lambda) I_\star(\lambda) \mathrm{d} \lambda = 4\pi \int \sigma_{\rm abs} (s, \lambda) B(\lambda, T) \mathrm{d} \lambda - \frac{\mathrm{d} m}{\mathrm{d} t}(s,T) L,
    \label{Eq. dust energy balance}
\end{equation}
where $\sigma_{\rm abs}$ is the absorption cross-section of the dust grain (which depends on its radius $s$ and on the wavelength $\lambda$), $I_\star$ is the stellar specific intensity (J/s/cm$^2$/\AA/sr), $B$ is the Planck function (assuming that the dust grain radiates like a black-body; same unit as $I_\star$), $T$ is the grain temperature, and $L$ is the dust latent heat of sublimation. The solid angle covered by the star ($\Omega$) is linked to the stellar distance, $d$, through:
\begin{equation}
    \Omega = 2\pi \left( 1 - \sqrt{1 - (R_\star/d)^2} \right),
    \label{Eq. Solid angle}
\end{equation}
and the mass loss $\frac{dm}{dt}$ is given by:
\begin{equation}
    \frac{\mathrm{d} m}{\mathrm{d} t} = - 4 \pi s^2 \alpha \sqrt{ \frac{m_{\rm gas}}{2 \pi k_B T}} \ p(T),
    \label{Eq. Dust sub rate}
\end{equation}
where $p(T)$ is the vapor pressure of the released gas, $m_{\rm gas}$ is the molecular mass of the sublimated molecule, $k_B$ is the Boltzmann constant, and $\alpha$ is a correction factor~$<1$. The vapor pressure $p(T)$ can be obtained from Clausius-Clapeyron's law:
\begin{equation}
    p(T) = p_{\infty} \exp \left(-\frac{m_{\rm gas} L}{k_B T} \right),
\end{equation}
where $p_\infty$ is the vapor pressure at infinite temperature. More complex $p-T$ relationships are also used \citep[e.g.][for water ice]{Lamy1974}. 

Equation \ref{Eq. dust energy balance} can be solved numerically to extract the equilibrium temperature $T$ of the grain, which in turn provides the mass loss rate (Eq. \ref{Eq. Dust sub rate}). For large grains ($s\geq 1~\mu$m), the cross section $\sigma_{\rm abs}$ is rather uniform throughout the UV/visible spectrum; as a result, the energy balance of the grain can be simply written as (neglecting the energy lost in sublimation):

\begin{equation}
    \Omega \pi s^2 \sigma T_{\rm eff}^4 = 4 \pi^2 s^2 \sigma T^4,
\end{equation}
with $\sigma$ the Stefan-Boltzmann constant and $T_{\rm eff}$ the stellar effective temperature. This yields: 
\begin{equation}
    T = T_{\rm eff} \left( \frac{\Omega}{4 \pi} \right) ^{\frac{1}{4}}.
\end{equation}
In this case, the sublimation rate of dust grains is similar to the sublimation rate of material directly at the surface of the nucleus (Sect. \ref{Sect. Volatile sublimation}). For smaller grains, however, the absorption cross section may vary by several orders of magnitude from visible to infrared wavelengths, leading to equilibrium temperatures that are very different from that of a macroscopic object, and thus to different sublimation rates. In this case, the exact resolution of Eq. \ref{Eq. dust energy balance} becomes mandatory, using cross sections derived from the Mie theory.

\begin{figure}[h]
\centering
    \includegraphics[scale = 0.18,     trim = 10 0 70 0,clip]{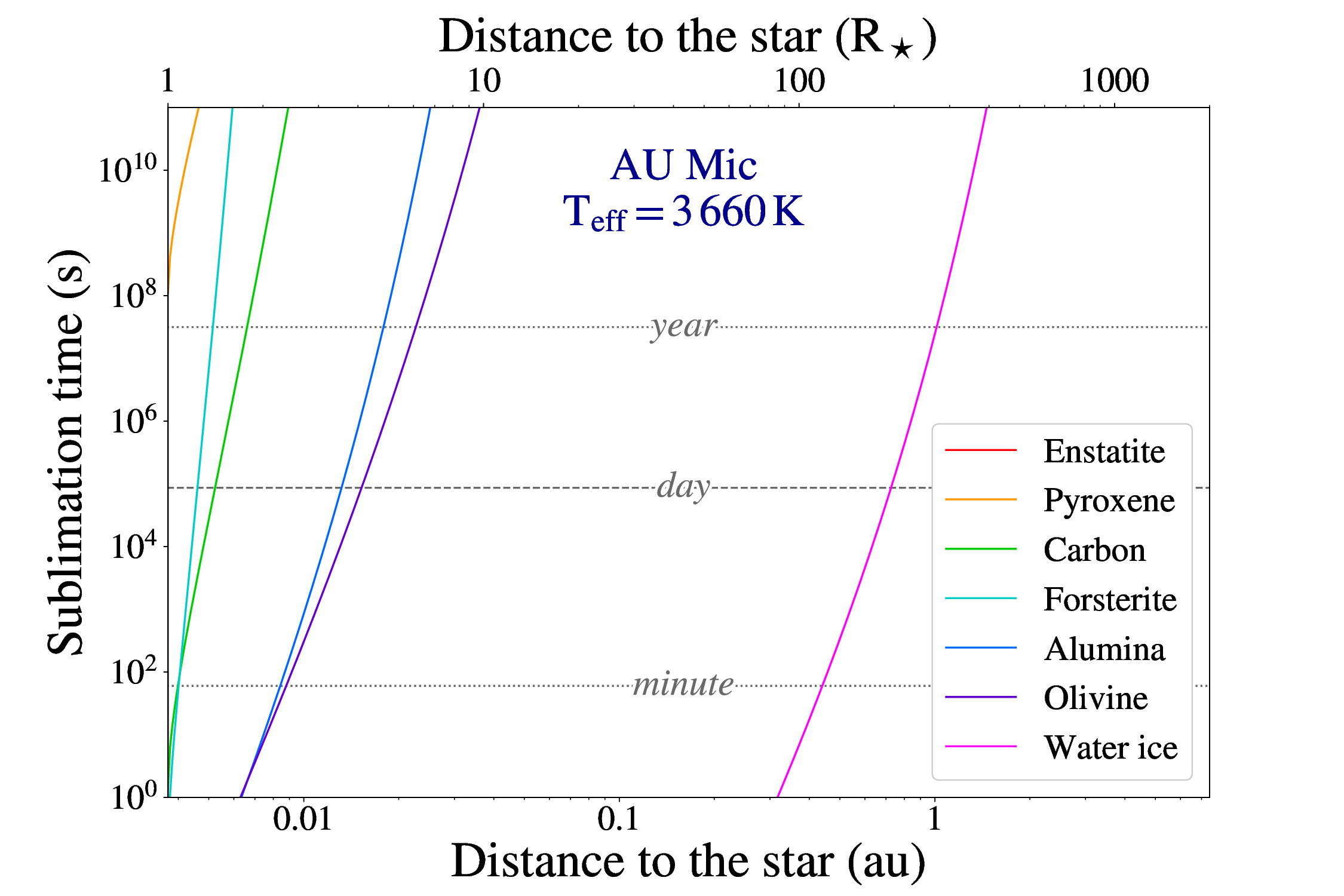}    
    \includegraphics[scale = 0.18,     trim = 30 0 60 0,clip]{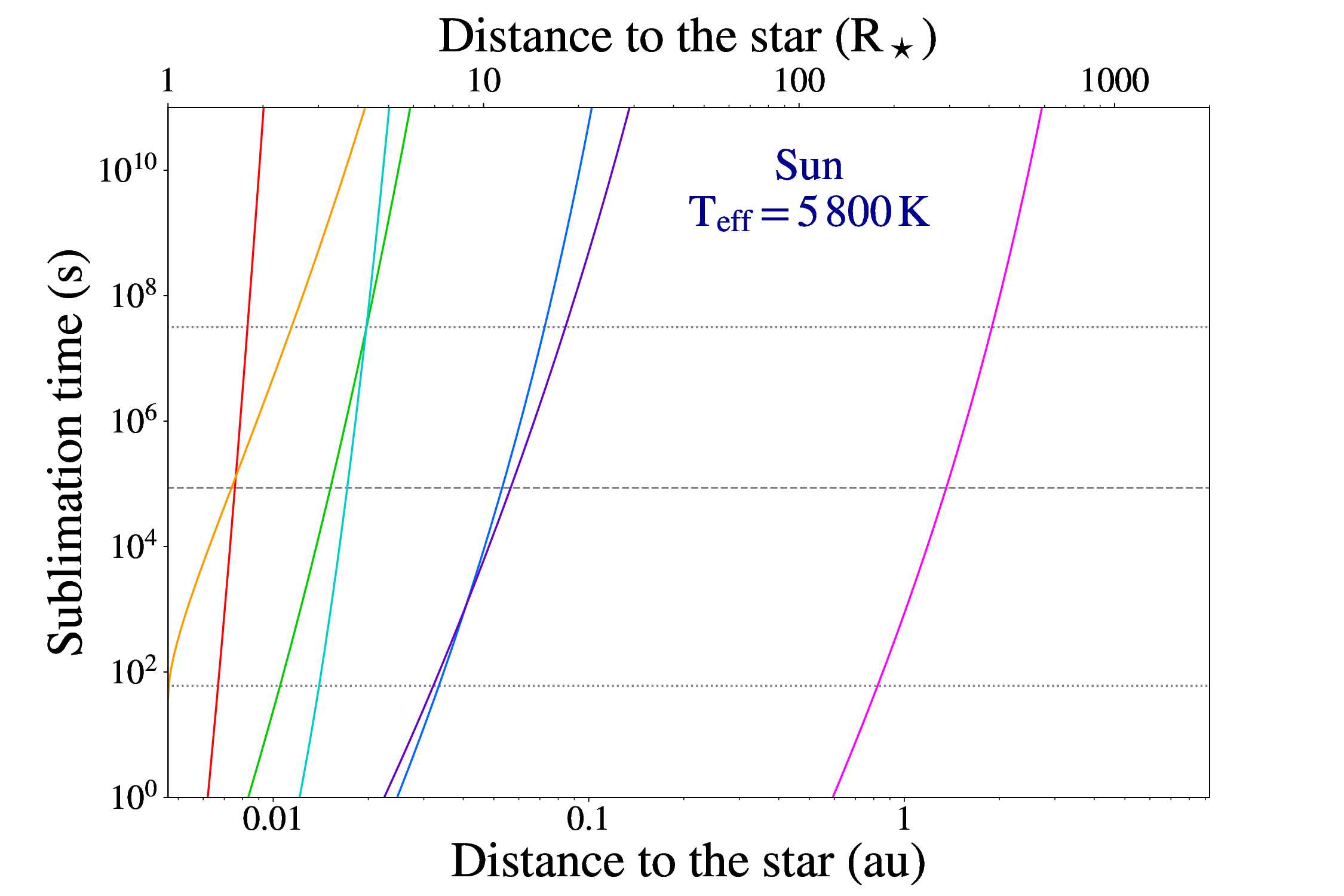}

    \vspace{0.2 cm}
    
    \includegraphics[scale = 0.18,     trim = 10 0 70 0,clip]{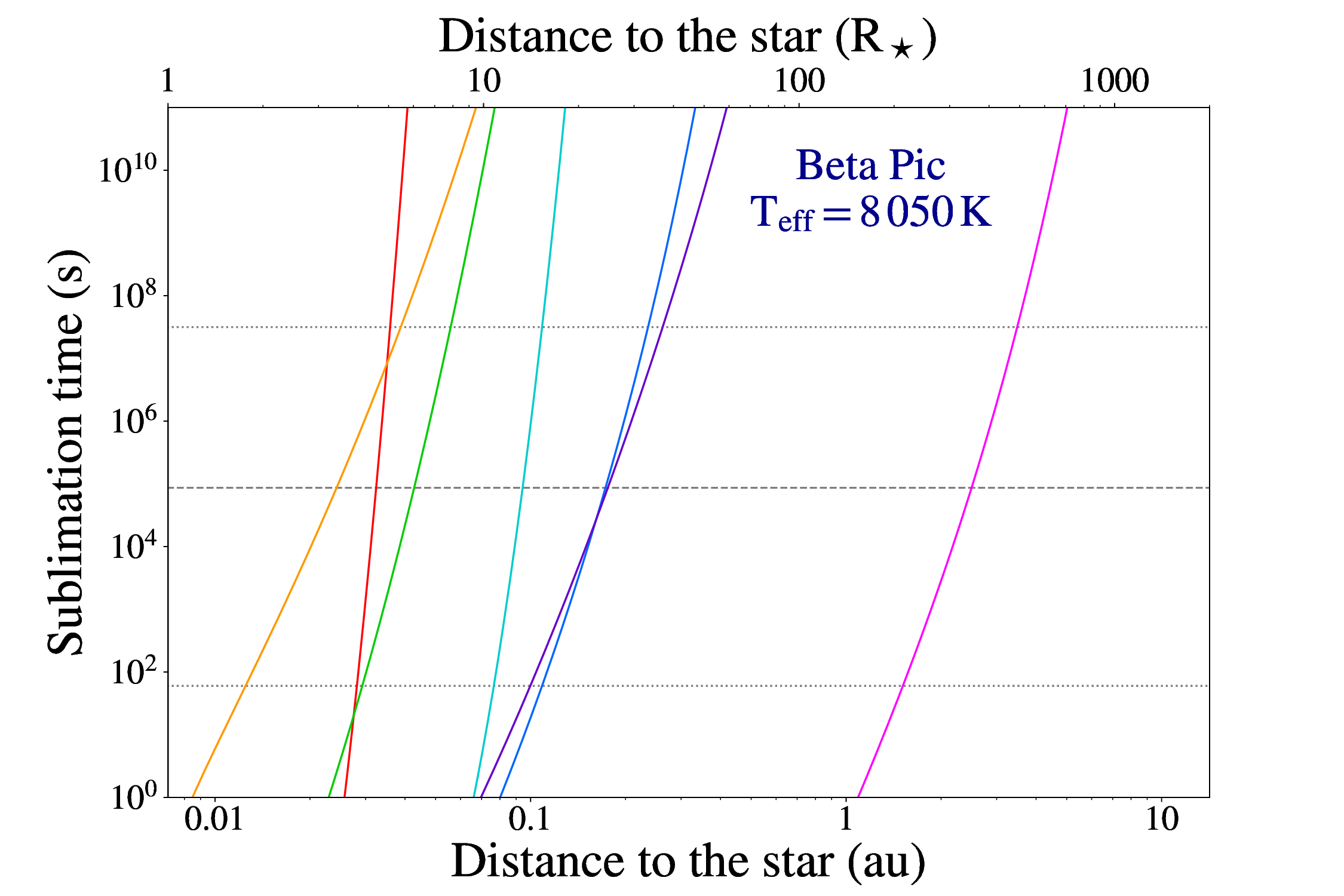}    
    \includegraphics[scale = 0.18,     trim = 30 0 60 0,clip]{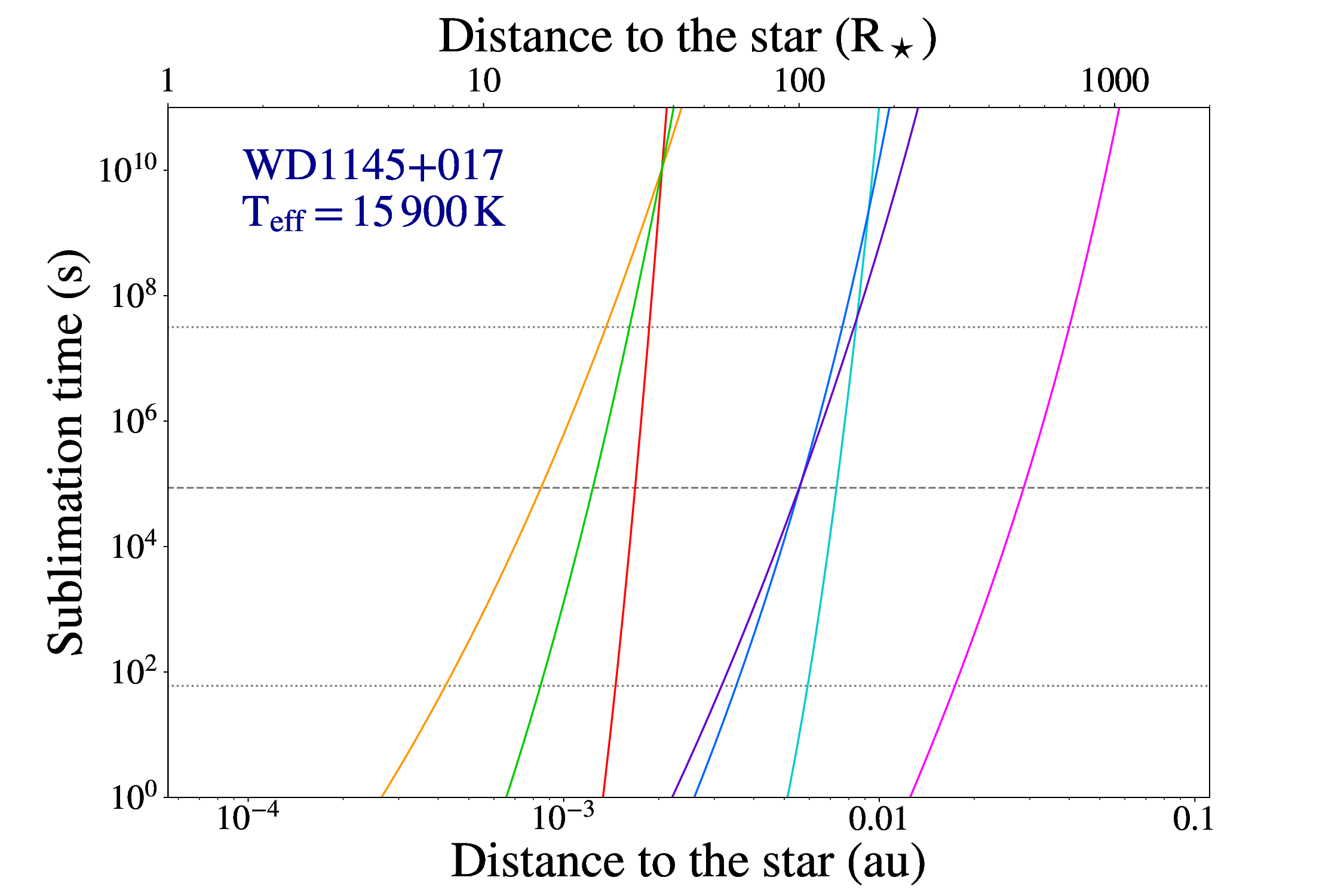}    
    
    \vspace{0.2 cm}
    
    \caption{\small Sublimation timescale of a 0.5 \textmu~m dust grain with various compositions around four different stars, as a function of the distance to the star. Dotted lines indicate a one-minute, one-day, and one-year times. Distances are expressed in stellar radii (upper axis; same scale for all stars) and au (lower axis; specific to each star).}
    \label{Fig. Dust sublim time vs dist}
\end{figure}

To illustrate the typical sublimation rate of dust in various stellar environments, Fig. \ref{Fig. Dust sublim time vs dist} provides the sublimation timescale (defined as $m/\frac{\mathrm{d} m}{\mathrm{d} t}$) of a 0.5 \textmu~m dust grain around our four case-study stars (AU~Mic, the Sun, {\bp}, and WD 1145+017; see the radius and temperature properties in Table \ref{table: case study stars}), for various compositions (olivine, pyroxene) and stellocentric distances. The grain cross sections were taken from \cite{budaj2015}, and the temperature-pressure relationships from \cite{vanlieshout2014} (forsterite, enstatite, alumina, amorphous carbon), \cite{Kimura2002} (olivine, pyroxene), and \cite{Lamy1974} (water ice). The grain radius, 0.5 \textmu~m, corresponds to the peak of the size distributions discussed in Sect. \ref{sec: dust escape}.

The ability of a given star to sublimate dust depends heavily on its effective temperature. For a $0.5\,\mu$m olivine grain orbiting \bp, the distance for which the sublimation time is 1 day is located around 0.18 au. Around the Sun and AU~Mic, this limit falls down to $\sim 0.06$ and $\sim 0.015$\,au respectively. Due to its small radius, dust sublimation around WD 1145+017 occurs within a very small region ($\sim$ 0.03\,au, about a solar radii). More detailed studies of dust sublimation around polluted white dwarfs can be found in \cite{Shestakova2022} and \cite{Steckloff2021}.

The sublimation rate also depends on the grain composition. While some minerals are sublimated at rather large stellocentric distances (e.g. olivines), others appear to be more resistant (e.g. enstatite). In general, a dust grain will be more resistant to sublimation when its cross section in the UV/visible (where the stellar flux is absorbed) is higher than in the IR (where the energy is re-emitted).

\begin{figure}[h]
\centering
    \includegraphics[scale = 0.18,     trim = 10 0 70 0,clip]{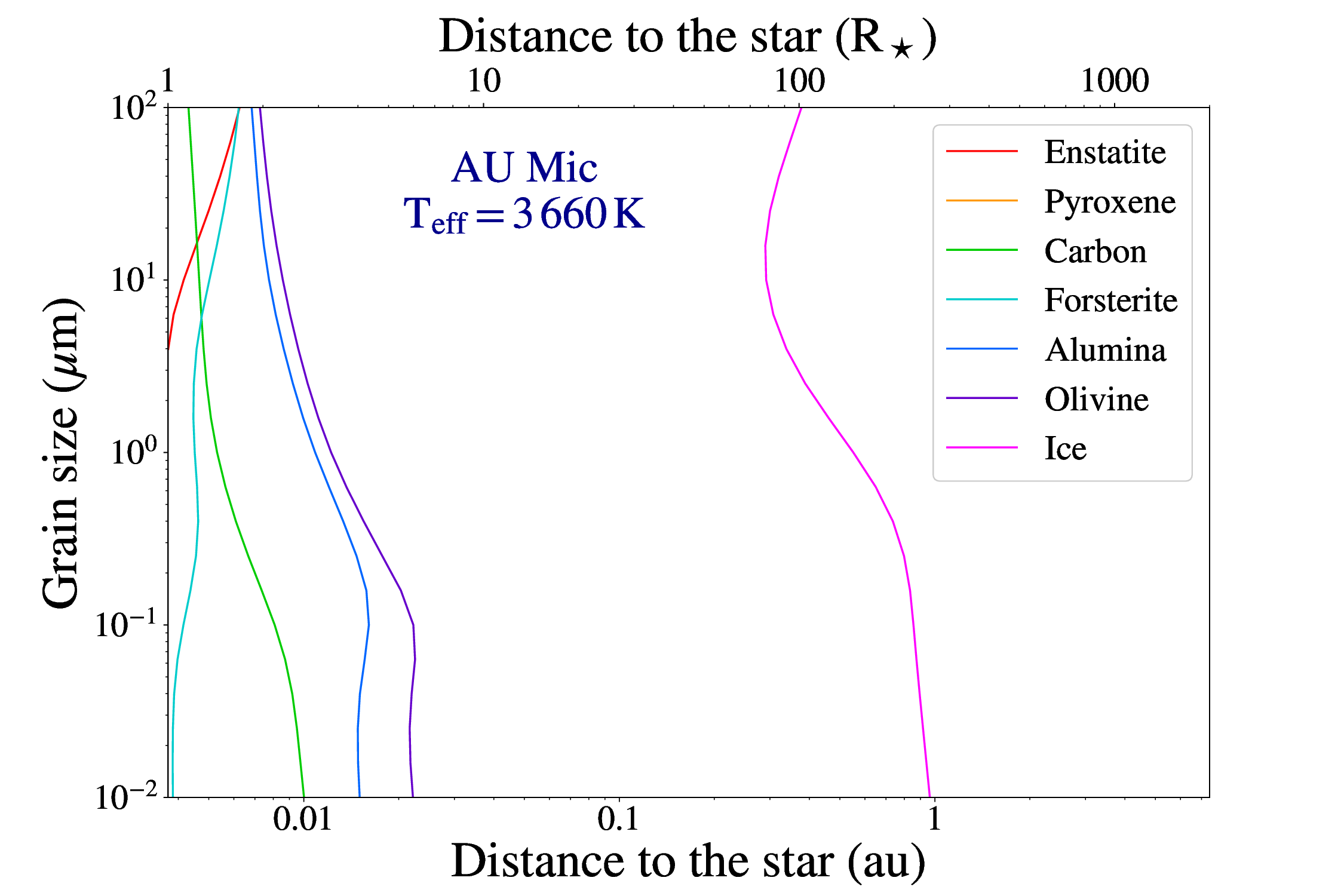}    
    \includegraphics[scale = 0.18,     trim = 25 0 60 0,clip]{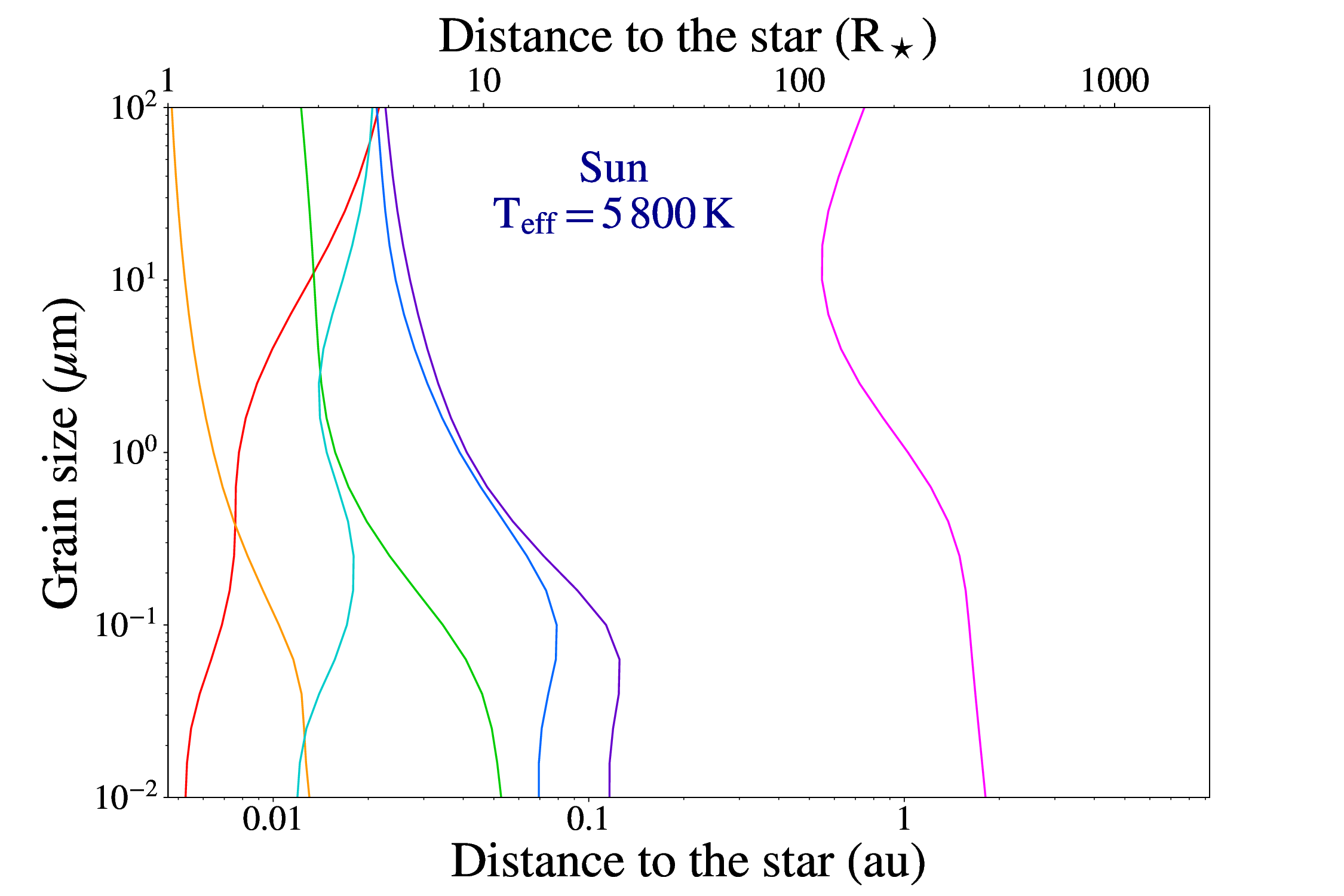}

    \vspace{0.2 cm}
    
    \includegraphics[scale = 0.18,     trim = 10 0 70 0,clip]{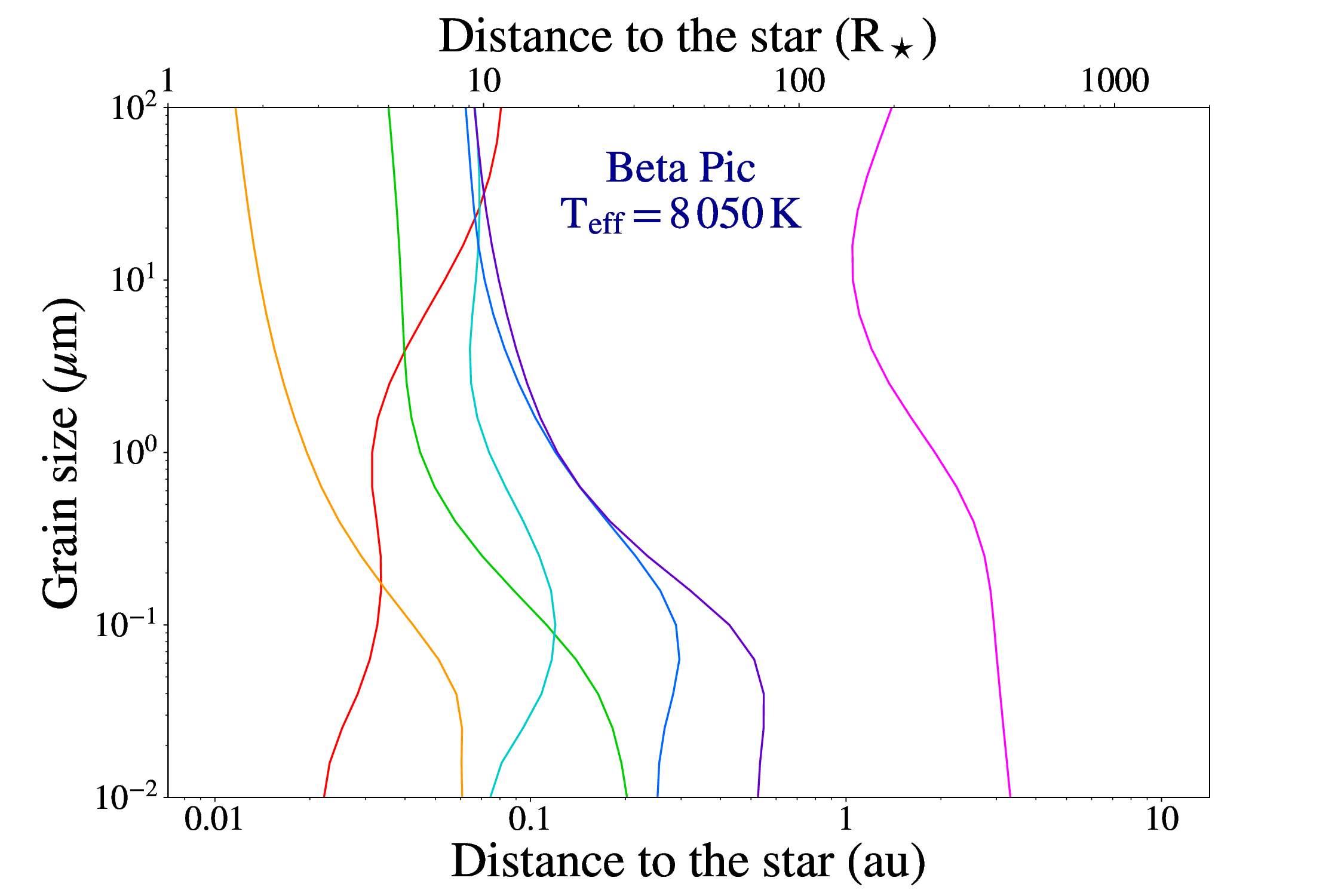}    
    \includegraphics[scale = 0.18,     trim = 25 0 60 0,clip]{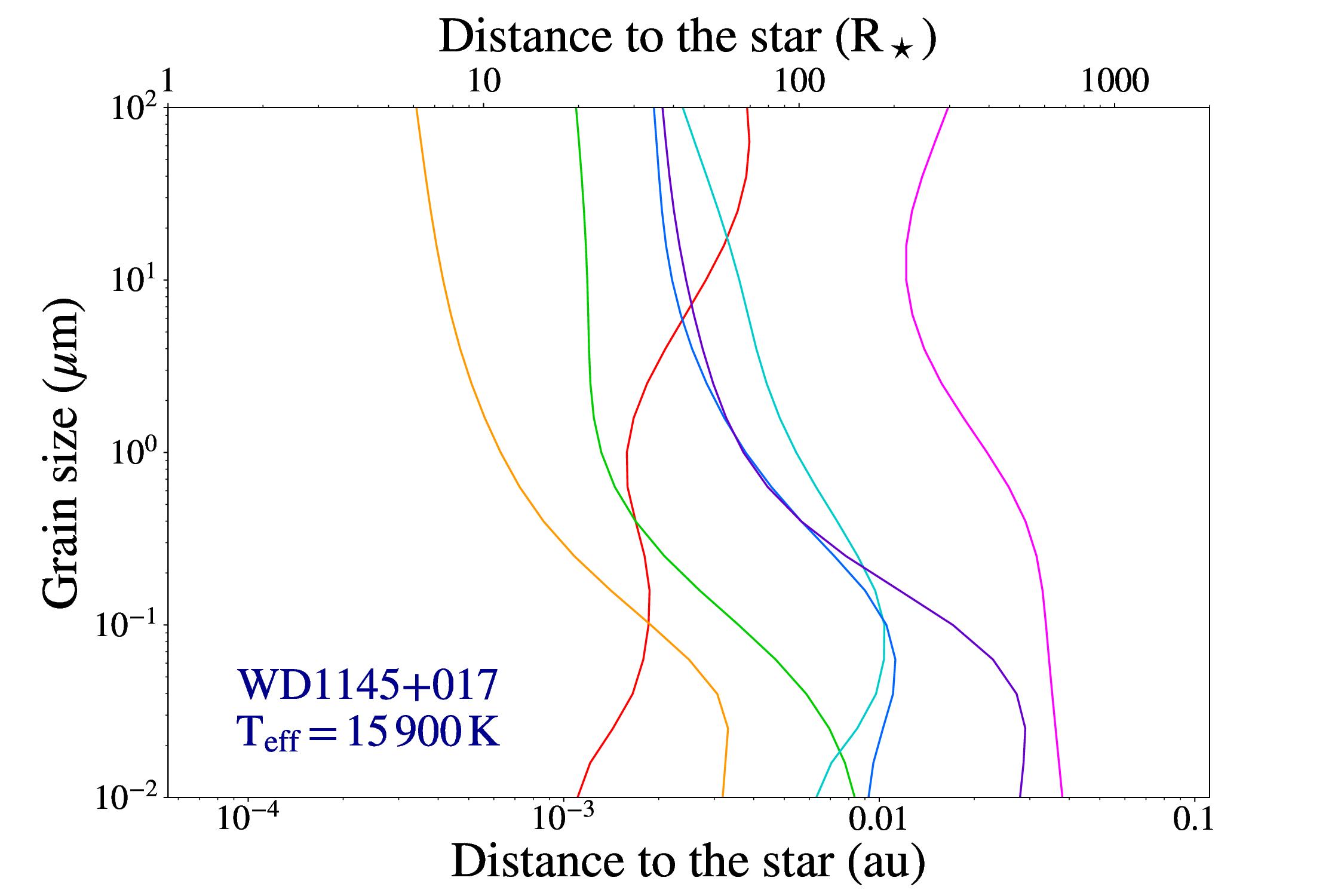}    

    \vspace{0.2 cm}
    
    \caption{\small Distance from the host star at which the sublimation timescale of a dust grain is 1 day, depending on the grain radius and composition. For all figures, the distance range is set to [1,\,2000]\,R$_\star$.}
    \label{Fig. Dust sublim dist vs radius}
\end{figure}

In a similar way, Fig \ref{Fig. Dust sublim dist vs radius} provides the sublimation distance (here defined as the stellar distance at which the sublimation timescale is 1 day), as a function of the grain radius. Although slight variations of the sublimation distance with grain size are observed, the main trend is that \bp\ is generally much more efficient at sublimating grains (below $\sim 0.1 - 0.5$ au, depending on the grain size) than the Sun or AU~Mic. This is consistent with the observation of large amounts of refractory ions (e.g. Fe$^+$, Ca$^+$...) in the tails of \bp\ exocomets.

We emphasize that the calculations presented here are simplistic, and only provide an approximate description of the behavior of dust grains around stars. In reality, dust grains may have inhomogeneous compositions (with various minerals taking various crystalline forms), and are likely not spherical (as assumed in the calculations of the cross-sections), resulting in potentially large variations in their sublimation rates and distances. For instance, \cite{Protopapa2018} showed that the sublimate rate of icy grains strongly varies with their size and purity. Grain porosity is also expected to influence the sublimation rate, although very few studies on this aspect are available \citep[e.g.,][]{Kossacki2021_ice_porosity}. Finally, it should be noted that young, dense dust discs may be optically thick, thereby reducing sublimation rates and shifting the sublimation lines closer to the star. 

The sublimation of dust grains produced by exocomets is related to the problem of `hot dust', which is detected around one-fifth of main-sequence stars \citep[e.g.][]{Ertel2016, Kirchschlager2017, Absil2021}. This dust is composed of sub-micron-size dust grains at a temperature above 1500~K \citep[][]{Defrere2012}, and can be detected in the near infrared using optical long-baseline interferometry. The nature of this dust is a mystery. These grains should rapidly sublimate or blow away under radiation pressure, and so far all attempts to dynamically explain hot dust have failed to completely reproduce the phenomenon \citep[e.g.][]{vanlieshout2014, Rieke2016, Sezestre2019, Kimura2020, Pearce2020}. This dust is most likely associated with exocomets, because some hot dust is variable which is best explained by stochastic cometary infall \citep[][]{Ertel2016}, and there is a tentative relation between hot-dust detections and circumstellar gas indicative of cometary activity \citep[][]{Rebollido_PhD}. However, cometary supply alone is unlikely to fully explain hot exozodiacal dust, or exozodis. Although star-grazing comets are promising candidates, the required comet inflow rates and dust ejecta size distributions conflict with current cometary and planetary system models. Additional dust-trapping mechanisms may work alongside cometary input to produce hot exozodis, but comets alone are insufficient to explain this phenomenon \citep[][]{Pearce2022}.

\FloatBarrier

\subsection{Dissociation and ionization of molecules}
\label{Sect. Dissociation}

Molecules released from cometary nuclei—such as H$_2$O and CO—may dissociate into lighter molecules (e.g., OH) or atoms (O, H), or they may undergo photoionization (e.g., forming H$_2$O$^+$), depending on the stellar flux they're exposed to. These processes are frequently observed in Solar System comets \citep[cf.][]{Feldman2004, Fortenberry2021}, and they are likely even more pronounced in the \bp\ system due to its star’s higher UV flux.

The dissociation rate of a molecule (e.g., H$_2$O) released by a comet can be calculated from its dissociation cross-section and the spectrum of its host star. This writes: 
\begin{equation}
    \mathrm{d}P_{\rm dis} = \Omega \int_{0}^{\infty} \frac{\lambda}{hc} I_{\star}(\lambda) \sigma_{\rm dis} (\lambda) \mathrm{d}\lambda \mathrm{d}t,
    \label{Eq. Dissociation}
\end{equation}
where $\mathrm{d}P_{\rm dis}$ is the dissociation probability within a time $\mathrm{d}t$, $\Omega$ is the solid angle covered by the host star (which depends on the distance between the particle and the star; Eq. \ref{Eq. Solid angle}), $I(\lambda)$ is the specific intensity of the stellar disc (J/s/cm$^2$/\AA/sr), $h$ is the Planck constant, $c$ is the speed of light, and $\sigma_{\rm dis}$ is the dissociation cross-section. For a given molecule, different sets of dissociation products are possible (e.g., H$_2$O~$\rightarrow$ H$_2$~+~O; H$_2$O$~\rightarrow$~H~+~OH, \ldots; cf. \citealt[][]{Combi2004}). Similarly, the ionization rate can be written: 
\begin{equation}
    \mathrm{d}P_{\rm ph} = \Omega \int_{0}^{\infty} \frac{\lambda}{hc} I_{\star}(\lambda) \sigma_{\rm ph} (\lambda) \mathrm{d}\lambda \mathrm{d}t,
    \label{Eq. Ionization}
\end{equation}
where $\mathrm{d}P_{\rm ph}$ is the photo-ionization probability within a time $\mathrm{d}t$ and $\sigma_{\rm ph}$ the photo-ionization cross-section. 

As an example, Table \ref{table: dissoc time case study stars} provides the typical dissociation and ionization times of key molecules (H$_2$O, HCN, CO$_2$, \ldots) at 0.2 au from our four case study stars \citep[corresponding to the typical transit distance of \bp\ exocomets,][]{Kiefer2014, Lecavelier2022}. Calculations were performed using the spectral distribution provided in Sect. \ref{Sect. intro}, and cross-sections from the Phidrates database \citep{Phidrates}. While around the Sun all molecules are able to survive $\sim$ 1 -- 10 hours, around \bp\ they are rapidly dissociated, particularly H$_2$O, HCN, and OH (lifetime $\sim 1$ s). As a result, it is little surprise that none of these species has been directly detected in \bp\ exocomets so far - only CO has been observed in the stable disc \citep{Roberge2000_betapic_CO, Dent_2014_betapic_CO}. The ionization timescale is generally higher than the dissociation timescale, although it can be dominant for specific species, like CO or C$_2$.

\begin{figure}[h!]
\centering
    \includegraphics[scale = 0.236,     trim = 75 0 60 60,clip]{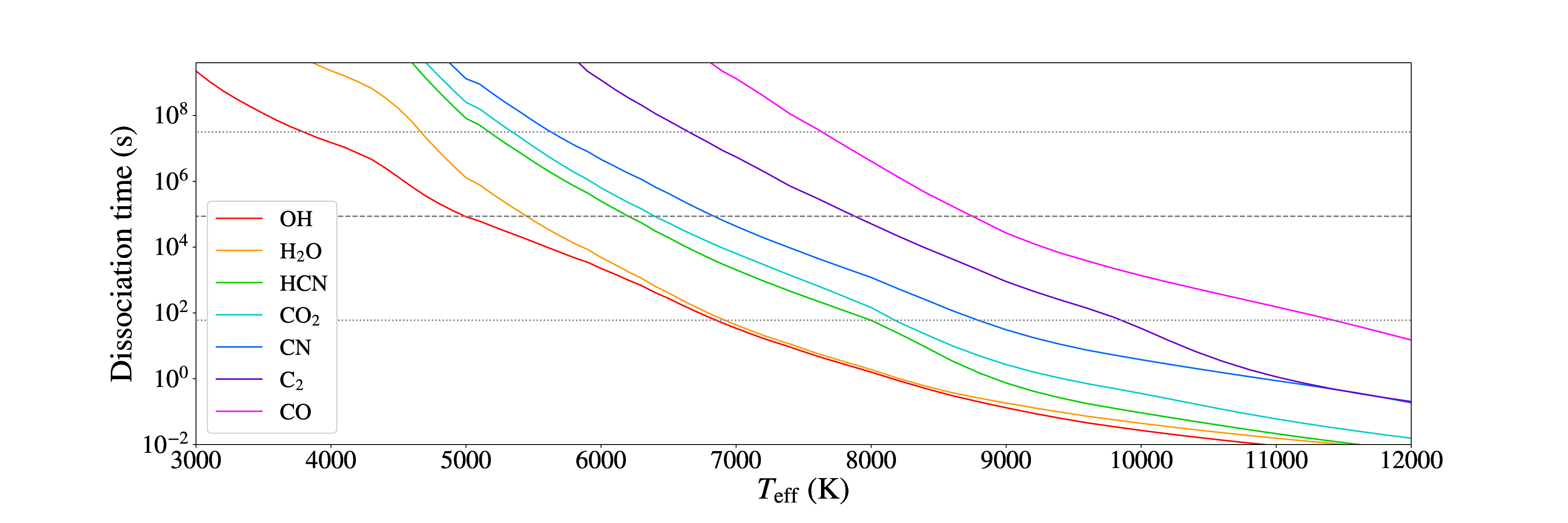}   
    \caption{\small Lifetimes (combining photo-dissociation and ionization) of various molecules at 0.2 au from their host star, depending on the stellar effective temperature. Dotted horizontal lines are identical to Fig. \ref{Fig. Dust sublim time vs dist}. Calculations were performed with synthetic spectra from the PHOENIX library, which are purely photospheric; as a result, these values should be considered as upper limits only.}
    \label{Fig. dissoc time vs Teff}
\end{figure}

More generally, Fig. \ref{Fig. dissoc time vs Teff} presents typical lifetimes for the studied molecules as a function of the stellar effective temperature, again at a distance of 0.2 au. These dissociation timescales were calculated using synthetic spectra from the PHOENIX library \citep{PHOENIX}, and again cross-sections from the Phidrates database. Since the PHOENIX spectra are purely photospheric, the dissociation timescales obtained are very likely overestimated, in particular for cool stars which should exhibit strong coronal and chromospheric emission. Nonetheless, Fig. \ref{Fig. dissoc time vs Teff} shows that stars hotter than $\sim 7000$~K should rapidly dissociate most molecules ($\tau_{\rm dissoc} \ll 1$ day), whereas around cooler stars they can survive significantly longer (as also shown in Table~\ref{table: dissoc time case study stars}). These short lifetimes of the most prominent fragment species observed in Solar System comets (such as OH, CN, C$_{2}$, etc; \cite{Feldman2004}) could explain why they have not been detected in exocometary systems yet.

\begin{table}[h!]
  \renewcommand{\arraystretch}{1.2}
    \begin{tabular}{ c c c c c c c c} 
    \hline
    \noalign{\smallskip}
    \hline
    \noalign{\smallskip}
    
    \textbf{Species} & Process &  WD 1145+017 &  AU~Mic & Sun & Beta Pic   \\
    
    \noalign{\smallskip}
    \hline
    \noalign{\smallskip}

     \multirow{2}{*}{OH} & Dissociation & 2.8$\times 10^{1}$ & 1.4$\times 10^{3}$ &  1.8$\times 10^{3}$  & 1.3$\times 10^{0}$\\
                         & Ionization   & 7.5$\times 10^{3}$ & 4.5$\times 10^{3}$ &  1.2$\times 10^{5}$  & 5.5$\times10^3$ \\
                         
    \noalign{\medskip}

    \multirow{2}{*}{HCN} & \multirow{2}{*}{"}  & 2.8$\times 10^{1}$ & 1.1$\times 10^{3}$ &  2.8$\times 10^{3}$  & 2.2$\times 10^{1}$   \\
                         &                     &  9.8$\times 10^{2}$ & 2.4$\times 10^{3}$ &  5.3$\times 10^{4}$  & 7.0$\times10^2$      \\

    \noalign{\medskip}

    \multirow{2}{*}{H$_2$O} &  \multirow{2}{*}{"}  & 4.4$\times 10^{1}$ & 2.1$\times 10^{3}$  &  4.0$\times 10^{3}$  & 1.5$\times 10^{0}$   \\
                            &                       & 8.8$\times 10^{2}$ & 3.6$\times 10^{3}$ &  7.6$\times 10^{4}$  & 1.1$\times 10^{3}$  \\
    \noalign{\medskip}

    \multirow{2}{*}{CN\,$^{\rm a}$} &  \multirow{2}{*}{"}  & 2.7$\times 10^{1}$ & 5.6$\times 10^{2}$ & 7.6$\times 10^{3}$  & 6.3$\times 10^{1}$   \\
                        &                      & -                  & -                  &  -                  &  -                   \\
    \noalign{\medskip}

    \multirow{2}{*}{CO$_2$} & \multirow{2}{*}{"}  & 3.5$\times 10^{1}$  &  1.7$\times 10^{3}$  & 2.1$\times 10^{4}$  &  1.0$\times 10^{2}$  \\
                            &                     & 1.1$\times 10^{3}$  &  2.0$\times 10^{3}$  & 4.0$\times 10^{4}$  &  7.1$\times 10^{2}$  \\
    \noalign{\medskip}

    \multirow{2}{*}{C$_2$} & \multirow{2}{*}{"}  & 3.5$\times 10^{2}$   &  1.5$\times 10^{4}$  & 2.4$\times 10^{5}$ & 6.4$\times 10^{3}$ \\
                           &                     &  2.2$\times 10^{2}$  &  1.4$\times 10^{3}$  & 2.6$\times 10^{4}$ & 2.4$\times10^2$  \\
    \noalign{\medskip}

    \multirow{2}{*}{CO} & \multirow{2}{*}{"}  & 4.1$\times 10^{2}$   &  5.7$\times 10^{3}$  & 6.7$\times 10^{4}$ &  6.3$\times 10^{2}$ \\
                        &                     &  1.6$\times 10^{3}$  &  2.9$\times 10^{3}$  & 6.5$\times 10^{4}$ &  1.1$\times 10^{3}$  \\

    \noalign{\smallskip}
    \hline
    \end{tabular} 

    \caption{\small Photo-dissociation and ionization timescales (in s) of common molecules around our four case-study stars, calculated at a stellocentric distance of 0.2 au.\\
    \footnotesize Notes. 
    $\boldsymbol{[}\textbf{a}\boldsymbol{]}$ Photo-ionization cross section for CN are not included in the Phidrates database. }

\label{table: dissoc time case study stars}
\end{table}

\subsection{Ionization of atoms and ions}
\label{Sect. Ionisation}

\subsubsection{Photo-ionization}

Around hot stars, efficient dust sublimation (Sect. \ref{Sect. Dust sublimation}) and molecular dissociation (Sect. \ref{Sect. Dissociation}) should lead to the production of large amounts of neutral atomic species, including O, C, and Fe. Depending on their ionization energies and on the intensity and distribution of the stellar flux, these species can be subsequently photo-ionized. The ionization rate of a given species can be calculated using its ionization cross section and the stellar luminosity, in a similar way as for molecules (Eq. \ref{Eq. Ionization}).

\begin{figure}[h!]
\centering
    \includegraphics[scale = 0.24,     trim = 75 0 60 50,clip]{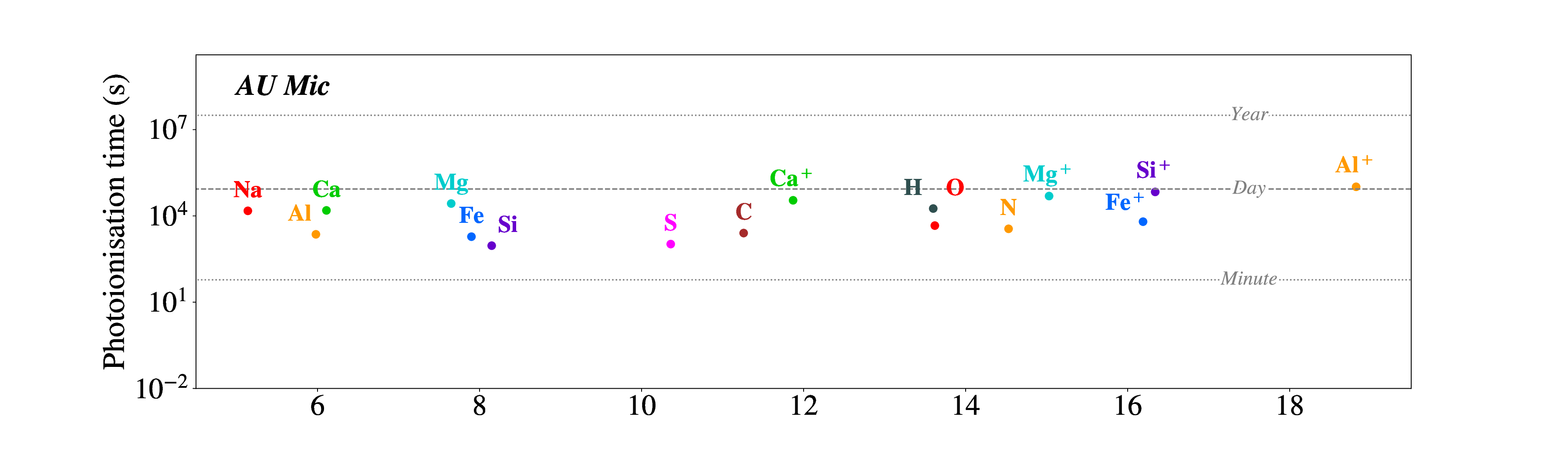}    
    \includegraphics[scale = 0.24,     trim = 75 0 60 60,clip]{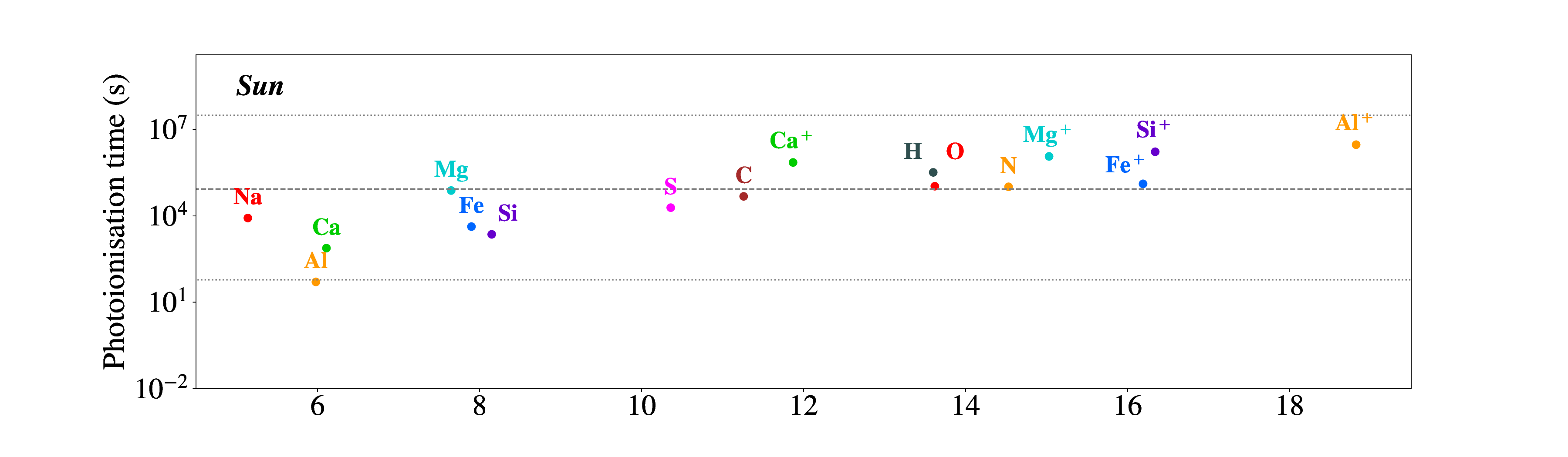}
    \includegraphics[scale = 0.24,     trim = 75 0 60 60,clip]{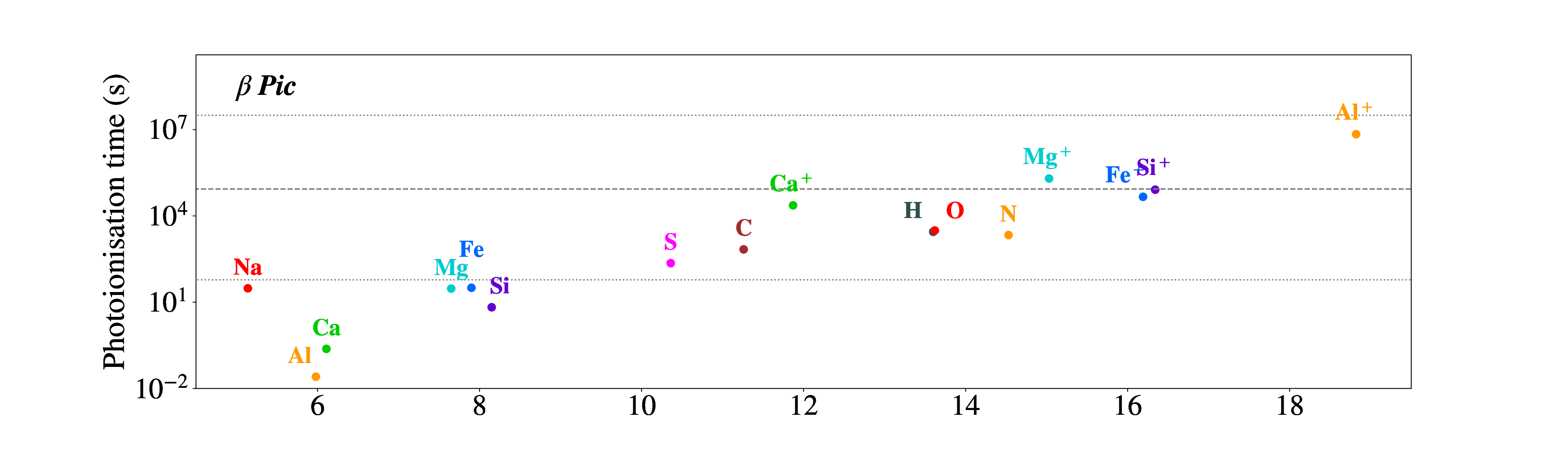}       
    \includegraphics[scale = 0.24,     trim = 75 0 60 60,clip]{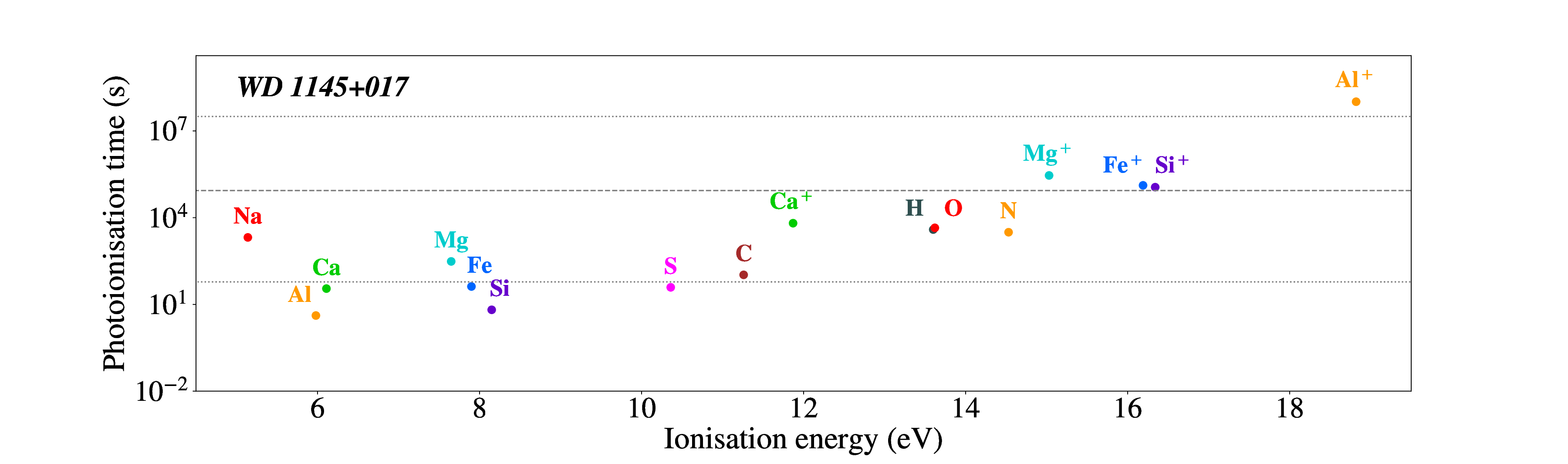}       

    \vspace{0.2cm}
    
    \caption{\small Photoionization timescales of various atoms and ions around our four case-study stars, at a distance of 0.2 au. Horizontal dotted lines indicates reference times of 1 minute, 1 day and 1 year.}
    \label{Fig. photo ion rate}
\end{figure}

As an example, Fig. \ref{Fig. photo ion rate} shows typical ionization times of various atomic and ionized species around our four case-study stars, at a stellocentric distance of 0.2~au. The calculations were performed using ionization cross sections from the Phidrates database, as was done for photo-dissociation, and the spectral distributions adopted in Sect. \ref{Sect. intro}. The largest differences between the four stars occur for neutral atoms: around the Sun and AU~Mic the photo-ionization timescales of Na, Fe, or Si are rather long ($\sim 10^4$~s, corresponding to a few days at 1 au), and they drop below 60~s around \bp, due to a much stronger photospheric continuum below 2000~\AA. This explains why refractory species observed around \bp\ are almost always ionized \citep[for instance, signatures in \fei\ or \nai\ are rare and extremely shallow;][]{Welsh2016_fe1_betapic, Hoeijmakers2025_betapic}, while neutral atoms such as Fe, Ni, and Na are commonly observed in the exospheres and tails of Solar System objects, even though they eventually become ionized, too \citep{Manfroid2021, Fulle2007}.   

For species with ionization threshold shorter than the Ly-$\alpha$ limit (10.2 eV $\sim 1216$ \AA), the difference in photo-ionization times is less important. Thanks to their relatively strong EUV fluxes, the Sun and AU~Mic are able to photo-ionize species such as C, C$^{+}$, H, or Fe$^+$, in 1 -- 10 days. These ionization rates should be sufficient to produce substantial quantities of highly charged ions within a single periastron, at least for star-grazing comets (e.g., Kreutz family; \citealt{Bryans2012}). In the solar system, observing such species is however difficult, because they are rapidly removed by radiation pressure and solar wind. Still, emission from C$^{+}$\ and C$^{2+}$\ was detected in the tails of the near-Sun comets C/2002~X5 (Kudo-Fujikawa) and 96P/Machholz at 0.19 and 0.12 au from the Sun, respectively \citep{Povich2003,Raymond2022}, as well as C$^{+}$, C$^{2+}$, O$^{+}$\ and O$^{2+}$\ in the distant tail of C/1996 B2 (Hyakutake) \citep{Gloeckler2000}, and O$^{3+}$\ in the far tail of C/2006 P1 (McNaught) \citep{Neugebauer2007}. These species are very likely produced by photoionization \citep[see table 3 of][]{Raymond2022}.

Similarly, \bp\ appears to be able to rapidly photoionise singly charged ions, such as Ca$^+$, Fe$^+$ or Si$^+$, owing to its strong chromospheric flux in the 500 - 900 \AA\ region \citep{Wu2025_betapic}. In particular, the lifetime of Al$^+$ at 0.2 au is about 80 days, which is of the same order at the timescale of a cometary periastron. Therefore, photo-ionization could explain the presence of Al$^{2+}$\ in \bp\ exocomets, along with collisional ionization \citep[][see Sect. \ref{Subsubsect. Gas compression}]{Beust1993}.

\begin{figure}[h!]
\centering
    \includegraphics[scale = 0.24,     trim = 75 0 60 60,clip]{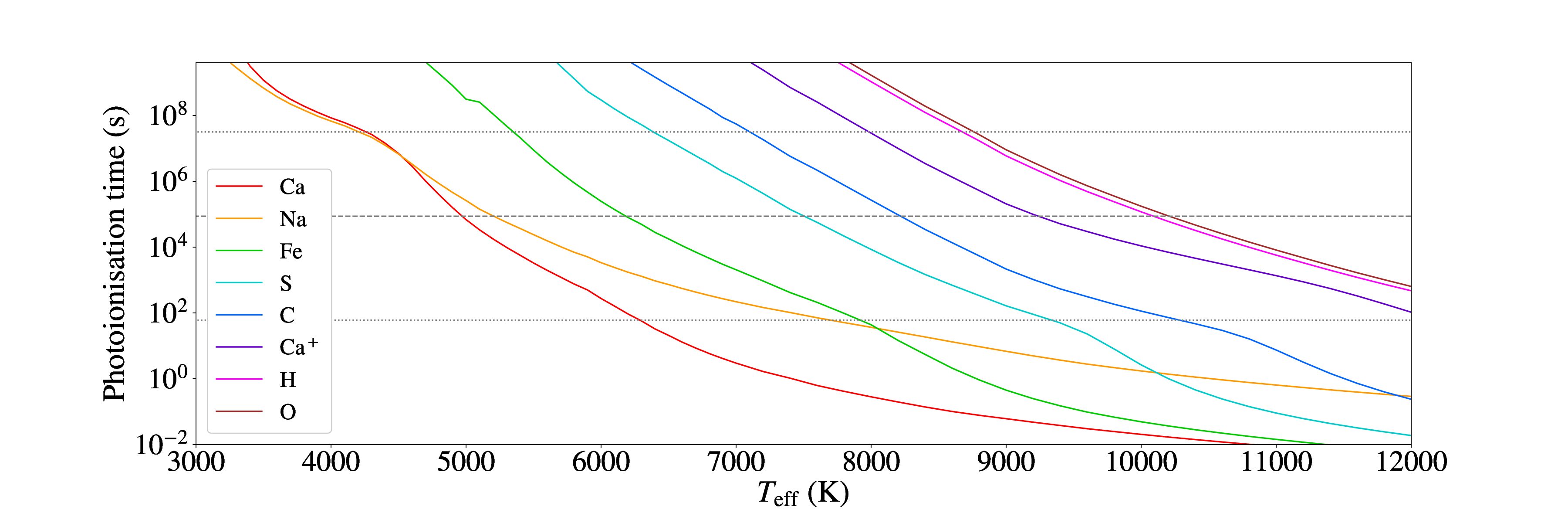}   
    \caption{\small Photoionization timescales of various species at 0.2 au from the host star, depending on the stellar effective temperature. Dotted horizontal lines are identical to Fig.~\ref{Fig. photo ion rate}. As for Fig.~\ref{Fig. dissoc time vs Teff}, purely photospheric fluxes were assumed.}
    \label{Fig. Photo ion time vs Teff}
\end{figure}

Fig. \ref{Fig. Photo ion time vs Teff} also provides the ionization timescales of the studied species as a function of the stellar effective temperature, using photospheric models from the PHOENIX library \citep{PHOENIX}. As for Fig. \ref{Fig. dissoc time vs Teff}, the PHOENIX spectra do not include chromospheric or coronal emission in the EUV/FUV; as a result, the ionization timescales provided here should be considered as upper limits. Yet, we notice that Ca$^+$, routinely detected in \bp\ exocomets thanks to the 390 nm \caii\ doublet, actually gets rapidly photo-ionized into Ca$^{2+}$ for stars hotter than $9\,000 - 10\,000$ K ($\sim$ A1V). This could considerably limit our ability to detect exocomet activity around such stars using optical spectroscopy.

Finally, we note that our calculations rely on the assumption that the atoms and ions are exposed to the unocculted stellar flux. In very young systems, however, discs may be optically thick to UV radiation \citep[mostly due to dust absorption; e.g.][]{Dullemond2010_protoplanetary_discs}, hence significantly suppressing photo-dissociation and photo-ionisation processes in the outer disc regions.

\subsubsection{Collisional ionization}

The different rates of photoionization found for the Sun and \bp\ (Fig. \ref{Fig. photo ion rate}) explain why neutral species, such as Na, Fe and Ni, are commonly detected in Solar System comets \citep{Manfroid2021} or planetary tails \citep{Wurz2022}, whereas in the \bp\ system these elements are primarily found in their ionized form \citep[see the review of][]{Strom2020}. However, the presence of highly ionized species (Al$^{2+}$, Si$^{3+}$, C$^{3+}$, ...) in \bp\ exocomets \citep{Deleuil1993} has long been puzzling, as an A5V star like \bp\ was not expected to emit sufficient EUV photons to produce species like C$^{3+}$. To resolve this problem, it has been proposed that these highly ionized species observed around \bp\ are produced through collisional ionization processes occurring within a dense and hot plasma environment.

Collisional ionization rates of various abundant elements were calculated in \cite{Arnaud1985}. The collisional ionization probability of a given atom or ion within $\mathrm{d}t$ can be written as: 

\begin{equation}
    \mathrm{d}P_{\rm col} = \frac{6.69 \cdot 10^{7}}{(k_B T_e)^{3/2}} \displaystyle \sum_{j} \frac{e^{-x_j}}{x_j} F_j(x_j) n_e \mathrm{d}t
    \label{Eq. DI rate}
\end{equation}
where the sum is performed over all the subshells $j$ of the initial ion, and where $x_j = I_j/(k_B T_e)$ with $I_j$ the energy of the sub-shell $j$. The electronic density $n_e$ and electronic temperature $T_e$ are expressed in cm$^{-3}$ and K, respectively. The function $F_j$ is described in \cite{Arnaud1985}; it depends on four coefficients $A_j$, $B_j$, $C_j$ and $D_j$ specific to the considered sub-shell.

Little literature exists on the typical temperature and density in exocometary tails. Based on the observed population of excited levels of Fe$^+$\ around \bp, \cite{Vrignaud2024b} recently showed that the electronic density in these environments must be below $10^7$ cm$^{-3}$ (see Sect. \ref{Subsect. Excitation}). Geometric estimates, based on the observed column density of Fe$^+$\ and on the typical size of the transiting clouds ($\sim 1$ R$_\star$) suggest electron densities of the order of $10^5 - 10^6$ cm$^{-3}$ \citep{Vrignaud2024b}. 

\begin{figure}[h!]
\centering
    \includegraphics[scale = 0.24,     trim = 75 0 60 60,clip]{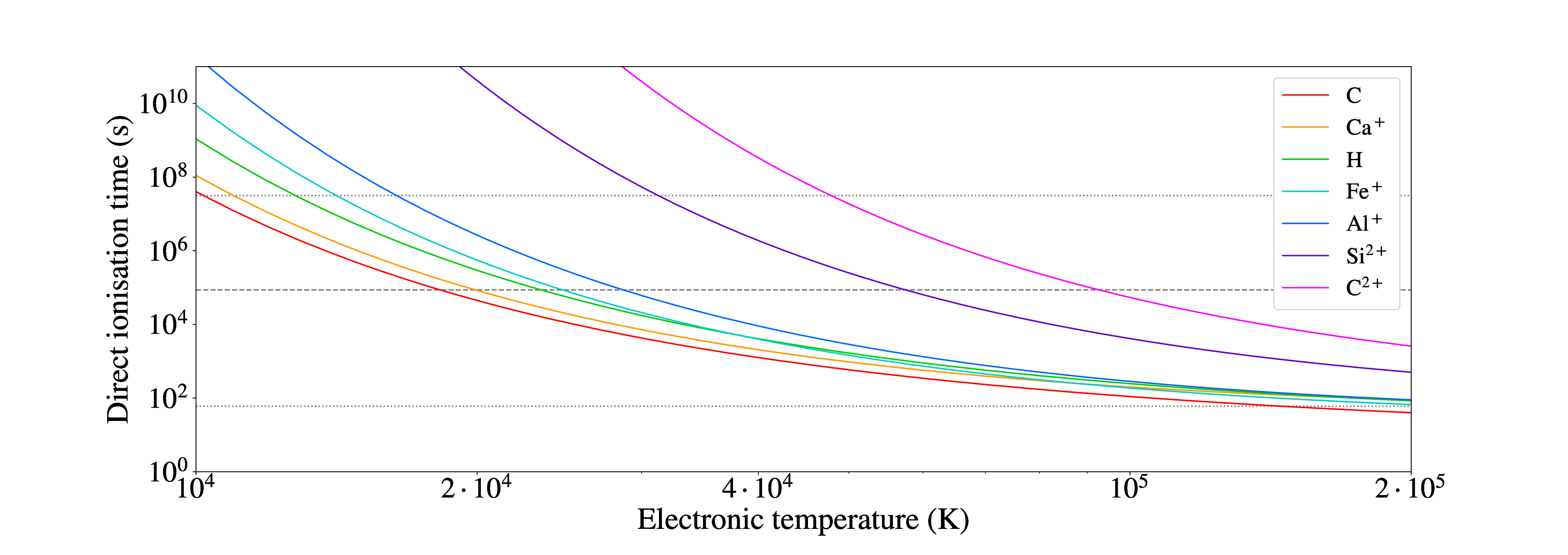}   
    \caption{\small Direct electron-impact ionization timescales of various species as a function of $T_e$, for an electron number density of $10^6$ cm$^{-3}$. Dotted horizontal lines are identical to Fig. \ref{Fig. photo ion rate}.}
    \label{Fig. DI rate vs T_e}
\end{figure}

Assuming an electronic density of $10^6$ cm$^{-3}$, Fig. \ref{Fig. DI rate vs T_e} provides the direct ionization times of a few species, as a function of the electronic temperature. We note that the various `highly ionized' species (Si$^{3+}$, C$^{3+}$, \ldots) observed in \bp\ exocomets are produced at temperatures of $\approx$ 100\,000 K. Such high temperatures could be reached when the cometary gas is compressed by radiation pressure \citep[][see Sect.~\ref{Subsubsect. Gas compression}]{Beust1993}.

\subsubsection{Recombination and ionization equilibrium}

Ions in cometary tails can also recombine with electrons to reproduce neutral or less ionized species. Two main processes exist: radiative recombination, which is fairly homogeneous among ions, and dielectric recombination, a resonant process that can dominate the recombination rate of specific ions. 

The radiative recombination probability of a given ion can be generally expressed as \citep[see][]{Shull1982}:

\begin{equation}
    \mathrm{d}P_{\rm rad} = A \left(\frac{T_e}{10^{4}\,{\rm K}} \right) ^{\eta} n_e \mathrm{d}t,
\end{equation}
with $T_e$ and $n_e$ the electronic temperature and density, and (A, $\eta$) parameters specific to the considered ion. Calculations in the following will be performed using the values of \cite{Shull1982} and \cite{Landini1991}. On the other hand, the dielectric recombination probability can be described by the following formula \citep[see for instance][]{Mazzotta1998}:

\begin{equation}
    \mathrm{d}P_{\rm dielec} = T^{-3/2} \displaystyle \sum_{j = 1}^{4} c_j \exp \left(- \frac{E_j}{k_BT_e} \right)       n_e \mathrm{d}t,
\end{equation}
where the parameters $c_j$ and $E_j$, $j = 1\dots4$, are specific to the studied ion.

\begin{figure}[h]
\centering
    \includegraphics[scale = 0.18,     trim = 10 0 70 30,clip]{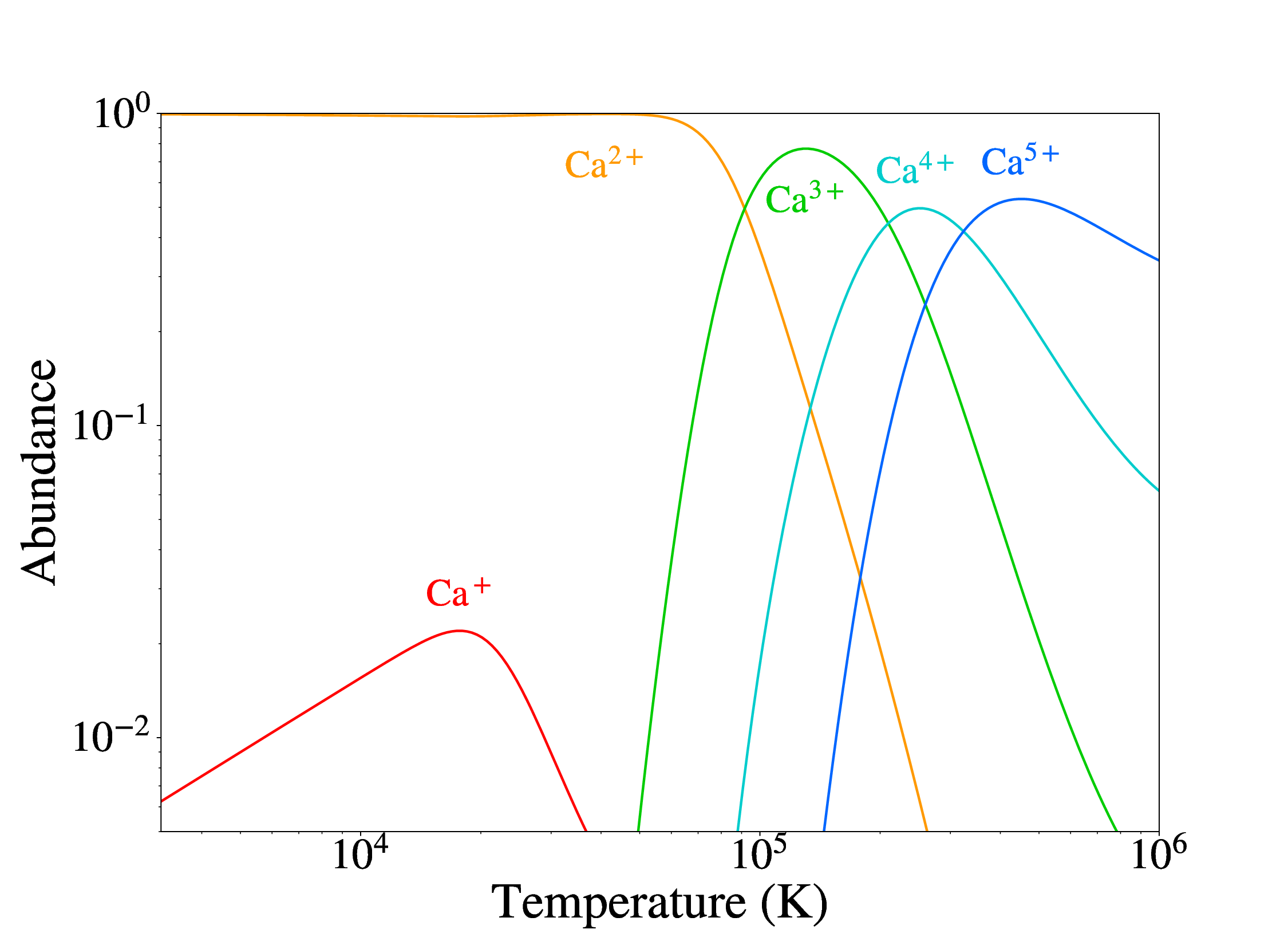}    
    \includegraphics[scale = 0.18,     trim = 15 0 60 30,clip]{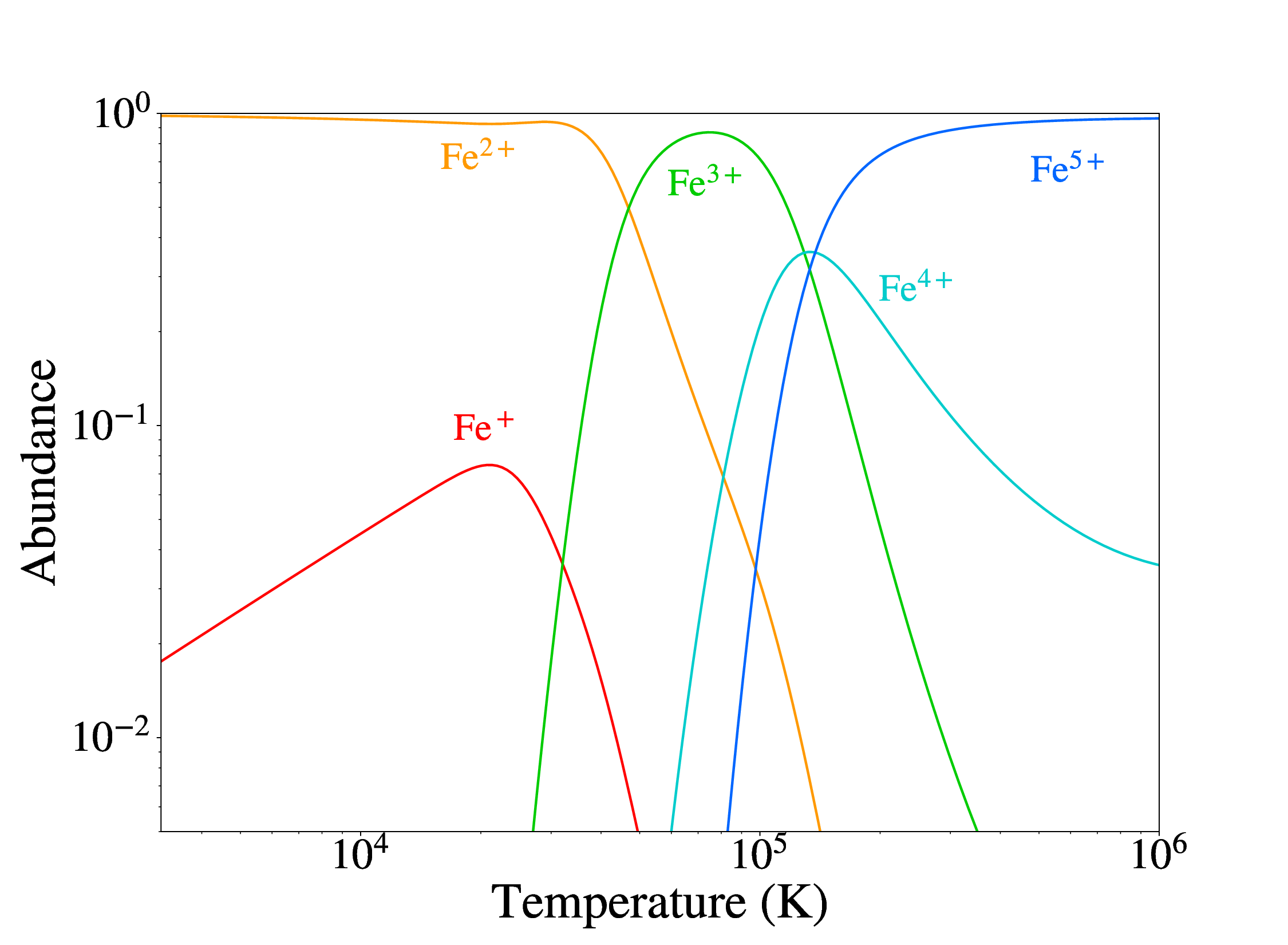}

    \includegraphics[scale = 0.18,     trim = 10 0 70 40,clip]{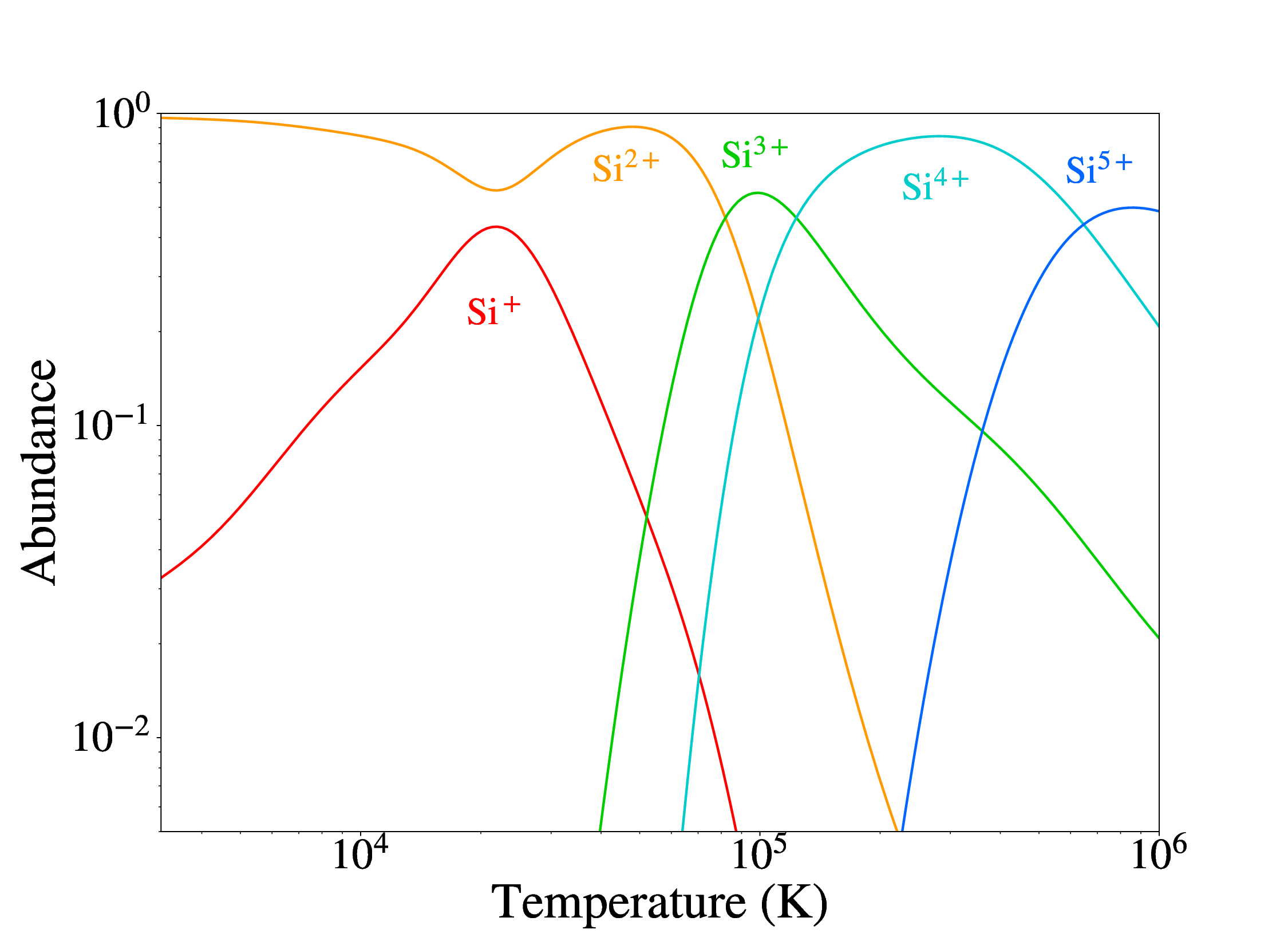}    
    \includegraphics[scale = 0.18,     trim = 15 0 60 40,clip]{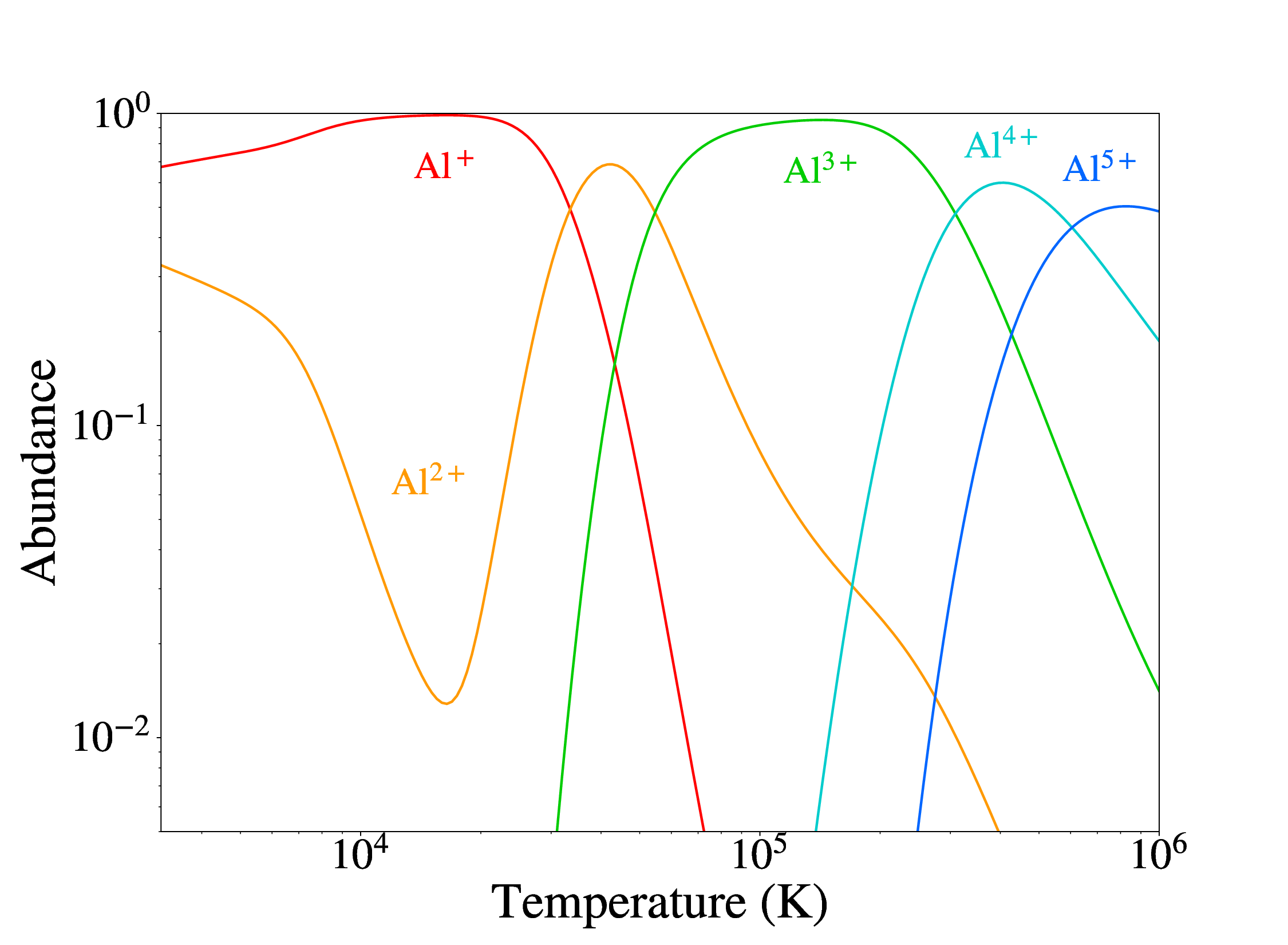}    

    \vspace{0.1 cm}
    
    \caption{\small Ionization equilibrium of a cometary plasma at 0.2 au from \bp, as a function of $T_e$, for different elements. An electronic density $n_e = 10^6$ cm$^{-3}$ was assumed.}
    \label{Fig. Ion equilibrium}
\end{figure}

Using the ionization and recombination processes described above, one can estimate the ionization state of a given element at equilibrium. This ionization state depends on the distance to the host star ($d$), on the electronic density ($n_e$), and on the electronic temperature ($T_e$). For instance, Fig. \ref{Fig. Ion equilibrium} provides the ionization equilibrium of Ca, Fe, Al and Si at 0.2 au from \bp, for temperatures in the range $3\cdot10^3 - 10^6$ K and an electronic density of $10^6$ cm$^{-3}$. Due to photo-ionization, the neutral fraction of these elements is always negligible. At low temperature ($T_e \leq 30\,000$~K), the dominant state can either be the singly-ionized form (e.g. Al$^+$) or the doubly-ionized (Fe$^{2+}$, Ca$^{2+}$, Si$^+$), depending on the photoionization rate of the singly-ionized form. At higher temperatures, strongly ionized species are produced: for instance, Si$^{3+}$\ (commonly observed in \bp\ exocomets) is associated with temperatures of $\sim10^5$~K. At such temperatures, the production of many other highly ionized species -- Ca$^{2+}$, Ca$^{3+}$, Fe$^{2+}$, Fe$^{3+}$, Al$^{3+}$, \ldots -- is expected. However, these species cannot be directly probed in \bp\ exocomets, due to the absence of optically-allowed transition at wavelengths accessible with the Hubble Space Telescope.

Note that the values shown in Fig. \ref{Fig. Ion equilibrium} should be regarded as rough estimates. In particular, the EUV spectrum of \bp\ -- which drives the photoionization of most atoms and ions -- was taken from \cite{Wu2025_betapic}. Their reconstruction relies on a simplified chromospheric model, constrained by observations in \ciii\ and \ovi\ emission lines with FUSE. The reliability of this spectrum, and therefore of our calculations, remains difficult to assess.

\FloatBarrier

\subsection{Excitation}
\label{Subsect. Excitation}

\begin{figure}[h!]
\centering
    \includegraphics[scale = 0.45,     trim = 20 0  20 10, clip]{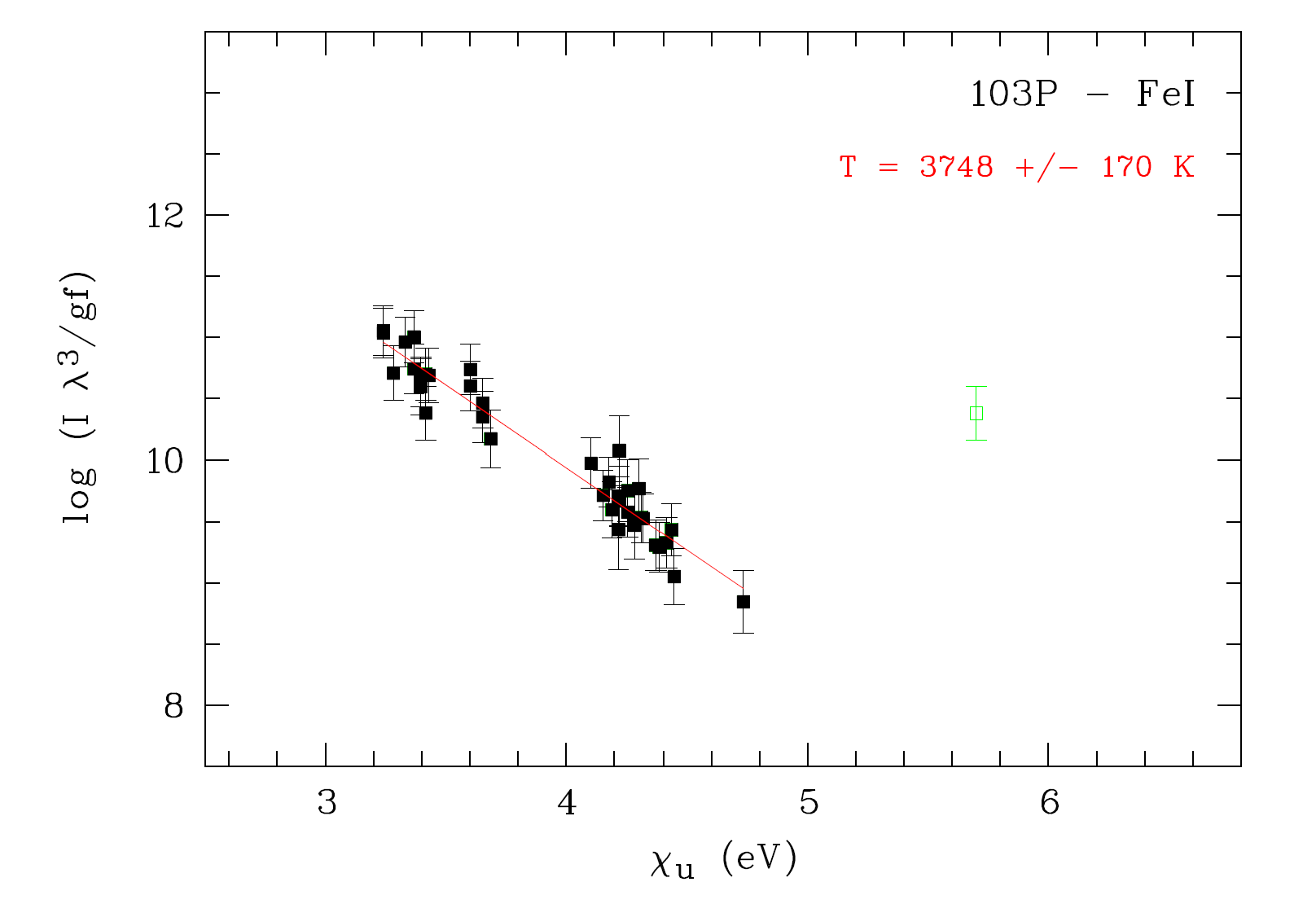}   
    \includegraphics[scale = 0.43,     trim = 45 45 40 50, clip]{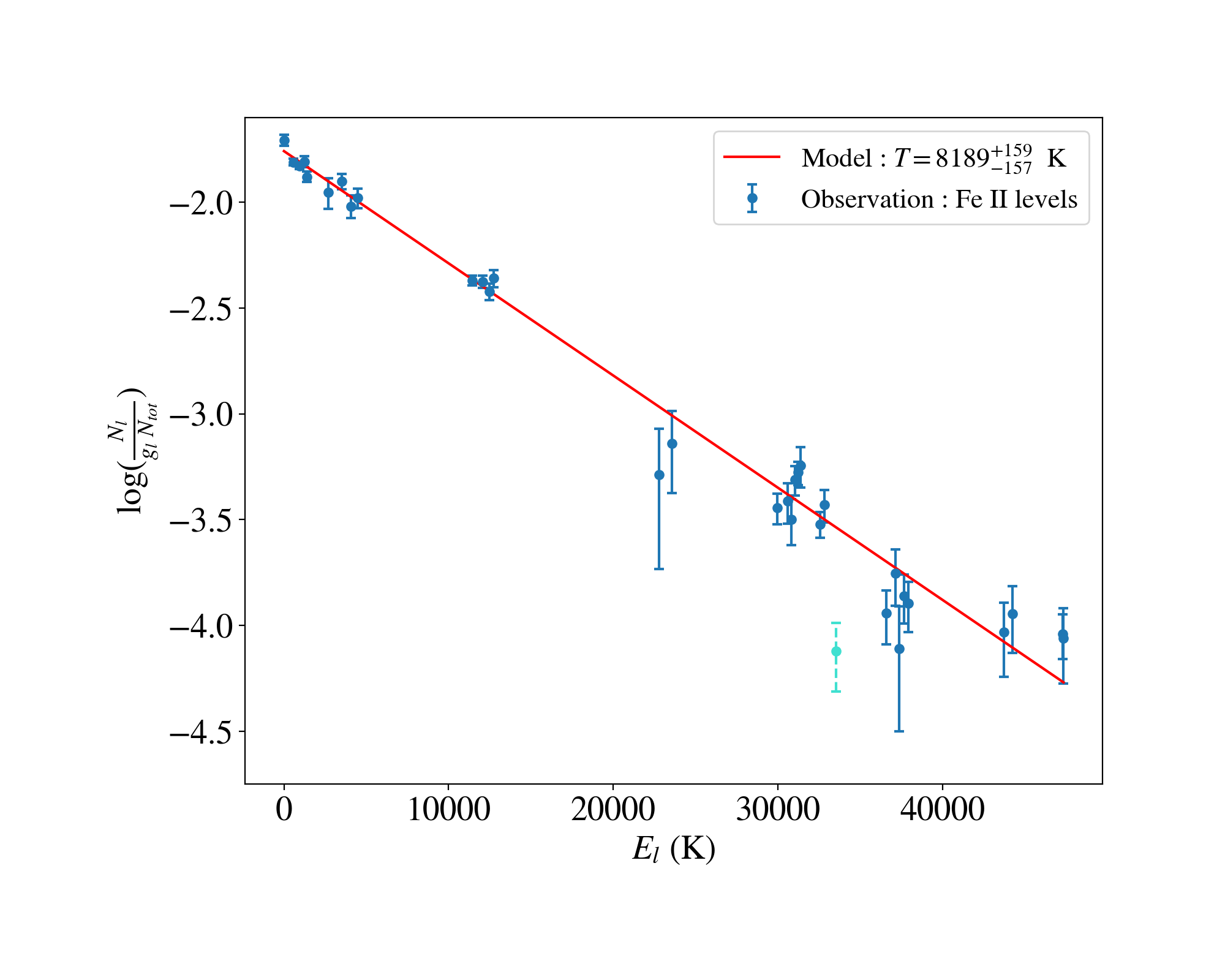}   
    \caption{\small Comparison of excitation diagrams of Fe and Fe$^+$ in emission spectra of comets and absorption spectra of \bp\ exocomets. For both diagrams, the y-axis shows the level abundance, and the x-axis shows the level energy. The slopes (red lines) are inversely proportional to the excitation temperature. \textbf{Left} : Excitation diagram of atomic Fe in 103P/Hartley 2, obtained through the analysis of the comet's emission lines \citep[taken from][]{Manfroid2021}. \textbf{Right} : Excitation diagram of Fe$^+$ in a \bp\ exocomet observed in 1997, derived from the exocomet's absorption signatures in \bp\ spectrum \citep[taken from][]{Vrignaud2024b}. }
    \label{Fig. excitation diagrams}
\end{figure}

The various molecules, atoms, and ions that make up cometary tails can get excited. The excitation state of those particles can then be probed, using either emission lines (when observing Solar System comets) or absorption lines (when studying transiting exocomets). Fig. \ref{Fig. excitation diagrams} provides two examples of excitation diagrams of species released by comets: the excitation diagram of atomic Fe in a solar comet (left), and the excitation diagram of Fe$^+$ in a \bp\ exocomet (right).

The observed excitation diagram can be linked to the local properties of the gas. Indeed, there are two main ways of exciting a particle: by collisions (e.g. with electrons) and by photo-absorption. Collisional excitation is effective at sufficiently high densities, and tends to bring the excitation temperature of the gas close to the kinetic temperature. On the other hand, photo-excitation becomes effective at close comet-to-star distance, and tends to align the excitation temperature of the gas with the black body temperature of the radiation field. 

In Solar System comets, the density of the outer coma is generally low enough  that photo-absorption dominates the excitation of the gas \citep{Bodewits2022}. As a result, the excitation temperatures of atomic Ni and Fe are generally found to be close to the solar effective temperature \citep[see also Fig. \ref{Fig. excitation diagrams}]{Manfroid2021}. However, in the case of Solar comets, the gas excitation is strongly affected by the gas radial velocity relative to the Sun. Comets with heliocentric radial velocities near 0 km~s$^{-1}$ generally show lower excitation temperatures and weaker emission lines, because the deep \fei\ and \nii\ absorption lines in the solar spectrum reduce the flux available to excite the main transitions of Fe and Ni \citep[Swings effect; see also][]{Swings1941,McKellar1944_swings, Furusho_2005_Na1_Swings}. On the other hand, for redshifted or blueshifted comets the flux received in \fei\ and \nii\ lines is higher, and the gas therefore tends to be more excited and to have stronger emission lines. 

A similar behavior is found in \bp\ exocomets \citep{Vrignaud2024b, Vrignaud2025}: the excitation temperature is generally found to be close to 8000 K (see Fig. \ref{Fig. excitation diagrams}), which corresponds to the stellar effective temperature. In this case, the Swings effect is not important, because the stellar spectrum is smoothed by the rapid rotation of the star (130 km~s$^{-1}$). 

The predictability of the excitation state in cometary atmospheres is very useful for estimating the total abundances of atoms and ions, as the population of all the excited levels of a given particle can then be inferred from measurements in a small subset of levels. As an example, \cite{Manfroid2021} analyzed the emission spectra of a large set of Solar System comets to measure the abundance of Fe and Ni in their atmospheres, and deduce their Ni/Fe ratios. For comets at $\geq 1$ au, this ratio is generally found to be super-solar, indicating that the nickel and iron atoms are not produced by direct sublimation of minerals, but instead by the dissociation of nickel carbonyls (e.g., Ni(CO)$_4$) and iron carbonyls (e.g., Fe(CO)$_5$), or by releasing Ni and Fe connected to PAHs \citep{Manfroid2021, Bromley2021}. By contrast, refractory ions (e.g. Fe$^+$, Ni$^+$, Si$^+$...) are generally found in solar abundances in \bp\ exocomets \citep{Vrignaud2025}, hinting that the refractories are produced from the complete sublimation of dust grains with solar composition. 

Finally, it should be emphasized that probing the excitation state of a comet relies on different approaches whether the comet is observed in our own Solar System (using emission spectroscopy) or around another star (using absorption spectroscopy). In the first case, only the abundance of excited, unstable levels (with very short lifetimes and thus strong emissivity) can be directly probed, allowing to build the excitation diagram of the gas only above a few eV (see for instance the left panel of Fig. \ref{Fig. excitation diagrams}). Absorption spectroscopy, however, is sensitive to the abundance of little excited, metastable states with long radiative lifetimes. In this case, only the low-energy part of the excitation diagram can be recovered (see the right panel of Fig. \ref{Fig. excitation diagrams}), which has the advantage of directly probing the column densities of the most populated levels.

\begin{figure}[h!]
\centering
    \includegraphics[scale = 0.38]{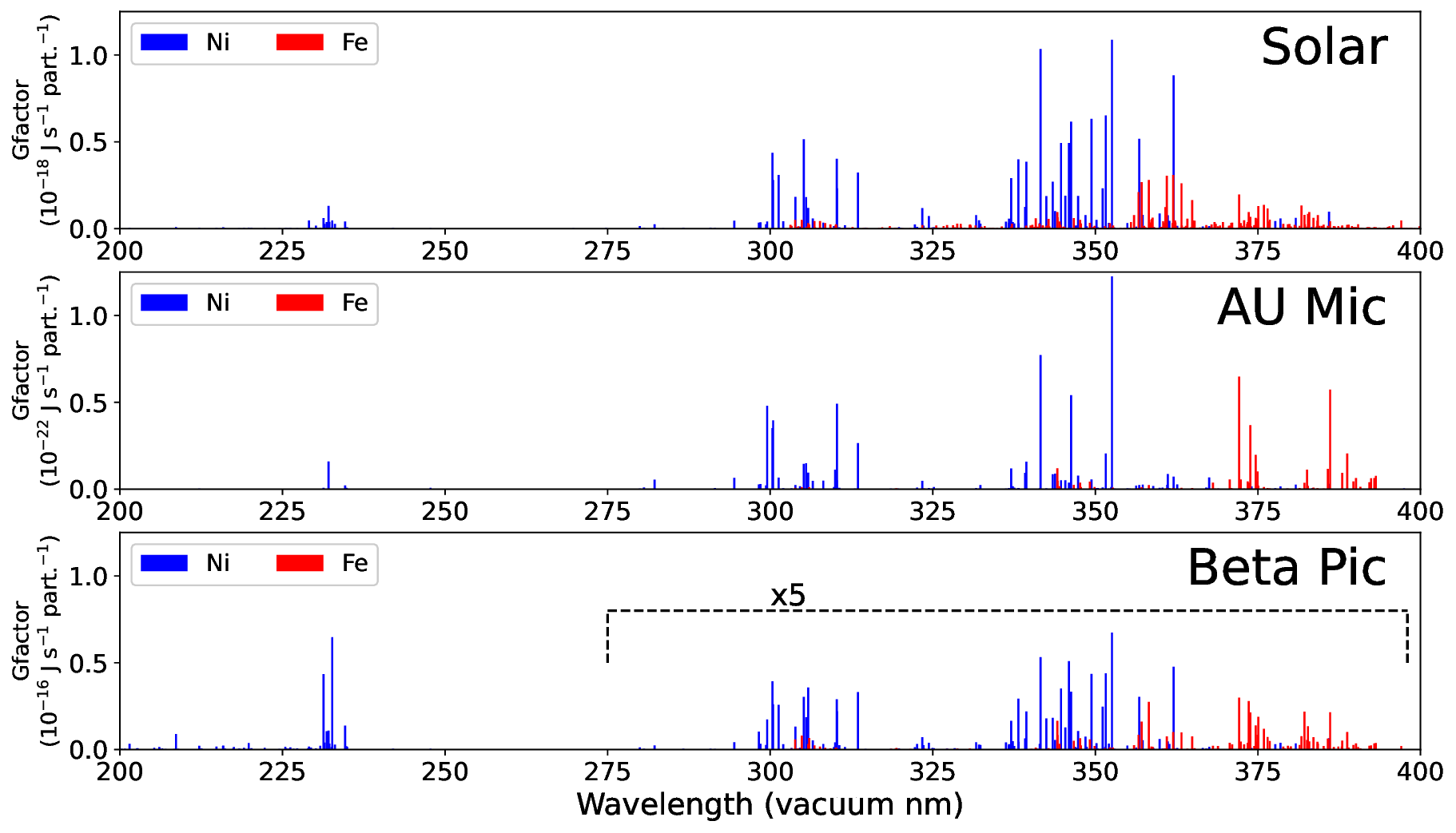}   
    \caption{\small Simulated photofluorescence emission spectra of Fe (red) and Ni (blue) for excitation by a solar spectrum (top), AU~Mic (middle), and \bp\ (bottom). Fluorescence model by \citet{Bromley2021}.}
    \label{Fig. FeNi}
\end{figure}

Detecting emission lines from exocomets appears to be difficult. To investigate what the spectra of neutral and ionized iron and nickel would look like around a given star, we used the photo-fluorescence model by \cite{Bromley2021}. Results for AU~Mic, the Sun and \bp\ are provided in Fig. \ref{Fig. FeNi} (using again synthetic spectra from the PHOENIX library, and assuming a stellocentric distance of 0.1 au). For \feii\ and \bp, fluxes per particles as high as $10^{-18}$ J/s can be expected. Multiplying this value by the typical number of Fe$^+$ ions in a \bp\ exocomet \citep[$\sim 10^{37}$][]{Vrignaud2024b} and assuming that the gas is observed from a distance of 19 pc (Sun - \bp\ distance), this translates into maximum intensities of $10^{-14}$ erg/s/cm$^2$/\AA, or about $10^{-4}$ of the \bp\ continuum in the Near-UV. This value is much below the typical noise level (1 \%) obtained with the STIS spectrograph (\emph{Space Telescope Imaging Spectrograph}, onboard the HST). In addition, emission features from exocomets close to the star would very likely be mixed with absorption features of exocomets transiting the star, making it difficult to distinguish between the two. 

\begin{figure}[h!]
\centering
    \includegraphics[scale = 0.38]{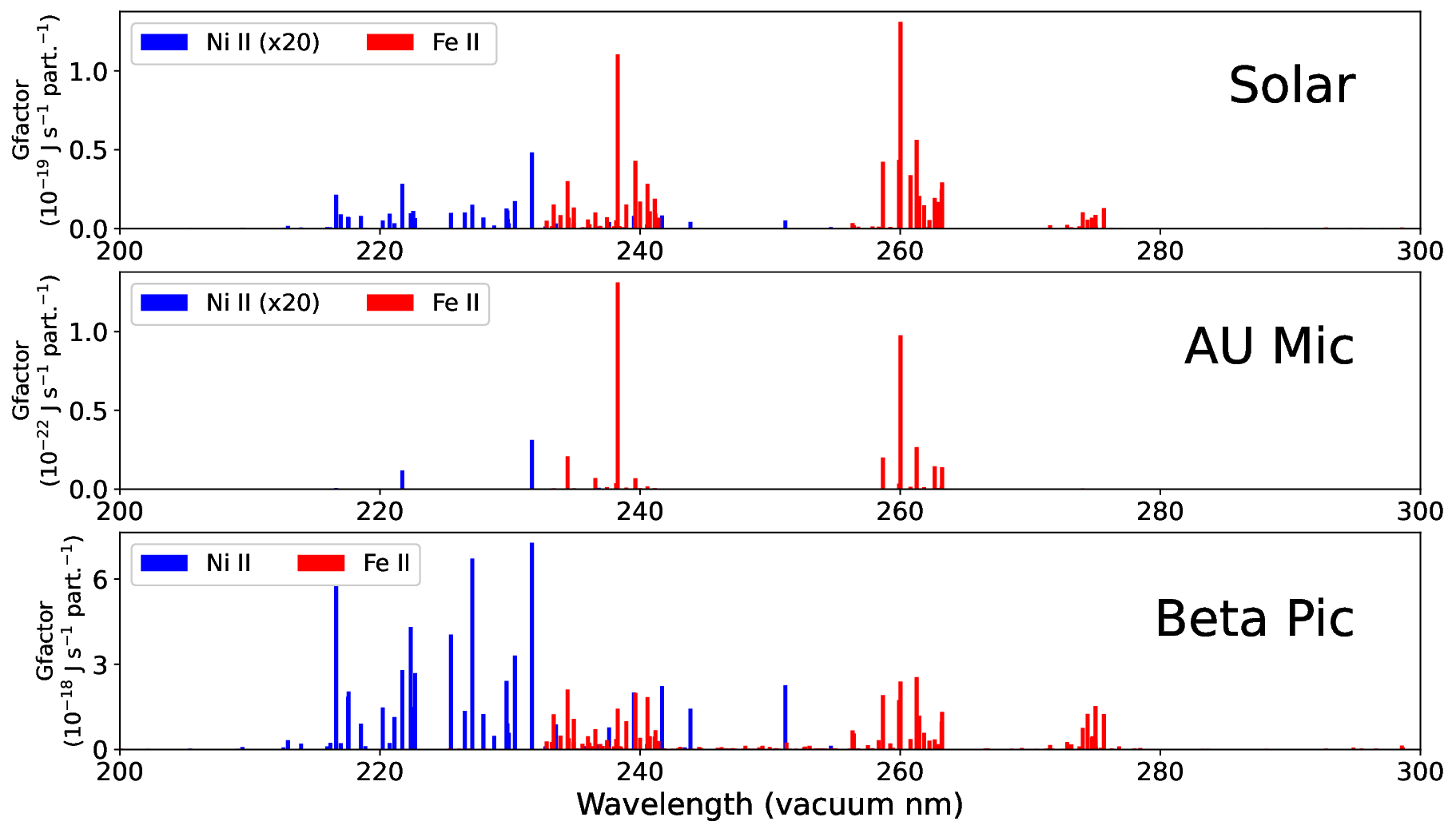}   
    \caption{\small Same as Fig. \ref{Fig. FeNi} for \feii\ and \niii. The line intensities of \niii\ were multiplied by 20 around the Sun and AU~Mic. Fluorescence model by \citet{Bromley2021}.}
    \label{Fig. Fe+Ni+}
\end{figure}


\section{Particle dynamics}

Gas and dust particles in cometary atmosphere and tails are subject to various forces that influence their trajectories, in addition to the gravitational attraction from the star. This section reviews the main mechanisms that affect the dynamics of dust (Sect. \ref{sec: Dust grain dynamics}) and gas (Sect. \ref{Sect. Gas dynamics}) released by exocomets, and how these mechanisms depend on the stellar environment.

\subsection{Dust grain dynamics}
\label{sec: Dust grain dynamics}

\subsubsection{Radiation pressure}
Once liberated from the comet nucleus's gravity, and decoupled from the cometary gas, radiation pressure and gravity are the two main forces acting on dust grains in the tails of Solar System comets. The two forces are given by:

\begin{equation}
\label{eq: rad pressure}
    F_r = \frac{Q_{pr}}{c}\left(\frac{L_{\odot}}{4 \pi r^2}\right) \pi R^2 ,
\end{equation}
and
\begin{equation}
\label{eq: gravity - dust}
    F_g = \frac{4}{3} \frac{GM_{\odot}}{r^2}\pi R^3 \rho ,
\end{equation}
where $Q_{pr}$ is the radiation pressure scattering efficiency factor, $r$ the heliocentric distance, $R$ the dust grain radius, $\rho$ the grain density, $L_{\odot}$ the solar luminosity, $M_{\odot}$ the solar mass, and $G$ the gravitational constant \citep{finson_theory_1968}.

The ratio of radiation pressure to gravitational forces \citep[$1 - \mu = F_r/F_g$, also often denoted $\beta$;][]{finson_theory_1968} determines whether a dust grain will remain on a bound orbit ($1 - \mu = \beta < 1$) or is swept out onto hyperbolic trajectories ($\beta > 1$). Since $F_r \propto R^2$ and $F_g \propto R^3$ for solid spherical grains, the $\beta$ value of a dust grain scales as $R^{-1}$, meaning that smaller grains are more easily pushed  away anti-sunward by radiation pressure than larger grains. For typical Solar System comets with visible dust tails, the radiation pressure forces acting on the grains in the tail are usually a few multiples larger than the magnitude of gravitational forces. To provide the reader with some context, the iconic comets C/2011 W3 (Lovejoy), C/2006 P1 (McNaught), and C/2014 Q1 (Pan-STARRS), which all entered the Near-Sun comet regime \citep[$q < 0.307$ au;][]{Jones2018}, released dust grains with $\beta$ values ranging from $\beta = [0.6, 2.5]$ \citep{sekanina_comet_2012}, $\beta = [0.2, 3.0]$ \citep{price_fine-scale_2019}, and $\beta \leq 0.2$ \citep{afghan_observations_2023} respectively. To estimate particle sizes, one can assume a density of $0.50$~g~cm$^{-3}$ and take the reciprocal of the beta values \citep{burns_radiation_1979}. For C/2011 W3 (Lovejoy), C/2006 P1 (McNaught), and C/2014 Q1 (PanSTARRS), the observed $\beta$ values correspond to estimated grain sizes $R = [0.4, 1.7]$ \textmu~m, $R = [0.3 , 5.0]$ \textmu~m, and $R > 5.0$ \textmu~m, respectively.

Whilst comets Lovejoy and McNaught produced visually spectacular dust tails with interesting features, comet C/1995 O1 (Hale-Bopp) could be a closer analogy to known exocomets, at least in terms of dust production. Comet Hale-Bopp was one of the most studied comets of the twentieth century and was easily visible with the naked eye, but unlike comets Lovejoy, McNaught, and PanSTARRS, Hale-Bopp's perihelion occurred much further out at $q \approx 0.9$ au. Analysis of Hale-Bopp images when it was at heliocentric distances between 1.02 au and 1.28 au concluded that dust grain sizes ranged from 0.3 \textmu~m to 8 \textmu~m \citep{kharchuk_model_2009}. This corresponds to an upper limit of $\beta \approx 3$ which is comparable to the upper $\beta$ values of comet McNaught. Furthermore, Hale-Bopp continued to remain active at greater distances and changes in the dust tail were observable for three years \citep{kramer_dynamical_2014}. Analysis by \cite{kramer_dynamical_2014} of Hale-Bopp beyond 20 au concluded that the observable dust tail consisted of grains with a typical $\beta$ value of 0.3. Due to its impressive dust production, Hale-Bopp has been used in exocomet studies to model transits in photometric light curves. \cite{lukyanyk_numerical_2024} showed that a comet similar to Hale-Bopp could be used to interpret a strong photometric transit observed in \bp\ light curve \citep{zieba2019}, assuming a transit distance of $\sim$~5~au and grains larger than 5 \textmu~m (see also Section~\ref{Sect. Solar System comets}).


The dependence of $\beta$ on radiation pressure and gravitational forces means that this parameter will vary with stellar luminosity and mass. Assuming that physical grain properties, such as size and density, remain unchanged, the $\beta$ value of a dust grain should scale as: 
\begin{equation}
    \frac{\beta_*}{\beta_\odot} = \frac{L_*}{L_\odot} \left( \frac{M_*}{M_\odot} \right)^{-1},
    \label{eq: beta scaling}
\end{equation}
where $\beta_\star$ is the $\beta$ factor of the grain around a given star, and $L_\star$ and $M_\star$ are the stellar luminosity and mass, respectively. Applying this equation to our case study stars, we find $\beta_\star/\beta_\odot$ = 0.17 for AU~Mic, 4.97 for \bp, and 0.02 for WD 1145+017. Thus, the $5$ \textmu~m grains ($\beta_\odot \approx 0.3$) used by \cite{lukyanyk_numerical_2024} to model $\upbeta$~Pic exocomets would possess typical $\beta$ values of $4.97 \times 0.3 \approx 1.5$ in the \bp\ system.


%

It should also be noted that, at small astrocentric distances, a dust grain is no longer exposed to the full area of the stellar disk: stellar photons impact the dust grains with a range of incident angles. Consequently, the typical $\beta$ calculation is no longer suitable for dust grains close to a star, as equation (\ref{eq: rad pressure}) assumes the star behaves as a point source for all heliocentric distances. Thus, an empirical geometric correction accounting for the deviation of radiation pressure forces from the inverse square law and the reduction in visible area of the disk is required. Such correction can be parameterized as:
\begin{equation}
    \log_{10}\left(\Gamma \right) =   \displaystyle \sum_{i = 0}^{4} a_i x^i,
    \label{Eq. Gamma}
\end{equation}
where $a_0 = -0.353$, $a_1 = 0.656$, $a_2 = -0.477$, $a_3 = 0.159$, $a_4 = -0.020$ are empirically derived, and $x = \log_{10}\left(r_h [R_*]\right)$ \citep{hanlon_empirical_2025}. This correction becomes necessary within $\sim 200$ $R_*$ ($\sim$ 1 au in the Solar System). It is particularly useful when modeling the dynamics of dust grains at very small stellar-centric distances, as the $\Gamma$ parameter can be multiplied by the standard $\beta$ factor to produce an effective $\beta$ value, given by $ \beta_\mathrm{eff} = \Gamma \beta$. For distances ranging from 6$R_*$ to 215$R_*$, $\Gamma$ varies from 0.86 to 1.00 \citep{hanlon_empirical_2025}. 

\begin{figure}
    \centering
    \includegraphics[width=0.9\linewidth]{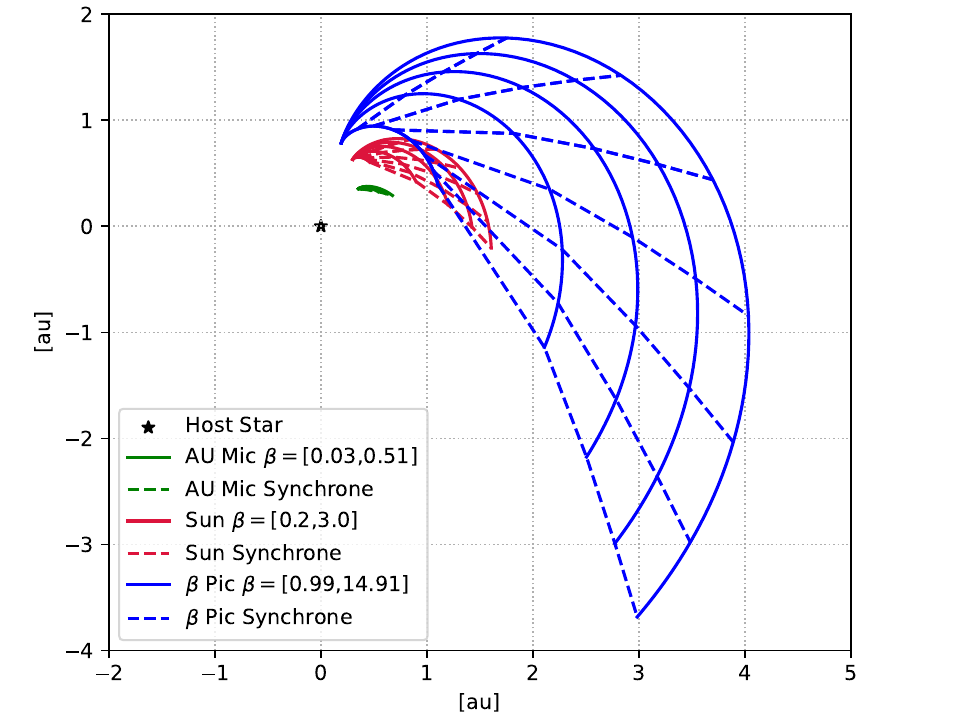}
    \caption{\small The simulated shape of an exocomet dust tail viewed above the plane of the orbit in the AU~Mic, Solar System, and $\upbeta$~Pic systems. The solid lines connect grains with the same $\beta$ values and the dashed lines connect grains of constant age. In each scenario the comet was simulated for 28 days and passed through a periastron of 0.50 au on day 14.}
    \label{fig: overlaid comets}
\end{figure}

Modeling of Solar System cometary dust tails often implements equations \ref{eq: rad pressure} and \ref{eq: gravity - dust}. As radiation pressure forces and gravity operate in opposite directions, the acceleration of a dust grain is given by:
\begin{equation}
\label{eq: grain accl}
    a = \frac{GM_*}{r^2}\left(1-\beta_\mathrm{eff} \right).
\end{equation}
Such an approach can provide an insight into where grains with different $\beta$ values are found within the dust tail. In such cases, lines connecting grains of equal $\beta$ are known as syndynes and lines connecting grains of the same age are referred to as synchrones \citep{finson_theory_1968}. Figure \ref{fig: overlaid comets} displays the simulated dust tail of a comet with the same grain properties as C/2006 P1 (McNaught) orbiting AU~Mic, the Sun, and $\upbeta$ Pic respectively. The tails were simulated by numerically modelling grain dynamics via the RK4 method with grains being assigned the Solar System $\beta$ values of Comet McNaught. For AU~Mic and $\upbeta$~Pic these were scaled using equation (\ref{eq: beta scaling}).  We note that the tail length and width increases rapidly with the stellar size and luminosity. These differences in shape arise from dust grains having a variety of starting conditions when ejected from the nucleus. In general, the orbital velocity of cometary nuclei increases with stellar mass, which in turn increases the initial velocity of an ejected dust grain. Coupled with variations in radiation pressure (which increases with stellar luminosity), this means that each star imparts a unique set of motion vectors to dust grains ejected from cometary nuclei. 
Around low-mass and low luminosity stars such as AU~Mic, orbital velocity and radiation pressure are significantly reduced compared to an exocomet at the same distance in the $\beta$ Pic system. On the other hand, dust grains around more massive and luminous stars have larger initial velocities and experience greater anti-sunward acceleration.  

Here, it should be noted that this form of analysis does not account for the dust production of the comet; a complete study of an exocometary dust tail would need to consider the dust production and distribution, as discussed in Sect. \ref{sec: dust escape}.

\subsubsection{Rocket Effect}
The sublimation of ices at the surface of a dust grain can produce a noticeable rocket effect that accelerates dust particles, due to the anisotropy of the vapor ejection (which is favorably released on the sunlit side of the grain). For spherical grains, the resulting acceleration is given by
\begin{equation}
    \label{eq: rocket effect}
    a_{\rm rocket} = \frac{3 \mu m_H Z v_{\rm th} f_{\rm ice}}{4 \rho_p a} = 1.1 \times 10^{-18}\frac{Z}{\rho_p a}, 
\end{equation}
where \textmu~ is the molecular weight of the sublimating ice, $m_H$ is the mass of hydrogen, $Z$ is the sublimation rate of the individual particle (molecules cm$^{-2}$\,s$^{-1}$), $v_{\rm th}$ is the mean thermal expansion speed of the gas, $f_{\rm ice}$ is the ice fraction of the particle, $\rho_p$ is the particle density, and $a$ is the particle radius \citep{kelley_distribution_2013}. All parameters are expressed in cgs units.

Similarly to how the magnitude of radiation pressure and gravitational forces are compared using a dimensionless ratio, $\alpha$ is used to compare the relative magnitude of rocket effect forces to gravitational forces such that \citep{reach_distribution_2009,kelley_distribution_2013}
\begin{equation}
    \label{eq: rocket:grav ratio}
    \alpha = 2.238 \times10^{26} \frac{3 \mu m_H Z\nu_{\rm th} f_{\rm ice}}{4 GM_\odot \rho_p a},
\end{equation}
where parameters are again expressed in cgs units. For a particle with a radius of 10 cm and assuming that only water ice is ejected, \cite{kelley_distribution_2013} found that $\alpha$ can be as high as 0.27, which is many orders of magnitude greater than the radiation-driven $\beta$ value of the dust particle ($\beta \approx 1 / \left(1\times 10^4 [\mu \text{m}]\right) = 1 \times 10^{-4}$).

\subsubsection{Charged grain motion} 
The Lorentz force is another force which can alter the structure of a cometary dust tail. In the Solar System, studies show correlations between the realignment of dust tail features and magnetic structures found in the solar wind \citep{price_fine-scale_2019,price_fine-scale_2023}. 

Since the Debye length (distance over which electrostatic effects persist) of a dust grain is significantly smaller than the typical distance between grains, charged dust grains can be considered as individual and isolated charges. As such, the Lorentz force acting on a dust grain is given by:
\begin{equation}
\label{eq:Lorentz force general}
    \vec{F}_L = q\left(\vec{E} + \vec{v_d} \times \vec{B}\right)
\end{equation}
where $v_d$ and $q$ are grain velocity and charge and $\vec{E}$, and $\vec{B}$ are the electric magnetic fields. The electric field `felt' by a dust grain is dependent on the velocity of the solar wind ($\vec{v}_\mathrm{sw}$) and its magnetic field such that $\vec{E} = -\vec{v}_\mathrm{sw} \times \vec{B}$. Substituting this into equation (\ref{eq:Lorentz force general}) leads to:
\begin{equation}
    \vec{F}_\mathrm{L} = q \left(-\vec{v}_\mathrm{sw} \times \vec{B} + \vec{v}_\mathrm{d} \times\vec{B} \right) = q \left(\vec{v}_\mathrm{rel} \times \vec{B} \right)
\end{equation}
where $\vec{v}_\mathrm{rel} = \vec{v}_\mathrm{d} - \vec{v}_\mathrm{sw}$. Further assuming $\vec{v}_\mathrm{sw} >> \vec{v_\mathrm{d}}$, the relative velocity term becomes $\vec{v_\mathrm{rel}} \approx - \vec{v}_\mathrm{sw}$, resulting in $\vec{F}_\mathrm{L} \approx -q\left(\vec{v}_\mathrm{sw} \times \vec{B}\right)$. 

From a Solar System perspective, Lorentz forces are associated with lofting (see section \ref{sec: dust escape}) and the realignment of fine-scale structures in the dust tails of highly active Solar System comets; these events being associated with Heliospheric Current Sheet (HCS) crossings \citep{price_fine-scale_2019,price_fine-scale_2023} where the polarity of the Interplanetary Magnetic Field (IMF) abruptly changes. Therefore, unlike plasma tails, the impact of the heliospheric magnetic field on the large-scale structure of a dust tail is relatively modest, having only been observed in few cases \citep{kramer_dynamical_2014,price_fine-scale_2019,price_fine-scale_2023}. In the AU~Mic system, however, the magnetic  environment is much stronger, with an average intensity of $550 \pm 30$ $G$  \citep[about 275 times more than the Sun]{donati_magnetic_2023}. Even though it is difficult to fully comment on the electrostatically-driven motion of exocometary dust grains in this system, the dusty tails of hypothetical AU~Mic exocomets could be severely disrupted from the interaction with their star's magnetic field. 

Regarding the \bp\ system, the stellar magnetic field is poorly constrained \citep[][proposes an upper limit field strength of 300 G, or $\sim 150 \times$ the solar value]{zwintz_revisiting_2019}. In this case, interactions with magnetic fields are usually not taken into account when modeling the light curves of exocomets. Finally, about 20 \% of white dwarfs have detectable magnetic fields \citep{Ferrario2015_white_dwarfs_magnetic_fields}. As a result, most studies about polluted white dwarfs don't include magnetic fields. However, it has been suggested that the lifetime of dusty debris discs around white dwarfs could be significantly reduced if the star is magnetized \citep[especially when $B>10$\,kG; see][]{Hogg2021_WD_magnetic_fields}, due to an increase of the accretion rate of the grains. 



\subsubsection{Poynting-Robertson drag}

The Poynting-Robertson (PR) effect arises when incident photons gradually decrease the angular momentum of a dust grain, causing it to spiral onto the surface of the host star \citep[e.g.][]{Ragot2003,klacka_poyntingrobertson_2014,klacka_times_2008}. The acceleration of a dust grain caused by radiation pressure and the PR effect is given by:
\begin{equation}
       \vec{a} = \Gamma \beta \frac{GM}{r^2}\left[ \left(1 - \frac{\vec{v}\cdot \vec{e}}{c}\right)\vec{e} - \frac{\vec{v}}{c}\right].
\end{equation}
where $r$, $\vec{v}$, and $\vec{e}$ are the distance from the Sun, the particle velocity, and the particle positional unit vector respectively \citep{klacka_poyntingrobertson_2014}, $\beta$ is the standard radiation pressure factor, and $\Gamma \leq 1$ is the correction factor given in Eq. \ref{Eq. Gamma}.

This form of the acceleration can be used to derive the secular, or time-averaged, change of orbital elements of dust grains. The result is a pair of coupled ordinary differential equations for the dust grain's semimajor axis, $a$ and its eccentricity, $e$ \citep{wyatt1950}:

\begin{equation}
\frac{\mathrm{d}a}{\mathrm{d}t} \approx -\left(\frac{1}{c^2}\right)
\frac{3\left(2+3e^2\right)L_{\star}}{16\pi R \rho a \left(1-e^2\right)^{3/2}}
\end{equation}

\begin{equation}
\frac{\mathrm{d}e}{\mathrm{d}t} \approx -\left(\frac{1}{c^2}\right)
\frac{15 e L_{\star}}{32\pi R \rho a^2 \sqrt{1-e^2}},
\end{equation}

\vspace{0.2 cm}

\noindent where $L_{\star}$ is the luminosity of the star, $c$ is the speed of light, and again $R$ and $\rho$ are the radius and density of the dust grain. The equations are exact when $\Gamma = 1$. 

Poynting-Robertson drag has a minimal effect on Solar System comets, as the timescale for a grain to spiral inward toward the Sun is significantly longer than the rapid motion changes induced by radiation pressure or solar wind interactions. However, this effect becomes more important when considering stars such as white dwarfs, which can be windless systems with minimal radiation pressure \citep[e.g.,][]{rafikov2011,malamud2021,veras2022,okuya2023}. In this case, dust grains will be slowly accreted towards the star, ultimately leading them to contaminate the white dwarf's photosphere.

\subsection{Gas/Ion dynamics}
\label{Sect. Gas dynamics}

We now investigate the various processes that drive the dynamics of gaseous particles (molecules, atoms, or ions) in exocometary tails. 

\subsubsection{Radiation pressure}
\label{Sect. radiation pressure}

Like dust grains, gas particles experience radiation pressure, which push them in the anti-stellar direction as they absorb photons. This pressure depends both on the stellar spectrum - the brighter the star, the stronger the radiation pressure - and on the spectral properties of the absorbing species.  

In the case where a gas particle is excited to a specific energy level and exhibits a single spectral line, its \(\beta\) value—defined as the ratio of radiation pressure to gravitational force—is given by \citep[e.g.,][]{Lagrange1995}:

\begin{equation}
    \beta = \frac{R_\star^2}{G m M_\star} \frac{1}{4 \pi \varepsilon_0} \frac{\pi e^2}{m_e c^2} \pi I_\star(\lambda) \frac{\lambda^2}{c} f,
\end{equation}
where $m$ is the mass of the absorbing particle, $M_\star$ is the stellar mass, $R_\star$ is the stellar radius, $\lambda$ is the wavelength of the transition, $I_\star$ is the stellar specific intensity per unit wavelength, $f$ is the oscillator strength of the line, $\varepsilon_0$ is the vacuum permittivity, $m_e$ is the electron mass, and $e$ is the elementary charge. The $\beta$ factor can also be expressed as a function of the stellar flux $\phi_\lambda$ received on Earth (in erg/s/cm$^{-2}$/\AA) and our distance to the star $d$, using: 
\begin{equation}
    \phi_\star(\lambda) = \left( \frac{R_\star}{d} \right) ^{-2} \pi I_\star(\lambda),
\end{equation}
which gives: 
\begin{equation}
    \beta = \frac{d^2}{G m M_\star} \frac{1}{4 \pi \varepsilon_0} \frac{\pi e^2}{m_e c^2} \frac{\lambda^2}{c} \phi_\star(\lambda) f.
    \label{Eq. Beta factor lit}
\end{equation}
\vspace{0.2 cm}

\noindent Numerically, Eq. \ref{Eq. Beta factor lit} becomes \citep[][]{Lagrange1995}: 
\begin{align}
\begin{aligned}
    \beta = 0.509 f \left( \frac{d}{\text{1 pc}} \right)^2  \left( \frac{m}{m_u} \right)^{-1}   \left( \frac{M_\star}{M_\odot} \right)^{-1}   \left( \frac{\lambda}{2000 \text{\AA}} \right)^{2} \times \\
    \vspace{0.5 cm}
    \frac{\phi_\star(\lambda)}{10^{-11} \text{\,erg\,cm}^{-2}\,\text{s}^{-1}\,\text{\AA}^{-1}} \ \ .
\label{Eq. Beta factor num}
\end{aligned}
\end{align}

In cases where the species has multiple spectral lines, the total $\beta$ value is obtained by summing Eq. \ref{Eq. Beta factor num} over all lines. If the species is also distributed over multiple excitation levels (as for instance Fe$^+$, Fig. \ref{Fig. excitation diagrams}), an effective $\beta$ value can be calculated by averaging the $\beta$ values across the different levels, weighted by their relative abundances. Assuming that the gas is characterized by an excitation temperature $T$, this is:

\begin{equation}
    \beta_{\rm eff} =  \frac{1}{Z(T)} \displaystyle  \sum_{\mathrm{lines}} \beta g e^{-\frac{E}{k_\mathrm{B} T}} , 
    \label{Eq. Beta factor total}
\end{equation}

\vspace{0.2 cm}

\noindent where $g$ and $E$ denote the lower level multiplicity and lower level energy of each line, $k_{\rm B}$ the Boltzmann constant, and $Z(T)$ the partition function. 

\begin{figure}[h!]
\centering
    \includegraphics[scale = 0.18,     trim = 10 0 60 60,clip]{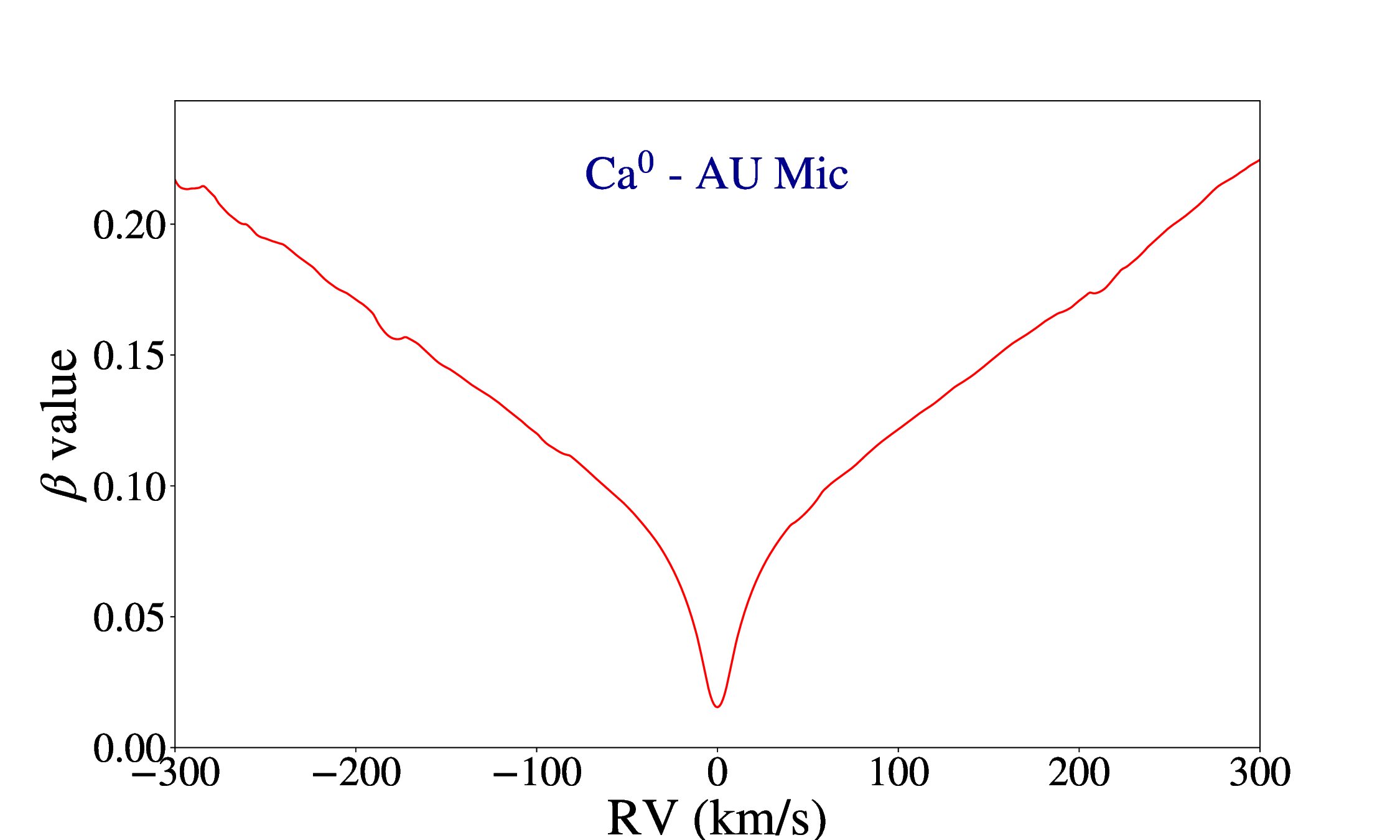}    
    \includegraphics[scale = 0.18,     trim = 10 0 50 60,clip]{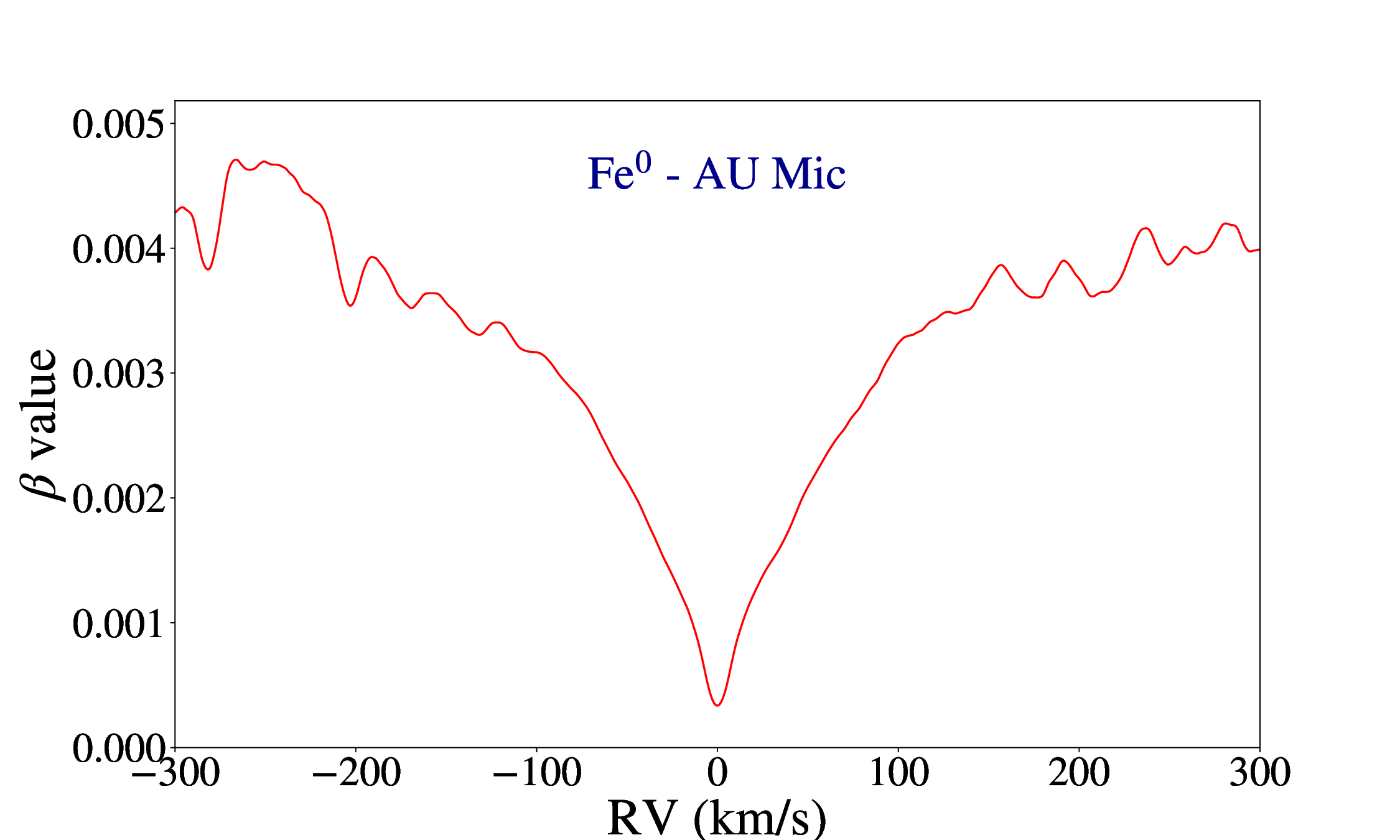}

    \vspace{0.1 cm}
    
    \includegraphics[scale = 0.18,     trim = 10 0 60 60,clip]{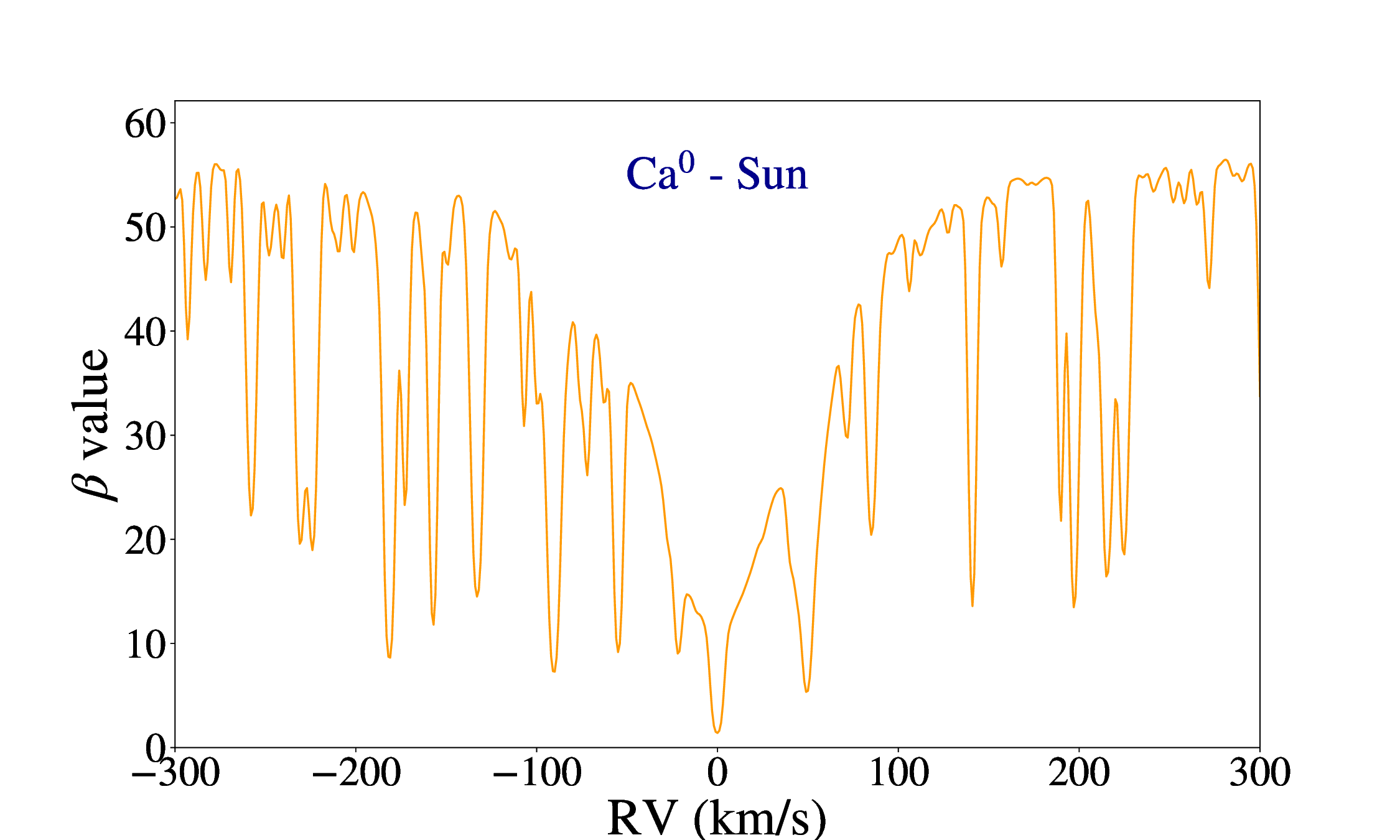}    
    \includegraphics[scale = 0.18,     trim = 10 0 50 60,clip]{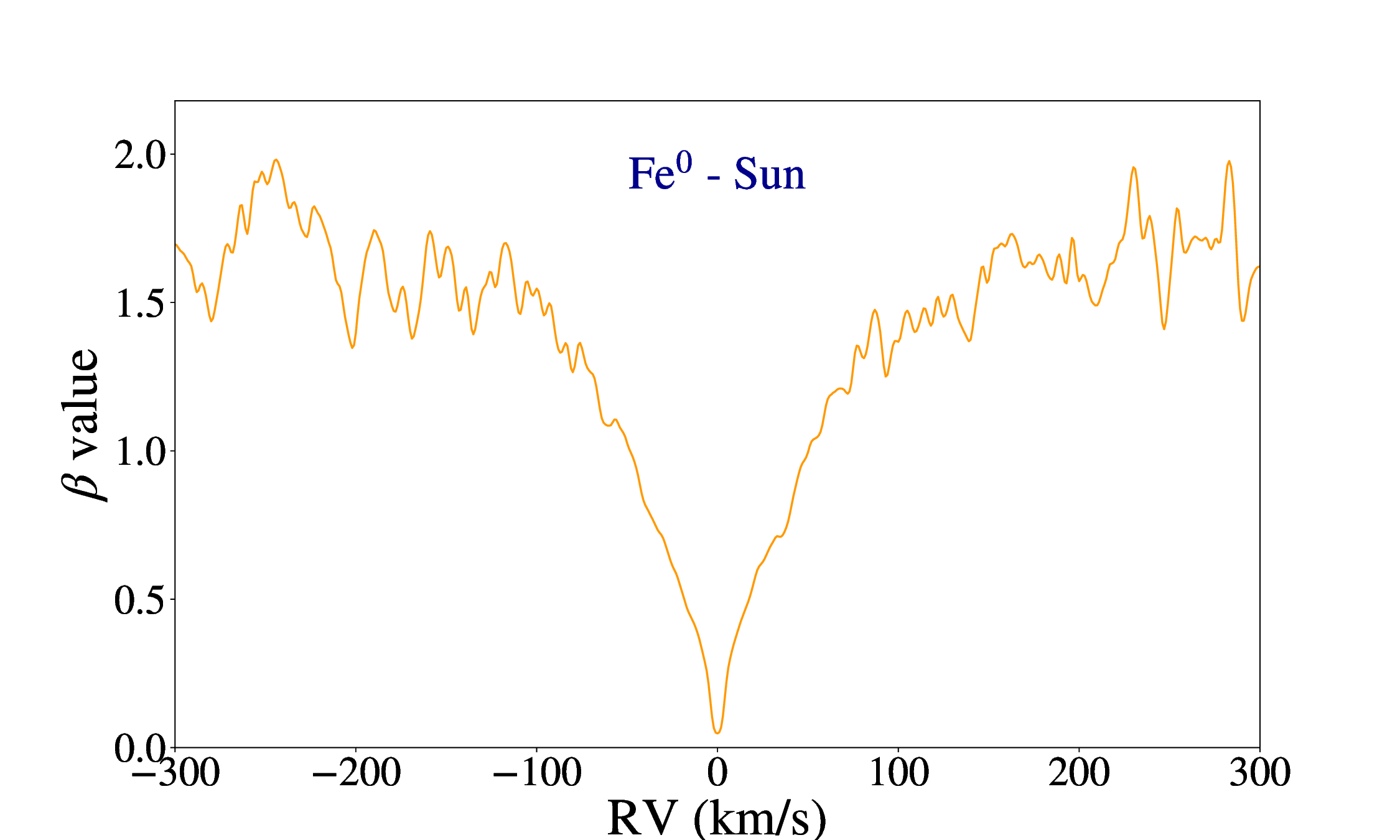}
    
    \vspace{0.1 cm}

    \includegraphics[scale = 0.18,     trim = 10 0 60 60,clip]{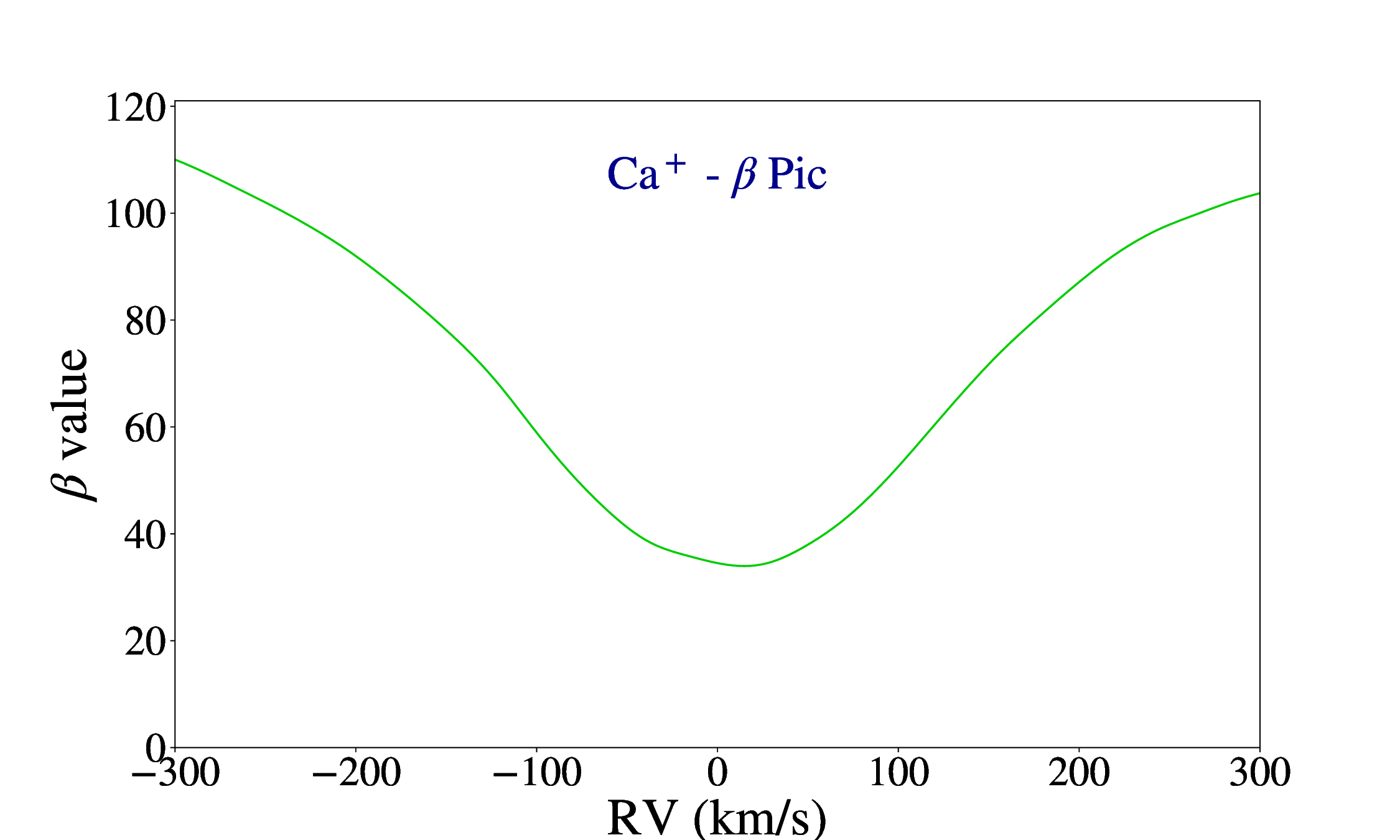}
    \includegraphics[scale = 0.18,     trim = 10 0 50 60,clip]{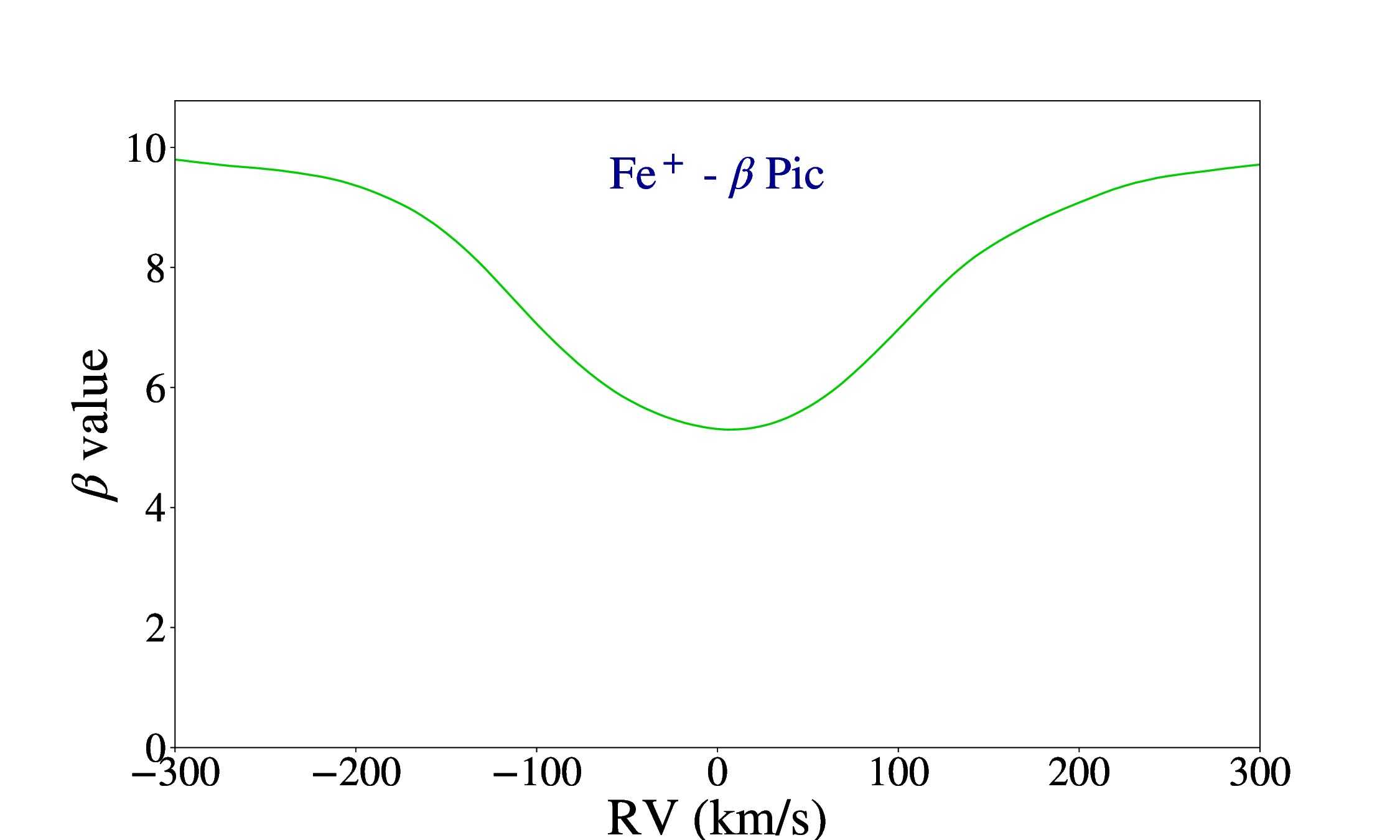}  
    
    \vspace{0.1 cm}

    \includegraphics[scale = 0.18,     trim = 10 0 60 60,clip]{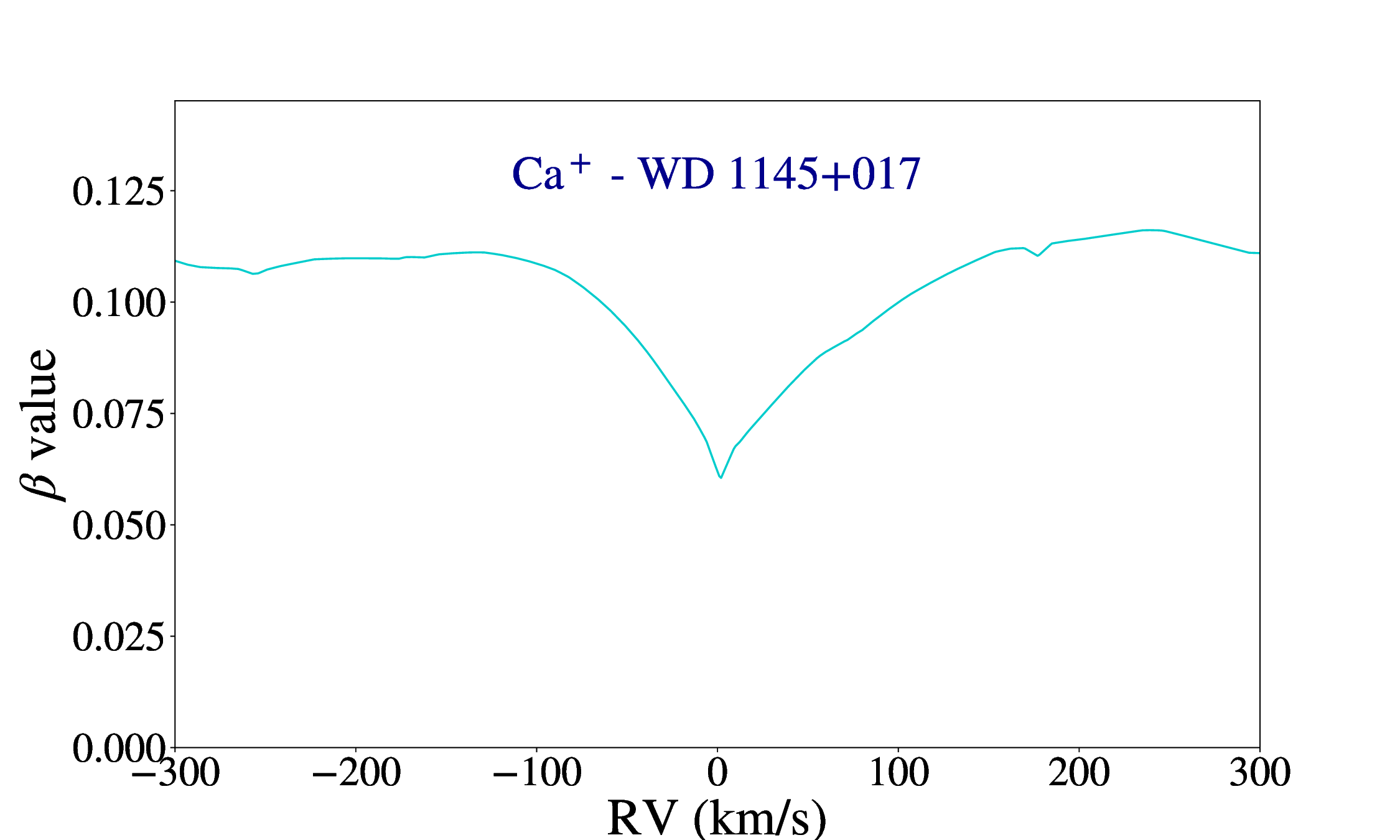}
    \includegraphics[scale = 0.18,     trim = 10 0 50 60,clip]{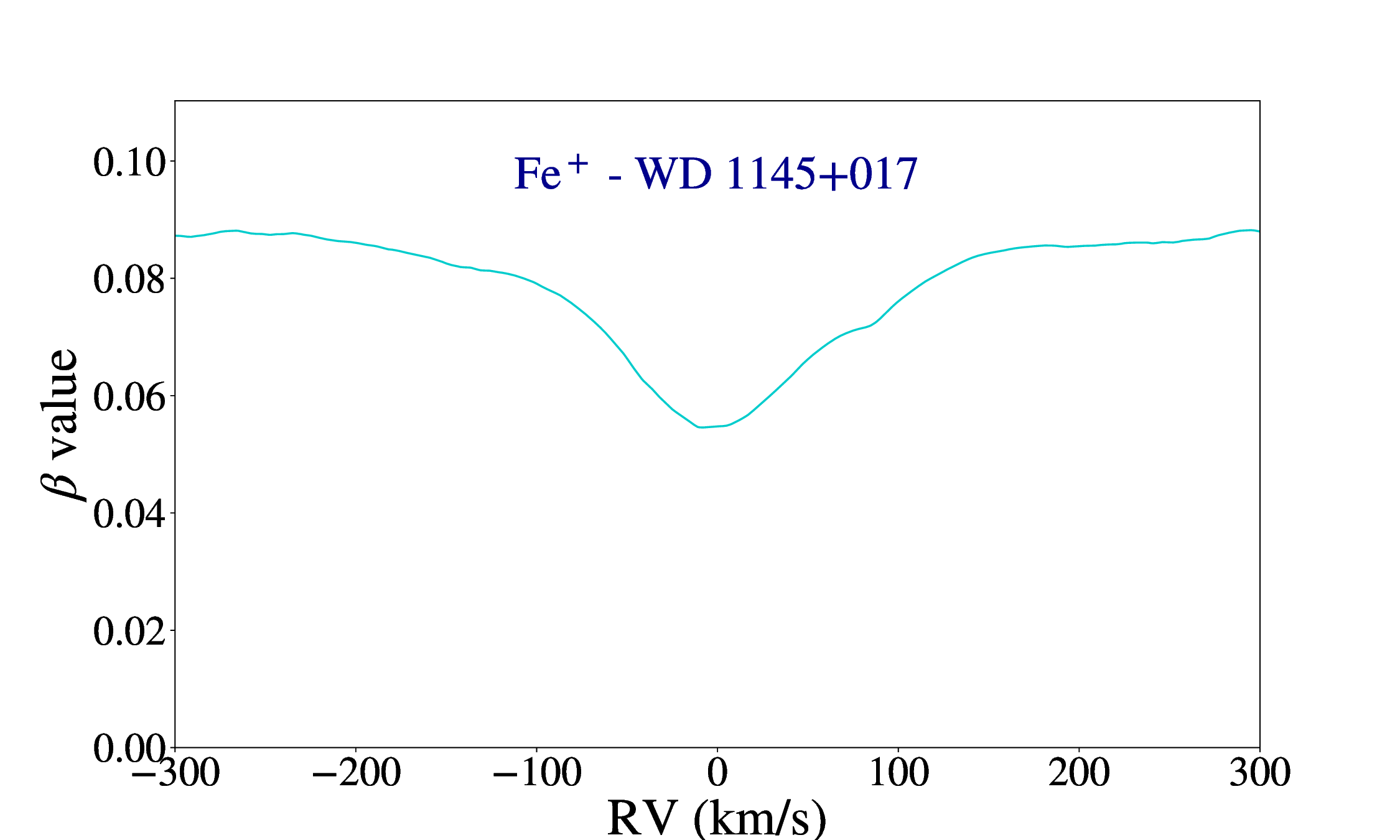}    

    \vspace{0.1 cm}

    \caption{\small Radiation pressure-to-gravity ratio of Ca and Fe around AU~Mic and the Sun, and of Fe$^+$ and Ca$^+$ around \bp\ and WD 1145+017, as a function of the radial velocity.}
    \label{Fig. Prad}
\end{figure}

Since the excitation temperature in cometary tails can be approximated by the stellar effective temperature (Sect. \ref{Subsect. Excitation}), Eq. \ref{Eq. Beta factor total} allows us to easily estimate the radiation pressure felt by a given atom or ion, around a given star. For instance, Fig. \ref{Fig. Prad} provide the $\beta$ values of neutral and ionized Ca and Fe, around our four case-study stars. For AU~Mic, \bp\ and~1145+017, the excitation temperature was set equal to the stellar effective temperature, following the results of \cite{Vrignaud2024b} for \bp. For the Sun, we used an excitation temperature of 4000~K, as found in \cite{Manfroid2021}.

The efficiency of radiation pressure strongly depends on the spectral type of the host star: while around AU~Mic the radiation pressure felt by Ca and Fe is rather weak ($\beta < 1$, Fig. \ref{Fig. Prad}), it becomes significant around the Sun ($\beta \approx 1$) and largely dominant around \bp\ ($\beta = 40-100$ for Ca$^+$, 5-10 for Fe$^+$). Due to its small size (0.012 R$_\odot$), the white dwarf WD 1145+017 exerts a much weaker radiation pressure ($\beta \sim 0.1$ for Ca$^+$\ and Fe$^+$). Such differences could induce strong contrasts between the dynamics of exocometary tails in those four systems, even though the motion of the most accelerated ions (e.g. Fe$^+$, Ca$^+$) could be slowed down by drag forces induced by other species (e.g. C$^+$; see for instance \cite{Brandeker2011_betapic_braking} and Sect. \ref{Sect. drag forces}).

The radiation pressure felt by a gas particle can  be highly sensitive to its radial velocity, particularly in the presence of absorption lines in the stellar spectrum. These absorption lines reduce the flux at the main transitions of the considered species, reducing the radiation pressure felt by the gas when its radial velocity is low. This effect is particularly strong for slowly rotating stars (e.g., the Sun or AU~Mic), where photospheric lines are very deep. For instance, the $\beta$ value of neutral Fe around the Sun can range from 0.05 to 1.5, depending on the radial velocity of the gas. A similar effect is observed for atomic sodium \citep{Spinrad1968_sodium_velocity, Brown1998_sodium_velocity_hale_bopp}. In contrast, this effect is less important in the \bp\ system due to the fast rotation of the host star (130 km~s$^{-1}$), even though the $\beta$ value of Ca$^+$\ still varies from $\sim 35$ at 0 km~s$^{-1}$ to $\sim 100$ for highly red or blue-shifted gas.

Finally, it should be noted that the radiation pressure acting on a given particle depends significantly on the location of its main spectral lines: species with lines further in the UV will generally feel a weaker radiation pressure than species with allowed transitions in the optical or NUV. This explains, for instance, the very weak radiation pressure felt by Si$^+$ around \bp\ ($\beta = 0.05 - 0.1$): the main \siii\ lines are all located below 1500 \AA, where the flux of \bp\ is rather weak. On the contrary, Fe$^+$\ ions experience a much larger pressure ($\beta = 5 - 10$) thanks to the strong \feii\ transitions in the NUV (2300--2800~\AA).

\subsubsection{Impact of stellar winds}

Stellar winds are continuous radial outflows emanating from the atmospheres of stars. In many cases, they play a significant role in the production and shaping of exocomet tails. The nature of these winds vary greatly in terms of density, speed, dynamic pressure (10$^{-10}$ - 10$^{1}$ Pa across all systems), and magnetic field, which in the Solar System is referred to as the interplanetary or heliospheric magnetic field. We refer the reader to \citet{Bennett2010} and \citet{Vidotto2021}, and references therein, for relevant reviews of this research field. In brief, stellar winds in extrasolar systems can differ dramatically from the collisionless solar, and the expansive studies undertaken of ion tails’ formation and interactions with the solar wind are therefore not always readily comparable with all exocomet systems, e.g. \citet{Strom2020}.

Stellar winds affect the dust and gas around exocomets in multiple ways. First, the wind interacts with the atmosphere of these bodies by picking up charged ions. Second, in the Solar System, the solar wind ($\sim 2 \times 10^{-14}$ M$_\odot$/yr) contributes a few percent to the production of ion tails in comets: charge exchange between solar wind protons and neutral gas species from the comae  produces new ions (CO$^+$, H$_2$O$^+$), which are then accelerated in the anti-solar direction \citep[the rest of the ion tails are produced from photo-ionization,][]{Combi2004}. In addition, coronal mass ejections (CME) can strongly increase the dissociation rates in comae, mostly through electron impact \citep[e.g.,][for 67P/Churyumov–Gerasimenko]{Noonan2018}.Third, charge exchange between highly charged ions in the solar wind and neutral gases surrounding comets results in highly characteristic X-ray emission. In the Solar System, comets produce up to 1~GW in X-rays \citep{Dennerl2012}, and similar interactions between stellar wind ions and interstellar neutrals have been detected around three main-sequence stars \citep{Kislyakova2024} 

Around \bp, studies indicate that the effect of the stellar wind \citep[$\sim 10^{-14}$ M$_\odot$/yr,][]{Bruhweiler1991} on ionic clouds is negligible, as their shape is primarily sculpted by radiation pressure \citep[][]{Beust2001}. Some other young stars, however, can exhibit strong mass loss rates. This is the case of Herbig stars (pre-main sequence stars younger than 10 Myr, with stellar type A or B), which exhibit stellar winds in the range $10^{-9} - 10^{-7}$ M$_\odot$/yr. Some of these stars have been hypothesized to host exocomets based on variable absorption signatures observed in refractory species \citep[e.g. HD 100546,][]{Grady1997}. However,  \cite{Beust2001} predicted that objects in such systems would be very difficult to detected, as their gaseous tails would be confined to compact regions by the intense stellar wind and would therefore cover only small fractions of the stellar disc. \cite{Beust2001} provided a maximum loss rate of $10^{-10}$ M$_\odot$/yr for the stellar wind above which no exocomets can be detected using absorption spectroscopy (e.g. in \mgii\ h \& k lines).  

Another interesting case is  AU~Mic, a young (23 Myr) M-dwarf surrounded by an edge-on  debris disc. Although the star has a relatively moderate stellar wind \citep[$1-2 \times 10^{-13}$ M$_\odot$/yr,][]{Alvarado2022}, it exhibits highly energetic flares and frequent coronal mass ejections, which release large amount of material at a daily frequency  \citep[$10^{18}$\,g = $10^{-15}$~M$_\odot$ or $10^{35}$\,erg for typical CME events, e.g.][]{Alvarado2022}. The effect of such events on potential exocomets has not been studied yet, but it is likely that they would significantly affect the shape and composition of comae and tails, by increasing the ionization rate and rapidly blowing the produced ions in the anti-stellar direction. 

As stars evolve beyond the main sequence, their mass loss rates increase dramatically during certain evolutionary phases. The red giant branch and asymptotic giant branch (AGB) phases are particularly significant, as mass loss rates can rise by several orders of magnitude. During the AGB phase, stellar winds can reach peak rates of $10^{-5} M_{\odot}/$yr, a value which is approximately 8 orders of magnitude higher than the current solar wind.  Such stellar winds would probably have a dramatic effect on exocomets, by rapidly sweeping away the evaporated material. Yet, the gas/ion interactions in this violent environment, coupled with orbital shifts of exocomets due to the stellar mass loss, not only could help explain nebular structures around young white dwarfs \citep{marshall2023}, but also help determine the distribution of reservoirs of planetary material which may eventually pollute the white dwarf \citep[see the accompanying chapter][]{Bannister2025SSR}).

\subsubsection{Drag forces in atomic and plasma tails}
\label{Sect. drag forces}

The various forces detailed above (radiation pressure, ion pick-up, stellar wind pressure, magnetic forces) are generally specific to the atom or ion they act on: two different ions will usually undergo different accelerations, depending on their masses, charges or spectral lines. Consequently, one may expect to observe a wide range of dynamics among the species released by exocomets, with, for example, species subject to greater radiation pressure expanding over larger distances. However, the motion of ions and atoms in cometary tails is also influenced by collisions with other particles, which induce drag forces and tend to couple their dynamics. These drag forces have been studied in previous studies of \bp\ exocomets, in particular \cite{Beust1989}, \cite{Beust1990} and \cite{Beust2001}, where further information on some of the effects below may be found.  \\

\textbf{Motion of particles in neutral gas}: A particle (of mass $m_1$) evolving in a field of neutral atoms of mass $m_2$ experiences a drag force, due to collisions with the field particles. This drag force can be described as \citep[see for instance][]{Kwok1975, Lagrange1998}: 

\begin{equation}
\vec{f}_{\rm col} = \frac{\mathrm{d} \vec{p}_1}{\mathrm{d} t}\Big|_{\rm col} = - k_{\rm col} \vec{v}_1, 
\label{Eq. drag force col}
\end{equation}
with $\vec{v}_1$ and $\vec{p}_1$ the velocity and momentum of the considered particle relative to the neutral gas, and $k_{\rm col}$ the drag coefficient, given by:
\begin{equation}
    k_{\rm col} = \pi a_2^2 \mu n_2\ \sqrt{v_T^2 + v_1^2},
\end{equation}
where $a_2$ is the radius of the field atoms, $n_2$ their volume density, $\mu = \frac{m_1 m_2}{m_1 + m_2}$, and $v_T$ the sound speed in the gas.

If the considered particle is an ion, it will also induce a dipole on the field atoms, increasing the dragging coefficient. This additional drag force can be expressed as \citep{McDaniel1964, Beust1990}: 
\begin{equation}
\vec{f}_{\rm dip} = \frac{\mathrm{d} \vec{p}_{1}}{\mathrm{d} t} = - k_{\rm dip} \vec{v}_1,
\label{Eq. drag force dip}
\end{equation}
with: 
\begin{equation}
    k_{\rm dip} = n_2 \pi \sqrt{\frac{4 \mu q_1^2 \alpha_2}{4 \pi \varepsilon_0}},
\end{equation}
and where $q_1$ is the charge of the ion and $\alpha_2$ the polarizability of the field atoms. For a gas with an arbitrary density, the total drag coefficient is obtained by summing Eqs. \ref{Eq. drag force col} and \ref{Eq. drag force dip} over all the different species making up the gas. 

\begin{figure}[h]
\centering
    \includegraphics[scale = 0.24,     trim = 75 15 60 60,clip]{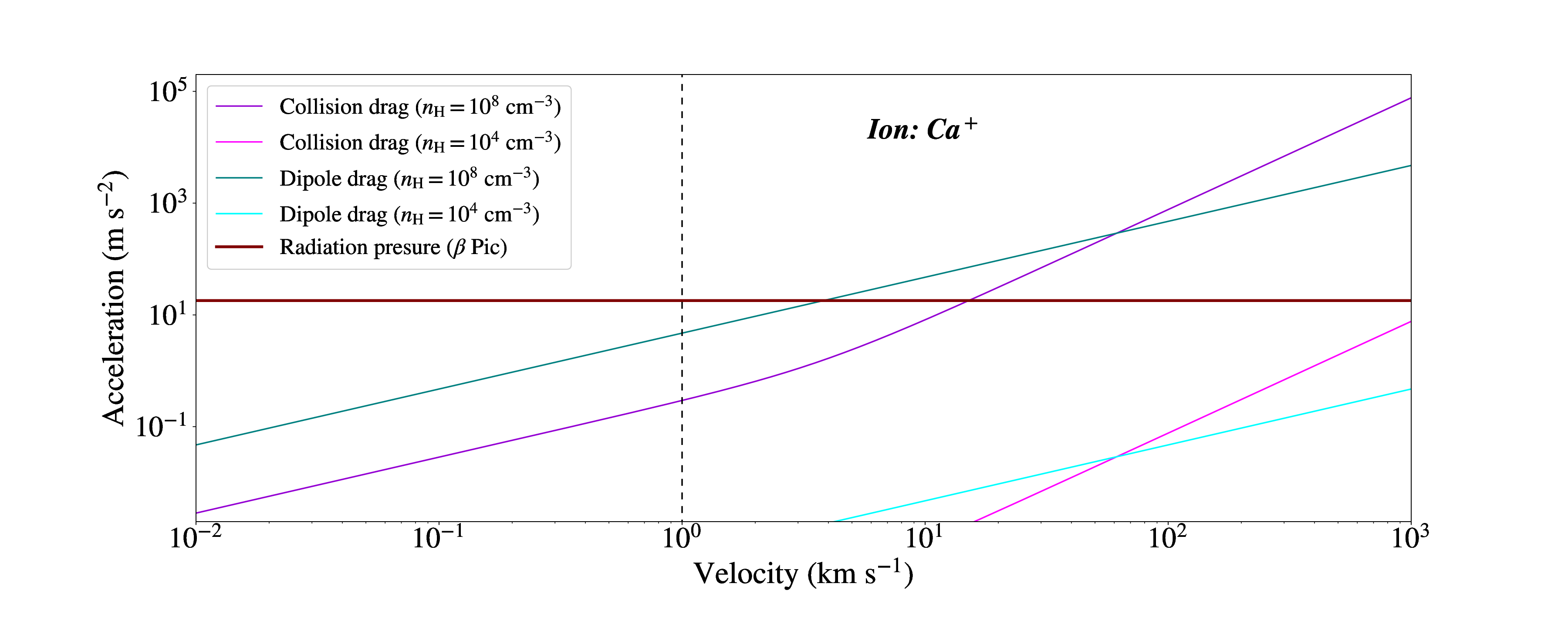}
    \includegraphics[scale = 0.24,     trim = 75 10 60 60,clip]{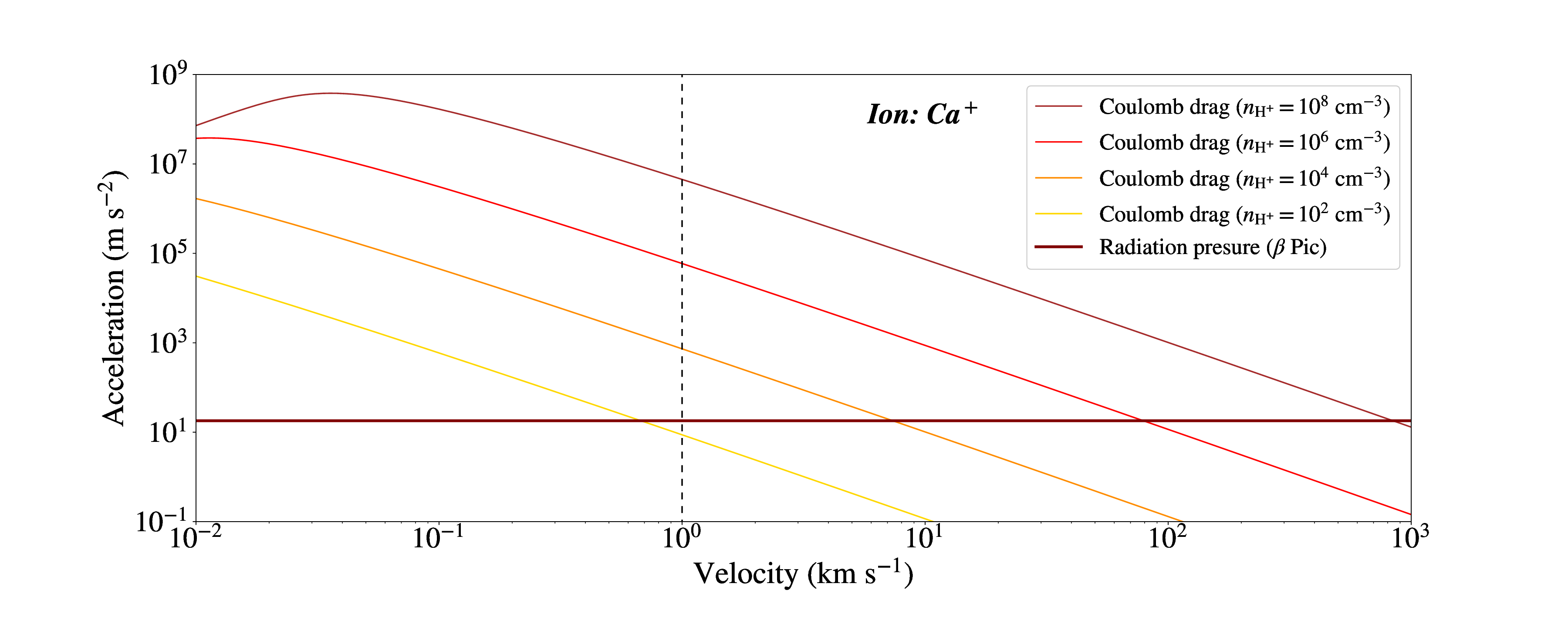}
    \caption{\small \textbf{Top}: Drag forces acting on Ca$^+$\ ions in a neutral gas of hydrogen, as a function of the Ca$^+$\ velocity relative to the gas. The black, brown line shows the radiation pressure felt by Ca$^+$\ ions at 20 R$_\star$ from \bp, using a $\beta$ value of 35 (zero-velocity gas). The vertical, dashed line indicates a velocity of 1 km~s$^{-1}$, typical of thermal movements. Note that only the norm of the acceleration is shown: for drag forces, the acceleration is opposed to the motion of the ion relative to the surrounding medium, and for radiation pressure the acceleration is directed in the anti-stellar direction. \newline    
    \textbf{Bottom}: Coulomb drag force acting of Ca$^+$\ ions in a plasma of ionized hydrogen, as a function of the Ca$^+$\ velocity relative to the plasma. An electronic temperature of 1000 K was assumed. The electronic density ($n_e$) was taken to be equal to the hydrogen density ($n_{\rm H^{+}}$). Horizontal and vertical lines are identical to the top panel.}
    \label{Fig. Drag forces Ca II}
\end{figure}

The intensities of the two drag forces are illustrated in Fig. \ref{Fig. Drag forces Ca II} for Ca$^+$\ (top panel), assuming a surrounding fluid of neutral hydrogen with densities of $10^{4}$ or $10^8$ cm$^{-3}$. We note that, for typical thermal or orbital velocities ($\sim$ km~s$^{-1}$), the contributions of \ref{Eq. drag force col} and \ref{Eq. drag force dip} are generally comparable, although dipole-induced interaction dominate for velocities below 50 km~s$^{-1}$. We also note that at low densities (e.g. $10^4$ cm$^{-3}$), dragging forces are rather weak, allowing the particles to accelerate to large velocities ($> 100$ km~s$^{-1}$) when undergoing strong radiation pressure \citep[see for instance][for the motion of Ca$^+$\ ions in atomic hydrogen gas]{Beust1989}. \\

\textbf{Motion of ions in plasma}: The motion of an ion in a dense, ionized exocometary tail was studied by \cite{Beust1990}. The interaction between a charged particle \#1 (of charge $q_1$ and mass $m_1$) and a field of particles \#2 (of charge $q_2$ and mass $m_2$) is described by the Coulomb potential. This interaction induces a drag force on particle \#1, which can be expressed as \citep[see][]{Beust1989, Beust2001}: 
\begin{equation}
    \vec{f}_{\rm coul} =\frac{\mathrm{d} \vec{p}_{1}}{\mathrm{d} t} = -2 \pi n_2 \mu \frac{C^2}{v_1^2} \ln \left( \frac{\lambda_{\rm D}^2 v_1^4}{C^2} + 1 \right)   \frac{\vec{v}_1}{v_1},
    \label{Eq. drag force coul}
\end{equation}
where $v_1$ is the velocity of the considered ion relative to the plasma, $n_2$ the volume density of particle \#2, $\mu = \frac{m_1 m_2}{m_1 + m_2}$, $C = \frac{1}{4 \pi \varepsilon_0} \frac{q_1 q_2}{\mu}$, and where $\lambda_D = \sqrt{\varepsilon_0 k_B T_e/(n_e e^2)}$ is the Debye length of the plasma, with $T_e$ the electronic temperature and $n_e$ the electronic density. 

Fig. \ref{Fig. Drag forces Ca II} (bottom panel) provides the norm of the drag force undergone by Ca$^+$\ ions in a plasma of ionized hydrogen of various densities, as a function of the Ca$^+$\ velocity relative to the plasma. The drag force is generally very strong, exceeding by several orders of magnitude the radiation pressure (which itself is very strong compared to other exocometary species). For typical thermal velocities ($\sim$ km~s$^{-1}$), the radiation pressure can overcome the drag force only if the local density is very low ($n_{\rm H^{+}} < 10^2$ cm$^{-3}$). Above this density, the dynamics of all ions in the plasma are coupled, and the plasma behaves as a single fluid. The effective radiation pressure acting on the gas is then the average of the radiation pressures felt by individual species (Ca$^+$, Fe$^+$...).

The ionization state and density of a cometary tail/atmosphere are thus key properties that influence the dynamics of individual particles. If the gas is neutral or diluted, drag forces are weak, and the dynamics of individual species rapidly decouple as they expand away from the nucleus. In this case, the motion of particles is set by their own radiation pressure and/or by the forces induced by the stellar wind. This is typically the case for the thin plasma tails of Solar System comets. On the contrary, in dense, ionized tails, drag forces efficiently couple the dynamics of all ionized species, maintaining them within the same fluid over large distances. This behavior has been observed for \bp\ exocomets \citep[][]{Vrignaud2024b}: cometary signatures in \feii, \niii, \mnii\ and \siii\ lines are very well correlated, showing that these ions can remain well-mixed over distances of several stellar radii despite undergoing different radiation pressures.

\subsubsection{Gas compression}
\label{Subsubsect. Gas compression}

The detection of highly ionized species (Al$^{2+}$, Si$^{3+}$, C$^{3+}$) in \bp\ exocomets has long remained puzzling, since the production of those ions is not expected from photo-ionization (Sect. \ref{Sect. Ionisation}). Much work have been conducted \citep[][]{Beust1989, Beust1990, Beust1993} to explain their presence in exocometary tail. The main conclusion of these studies is that highly ionized species are produced from collisional ionization in a dense, hot 'bow shock', located at the front of the comae. This shock is created by a combined effect of radiation pressure --- which accelerates the gas in the anti-stellar direction, Sect. \ref{Sect. radiation pressure} --- and drag forces with a neutral medium --- which resist to the movement of ions, Sect. \ref{Sect. drag forces}. These mechanisms strongly compress the gas, allowing it to reach high temperatures \citep[$>$ 30\,000 K;][]{Beust1993} at which Al$^{+}$\ is rapidly ionized into Al$^{2+}$.

In \cite{Beust1993}, the dynamics of cometary ions is controlled by radiation pressure and drag forces. The drag forces are assumed to be induced by a surrounding cloud of neutral hydrogen (see Eqs. \ref{Eq. drag force col}, \ref{Eq. drag force dip}, and the first panel of Fig. \ref{Fig. Drag forces Ca II}). An estimate for the size of the bow shock (which depend on the considered ion, e.g. Ca$^+$) is then given by:

\begin{equation}
    r = \sqrt{\frac{F v_e}{\beta g} \left(1 + \frac{a_2}{a_1} \right) },
\end{equation}
with $F$ a factor characterizing the sensitivity of the studied ion to collisions with neutral hydrogen, $v_e$ the expansion velocity of the hydrogen cloud, $\beta$ the radiation pressure to gravity ratio (Sect. \ref{Sect. radiation pressure}), $g$ the stellar gravity acting on the comet, and $a_1$, $a_2$ the sizes of the two layers used to model the shock front. This formula was used by \cite{Kiefer2014} to infer basic properties of \bp\ cometary nuclei from their covering factor in \caii\ lines. 


It should be noted, however, that the model proposed by \cite{Beust1993} could be refined. In particular, the assumption that ion trajectories (Ca$^+$, Mg$^+$, ...) are primarily slowed by collisions with neutral hydrogen is debatable, as observations indicate that most of the gas in \bp\ exocomets is ionized, implying significantly stronger drag forces than previously considered (Sect. \ref{Sect. drag forces}). In fact, the observed tendency of ions to remain well-mixed in exocometary tails \citep[][]{Vrignaud2024b} suggests that ion-ion interactions play a dominant role in the dynamics of cometary tails. In addition, the role of photoionisation was probably under-estimated in the study of \cite{Beust1993}, as our calculations showed that this mechanism can also produce Al$^{2+}$ if the comet is sufficiently close to the star. Despite these limitations, the \cite{Beust1993} model provides a valuable foundation for developing more accurate descriptions of the physics of \bp ~exocomets.

\section{Case studies}
\label{sec: case study stars}

This section provides a comparative analysis of comet-like activity in four different stellar environments: the Sun, AU~Mic, \bp, and WD 1145+017. A summary of the main properties of (exo)comets in the four systems is also presented in Table \ref{Tab. Differences between systems}, allowing for a direct comparison.

\subsection{Solar System comets}
\label{Sect. Solar System comets}

The observable characteristics of comets in our own Solar System are readily discussed in many works, notably in {\it Comets III} \citep[see also the reviews of][]{Biver2022,Bodewits2022,Engrand2023,Knight2024}. Here, we consider them from a different perspective: if we were to study the Sun from another Solar System, what would be the potential observable signatures of our comets and other small bodies?

The most obvious candidates for consideration are the various populations of near-Sun comets \citep[cf. reviews by][]{Marsden2005,Jones2018}, since these are nominally most analogous to the \bp\ exocomets: they share similar periastron distance (a few $1-10$ stellar radii) and produce large amounts of refractory gas through dust sublimation \citep[e.g., ][]{Iseli2002_sungrazing}. More than 5000 near-Sun comets have been discovered to date; however, nearly all are thought to be 10s of meters in size or smaller \citep[e.g.,][]{Sekanina2003,Knight2010}, far too small for remote detection. The vast majority of these comets are members of the Kreutz group \citep{Battams2017}, a group of sungrazing comets created by hierarchical fragmentation of a single parent body near the Sun, likely within the last few millenia \citep[e.g.,][]{Marsden1989}. Over the last two centuries, several spectacularly bright members of the group have been detected whose brightness implied radii of order 1~km or more, such as Ikeya-Seki \citep[\cite{Preston1967}; see also][]{Sekanina2002,Knight2010}, suggesting that the progenitor body may have been of order 10~km in radius. 

Among comets with perihelion distances near Earth, C/1995 O1 (Hale-Bopp) offers the most optimistic test case for what might be detectable as an exocomet. Hale-Bopp's 37~km radius \citep{Szabo2012} is among the largest known \citep[see review by][]{Knight2024} and it had the highest measured activity near 1~au, reaching production rates of $\sim$10$^{31}$ molecules/s of H$_2$O and $\sim$10$^{30}$ molecules/s of CO \citep[e.g.,][]{Biver2002,Schleicher2024} and $Af{\rho} > 10^6$ cm \citep{Womack2021}. The quantity $Af\rho$ is physically meaningless but is widely used as a proxy for dust production. It is measured within a circular aperture of radius $\rho$, and is the product of the albedo, $A$, of grains in the coma, the filling factor, $f$, of grains in the aperture, and $\rho$ \citep{A'Hearn1984}. Given that an $Af{\rho}$ value of 1000~cm is approximately equivalent to one metric ton of dust production per second \citep[C.~Arpigny as quoted by][]{A'Hearn1995}, this translates into a dust production rate of $>10^6$~kg/s for Hale-Bopp.

In an early investigation of photometric transit detection capability, \citet{Lecavelier1999_photometry} found that comets with dust production rates as low as $\sim$10$^3$ kg/s could be detectable from highly favorable viewing geometry due to the optical depth along their tail (which must exceed $\sim 10^{-4}$ for detection). Below, we briefly consider a different criterion: what could be detectable from any viewing geometry? \citet{Lecavelier2022} identified exocomet transits in the $\upbeta$~Pic system using the Transiting Exoplanet Survey Telescope (TESS) with transit depths shallower than $2{\times}10^{-4}$. Taking this as a representative detection threshold and multiplying by ${\pi}R_\odot^2$ yields the cross-section of dust along a column towards the Sun that a comet would need to produce in order to potentially be detected as a transit. Approximating comet dust as spheres of radius 1~\textmu~m and assuming a grain density of 500 kg/m$^3$ \citep[consistent with Rosetta's measurement of 67P's bulk nucleus;][]{Jorda2016}, this would require ${\sim}2{\times}10^{11}$~kg of dust, or about $\sim$9~hr of Hale-Bopp's peak activity.  
Assuming that grains remain within the column for 24~hr (likely an underestimate given that the column would have the diameter of the Sun), Hale-Bopp would have been detectable within $\sim$1.5 au of the Sun for a duration of $\sim$5 months. Alternatively, this mass of dust is equivalent to the complete dissolution of a nucleus of radius $\sim$600~m (assuming a dust-to-gas mass ratio of 1), implying that many of the brightest Kreutz comets might have been detectable as single events. This argument is obviously an over simplification, but demonstrates that it is at least plausible that some comets in our current Solar System could be detected as exocomets. More realistic models \citep[e.g.,][]{Lecavelier1999_photometry} might include a grain size distribution, different densities for grains versus the nucleus, variable dust-to-gas ratio, radiation pressure, forward-scattering, grain composition, etc. but are beyond the scope of this paper. See \citet{Kalman2024} for a discussion of recent modeling efforts.


Since detection of exocomets by pollution of the star's atmosphere has only been detected for white dwarfs, it can be expected that this mechanism would be irrelevant for extra-solar detection of our system's comets. In the interest of completeness, we note that such an event would be extremely rare in the current Solar System. No comets or asteroids have been observed to impact the Sun, although a few of the poorly observed, tiny Kreutz sungrazers have had nominal orbits with perihelion distance below 1~R$_\odot$ but none appeared to survive to this point. See Section 1.3.3 of \citet{Jones2018} for additional discussion of these so-called ``sundivers.''

The current distribution of comet orbits makes extra-solar detection less favorable. The majority of comets are thought to be isotropically distributed throughout the Oort cloud \citep[see the recent review by][]{Kaib2022}. When perturbed into the inner Solar System, they generally retain their near-random inclination distribution. Thus, no view of the Solar System would yield a preferred orientation for transits of comets arriving from the Oort cloud. Since most short period comets have orbits with inclinations below 20$^\circ$ \citep{Kaib2022} and these comets have much shorter orbital periods than Oort cloud comets, there is a clear preference for edge-on observations. However, there are no current short-period comets with activity anywhere near that of Hale-Bopp; 1P/Halley, itself an outlier in the population, is about an order of magnitude less active \citep[e.g.,][]{schleicher_narrowband_1998}. There are a small number of known Centaurs with sizes comparable to Hale-Bopp \citep[e.g., 29P/Schwassmann-Wachmann~1;][]{Schambeau2015}. Centaurs are objects currently orbiting on low inclination orbits in the outer planet region, but which may evolve into or have already been on orbits that reach the inner Solar System \cite[cf.][]{Sarid2019}. Thus, at certain times in the Solar System's history, there might be a regularly detectable object, but such periods are likely infrequent. Furthermore, many of these objects will be thermally processed during their long dynamical evolution into such an orbit \citep{Gkotsinas2022}, potentially suppressing activity as compared to a comparably sized Oort cloud comet reaching the same perihelion distance.



\subsection{$\upbeta$~Pictoris}

\bp\ was the first star discovered to be an exocomet system, using observations of variable features in \caii\ H\&K lines \citep[][]{Ferlet1987} attributed to the transit of large cometary-like tails in from of the star. Since then, many studies \citep[e.g.][]{Beust1990, Beust1993, Beust1998} have tried to model and simulate these exocomets, in order to demonstrate that they are a valid scenario to explain the observed events, and to better constrain what physical physical properties these objects should have. A full review on this system is provided in the accompanying chapter Lu et al. (submitted). 

The main differences between Solar System comets and those observed around \bp\ lie in (1) the very high production rates and (2) the high level of ionization of the  surrounding gas. Concerning the production rates of \bp\ exocomets, studies \citep{Beust1989} have shown that comets in the \bp\ system could release hydrogen at rates of $\sim 10^{33}$ atoms/s, or about $1000 \times$ higher than Solar System comets. This is mostly due to the higher luminosity ($\sim 9\,L_\odot$) and temperature ($\sim 8000$ K) of \bp\ compared to the Sun, as well as the close proximity \citep[$10-20\,R_\star$,][]{Kiefer2014} at which the comets are observed (the production rate of a comet typically scales as $r^{-2}$, see Sect. 2). As a result, the total amount of gas found in a \bp\ cometary tail typically corresponds to the full evaporation of a $\sim$ km$^3$ solid body \citep{Lagrange1996, Vrignaud2024a}. Dust production rates are also significant. \cite{Lecavelier2022} analyzed 30 photometric events over a five-month observation period with the TESS satellite (average of 1 event every 5 days), and found dust production rates at 1 au ranging between $10^5$ and $10^6$ kg/s. As discussed in the previous subsection, these rates compare with that of Hale-Bopp, which was an exceptionally active comet by Solar System standards. 

Such high production rates may appear contradictory with the dynamical mechanisms proposed to explain the presence of star-grazing comets around \bp\ (i.e., orbital perturbation from \bp\ b and c that slowly elongate the orbits of minor bodies; see the accompanying chapters Mustill et al. and Lu et al. (submitted)), which act over long timescales. However, recent simulations by \cite{Beust2024}, incorporating planetary perturbations and estimates of mass loss at each periastron passage, have helped resolve this apparent contradiction.  The study suggested that a likely reservoir of exocomets exists at $\sim 1$au from the star, corresponding to high-order mean motion resonances with \bp\ c \citep[see the accompanying chapter][for further details]{Bannister2025SSR}).

\begin{figure}[h!]
    \centering
    \includegraphics[scale = 0.18, trim =  50 0 40 0   ]{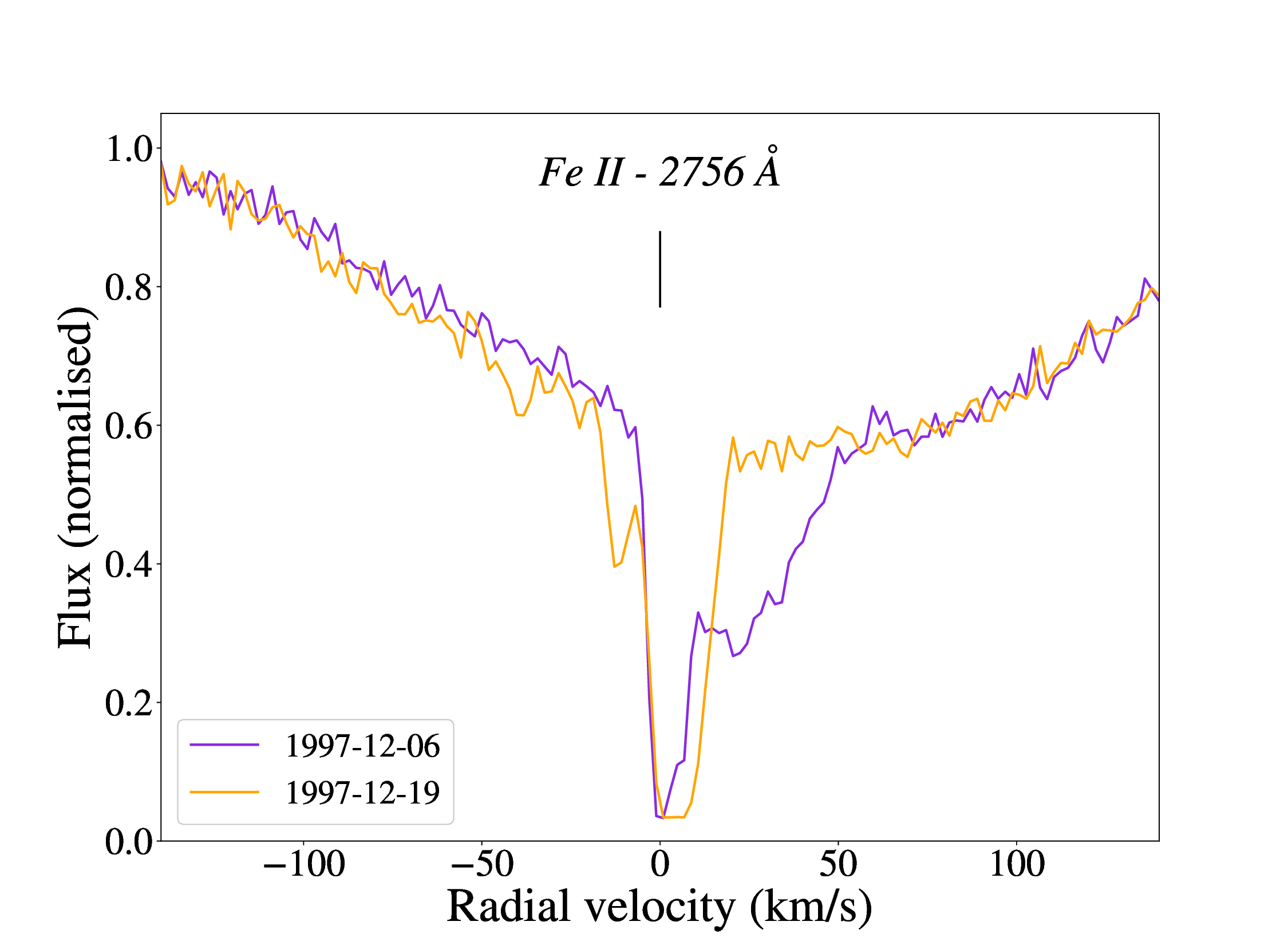}
    \includegraphics[scale = 0.18, trim =  70 0 70 0   ]{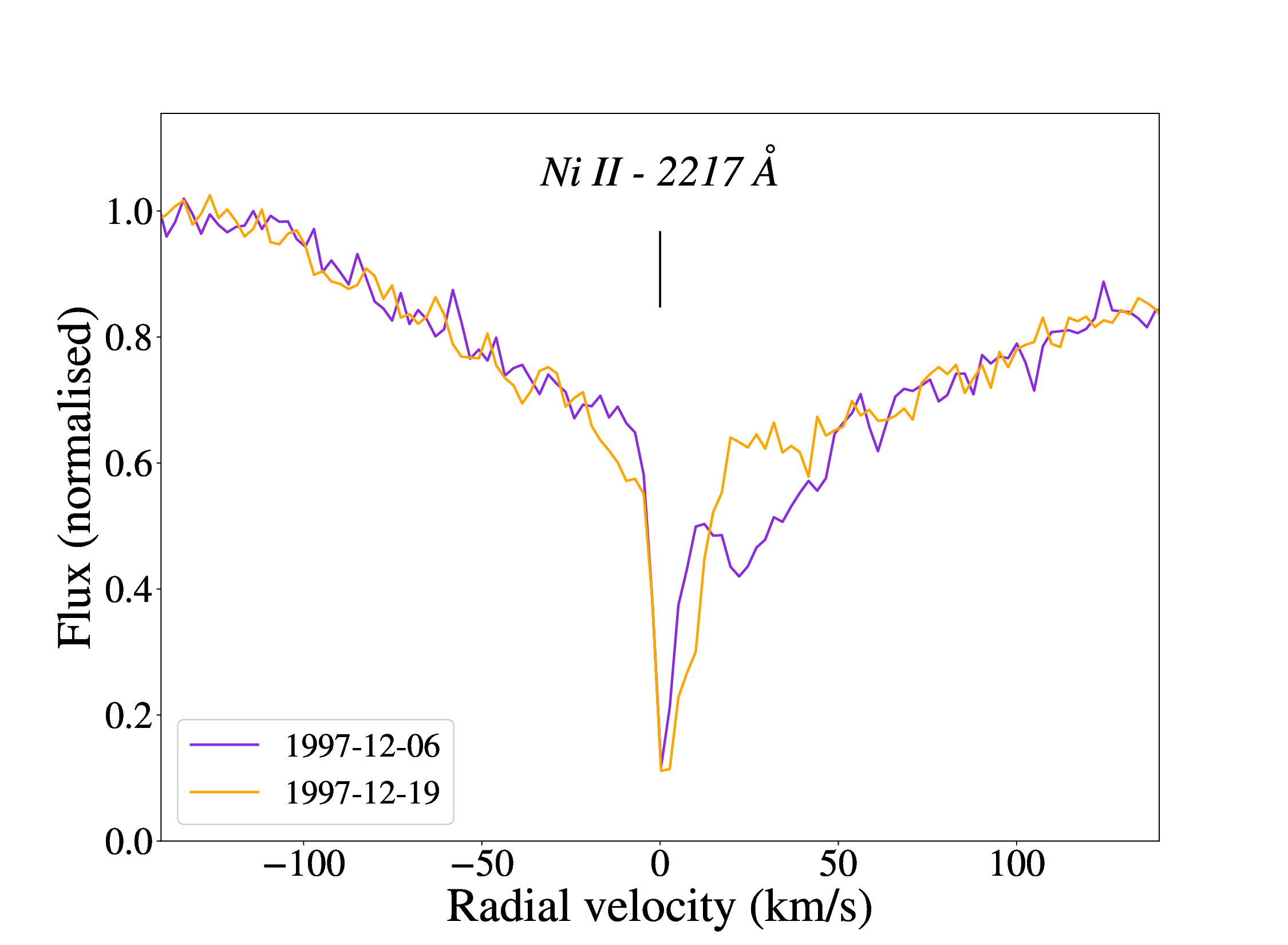}
    \caption{\small Comparison of a \feii\ line (2756 \AA, left) and \niii\ line (2217 \AA, right) in \bp\ spectrum, as observed at two epochs in 1997 with  HST \citep[program 7512;][]{Lagrange1997hst}. A red shifted absorption (between roughly +10 and +50 km~s$^{-1}$) attributed to a transiting exocomet is detected in the December 6, 1997 observation, in both lines. The similarity of the two absorption features suggests that Fe$^+$\ and Ni$^+$\ are well-mixed in the gaseous tail, despite undergoing different radiation pressure.}
    \label{fig: Comparison Fe II - Ni II}
\end{figure}

The ionization level of the gas in the \bp\ system also distinguishes these comets from Solar System comets. As discussed  in Sect. \ref{Sect. Ionisation}, photo-ionization occurs rather slow for Solar System comets;  for instance, at 1 au, the typical timescale for atomic photoionization is approximately 10 days \citep{Huebner1992}, allowing Fe and Ni to be detected \citep{Manfroid2021}. In contrast, around \bp, many abundant elements (e.g., Fe, Na, Ca) are photo-ionized almost instantaneously, within seconds or minutes. The ionization can also be enhanced by electronic collisions, producing highly charged species such as Al$^{2+}$, Si$^{3+}$, and C$^{3+}$. As a result, nearly all the gas that is released by \bp\ exocomets is in the form of a dense plasma ($n_e \sim 10^6$ cm$^{-3}$ for the most active cases). This leads to a markedly different dynamical behavior compared to Solar System comets: instead of distinct, radiation-driven ion tails, the plasma in \bp\ cometary tails undergoes self-braking. Species subject to weaker radiation pressure (e.g., O$^{+}$, C$^{+}$) act to slow down those experiencing stronger acceleration (e.g., Ca$^+$, Fe$^+$). Consequently, the various ions remain well-mixed rather than separating by mass or charge.  This behavior, which is illustrated on Fig. \ref{fig: Comparison Fe II - Ni II}, was highlighted in \cite{Vrignaud2024b} and is discussed further in Section~\ref{Sect. drag forces}.

Future models of ion dynamics around \bp\  will aim to incorporate key physical processes discussed in this paper. These models should account for sublimation rates (Sect. \ref{Sect. Dust sublimation}), the effects of radiation pressure (Sect. \ref{Sect. radiation pressure}), and drag forces within the plasma environment (Sect. \ref{Sect. drag forces}). Additionally, upcoming HST observations planned for 2025 \citep{Vrignaud2024_HST} will offer new perspective on \bp\ exocomets, helping to refine theoretical predictions and improve our understanding of their behavior.

\subsection{AU~Mic}
AU~Mic is a nearby \citep[$\approx 10 $ pc;][]{gaia_collaboration_gaia_2018}, young \citep[$18_{-2.4}^{+2.0}$ Myr;][]{miret-roig_dynamical_2020} M1 dwarf, known for its planetary system, which consists of two confirmed warm Neptunes and two other planetary candidates \citep{donati_magnetic_2023}. The mass estimates for the two confirmed planets range from 10.2 - 14.3 $M_\oplus$ and 14.2 - 34.9 $M_\oplus$ \citep{donati_magnetic_2023,plavchan_planet_2020,klein_one_2022, Wittrock_2023_AU_mic_d} and orbit at a distance of 0.07 au \citep{plavchan_planet_2020} and 0.11 au \citep{zicher_one_2022}, respectively. The system also hosts an edge-on debris disk which extends from 50 to 210 au \citep{kalas_discovery_2004} and an inner dusty belt between 8.8 and 43 au \citep{macgregor_millimeter_2013}. As one of the most extensively studied debris disks around an M dwarf \citep[e.g.,][]{grady_eroding_2020, arnold_stumbling_2022, lawson_jwstnircam_2023, boccaletti_observations_2018}, AU~Mic presents a compelling target for future searches for cometary activity \citep{janson_imaging_2023}. The system’s edge-on orientation, similar to that of \bp, could facilitate the detection of transiting or spectroscopically identifiable comet-like bodies. However, AU~Mic's low luminosity ($L = 0.10~L_{\odot}$) would limit the dust production compared to a Solar System comet, and liberated dust grains would experience smaller radiation pressure than its Solar System counterpart (see Table \ref{table: case study stars}). The shape of a hypothetical AU~Mic comet dust tail is displayed in Fig. \ref{fig: overlaid comets}. Here, it can be seen the dust tail is significantly smaller in size than the Solar System and $\upbeta$~Pic counterparts. This suggests that the photometric detection of transiting exocomet dust tails in the AU~Mic system could be elusive.


M dwarf stars are known to be highly active and TESS light curve data of AU~Mic shows frequent flares occurring regularly  \citep{gilbert_flares_2022}. Space weather modeling suggested it is unlikely planetary atmospheres could survive in such conditions \citep{alvarado-gomez_simulating_2022}. Given that even coronal mass ejections have been observed to disrupt the plasma and dust environments of Solar System comets \citep[e.g.,][]{edberg_cme_2016,jia_study_2009,jian-chun_disconnection_2011,kuchar_observations_2008}, it is reasonable to suggest AU~Mic comets would also be altered by extreme space weather events. Consequently, the combination of relatively low $\beta$ values, poor dust production, and intense space weather would probably undermine the possibility of cometary tails to form and become detectable in the AU~Mic system.

\subsection{WD 1145+017}

Cometary bodies in white dwarf planetary systems have likely already been detected, both directly and indirectly, with more discoveries expected in the future. These exocomets, however, are observed during the late stages of their destruction, or after they have been completely destroyed. Hence, when compared to main-sequence stars, the observational signatures of exocomets around white dwarfs are fundamentally different, as these objects are observed through active tidal breakup and chemical signatures in the stellar photosphere, rather than through the transit of sublimation-driven tails.

One method of detecting exocomets around white dwarfs is by looking in the atmospheres of the white dwarfs themselves. Because of these stars' crushingly high density, the photospheres of most white dwarfs could support only primordial hydrogen and helium. Consequently, any metals detected are transient features from remnant planetary systems \citep{bonsor2024,williams2024,xu2024}. Most of these accreted materials resemble dry, carbon-poor asteroidal material. However, certain detections include elevated levels of carbon and volatile elements—key ingredients for water-rich bodies—suggesting contributions from icy or more primitive objects and highlighting the diverse compositions of small bodies that once orbited the star \citep{farihi2013,raddi2015,gentilefusillo2017,xu2017,hoskin2020,izquierdo2021}. 

An alternative and indirect method to detect exocomets within white dwarf atmospheres is to consider their hydrogen budget as a function of time: as exocomets deposit hydrogen in their atmospheres, this hydrogen cannot escape through winds or sink into the core. Therefore, a model for exocometary accretion as a function of time \citep[e.g.][]{oconnor2023,pham2024}, combined with a large enough observational sample of hydrogen in white dwarfs \citep[e.g.][]{raddi2015}, could pose a statistical argument for the detection of these exocomets \citep{veras2014b}.

\begin{figure}[h!]
    \centering
    \includegraphics[scale = 0.77]{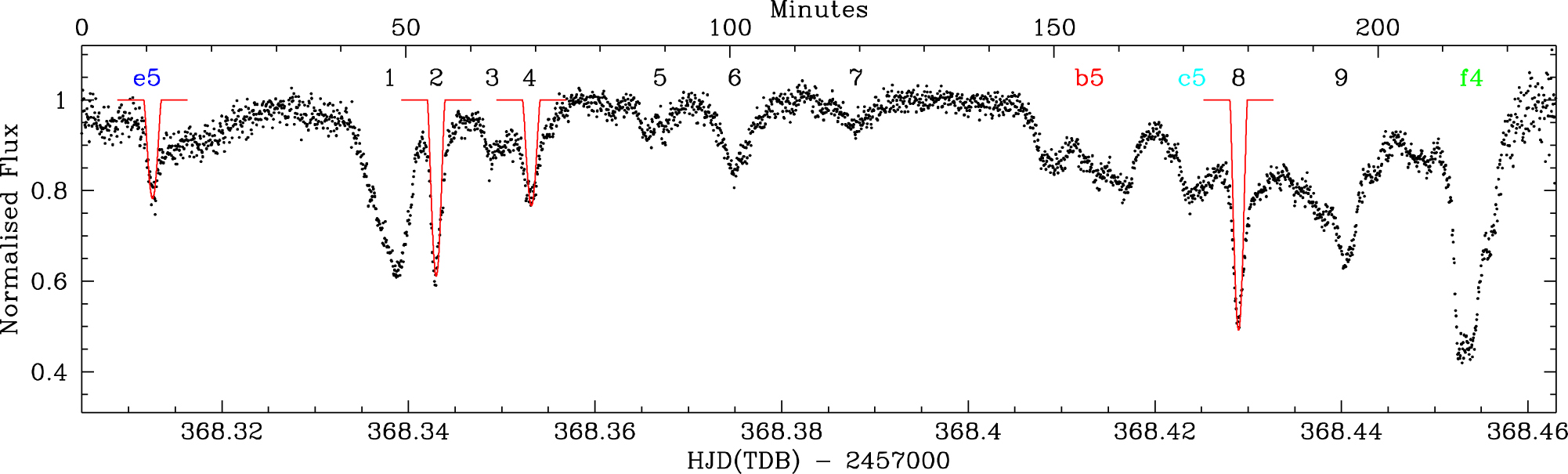}
    \caption{\small Photometry of WD 1145+017 obtained with the TNT/ULTRASPEC camera \citep[reproduction of Fig. 2 of][]{gansicke2016}. The light curve shows the transit of numerous debris over a single night. The alphanumeric labels specify features of interest. These features change shape with subsequent observations, and showcase intricate substructure which is likely associated with active and complex breakup and sublimation of a parent body which may be an exocomet, exo-asteroid, or some hybrid of the two.}
    \label{fig:WD1145Breakup}
\end{figure}

Despite the above considerations, perhaps the most striking observational manifestation of an exocomet orbiting a white dwarf is a photometric transit curve shaped by active sublimation. The first minor planet found to be disrupting around a white dwarf broke up around WD~1145+017 \citep{vanderburg2015}. The minor planet's orbital period was approximately 4.5 hours, and broke up over the course of about 7 years (since first detection in 2015); activity around that star has now, as of 2024, largely ceased \citep{aungwerojwit2024}. The breakup was observed through complex and dynamic photometric transit curves, which differ significantly from those produced by exoplanets transiting their stars and instead resemble exocomet transits. The photometric curves for WD~1145+107 (e.g. Fig. \ref{fig:WD1145Breakup}) are full of structures which indicate ongoing breakup and a potential collisional cascade. How exactly this breakup is linked to sublimation (similarly to comets) as opposed to tidal fragmentation remains an outstanding question, and requires further modeling \citep{veras2017,xu2018,duvvuri2020}, especially as many additional systems like WD~1145+017 are now increasing our sample of disrupted exo-minor planets (whether they be exo-asteroids or exocomets) \citep{guidry2021,malamud2024}.

\FloatBarrier

\begin{sidewaystable}
\centering
\begin{tabularx}{0.998\textwidth}{ l l l l  }    
    \hline 
    \noalign{\smallskip}
    \hline 
    \noalign{\smallskip}
    \textbf{System} & Dust grains & Volatiles & Refractory atoms \& ions \\
    \noalign{\smallskip}
    \hline
    \noalign{\medskip}

                 & Released by volatile sublimation          &    Dominant species : H$_2$O, CO, CO$_2$                 & Complex production mechanisms: \\
                 & Dust tails shaped by radiation              &  Slow photo-dissociation ($\approx 1$ day                & carbonyls sublimation ($\geq 1$ au),  \\
    \textbf{Sun} & pressure                                    &  at 1 au)                                                &  dust sublimation ($\leq 0.1$ au)\\
                 & Power-law size distribution                 &  Ionization by solar EUV photons                         & Slow photoionization ($\approx 1-100$ day \\
                 &                                             &   and solar wind   (H$_2$O$^+$, CO$^+$...)               & at 1 au) \\
    
    \noalign{\medskip}
    \hdashline
    \noalign{\medskip}

                                  & Large production ($\dot M_{\rm 1\, au} \geq 10^5$ kg/s) &  High loss rate ($\sim 10^{33}$ atoms/s)      & Rapid production from grain sublimation\\
    \multirow{2}{*}{\textbf{\bp}} & Rapid sublimation below 0.2 au                          &  Rapid dissociation: CO, H$_2$O... are        & Rapid ionization of neutrals (Fe, Ni, Mg..)\\
                                  & Large radiation pressure ($\beta > 1)$                  &  not detected in the variable gas             & High plasma density: species remain\\
                                  &                                                         &  (although CO is present in the disc)         & well-mixed\\

    \noalign{\medskip}
    \hdashline
    \noalign{\medskip}

                    & Weaker volatile sublimation could               & Smaller volatile outgassing                 &   Low production from dust sublimation\\
                    & induce smaller dust activity                    & Higher photo-ionization rates due to        &   Photoionisation of neutrals and ions is\\
    \textbf{AU~Mic} & Smaller radiation pressure : more               & stronger EUV flux                          &   quicker than in the Solar System\\
                    & compact dust tails                              & Stellar flares accelerate ionization        &   Tail structure affected by flare events \\
                    &                                                 & and affects tail structure                 & \\
    
    \noalign{\medskip}
    \hdashline
    \noalign{\medskip}

                         & Observations at the last stages of           & Volatiles are quickly dissociated, and        & Detected from atmospheric pollution\\
                         & destruction                                  & accreted to the surface of the WD.            & (e.g. Fe II, Si II...) \\
    \textbf{WD 1145+17}  & Dust release from tidal disruption           & Gas detected from atmospheric                 & Observations suggest that the accreted \\
                         & Low radiation pressure ; grain dynamics      & pollution (e.g. C, N, O)                      & material is mostly rocky\\
                         & affected by Poyting-Roberston drag          & & \\

    \noalign{\smallskip}
    \hline
\end{tabularx} 
\captionsetup{width=\textwidth}
\caption{\small Recapitulation of the main properties of exocomets in our four case-study systems. The three columns discuss the observed or expected behavior of dust grains, volatiles (e.g. H$_2$O, CO$_2$, ...) and refractory species (e.g. Fe, Fe$^+$, ...).}
\label{Tab. Differences between systems}
\end{sidewaystable}

\subsection{A bias towards A-type stars?}
\label{Sect. Discusion}

As seen in the previous sections, the physical processes influencing comets can vary greatly depending on the type of stellar environment to which they are exposed. The stellar properties therefore determine the ability of a minor body to produce tails on close approach to its host star, and of these tails to become detectable (either in spectroscopy or photometry) from Earth.

The primary technique for detecting exocomets is \textbf{spectroscopy}, which is sensitive to gaseous and ionized species. The simplest method involves searching for time variable absorption in optically allowed transitions in the visible spectrum, such as \cai, \caii, or \nai\ lines. However, this technique has proven to be heavily biased toward A-type stars, such as \bp. This bias can be attributed to several factors.

On the one hand, exocomets around cool stars will tend to have less activity, due to a lower stellar flux. Indeed, the production rate of a comet (for a given stellar distance) typically scales as $L_\star$, meaning that comets around the Sun and AU~Mic will generally produce 10$\times$ and 100$\times$ (respectively) less material than \bp\ exocomets.  In addition, sublimation lines around cool stars are further in: while refractory dust starts sublimating at $\sim 0.2$~au around \bp\ (Sect. \ref{Sect. Dust sublimation}), this limit falls down to $\sim 0.06$~au around the Sun - a region that is rarely reached by large Solar System comets. This considerably limits our ability to detect exocomets around cool stars, at least when the targeted species are Ca or Na. If we assume that the sublimation lines of refractories have to be located sufficiently far way from the star, say 0.1 au, for exocomets to produce significant amounts of refractory gas, we find that the minimum effective temperature above which exocomets can be detected in optical spectroscopy is around 7000 K, corresponding to a $\sim$ F1V type.

On the other hand, strong stellar luminosities may undermine the detectability of potential exocomets. In particular, we saw in Sect. \ref{Sect. Ionisation} that, above $\sim 9\,500$ K ($\approx$ A1V), the photospheric flux starts to quickly photo-ionize Ca$^+$\ into Ca$^{2+}$, limiting our ability to detect exocomets using optical spectral lines like \caii. Detecting exocomets around O-B stars would therefore require targeting lines from more resistant species in the UV (e.g \cii, \aliii, \siiv, \civ...), which is observationally challenging. As pointed out by \cite{Rebollido_PhD}, another limitation of O-B stars is that ice lines are located at very large stellocentric distances ($\geq 100$ au, Fig. \ref{fig:sublimationlines MS}). Exocomets in those systems would thus start to show activity very far from the star, where the transit probability is extremely low, and could be fully destroyed before reaching the $\sim$ 1~au region (where the transit - and thus detection - probability becomes significant).

Even though the proposed thresholds are only approximate, they suggest that the favored $T_{\rm eff}$ for detecting exocomets in visible-light spectroscopy is $\sim 7000 - 9500$ K, or F1V~-~A1V. A-type stars thus naturally appear to be the most favored systems for exocomet observations, explaining why most previous detections - or hints of detection - have been obtained around such stars \citep[][]{Rebollido_PhD}. 

Detecting exocometary gas around cooler stars may instead require focusing on spectral lines other than \caii\ or \nai, since these are mainly produced from dust sublimation. For instance, the infrared bands of H$_2$O or CO$_2$ may represent better targets, in the case where the gas in not strongly dissociation nor ionized. These bands are accessible with the JWST, which could allows us to detect exocomets around Sun-like stars. 

The other primary method for detecting exocomets is \textbf{photometry}, which relies on observing the transit of the dust tail. This technique could, a priori, allow for the detection of exocomets across a broader range of stellar effective temperatures, as dust activity can be triggered even around cool stars. For instance, we saw in Sect. \ref{Sect. Solar System comets} that the transit of a Hale-Bopp-like comet around a Sun-like star would be detectable in photometry, if the observation conditions are favorable. Yet, very little photometric detections of exocomets have been made, the clearest being \bp\ \citep[A5V, $T_{\rm eff} = 8000$ K; see][]{zieba2019} and KIC 3542116 \& 11084727 \citep[F2V, $T_{\rm eff} = 6800$ K; see][]{Rappaport2018}. The existence of some bias regarding photometric detections thus remains uncertain. It is however possible that hotter stars, such as A-type stars, are also favored, as their higher luminosities would naturally trigger larger dust productions.


\section{Conclusion}
Exocomets provide a direct means of probing the composition and dynamical evolution of planetary systems beyond our own. Their activity, likely driven by sublimation, desorption, sputtering, or impacts, reveals how small bodies respond to stellar irradiation, winds, and magnetic fields under conditions far more diverse than those in the Solar System. These processes, when viewed across different stellar environments, reveal not only the diversity of small-body evolution but also the shared physical mechanisms that shape planetary systems throughout the Galaxy.

By comparing the Sun, \bp, AU~Mic, and WD 1145+017, we highlight that stellar luminosity and activity fundamentally govern the onset and efficiency of mass loss. Around hot A-type stars like \bp, high radiation pressure and elevated temperatures shift the ice lines outward, allowing CO- and CO$_2$-driven activity far from the star. Around active M dwarfs like AU~Mic, strong stellar winds and flares may dominate material erosion and dust removal. Around white dwarfs such as WD 1145+017, the remnants of small bodies accrete onto the stellar surface, producing the metal-polluted atmospheres that reveal the composition of disrupted planetary material.

The morphology and composition of exocometary tails can reveal valuable diagnostics of the local environment. The interplay between radiation pressure, stellar winds, and magnetic fields determines the balance between gas and dust, the ionization structure of the coma, and the persistence of charged species. In systems like \bp, the dense, ionized plasma tails exhibit strong coupling between ions of differing radiation pressure, leading to self-braking and compositional mixing—behaviors unseen in Solar System comets. Around low-luminosity M-dwarfs, faint dust emission and intense space weather may render transiting tails nearly undetectable.

Future progress will depend on coordinated, multi-wavelength observations coupled with advances in thermophysical and dynamical modeling. UV and optical spectroscopy will continue to constrain gas composition and plasma behavior, while infrared measurements with facilities such as JWST and Ariel will probe volatile release and dust-grain properties. High-resolution time-domain surveys will expand the census of exocometary systems and allow direct measurements of variability and mass-loss rates. By linking these observations with detailed modeling of sublimation, ion dynamics, and radiative transfer, we can reconstruct how small bodies evolve under different stellar conditions and how their material is redistributed into circumstellar environments.

\backmatter

\bmhead{Acknowledgements}

We gratefully acknowledge the International Space Science Insitute, ISSI, Bern, for supporting  and hosting the workshop on ``Exocomets: Bridging our Understanding of Minor Bodies in Solar and Exoplanetary Systems'', during which this work was initiated in July 2024. \newline
We would like to thank Dr. Carl Schmidt (BU) for valuable discussions and feedback on release mechanisms.\newline
We thank Dr. Yankin Wu for kindly sharing the panchromatic spectral model of \bp\ presented in \cite{Wu2025_betapic}. \newline
We thank Dr. Steven Bromley for sending the fluorescence spectra of Fe and Ni around our case study stars.\newline
T.V. acknowledges funding from the Centre National d’\'Etudes Spatiales (CNES).\newline
D.Z.S. is supported by an NSF Astronomy and Astrophysics Postdoctoral Fellowship under award AST-2303553. This research award is partially funded by a generous gift of Charles Simonyi to the NSF Division of Astronomical Sciences. The award is made in recognition of significant contributions to Rubin Observatory’s Legacy Survey of Space and Time. \newline
T.D.P. is supported by a UKRI Stephen Hawking Fellowship.\newline
J.H. acknowledges support by the UK Science and Technology Facilities Council (STFC) through grant number ST/Y509784/1. \\

\noindent The views expressed in this article are those of the authors and do not reflect the official policy or position of the U.S. Naval Academy, Department of the Navy, the Department of Defense, or the U.S. Government.

\bmhead{Conflict of interest/Competing interests}
Not applicable.
\bigskip

\bibliography{sn-bibliography}

@ARTICLE{Asphaug1996,
       author = {{Asphaug}, Erik and {Benz}, Willy},
        title = "{Size, Density, and Structure of Comet Shoemaker-Levy 9 Inferred from the Physics of Tidal Breakup}",
      journal = {\icarus},
         year = 1996,
        month = jun,
       volume = {121},
       number = {2},
        pages = {225-248},
          doi = {10.1006/icar.1996.0083},
       adsurl = {https://ui.adsabs.harvard.edu/abs/1996Icar..121..225A}
}

@ARTICLE{Snodgrass2025,
       author = {{Snodgrass}, Colin and {Holt}, Carrie E. and {Kelley}, Michael S.~P. and {Opitom}, Cyrielle and {Guilbert-Lepoutre}, Aur{\'e}lie and {Knight}, Matthew M. and {Kokotanekova}, Rosita and {Jehin}, Emmanuel and {Mazzotta Epifani}, Elena and {Migliorini}, Alessandra and {Tubiana}, Cecilia and {Micheli}, Marco and {Farnocchia}, Davide},
        title = "{First JWST spectrum of distant activity in long-period comet C/2024 E1 (Wierzchos)}",
      journal = {\mnras},
         year = 2025,
        month = jul,
       volume = {541},
       number = {1},
        pages = {L8-L13},
          doi = {10.1093/mnrasl/slaf046},
archivePrefix = {arXiv},
       eprint = {2503.14071},
 primaryClass = {astro-ph.EP},
       adsurl = {https://ui.adsabs.harvard.edu/abs/2025MNRAS.541L...8S}
}

@ARTICLE{Parhi2025,
       author = {{Parhi}, Adam and {Prialnik}, Dina},
        title = "{Combined Orbital and Thermal Evolution of Oort Cloud Comets}",
      journal = {arXiv e-prints},
         year = 2025,
        month = oct,
          eid = {arXiv:2510.26549},
        pages = {arXiv:2510.26549},
          doi = {10.48550/arXiv.2510.26549},
archivePrefix = {arXiv},
       eprint = {2510.26549},
 primaryClass = {astro-ph.EP},
       adsurl = {https://ui.adsabs.harvard.edu/abs/2025arXiv251026549P}
}

@ARTICLE{Sekanina1994,
       author = {{Sekanina}, Z. and {Chodas}, P.~W. and {Yeomans}, D.~K.},
        title = "{Tidal disruption and the appearance of periodic comet Shoemaker-Levy 9.}",
      journal = {\aap},
         year = 1994,
        month = sep,
       volume = {289},
        pages = {607-636},
       adsurl = {https://ui.adsabs.harvard.edu/abs/1994A&A...289..607S}
}

@INCOLLECTION{Knight2024,
       author = {{Knight}, Matthew M. and {Kokotanekova}, Rosita and {Samarasinha}, Nalin H.},
        title = "{Physical and Surface Properties of Comet Nuclei from Remote Observations}",
    booktitle = {Comets III},
         year = 2024,
       editor = {{Meech}, Karen. J. and {Combi}, Michael. R. and {Bockel{\'e}e-Morvan}, Dominique and {Raymodn}, Sean. N. and {Zolensky}, Michael. E.},
        pages = {361-404},
          doi = {10.2458/azu_uapress_9780816553631-ch012},
       adsurl = {https://ui.adsabs.harvard.edu/abs/2024come.book..361K}
}

@ARTICLE{Weaver1994,
       author = {{Weaver}, H.~A. and {Feldman}, P.~D. and {A'Hearn}, M.~F. and {Arpigny}, C. and {Brown}, R.~A. and {Helin}, E.~F. and {Levy}, D.~H. and {Marsden}, B.~G. and {Meech}, K.~J. and {Larson}, S.~M. and {Noll}, K.~S. and {Scotti}, J.~V. and {Sekanina}, Z. and {Shoemaker}, C.~S. and {Shoemaker}, E.~M. and {Smith}, T.~E. and {Storrs}, A.~D. and {Yeomans}, D.~K. and {Zellner}, B.},
        title = "{Hubble Space Telescope Observations of Comet P/Shoemaker-Levy 9 (1993e)}",
      journal = {Science},
         year = 1994,
        month = feb,
       volume = {263},
       number = {5148},
        pages = {787-791},
          doi = {10.1126/science.263.5148.787},
       adsurl = {https://ui.adsabs.harvard.edu/abs/1994Sci...263..787W}
}

@ARTICLE{Hammel1995,
       author = {{Hammel}, H.~B. and {Beebe}, R.~F. and {Ingersoll}, A.~P. and {Orton}, G.~S. and {Mills}, J.~R. and {Simon}, A.~A. and {Chodas}, P. and {Clarke}, J.~T. and {de Jong}, E. and {Dowling}, T.~E. and {Harrington}, J. and {Huber}, L.~F. and {Karkoschka}, E. and {Santori}, C.~M. and {Toigo}, A. and {Yeomans}, D. and {West}, R.~A.},
        title = "{HST Imaging of Atmospheric Phenomena Created by the Impact of Comet Shoemaker-Levy 9}",
      journal = {Science},
         year = 1995,
        month = mar,
       volume = {267},
       number = {5202},
        pages = {1288-1296},
          doi = {10.1126/science.7871425},
       adsurl = {https://ui.adsabs.harvard.edu/abs/1995Sci...267.1288H}
}

@ARTICLE{Bottke2005,
       author = {{Bottke}, William F. and {Durda}, Daniel D. and {Nesvorn{\'y}}, David and {Jedicke}, Robert and {Morbidelli}, Alessandro and {Vokrouhlick{\'y}}, David and {Levison}, Hal},
        title = "{The fossilized size distribution of the main asteroid belt}",
      journal = {\icarus},
         year = 2005,
        month = may,
       volume = {175},
       number = {1},
        pages = {111-140},
          doi = {10.1016/j.icarus.2004.10.026},
       adsurl = {https://ui.adsabs.harvard.edu/abs/2005Icar..175..111B}
}

@INCOLLECTION{Nesvorny2015,
       author = {{Nesvorn{\'y}}, D. and {Bro{\v{z}}}, M. and {Carruba}, V.},
        title = "{Identification and Dynamical Properties of Asteroid Families}",
    booktitle = {Asteroids IV},
         year = 2015,
       editor = {{Michel}, Patrick and {DeMeo}, Francesca E. and {Bottke}, William F.},
        pages = {297-321},
          doi = {10.2458/azu_uapress_9780816532131-ch016},
       adsurl = {https://ui.adsabs.harvard.edu/abs/2015aste.book..297N}
}

@ARTICLE{Bottke2017,
       author = {{Bottke}, William F. and {Norman}, Marc D.},
        title = "{The Late Heavy Bombardment}",
      journal = {Annual Review of Earth and Planetary Sciences},
         year = 2017,
        month = aug,
       volume = {45},
       number = {1},
        pages = {619-647},
          doi = {10.1146/annurev-earth-063016-020131},
       adsurl = {https://ui.adsabs.harvard.edu/abs/2017AREPS..45..619B}
}

@ARTICLE{Wetherill1975,
       author = {{Wetherill}, G.~W.},
        title = "{Late heavy bombardment of the moon and terrestrial planets.}",
      journal = {Lunar and Planetary Science Conference Proceedings},
         year = 1975,
        month = jan,
       volume = {2},
        pages = {1539-1561},
       adsurl = {https://ui.adsabs.harvard.edu/abs/1975LPSC....6.1539W}
}

@ARTICLE{Strom2005,
       author = {{Strom}, Robert G. and {Malhotra}, Renu and {Ito}, Takashi and {Yoshida}, Fumi and {Kring}, David A.},
        title = "{The Origin of Planetary Impactors in the Inner Solar System}",
      journal = {Science},
         year = 2005,
        month = sep,
       volume = {309},
       number = {5742},
        pages = {1847-1850},
          doi = {10.1126/science.1113544},
archivePrefix = {arXiv},
       eprint = {astro-ph/0510200},
 primaryClass = {astro-ph},
       adsurl = {https://ui.adsabs.harvard.edu/abs/2005Sci...309.1847S}
}

@ARTICLE{Strom2015,
       author = {{Strom}, Robert G. and {Renu}, Malhotra and {Xiao}, Zhi-Yong and {Ito}, Takashi and {Yoshida}, Fumi and {Ostrach}, Lillian R.},
        title = "{The inner solar system cratering record and the evolution of impactor populations}",
      journal = {Research in Astronomy and Astrophysics},
         year = 2015,
        month = mar,
       volume = {15},
       number = {3},
          eid = {407},
        pages = {407},
          doi = {10.1088/1674-4527/15/3/009},
archivePrefix = {arXiv},
       eprint = {1407.4521},
 primaryClass = {astro-ph.EP},
       adsurl = {https://ui.adsabs.harvard.edu/abs/2015RAA....15..407S}
}

@ARTICLE{Bryans2012,
       author = {{Bryans}, P. and {Pesnell}, W.~D.},
        title = "{The Extreme-ultraviolet Emission from Sun-grazing Comets}",
      journal = {\apj},
         year = 2012,
        month = nov,
       volume = {760},
       number = {1},
          eid = {18},
        pages = {18},
          doi = {10.1088/0004-637X/760/1/18},
archivePrefix = {arXiv},
       eprint = {1209.5708},
 primaryClass = {astro-ph.SR},
       adsurl = {https://ui.adsabs.harvard.edu/abs/2012ApJ...760...18B}
}

@ARTICLE{Hui2018,
       author = {{Hui}, Man-To and {Jewitt}, David and {Clark}, David},
        title = "{Pre-discovery Observations and Orbit of Comet C/2017 K2 (PANSTARRS)}",
      journal = {\aj},
         year = 2018,
        month = jan,
       volume = {155},
       number = {1},
          eid = {25},
        pages = {25},
          doi = {10.3847/1538-3881/aa9be1},
archivePrefix = {arXiv},
       eprint = {1711.06355},
 primaryClass = {astro-ph.EP},
       adsurl = {https://ui.adsabs.harvard.edu/abs/2018AJ....155...25H}
}

@INPROCEEDINGS{Bennett2010,
       author = {{Bennett}, P.~D.},
        title = "{Chromospheres and Winds of Red Supergiants: An Empirical Look at Outer Atmospheric Structure}",
    booktitle = {Hot and Cool: Bridging Gaps in Massive Star Evolution},
         year = 2010,
       editor = {{Leitherer}, C. and {Bennett}, P.~D. and {Morris}, P.~W. and {Van Loon}, J. Th.},
       series = {Astronomical Society of the Pacific Conference Series},
       volume = {425},
        month = jun,
        pages = {181},
          doi = {10.48550/arXiv.1004.1853},
archivePrefix = {arXiv},
       eprint = {1004.1853},
 primaryClass = {astro-ph.SR},
       adsurl = {https://ui.adsabs.harvard.edu/abs/2010ASPC..425..181B}
}

@ARTICLE{Vidotto2021,
       author = {{Vidotto}, Aline A.},
        title = "{The evolution of the solar wind}",
      journal = {Living Reviews in Solar Physics},
         year = 2021,
        month = dec,
       volume = {18},
       number = {1},
          eid = {3},
        pages = {3},
          doi = {10.1007/s41116-021-00029-w},
archivePrefix = {arXiv},
       eprint = {2103.15748},
 primaryClass = {astro-ph.SR},
       adsurl = {https://ui.adsabs.harvard.edu/abs/2021LRSP...18....3V}
}

@ARTICLE{Killen2012_moon,
       author = {{Killen}, R.~M. and {Hurley}, D.~M. and {Farrell}, W.~M.},
        title = "{The effect on the lunar exosphere of a coronal mass ejection passage}",
      journal = {Journal of Geophysical Research (Planets)},
         year = 2012,
        month = mar,
       volume = {117},
          eid = {E00K02},
        pages = {E00K02},
          doi = {10.1029/2011JE004011},
       adsurl = {https://ui.adsabs.harvard.edu/abs/2012JGRE..117.0K02K}
}

@ARTICLE{Phillips1977,
       author = {{Phillips}, E. and {Lee}, L.~C. and {Judge}, D.~L.},
        title = "{Absolute photoabsorption cross sections for H2O and D2O from lambda 180-790 A}",
      journal = {\jqsrt},
         year = 1977,
        month = sep,
       volume = {18},
        pages = {309-313},
          doi = {10.1016/0022-4073(77)90061-9},
       adsurl = {https://ui.adsabs.harvard.edu/abs/1977JQSRT..18..309P}
}

@ARTICLE{Ragot2003,
       author = {{Ragot}, B.~R. and {Kahler}, S.~W.},
        title = "{Interactions of Dust Grains with Coronal Mass Ejections and Solar Cycle Variations of the F-Coronal Brightness}",
      journal = {\apj},
         year = 2003,
        month = sep,
       volume = {594},
       number = {2},
        pages = {1049-1059},
          doi = {10.1086/377076},
       adsurl = {https://ui.adsabs.harvard.edu/abs/2003ApJ...594.1049R}
}

@ARTICLE{Neugebauer2007,
       author = {{Neugebauer}, M. and {Gloeckler}, G. and {Gosling}, J.~T. and {Rees}, A. and {Skoug}, R. and {Goldstein}, B.~E. and {Armstrong}, T.~P. and {Combi}, M.~R. and {M{\"a}kinen}, T. and {McComas}, D.~J. and {von Steiger}, R. and {Zurbuchen}, T.~H. and {Smith}, E.~J. and {Geiss}, J. and {Lanzerotti}, L.~J.},
        title = "{Encounter of the Ulysses Spacecraft with the Ion Tail of Comet MCNaught}",
      journal = {\apj},
         year = 2007,
        month = oct,
       volume = {667},
       number = {2},
        pages = {1262-1266},
          doi = {10.1086/521019},
       adsurl = {https://ui.adsabs.harvard.edu/abs/2007ApJ...667.1262N}
}

@ARTICLE{Povich2003,
       author = {{Povich}, Matthew S. and {Raymond}, John C. and {Jones}, Geraint H. and {Uzzo}, Michael and {Ko}, Yuan-Kuen and {Feldman}, Paul D. and {Smith}, Peter L. and {Marsden}, Brian G. and {Woods}, Thomas N.},
        title = "{Doubly Ionized Carbon Observed in the Plasma Tail of Comet Kudo-Fujikawa}",
      journal = {Science},
         year = 2003,
        month = dec,
       volume = {302},
       number = {5652},
        pages = {1949-1952},
          doi = {10.1126/science.1092142},
       adsurl = {https://ui.adsabs.harvard.edu/abs/2003Sci...302.1949P}
}

@ARTICLE{Yakshinskiy1999,
       author = {{Yakshinskiy}, B.~V. and {Madey}, T.~E.},
        title = "{Photon-stimulated desorption as a substantial source of sodium in the lunar atmosphere}",
      journal = {\nat},
         year = 1999,
        month = aug,
       volume = {400},
       number = {6745},
        pages = {642-644},
          doi = {10.1038/23204},
       adsurl = {https://ui.adsabs.harvard.edu/abs/1999Natur.400..642Y}
}

@inbook{McClintock2018, place={Cambridge}, series={Cambridge Planetary Science}, title={Observations of Mercury’s Exosphere: Composition and Structure}, booktitle={Mercury: The View after MESSENGER}, publisher={Cambridge University Press}, author={McClintock, William E. and Cassidy, Timothy A. and Merkel, Aimee W. and Killen, Rosemary M. and Burger, Matthew H. and Vervack, Ronald J.}, editor={Solomon, Sean C. and Nittler, Larry R. and Anderson, Brian J.Editors}, year={2018}, pages={371–406}, collection={Cambridge Planetary Science}}

@article{Pinto2023, 
year = {2023}, 
title = {{First Detection of CO2 Emission in a Centaur: JWST NIRSpec Observations of 39P/Oterma}}, 
author = {Pinto, O. Harrington and Kelley, M. S. P. and Villanueva, G. L. and Womack, M. and Faggi, S. and McKay, A. and DiSanti, M. A. and Schambeau, C. and Fernandez, Y. and Bauer, J. and Feaga, L. and Wierzchos, K.}, 
journal = {The Planetary Science Journal}, 
doi = {10.3847/psj/acf928}, 
pages = {208}, 
number = {11}, 
volume = {4}
}

@article{Eichhorn1978,
  author    = {G. Eichhorn},
  title     = {Time-resolved temperature measurements in hypervelocity dust impact},
  journal   = {Planetary and Space Science},
  year      = {1978},
  volume    = {26},
  number    = {5},
  pages     = {463--467},
  doi       = {10.1016/0032-0633(78)90006-0},
  url       = {https://www.sciencedirect.com/science/article/pii/0032063378900060}
}

@incollection{Combi2004, 
year = {2004}, 
title = {{Gas dynamics and kinetics in the cometary coma: theory and observations}}, 
author = {Combi, Michael R and Harris, Walter M and Smyth, William H}, 
editor = {Weaver, Harold A and Keller, Horst Uwe and  and Festou, M}, 
url = {http://adsabs.harvard.edu/cgi-bin/nph-data\_query?bibcode=2004come.book..523C\&link\_type=ABSTRACT}, 
urldate = {0}, 
pages = {523}, 
series = {Comets II}, 
publisher = {University of Arizona Press}
}

@incollection{Feldman2004, 
year = {2004}, 
title = {{Spectroscopic Investigations of Fragment Species in the Coma}}, 
author = {Feldman, paul D and Cochran, Anita L and Combi, Michael R}, 
editor = {Weaver, Harold A and Festou, M and  and Keller, Horst Uwe}, 
urldate = {0}, 
pages = {425}, 
series = {Comets II}, 
publisher = {University of Arizona Press}, 
month = {02}
}

@article{Pokorny2018, 
year = {2018}, 
title = {{A Comprehensive Model of the Meteoroid Environment around Mercury}}, 
author = {Pokorný, Petr and Sarantos, Menelaos and Janches, Diego}, 
journal = {The Astrophysical Journal}, 
issn = {0004-637X}, 
doi = {10.3847/1538-4357/aad051}, 
eprint = {1807.02749}, 
pages = {31}, 
number = {1}, 
volume = {863}
}

@ARTICLE{Kislyakova2024,
       author = {{Kislyakova}, K.~G. and {G{\"u}del}, M. and {Koutroumpa}, D. and {Carter}, J.~A. and {Lisse}, C.~M. and {Boro Saikia}, S.},
        title = "{X-ray detection of astrospheres around three main-sequence stars and their mass-loss rates}",
      journal = {Nature Astronomy},
         year = 2024,
        month = may,
       volume = {8},
        pages = {596-605},
          doi = {10.1038/s41550-024-02222-x},
archivePrefix = {arXiv},
       eprint = {2404.14980},
 primaryClass = {astro-ph.SR},
       adsurl = {https://ui.adsabs.harvard.edu/abs/2024NatAs...8..596K}
}

@article{Fortenberry2021, 
year = {2021}, 
title = {{Knowledge Gaps in the Cometary Spectra of Oxygen-bearing Molecular Cations}}, 
author = {Fortenberry, Ryan C. and Bodewits, Dennis and Pierce, Donna M.}, 
journal = {The Astrophysical Journal Supplement Series}, 
issn = {0067-0049}, 
doi = {10.3847/1538-4365/ac0cfd}, 
pages = {6}, 
number = {1}, 
volume = {256}
}

@article{Dennerl2012, 
year = {2012}, 
title = {{Solar system X‐rays from charge exchange processes}}, 
author = {Dennerl, Konrad and Lisse, Carey M and Bhardwaj, Anil and Christian, Damian J and Wolk, Scott J and Bodewits, Dennis and Zurbuchen, T H and Combi, Michael R and Lepri, Susan T}, 
journal = {Astronomische Nachrichten}, 
doi = {10.1002/asna.201211663}, 
pages = {324 -- 334}, 
number = {4}, 
volume = {333}, 
language = {English}, 
month = {04}
}

@ARTICLE{Szalay2018,
       author = {{Szalay}, J.~R. and {Poppe}, A.~R. and {Agarwal}, J. and {Britt}, D. and {Belskaya}, I. and {Hor{\'a}nyi}, M. and {Nakamura}, T. and {Sachse}, M. and {Spahn}, F.},
        title = "{Dust Phenomena Relating to Airless Bodies}",
      journal = {\ssr},
         year = 2018,
        month = aug,
       volume = {214},
       number = {5},
          eid = {98},
        pages = {98},
          doi = {10.1007/s11214-018-0527-0},
       adsurl = {https://ui.adsabs.harvard.edu/abs/2018SSRv..214...98S}
}

@ARTICLE{Cassidy2021,
       author = {{Cassidy}, T.~A. and {Schmidt}, C.~A. and {Merkel}, A.~W. and {Jasinski}, J.~M. and {Burger}, M.~H.},
        title = "{Detection of Large Exospheric Enhancements at Mercury due to Meteoroid Impacts}",
      journal = {\psj},
         year = 2021,
        month = oct,
       volume = {2},
       number = {5},
          eid = {175},
        pages = {175},
          doi = {10.3847/PSJ/ac1a19},
       adsurl = {https://ui.adsabs.harvard.edu/abs/2021PSJ.....2..175C}
}

@article{Killen2016, 
year = {2016}, 
title = {{Pathways for energization of Ca in Mercury’s exosphere}}, 
author = {Killen, Rosemary M.}, 
journal = {Icarus}, 
issn = {0019-1035}, 
doi = {10.1016/j.icarus.2015.12.035}, 
pages = {32--36}, 
volume = {268}
}

@article{Stern1999,
  author = {Stern, S. A.},
  title = {The lunar atmosphere: History, status, current problems, and context},
  journal = {Reviews of Geophysics},
  volume = {37},
  number = {4},
  pages = {453-491},
  year = {1999},
  doi = {10.1029/1999RG900005}
}

@article{Nie2024, 
year = {2024}, 
title = {{Lunar soil record of atmosphere loss over eons}}, 
author = {Nie, Nicole X. and Dauphas, Nicolas and Zhang, Zhe J. and Hopp, Timo and Sarantos, Menelaos}, 
journal = {Science Advances}, 
doi = {10.1126/sciadv.adm7074}, 
pmid = {39093970}, 
pmcid = {PMC11296337}, 
pages = {eadm7074}, 
number = {31}, 
volume = {10}
}

@ARTICLE{Collette2014,
       author = {{Collette}, A. and {Sternovsky}, Z. and {Horanyi}, M.},
        title = "{Production of neutral gas by micrometeoroid impacts}",
      journal = {\icarus},
         year = 2014,
        month = jan,
       volume = {227},
        pages = {89-93},
          doi = {10.1016/j.icarus.2013.09.009},
       adsurl = {https://ui.adsabs.harvard.edu/abs/2014Icar..227...89C}
}

@article{Johnson1991,
  author = {Johnson, R. E. and Baragiola, R. A.},
  title = {Lunar surface: Sputtering and secondary ion mass spectrometry},
  journal = {Geophysical Research Letters},
  volume = {18},
  number = {11},
  pages = {2063-2066},
  year = {1991},
  doi = {10.1029/91GL02664}
}

@article{Cassidy2005,
  author = {Cassidy, T. A. and Johnson, R. E.},
  title = {Monte Carlo model of sputtering and other ejection processes within a regolith},
  journal = {Icarus},
  volume = {176},
  number = {2},
  pages = {499-507},
  year = {2005},
  doi = {10.1016/j.icarus.2005.02.013}
}

@book{Tielens2005,
  author = {Tielens, A. G. G. M.},
  title = {The Physics and Chemistry of the Interstellar Medium},
  publisher = {Cambridge University Press},
  year = {2005},
  doi = {10.1017/CBO9780511810076}
}

@article{Plainaki2012,
  author = {Plainaki, C. and Milillo, A. and Mura, A. and Orsini, S. and Massetti, S.},
  title = {Exospheric O2 at Europa: The result of radiolytic, thermal, and ion-sputtering processes},
  journal = {Planetary and Space Science},
  volume = {88},
  pages = {42-52},
  year = {2012},
  doi = {10.1016/j.pss.2013.07.011}
}

@article{Killen2012_mercury,
  author = {Killen, R. M. and Sarantos, M. and Potter, A. E. and Reiff, P.},
  title = {The effect of solar wind variability on the Mercury exosphere},
  journal = {Icarus},
  volume = {209},
  number = {1},
  pages = {75-87},
  year = {2012},
  doi = {10.1016/j.icarus.2010.08.005}
}

@article{Vorburger2012,
  author = {Vorburger, A. and Wurz, P. and Barabash, S. et al.},
  title = {Energetic neutral atom imaging of surface sputtering on the Moon},
  journal = {Journal of Geophysical Research: Planets},
  volume = {117},
  year = {2012},
  doi = {10.1029/2011JE004092}
}

@article{Poppe2016,
  author = {Poppe, A. R. and Curry, S. M. and Fatemi, S. et al.},
  title = {Solar wind interaction with the Moon: A synthesis of disparate measurements},
  journal = {Journal of Geophysical Research: Space Physics},
  volume = {121},
  pages = {4129-4143},
  year = {2016},
  doi = {10.1002/2016JA022529}
}

@article{Johnson2013,
  author = {Johnson, R. E.},
  title = {Sputtering and composition of planetary surfaces},
  journal = {Reviews of Modern Physics},
  volume = {85},
  pages = {801-812},
  year = {2013},
  doi = {10.1103/RevModPhys.85.801}
}

@article{Sigmund1981,
  author = {Sigmund, P.},
  title = {Sputtering by ion bombardment: Theoretical concepts},
  journal = {Sputtering by Particle Bombardment I, Springer, Berlin, Heidelberg},
  pages = {9-71},
  year = {1981},
  doi = {10.1007/3540111347}
}

@article{Janches2021, 
year = {2021}, 
title = {{Meteoroids as One of the Sources for Exosphere Formation on Airless Bodies in the Inner Solar System}}, 
author = {Janches, Diego and Berezhnoy, Alexey A. and Christou, Apostolos A. and Cremonese, Gabriele and Hirai, Takayuki and Horányi, Mihály and Jasinski, Jamie M. and Sarantos, Menelaos}, 
journal = {Space Science Reviews}, 
issn = {0038-6308}, 
doi = {10.1007/s11214-021-00827-6}, 
pages = {50}, 
number = {4}, 
volume = {217}
}

@article{Shi1995,
  author = {Shi, J. and Baragiola, R. A. and Grosjean, D. E. and Johnson, R. E.},
  title = {Sputtering of water ice surfaces and the production of extended atmospheres},
  journal = {Journal of Geophysical Research: Planets},
  volume = {100},
  pages = {26387-26395},
  year = {1995},
  doi = {10.1029/95JE02822}
}

@article{Fama2008,
  author = {Famá, M. and Shi, J. and Baragiola, R. A.},
  title = {Sputtering of ice by low-energy ions},
  journal = {Surface Science},
  volume = {602},
  number = {1},
  pages = {156-161},
  year = {2008},
  doi = {10.1016/j.susc.2007.10.011}
}

@article{Farnham2009, 
year = {2009}, 
title = {{Coma morphology of Jupiter-family comets}}, 
author = {Farnham, Tony L.}, 
journal = {Planetary and Space Science}, 
issn = {0032-0633}, 
doi = {10.1016/j.pss.2009.02.008}, 
pages = {1192--1217}, 
number = {10}, 
volume = {57}
}

@article{Luspay2015, 
year = {2015}, 
title = {{Composition-dependent outgassing of comet 67P/Churyumov-Gerasimenko from ROSINA/DFMS. Implications for nucleus heterogeneity?}}, 
author = {Luspay-Kuti, A and Hassig, M and Fuselier, S A and Mandt, K E and Altwegg, K and Balsiger, H and Gasc, S and Jackel, A and Roy, L Le and Rubin, M and Tzou, C Y and Wurz, P and Mousis, O and Dhooghe, F and Berthelier, J J and Fiethe, B and Gombosi, T.\textbackslashtextasciitildeI. and Mall, Urs A}, 
journal = {Astronomy and Astrophysics}, 
doi = {10.1051/0004-6361/201526205}, 
pages = {A4}, 
volume = {583}
}

@article{Combi2020, 
year = {2020}, 
title = {{The surface distributions of the production of the major volatile species, H2O, CO2, CO and O2, from the nucleus of comet 67P/Churyumov-Gerasimenko …}}, 
author = {Combi, Michael R and Shou, Y and Fougere, N and Tenishev, V and 2020}, 
journal = {Icarus}, 
doi = {10.1016/j.icarus.2019.113421}, 
pages = {113421}, 
volume = {335}
}

@misc{Jorda2008, 
year = {2008}, 
author = {Jorda, Laurent and Crovisier, Jacques and Green, D W E}, 
title = {{The Correlation Between Visual Magnitudes and Water Production Rates}}, 
url = {http://adsabs.harvard.edu/cgi-bin/nph-data\_query?bibcode=2008LPICo1405.8046J\&link\_type=ARTICLE}, 
urldate = {0}
}

@article{Meech2017, 
year = {2017}, 
title = {{CO-driven Activity in Comet C/2017 K2 (PANSTARRS)}}, 
author = {Meech, Karen J and Kleyna, Jan T and Hainaut, Olivier R and Micheli, Marco and Bauer, James and Denneau, Larry and Keane, Jacqueline V and Stephens, Haynes and Jedicke, Robert and Wainscoat, Richard and Weryk, Robert and Flewelling, Heather and Schunova-Lilly, Eva and Magnier, Eugene and Chambers, Kenneth C}, 
journal = {Astrophysical Journal Letters}, 
doi = {10.3847/2041-8213/aa921f}, 
pages = {L8}, 
number = {1}, 
volume = {849}
}

@article{Faggi2024, 
year = {2024}, 
title = {{Heterogeneous outgassing regions identified on active centaur 29P/Schwassmann–Wachmann 1}}, 
author = {Faggi, Sara and Villanueva, Geronimo L. and McKay, Adam and Pinto, Olga Harrington and Kelley, Michael S. P. and Bockelée-Morvan, Dominique and Womack, Maria and Schambeau, Charles A. and Feaga, Lori and DiSanti, Michael A. and Bauer, James M. and Biver, Nicolas and Wierzchos, Kacper and Fernandez, Yanga R.}, 
journal = {Nature Astronomy}, 
doi = {10.1038/s41550-024-02319-3}, 
pages = {1--9}
}

@ARTICLE{Bodewits2016,
       author = {{Bodewits}, D. and {Lara}, L.~M. and {A'Hearn}, M.~F. and {La Forgia}, F. and {Gicquel}, A. and {Kovacs}, G. and {Knollenberg}, J. and {Lazzarin}, M. and {Lin}, Z. -Y. and {Shi}, X. and {Snodgrass}, C. and {Tubiana}, C. and {Sierks}, H. and {Barbieri}, C. and {Lamy}, P.~L. and {Rodrigo}, R. and {Koschny}, D. and {Rickman}, H. and {Keller}, H.~U. and {Barucci}, M.~A. and {Bertaux}, J. -L. and {Bertini}, I. and {Boudreault}, S. and {Cremonese}, G. and {Da Deppo}, V. and {Davidsson}, B. and {Debei}, S. and {De Cecco}, M. and {Fornasier}, S. and {Fulle}, M. and {Groussin}, O. and {Guti{\'e}rrez}, P.~J. and {G{\"u}ttler}, C. and {Hviid}, S.~F. and {Ip}, W. -H. and {Jorda}, L. and {Kramm}, J. -R. and {K{\"u}hrt}, E. and {K{\"u}ppers}, M. and {L{\'o}pez-Moreno}, J.~J. and {Marzari}, F. and {Naletto}, G. and {Oklay}, N. and {Thomas}, N. and {Toth}, I. and {Vincent}, J. -B.},
        title = "{Changes in the Physical Environment of the Inner Coma of 67P/Churyumov-Gerasimenko with Decreasing Heliocentric Distance}",
      journal = {\aj},
         year = 2016,
        month = nov,
       volume = {152},
       number = {5},
          eid = {130},
        pages = {130},
          doi = {10.3847/0004-6256/152/5/130},
archivePrefix = {arXiv},
       eprint = {1607.05632},
 primaryClass = {astro-ph.EP},
       adsurl = {https://ui.adsabs.harvard.edu/abs/2016AJ....152..130B}
}

@ARTICLE{Welsh2006,
       author = {{Welsh}, B.~Y. and {Wheatley}, J. and {Browne}, S.~E. and {Siegmund}, O.~H.~W. and {Doyle}, J.~G. and {O'Shea}, E. and {Antonova}, A. and {Forster}, K. and {Seibert}, M. and {Morrissey}, P. and {Taroyan}, Y.},
        title = "{GALEX high time-resolution ultraviolet observations of dMe flare events}",
      journal = {\aap},
         year = 2006,
        month = nov,
       volume = {458},
       number = {3},
        pages = {921-930},
          doi = {10.1051/0004-6361:20065304},
archivePrefix = {arXiv},
       eprint = {astro-ph/0608254},
 primaryClass = {astro-ph},
       adsurl = {https://ui.adsabs.harvard.edu/abs/2006A&A...458..921W}
}

@ARTICLE{Jewitt2009,
       author = {{Jewitt}, David},
        title = "{The Active Centaurs}",
      journal = {\aj},
         year = 2009,
        month = may,
       volume = {137},
       number = {5},
        pages = {4296-4312},
          doi = {10.1088/0004-6256/137/5/4296},
archivePrefix = {arXiv},
       eprint = {0902.4687},
 primaryClass = {astro-ph.EP},
       adsurl = {https://ui.adsabs.harvard.edu/abs/2009AJ....137.4296J}
}

@ARTICLE{GuilbertLepoutre2012,
       author = {{Guilbert-Lepoutre}, Aur{\'e}lie},
        title = "{Survival of Amorphous Water Ice on Centaurs}",
      journal = {\aj},
         year = 2012,
        month = oct,
       volume = {144},
       number = {4},
          eid = {97},
        pages = {97},
          doi = {10.1088/0004-6256/144/4/97},
       adsurl = {https://ui.adsabs.harvard.edu/abs/2012AJ....144...97G}
}

@ARTICLE{DelloRusso2016,
       author = {{Dello Russo}, Neil and {Kawakita}, Hideyo and {Vervack}, Ronald J. and {Weaver}, Harold A.},
        title = "{Emerging trends and a comet taxonomy based on the volatile chemistry measured in thirty comets with high-resolution infrared spectroscopy between 1997 and 2013}",
      journal = {\icarus},
         year = 2016,
        month = nov,
       volume = {278},
        pages = {301-332},
          doi = {10.1016/j.icarus.2016.05.039},
       adsurl = {https://ui.adsabs.harvard.edu/abs/2016Icar..278..301D}
}

@ARTICLE{Luspay2018,
       author = {{Luspay-Kuti}, Adrienn and {Mousis}, Olivier and {Lunine}, Jonathan I. and {Ellinger}, Yves and {Pauzat}, Fran{\c{c}}oise and {Raut}, Ujjwal and {Bouquet}, Alexis and {Mandt}, Kathleen E. and {Maggiolo}, Romain and {Ronnet}, Thomas and {Brugger}, Bastien and {Ozgurel}, Ozge and {Fuselier}, Stephen A.},
        title = "{Origin of Molecular Oxygen in Comets: Current Knowledge and Perspectives}",
      journal = {\ssr},
         year = 2018,
        month = dec,
       volume = {214},
       number = {8},
          eid = {115},
        pages = {115},
          doi = {10.1007/s11214-018-0541-2},
       adsurl = {https://ui.adsabs.harvard.edu/abs/2018SSRv..214..115L}
}

@ARTICLE{Biver2022,
       author = {{Biver}, Nicolas and {Dello Russo}, Neil and {Opitom}, Cyrielle and {Rubin}, Martin},
        title = "{Chemistry of comet atmospheres}",
      journal = {arXiv e-prints},
         year = 2022,
        month = jul,
          eid = {arXiv:2207.04800},
        pages = {arXiv:2207.04800},
          doi = {10.48550/arXiv.2207.04800},
archivePrefix = {arXiv},
       eprint = {2207.04800},
 primaryClass = {astro-ph.EP},
       adsurl = {https://ui.adsabs.harvard.edu/abs/2022arXiv220704800B}
}

@ARTICLE{harringtonpinto2022,
       author = {{Harrington Pinto}, Olga and {Womack}, Maria and {Fernandez}, Yanga and {Bauer}, James},
        title = "{A Survey of CO, CO$_{2}$, and H$_{2}$O in Comets and Centaurs}",
      journal = {\psj},
         year = 2022,
        month = nov,
       volume = {3},
       number = {11},
          eid = {247},
        pages = {247},
          doi = {10.3847/PSJ/ac960d},
archivePrefix = {arXiv},
       eprint = {2209.09985},
 primaryClass = {astro-ph.EP},
       adsurl = {https://ui.adsabs.harvard.edu/abs/2022PSJ.....3..247H}
}

@ARTICLE{Gkotsinas2022,
       author = {{Gkotsinas}, Anastasios and {Guilbert-Lepoutre}, Aur{\'e}lie and {Raymond}, Sean N. and {Nesvorny}, David},
        title = "{Thermal Processing of Jupiter-family Comets during Their Chaotic Orbital Evolution}",
      journal = {\apj},
         year = 2022,
        month = mar,
       volume = {928},
       number = {1},
          eid = {43},
        pages = {43},
          doi = {10.3847/1538-4357/ac54ac},
archivePrefix = {arXiv},
       eprint = {2202.06685},
 primaryClass = {astro-ph.EP},
       adsurl = {https://ui.adsabs.harvard.edu/abs/2022ApJ...928...43G}
}

@ARTICLE{Sarid2019,
       author = {{Sarid}, G. and {Volk}, K. and {Steckloff}, J.~K. and {Harris}, W. and {Womack}, M. and {Woodney}, L.~M.},
        title = "{29P/Schwassmann-Wachmann 1, A Centaur in the Gateway to the Jupiter-family Comets}",
      journal = {\apjl},
         year = 2019,
        month = sep,
       volume = {883},
       number = {1},
          eid = {L25},
        pages = {L25},
          doi = {10.3847/2041-8213/ab3fb3},
archivePrefix = {arXiv},
       eprint = {1908.04185},
 primaryClass = {astro-ph.EP},
       adsurl = {https://ui.adsabs.harvard.edu/abs/2019ApJ...883L..25S}
}

@phdthesis{Bodewits2010,
title = "Dynamics of highly charged ions interacting with surfaces",
author = "Erwin Bodewits",
note = "Relation: http://www.rug.nl/ Rights: University of Groningen",
year = "2010",
language = "English",
isbn = "9789036746731",
publisher = "s.n.",
}

@ARTICLE{Schambeau2015,
       author = {{Schambeau}, Charles A. and {Fern{\'a}ndez}, Yanga R. and {Lisse}, Carey M. and {Samarasinha}, Nalin and {Woodney}, Laura M.},
        title = "{A new analysis of Spitzer observations of Comet 29P/Schwassmann-Wachmann 1}",
      journal = {\icarus},
         year = 2015,
        month = nov,
       volume = {260},
        pages = {60-72},
          doi = {10.1016/j.icarus.2015.06.038},
archivePrefix = {arXiv},
       eprint = {1506.07037},
 primaryClass = {astro-ph.EP},
       adsurl = {https://ui.adsabs.harvard.edu/abs/2015Icar..260...60S}
}

@ARTICLE{Kaib2022,
       author = {{Kaib}, Nathan A. and {Volk}, Kathryn},
        title = "{Dynamical Population of Comet Reservoirs}",
      journal = {arXiv e-prints},
         year = 2022,
        month = may,
          eid = {arXiv:2206.00010},
        pages = {arXiv:2206.00010},
          doi = {10.48550/arXiv.2206.00010},
archivePrefix = {arXiv},
       eprint = {2206.00010},
 primaryClass = {astro-ph.EP},
       adsurl = {https://ui.adsabs.harvard.edu/abs/2022arXiv220600010K}
}

@ARTICLE{Lecavelier1999_photometry,
       author = {{Lecavelier Des Etangs}, A. and {Vidal-Madjar}, A. and {Ferlet}, R.},
        title = "{Photometric stellar variation due to extra-solar comets}",
      journal = {\aap},
         year = 1999,
        month = mar,
       volume = {343},
        pages = {916-922},
          doi = {10.48550/arXiv.astro-ph/9812381},
archivePrefix = {arXiv},
       eprint = {astro-ph/9812381},
 primaryClass = {astro-ph},
       adsurl = {https://ui.adsabs.harvard.edu/abs/1999A&A...343..916L}
}

@ARTICLE{Engrand2023,
       author = {{Engrand}, Cecile and {Lasue}, J{\'e}r{\'e}mie and {Wooden}, Diane H. and {Zolensky}, Mike E.},
        title = "{Chemical and physical properties of cometary dust}",
      journal = {arXiv e-prints},
         year = 2023,
        month = may,
          eid = {arXiv:2305.03417},
        pages = {arXiv:2305.03417},
          doi = {10.48550/arXiv.2305.03417},
archivePrefix = {arXiv},
       eprint = {2305.03417},
 primaryClass = {astro-ph.EP},
       adsurl = {https://ui.adsabs.harvard.edu/abs/2023arXiv230503417E}
}

@ARTICLE{Pollack1987,
       author = {{Pollack}, J.~B. and {Kasting}, J.~F. and {Richardson}, S.~M. and {Poliakoff}, K.},
        title = "{The case for a wet, warm climate on early Mars}",
      journal = {\icarus},
         year = 1987,
        month = aug,
       volume = {71},
       number = {2},
        pages = {203-224},
          doi = {10.1016/0019-1035(87)90147-3},
       adsurl = {https://ui.adsabs.harvard.edu/abs/1987Icar...71..203P}
}

@ARTICLE{Howe2020,
       author = {{Howe}, Alex R. and {Adams}, Fred C. and {Meyer}, Michael R.},
        title = "{Survival of Primordial Planetary Atmospheres: Photodissociation-driven Mass Loss}",
      journal = {\apj},
         year = 2020,
        month = may,
       volume = {894},
       number = {2},
          eid = {130},
        pages = {130},
          doi = {10.3847/1538-4357/ab620c},
archivePrefix = {arXiv},
       eprint = {1912.08820},
 primaryClass = {astro-ph.EP},
       adsurl = {https://ui.adsabs.harvard.edu/abs/2020ApJ...894..130H}
}

@ARTICLE{Jorda2016,
       author = {{Jorda}, L. and {Gaskell}, R. and {Capanna}, C. and {Hviid}, S. and {Lamy}, P. and {{\v{D}}urech}, J. and {Faury}, G. and {Groussin}, O. and {Guti{\'e}rrez}, P. and {Jackman}, C. and {Keihm}, S.~J. and {Keller}, H.~U. and {Knollenberg}, J. and {K{\"u}hrt}, E. and {Marchi}, S. and {Mottola}, S. and {Palmer}, E. and {Schloerb}, F.~P. and {Sierks}, H. and {Vincent}, J. -B. and {A'Hearn}, M.~F. and {Barbieri}, C. and {Rodrigo}, R. and {Koschny}, D. and {Rickman}, H. and {Barucci}, M.~A. and {Bertaux}, J.~L. and {Bertini}, I. and {Cremonese}, G. and {Da Deppo}, V. and {Davidsson}, B. and {Debei}, S. and {De Cecco}, M. and {Fornasier}, S. and {Fulle}, M. and {G{\"u}ttler}, C. and {Ip}, W. -H. and {Kramm}, J.~R. and {K{\"u}ppers}, M. and {Lara}, L.~M. and {Lazzarin}, M. and {Lopez Moreno}, J.~J. and {Marzari}, F. and {Naletto}, G. and {Oklay}, N. and {Thomas}, N. and {Tubiana}, C. and {Wenzel}, K. -P.},
        title = "{The global shape, density and rotation of Comet 67P/Churyumov-Gerasimenko from preperihelion Rosetta/OSIRIS observations}",
      journal = {\icarus},
         year = 2016,
        month = oct,
       volume = {277},
        pages = {257-278},
          doi = {10.1016/j.icarus.2016.05.002},
       adsurl = {https://ui.adsabs.harvard.edu/abs/2016Icar..277..257J}
}

@ARTICLE{Kalman2024,
       author = {{K{\'a}lm{\'a}n}, Szil{\'a}rd and {Szab{\'o}}, Gyula M. and {Kiss}, Csaba},
        title = "{Exocomet Models in Transit: Light Curve Morphology in the Optical{\textemdash}Near Infrared Wavelength Range}",
      journal = {\pasp},
         year = 2024,
        month = aug,
       volume = {136},
       number = {8},
          eid = {084401},
        pages = {084401},
          doi = {10.1088/1538-3873/ad4fe3},
archivePrefix = {arXiv},
       eprint = {2405.13663},
 primaryClass = {astro-ph.EP},
       adsurl = {https://ui.adsabs.harvard.edu/abs/2024PASP..136h4401K}
}

@ARTICLE{Bahr2012,
       author = {{Bahr}, D.~A. and {Baragiola}, R.~A.},
        title = "{Photodesorption of Solid CO$_{2}$ by Ly{\ensuremath{\alpha}}}",
      journal = {\apj},
         year = 2012,
        month = dec,
       volume = {761},
       number = {1},
          eid = {36},
        pages = {36},
          doi = {10.1088/0004-637X/761/1/36},
       adsurl = {https://ui.adsabs.harvard.edu/abs/2012ApJ...761...36B}
}

@article{Yakshinskiy2000, 
year = {2000}, 
title = {{Desorption induced by electronic transitions of Na from SiO2: relevance to tenuous planetary atmospheres}}, 
author = {Yakshinskiy, B.V. and Madey, T.E.}, 
journal = {Surface Science}, 
issn = {0039-6028}, 
doi = {10.1016/s0039-6028(00)00022-4}, 
pages = {160--165}, 
number = {1-3}, 
volume = {451}
}

@article{Fillion2022, 
year = {2022}, 
title = {{Vacuum-UV Photodesorption from Compact Amorphous Solid Water: Photon Energy Dependence, Isotopic and Temperature Effects}}, 
author = {Fillion, Jean-Hugues and Dupuy, Rémi and Féraud, Géraldine and Romanzin, Claire and Philippe, Laurent and Putaud, Thomas and Baglin, Vincent and Cimino, Roberto and Marie-Jeanne, Patrick and Jeseck, Pascal and Michaut, Xavier and Bertin, Mathieu}, 
journal = {ACS Earth and Space Chemistry}, 
issn = {2472-3452}, 
doi = {10.1021/acsearthspacechem.1c00302}, 
eprint = {2103.15435}, 
pages = {100--115}, 
number = {1}, 
volume = {6}
}

@ARTICLE{Lecavelier2022,
       author = {{Lecavelier des Etangs}, Alain and {Cros}, Lucie and {H{\'e}brard}, Guillaume and {Martioli}, Eder and {Duquesnoy}, Marc and {Kenworthy}, Matthew A. and {Kiefer}, Flavien and {Lacour}, Sylvestre and {Lagrange}, Anne-Marie and {Meunier}, Nad{\`e}ge and {Vidal-Madjar}, Alfred},
        title = "{Exocomets size distribution in the {\ensuremath{\beta}} Pictoris planetary system}",
      journal = {Scientific Reports},
         year = 2022,
        month = apr,
       volume = {12},
          eid = {5855},
        pages = {5855},
          doi = {10.1038/s41598-022-09021-2},
archivePrefix = {arXiv},
       eprint = {2204.13618},
 primaryClass = {astro-ph.EP},
       adsurl = {https://ui.adsabs.harvard.edu/abs/2022NatSR..12.5855L}
}

@ARTICLE{Manfroid2021,
       author = {{Manfroid}, J. and {Hutsem{\'e}kers}, D. and {Jehin}, E.},
        title = "{Iron and nickel atoms in cometary atmospheres even far from the Sun}",
      journal = {\nat},
         year = 2021,
        month = may,
       volume = {593},
       number = {7859},
        pages = {372-374},
          doi = {10.1038/s41586-021-03435-0},
       adsurl = {https://ui.adsabs.harvard.edu/abs/2021Natur.593..372M}
}

@ARTICLE{Kelley2023,
       author = {{Kelley}, Michael S.~P. and {Hsieh}, Henry H. and {Bodewits}, Dennis and {Saki}, Mohammad and {Villanueva}, Geronimo L. and {Milam}, Stefanie N. and {Hammel}, Heidi B.},
        title = "{Spectroscopic identification of water emission from a main-belt comet}",
      journal = {\nat},
         year = 2023,
        month = jul,
       volume = {619},
       number = {7971},
        pages = {720-723},
          doi = {10.1038/s41586-023-06152-y},
       adsurl = {https://ui.adsabs.harvard.edu/abs/2023Natur.619..720K}
}

@ARTICLE{Collings2004,
       author = {{Collings}, Mark P. and {Anderson}, Mark A. and {Chen}, Rui and {Dever}, John W. and {Viti}, Serena and {Williams}, David A. and {McCoustra}, Martin R.~S.},
        title = "{A laboratory survey of the thermal desorption of astrophysically relevant molecules}",
      journal = {\mnras},
         year = 2004,
        month = nov,
       volume = {354},
       number = {4},
        pages = {1133-1140},
          doi = {10.1111/j.1365-2966.2004.08272.x},
       adsurl = {https://ui.adsabs.harvard.edu/abs/2004MNRAS.354.1133C}
}

@ARTICLE{Killen2007,
       author = {{Killen}, Rosemary and {Cremonese}, Gabrielle and {Lammer}, Helmut and {Orsini}, Stefano and {Potter}, Andrew E. and {Sprague}, Ann L. and {Wurz}, Peter and {Khodachenko}, Maxim L. and {Lichtenegger}, Herbert I.~M. and {Milillo}, Anna and {Mura}, Alessandro},
        title = "{Processes that Promote and Deplete the Exosphere of Mercury}",
      journal = {\ssr},
         year = 2007,
        month = oct,
       volume = {132},
       number = {2-4},
        pages = {433-509},
          doi = {10.1007/s11214-007-9232-0},
       adsurl = {https://ui.adsabs.harvard.edu/abs/2007SSRv..132..433K}
}

@article{Filacchione2016, 
year = {2016}, 
title = {{Seasonal exposure of carbon dioxide ice on the nucleus of comet 67P/Churyumov-Gerasimenko}}, 
author = {Filacchione, G. and Raponi, A. and Capaccioni, F. and Ciarniello, M. and Tosi, F. and Capria, M. T. and Sanctis, M. C. De and Migliorini, A. and Piccioni, G. and Cerroni, P. and Barucci, M. A. and Fornasier, S. and Schmitt, B. and Quirico, E. and Erard, S. and Bockelee-Morvan, D. and Leyrat, C. and Arnold, G. and Mennella, V. and Ammannito, E. and Bellucci, G. and Benkhoff, J. and Bibring, J. P. and Blanco, A. and Blecka, M. I. and Carlson, R. and Carsenty, U. and Colangeli, L. and Combes, M. and Combi, M. and Crovisier, J. and Drossart, P. and Encrenaz, T. and Federico, C. and Fink, U. and Fonti, S. and Fulchignoni, M. and Ip, W.-H. and Irwin, P. and Jaumann, R. and Kuehrt, E. and Langevin, Y. and Magni, G. and McCord, T. and Moroz, L. and Mottola, S. and Palomba, E. and Schade, U. and Stephan, K. and Taylor, F. and Tiphene, D. and Tozzi, G. P. and Beck, P. and Biver, N. and Bonal, L. and Combe, J.-Ph. and Despan, D. and Flamini, E. and Formisano, M. and Frigeri, A. and Grassi, D. and Gudipati, M. S. and Kappel, D. and Longobardo, A. and Mancarella, F. and Markus, K. and Merlin, F. and Orosei, R. and Rinaldi, G. and Cartacci, M. and Cicchetti, A. and Hello, Y. and Henry, F. and Jacquinod, S. and Reess, J. M. and Noschese, R. and Politi, R. and Peter, G.}, 
journal = {Science}, 
issn = {0036-8075}, 
doi = {10.1126/science.aag3161}, 
pmid = {27856846}, 
url = {https://www.science.org/doi/10.1126/science.aag3161}, 
pages = {1563--1566}, 
number = {6319}, 
volume = {354}
}

@ARTICLE{Luspay2022,
       author = {{Luspay-Kuti}, Adrienn and {Mousis}, Olivier and {Pauzat}, Fran{\c{c}}oise and {Ozgurel}, Ozge and {Ellinger}, Yves and {Lunine}, Jonathan I. and {Fuselier}, Stephen A. and {Mandt}, Kathleen E. and {Trattner}, Karlheinz J. and {Petrinec}, Steven M.},
        title = "{Dual storage and release of molecular oxygen in comet 67P/Churyumov-Gerasimenko}",
      journal = {Nature Astronomy},
         year = 2022,
        month = mar,
       volume = {6},
        pages = {724-730},
          doi = {10.1038/s41550-022-01614-1},
       adsurl = {https://ui.adsabs.harvard.edu/abs/2022NatAs...6..724L}
}

@book{Huebner2006, 
year = {2006}, 
title = {{Heat and Gas Diffusion in Comet Nuclei}}, 
author = {Huebner, Walter F and Benkhoff, J and Capria, M T and Coradini, A and Sanctis, Maria C De and Orosei, R and Prialnik, D}, 
isbn = {475371229}, 
urldate = {0}, 
series = {International Space Science Institute}, 
publisher = {International Space Science Institute}, 
language = {English}
}

@article{Zhang2023, 
year = {2023}, 
title = {{Sodium Brightening of (3200) Phaethon near Perihelion}}, 
author = {Zhang, Qicheng and Battams, Karl and Ye, Quanzhi and Knight, Matthew M. and Schmidt, Carl A.}, 
journal = {The Planetary Science Journal}, 
doi = {10.3847/psj/acc866}, 
eprint = {2303.17625}, 
pages = {70}, 
number = {4}, 
volume = {4}
}

@ARTICLE{Kruczkiewicz2024,
       author = {{Kruczkiewicz}, F. and {Dulieu}, F. and {Ivlev}, A.~V. and {Caselli}, P. and {Giuliano}, B.~M. and {Ceccarelli}, C. and {Theul{\'e}}, P.},
        title = "{Comprehensive laboratory constraints on thermal desorption of interstellar ice analogues}",
      journal = {\aap},
         year = 2024,
        month = jun,
       volume = {686},
          eid = {A236},
        pages = {A236},
          doi = {10.1051/0004-6361/202346948},
archivePrefix = {arXiv},
       eprint = {2404.02695},
 primaryClass = {astro-ph.GA},
       adsurl = {https://ui.adsabs.harvard.edu/abs/2024A&A...686A.236K}
}

@article{Saki2024, 
year = {2024}, 
title = {{Parent Volatile Outgassing Associations in Cometary Nuclei: Synthesizing Rosetta Measurements and Ground-based Observations}}, 
author = {Saki, Mohammad and Bodewits, Dennis and Bonev, Boncho P. and Russo, Neil Dello and Luspay-Kuti, Adrienn and Noonan, John W. and Combi, Michael. R. and Shou, Yinsi}, 
journal = {The Planetary Science Journal}, 
doi = {10.3847/psj/ad118f}, 
pages = {70}, 
number = {3}, 
volume = {5}
}

@ARTICLE{Hsieh2006,
       author = {{Hsieh}, Henry H. and {Jewitt}, David},
        title = "{A Population of Comets in the Main Asteroid Belt}",
      journal = {Science},
         year = 2006,
        month = apr,
       volume = {312},
       number = {5773},
        pages = {561-563},
          doi = {10.1126/science.1125150},
       adsurl = {https://ui.adsabs.harvard.edu/abs/2006Sci...312..561H}
}

@ARTICLE{Marsden1989,
       author = {{Marsden}, B.~G.},
        title = "{The Sungrazing Comet Group. II.}",
      journal = {\aj},
         year = 1989,
        month = dec,
       volume = {98},
        pages = {2306},
          doi = {10.1086/115301},
       adsurl = {https://ui.adsabs.harvard.edu/abs/1989AJ.....98.2306M}
}

@ARTICLE{Womack2021,
       author = {{Womack}, M. and {Curtis}, O. and {Rabson}, D.~A. and {Harrington Pinto}, O. and {Wierzchos}, K. and {Cruz Gonzalez}, S. and {Sarid}, G. and {Mentzer}, C. and {Lastra}, N. and {Pichette}, N. and {Ruffini}, N. and {Cox}, T. and {Rivera}, I. and {Micciche}, A. and {Jackson}, C. and {Homich}, A. and {Tollison}, A. and {Reed}, S. and {Zilka}, J. and {Henning}, B. and {Spinar}, M. and {Escoto}, S. Rosslyn and {Erdahl}, T. and {Goldschen-Ohm}, Marcel P. and {Uhl}, W.~T.},
        title = "{The Visual Lightcurve of Comet C/1995 O1 (Hale-Bopp) from 1995 to 1999}",
      journal = {\psj},
         year = 2021,
        month = feb,
       volume = {2},
       number = {1},
          eid = {17},
        pages = {17},
          doi = {10.3847/PSJ/abd32c},
archivePrefix = {arXiv},
       eprint = {2008.06761},
 primaryClass = {astro-ph.EP},
       adsurl = {https://ui.adsabs.harvard.edu/abs/2021PSJ.....2...17W}
}

@ARTICLE{Biver2002,
       author = {{Biver}, Nicolas and {Bockel{\'e}e-Morvan}, Dominique and {Colom}, Pierre and {Crovisier}, Jacques and {Henry}, Florence and {Lellouch}, Emmanuel and {Winnberg}, Anders and {Johansson}, Lars E.~B. and {Gunnarsson}, Marcus and {Rickman}, Hans and {Rantakyr{\"o}}, Fredrik and {Davies}, John K. and {Dent}, William R.~F. and {Paubert}, Gabriel and {Moreno}, Rapha{\"e}l and {Wink}, J{\"o}rn and {Despois}, Didier and {Benford}, Dominic J. and {Gardner}, Matt and {Lis}, Dariusz C. and {Mehringer}, David and {Phillips}, Thomas G. and {Rauer}, Heike},
        title = "{The 1995{\textendash}2002 Long-Term Monitoring of Comet C/1995 O1 (HALE{\textendash}BOPP) at Radio Wavelength}",
      journal = {Earth Moon and Planets},
         year = 2002,
        month = jun,
       volume = {90},
       number = {1},
        pages = {5-14},
          doi = {10.1023/A:1021599915018},
       adsurl = {https://ui.adsabs.harvard.edu/abs/2002EM&P...90....5B}
}

@ARTICLE{Schleicher2024,
       author = {{Schleicher}, David G. and {Birch}, Peter V. and {Farnham}, Tony L. and {Bair}, Allison N.},
        title = "{The Extreme Activity in Comet Hale-Bopp (C/1995 O1): Investigations of Extensive, Narrowband Photoelectric Photometry}",
      journal = {arXiv e-prints},
         year = 2024,
        month = sep,
          eid = {arXiv:2409.13005},
        pages = {arXiv:2409.13005},
          doi = {10.48550/arXiv.2409.13005},
archivePrefix = {arXiv},
       eprint = {2409.13005},
 primaryClass = {astro-ph.EP},
       adsurl = {https://ui.adsabs.harvard.edu/abs/2024arXiv240913005S}
}

@ARTICLE{Sekanina2002,
       author = {{Sekanina}, Zdenek},
        title = "{Statistical Investigation and Modeling of Sungrazing Comets Discovered with the Solar and Heliospheric Observatory}",
      journal = {\apj},
         year = 2002,
        month = feb,
       volume = {566},
       number = {1},
        pages = {577-598},
          doi = {10.1086/324335},
       adsurl = {https://ui.adsabs.harvard.edu/abs/2002ApJ...566..577S}
}

@ARTICLE{Sekanina2003,
       author = {{Sekanina}, Zdenek},
        title = "{Erosion Model for the Sungrazing Comets Observed with the Solar and Heliospheric Observatory}",
      journal = {\apj},
         year = 2003,
        month = nov,
       volume = {597},
       number = {2},
        pages = {1237-1265},
          doi = {10.1086/378192},
       adsurl = {https://ui.adsabs.harvard.edu/abs/2003ApJ...597.1237S}
}

@ARTICLE{Szabo2012,
       author = {{Szab{\'o}}, Gyula M. and {Kiss}, L{\'a}szl{\'o} L. and {P{\'a}l}, Andr{\'a}s and {Kiss}, Csaba and {S{\'a}rneczky}, Kriszti{\'a}n and {Juh{\'a}sz}, Attila and {Hogerheijde}, Michiel R.},
        title = "{Evidence for Fresh Frost Layer on the Bare Nucleus of Comet Hale-Bopp at 32 AU Distance}",
      journal = {\apj},
         year = 2012,
        month = dec,
       volume = {761},
       number = {1},
          eid = {8},
        pages = {8},
          doi = {10.1088/0004-637X/761/1/8},
archivePrefix = {arXiv},
       eprint = {1210.2785},
 primaryClass = {astro-ph.EP},
       adsurl = {https://ui.adsabs.harvard.edu/abs/2012ApJ...761....8S}
}

@article{Pitjeva2021, 
year = {2021}, 
title = {{Estimates of the change rate of solar mass and gravitational constant based on the dynamics of the Solar System}}, 
author = {Pitjeva, E. V. and Pitjev, N. P. and Pavlov, D. A. and Turygin, C. C.}, 
journal = {Astronomy \& Astrophysics}, 
issn = {0004-6361}, 
doi = {10.1051/0004-6361/202039893}, 
eprint = {2201.09804}, 
pages = {A141}, 
volume = {647}
}

@ARTICLE{Knight2010,
       author = {{Knight}, Matthew M. and {A'Hearn}, Michael F. and {Biesecker}, Douglas A. and {Faury}, Guillaume and {Hamilton}, Douglas P. and {Lamy}, Philippe and {Llebaria}, Antoine},
        title = "{Photometric Study of the Kreutz Comets Observed by SOHO from 1996 to 2005}",
      journal = {\aj},
         year = 2010,
        month = mar,
       volume = {139},
       number = {3},
        pages = {926-949},
          doi = {10.1088/0004-6256/139/3/926},
       adsurl = {https://ui.adsabs.harvard.edu/abs/2010AJ....139..926K}
}

@ARTICLE{Battams2017,
       author = {{Battams}, Karl and {Knight}, Matthew M.},
        title = "{SOHO comets: 20 years and 3000 objects later}",
      journal = {Philosophical Transactions of the Royal Society of London Series A},
         year = 2017,
        month = may,
       volume = {375},
       number = {2097},
          eid = {20160257},
        pages = {20160257},
          doi = {10.1098/rsta.2016.0257},
archivePrefix = {arXiv},
       eprint = {1611.02279},
 primaryClass = {astro-ph.EP},
       adsurl = {https://ui.adsabs.harvard.edu/abs/2017RSPTA.37560257B}
}

@ARTICLE{Marsden2005,
       author = {{Marsden}, Brian G.},
        title = "{Sungrazing Comets}",
      journal = {\araa},
         year = 2005,
        month = sep,
       volume = {43},
       number = {1},
        pages = {75-102},
          doi = {10.1146/annurev.astro.43.072103.150554},
       adsurl = {https://ui.adsabs.harvard.edu/abs/2005ARA&A..43...75M}
}

@article{Itikawa2005, 
year = {2005}, 
title = {{Cross Sections for Electron Collisions with Water Molecules}}, 
author = {Itikawa, Yukikazu and Mason, Nigel}, 
journal = {Journal of Physical and Chemical Reference Data}, 
doi = {10.1063/1.1799251}, 
pages = {1 -- 22}, 
number = {1}, 
volume = {34}
}

@article{Cravens1986, 
year = {1986}, 
title = {{Vibrational and rotational cooling of electrons by water vapor}}, 
author = {Cravens, T.E. and Korosmezey, A.}, 
journal = {Planetary and Space Science}, 
issn = {0032-0633}, 
doi = {10.1016/0032-0633(86)90005-x}, 
pages = {961--970}, 
number = {10}, 
volume = {34}
}

@INCOLLECTION{Bodewits2022,
       author = {{Bodewits}, D. and {Bonev}, B.~P. and {Cordiner}, M.~A. and {Villanueva}, G.~L.},
        title = "{Radiative Processes as Diagnostics of Cometary Atmospheres}",
    booktitle = {Comets III},
         year = 2024,
       editor = {{Meech}, Karen. J. and {Combi}, Michael. R. and {Bockel{\'e}e-Morvan}, Dominique and {Raymodn}, Sean. N. and {Zolensky}, Michael. E.},
        pages = {407-432},
          doi = {10.2458/azu_uapress_9780816553631-ch013},
       adsurl = {https://ui.adsabs.harvard.edu/abs/2024come.book..407B}
}

@article{Wurz2022, 
year = {2022}, 
title = {{Particles and Photons as Drivers for Particle Release from the Surfaces of the Moon and Mercury}}, 
author = {Wurz, P. and Fatemi, S. and Galli, A. and Halekas, J. and Harada, Y. and Jäggi, N. and Jasinski, J. and Lammer, H. and Lindsay, S. and Nishino, M. N. and Orlando, T. M. and Raines, J. M. and Scherf, M. and Slavin, J. and Vorburger, A. and Winslow, R.}, 
journal = {Space Science Reviews}, 
issn = {0038-6308}, 
doi = {10.1007/s11214-022-00875-6}, 
pages = {10}, 
number = {3}, 
volume = {218}
}

@techreport{Rubin2019, 
year = {2019}, 
author = {Rubin, Martin and Bekaert, David V and Broadley, Michael W and Drozdovskaya, Maria N and Wampfler, Susanne F}, 
title = {{Volatile Species in Comet 67P/Churyumov-Gerasimenko: Investigating the Link from the ISM to the Terrestrial Planets}}, 
url = {message:\%3C04697064-9816-46EE-B2DC-1949694A9119@gmail.com\%3E}, 
urldate = {0}, 
pages = {1792--1811}, 
language = {English}, 
month = {08}
}

@article{Bodewits2011, 
year = {2011}, 
title = {{Collisional Excavation of Asteroid (596) Scheila}}, 
author = {Bodewits, Dennis and Kelley, Michael S and Li, Jian-Yang and Landsman, Wayne B and Besse, Sebastien and A'Hearn, Michael F}, 
journal = {Astrophysical Journal Letters}, 
doi = {10.1088/2041-8205/733/1/l3}, 
pages = {L3}, 
number = {1}, 
volume = {733}, 
month = {04}
}

@article{Bromley2021, 
year = {2021}, 
title = {{Atomic Iron and Nickel in the Coma of C/1996 B2 (Hyakutake): Production Rates, Emission Mechanisms, and Possible Parents}}, 
author = {Bromley, S. J. and Neff, B. and Loch, S. D. and Marler, J. P. and Országh, J. and Venkataramani, K. and Bodewits, D.}, 
journal = {The Planetary Science Journal}, 
doi = {10.3847/psj/ac2dff}, 
pages = {228}, 
number = {6}, 
volume = {2}
}

@article{Jones2018, 
year = {2018}, 
title = {{The Science of Sungrazers, Sunskirters, and Other Near-Sun Comets}}, 
author = {Jones, Geraint H and Knight, Matthew Manning and Battams, Karl and Boice, Daniel C and Brown, John and Giordano, Silvio and Raymond, John and Snodgrass, Colin and Steckloff, Jordan K and Weissman, Paul and Fitzsimmons, Alan and Lisse, Carey M and Opitom, Cyrielle and Birkett, Kimberley S and Bzowski, Maciej and Decock, Alice and Mann, Ingrid and Ramanjooloo, Yudish and McCauley, Patrick}, 
journal = {Space Science Reviews}, 
doi = {10.1007/s11214-017-0446-5}, 
pages = {\#20}, 
number = {1}, 
volume = {214}
}

@INCOLLECTION{Meech2005,
       author = {{Meech}, K.~J. and {Svoren}, J.},
        title = "{Using cometary activity to trace the physical and chemical evolution of cometary nuclei}",
    booktitle = {Comets II},
         year = 2004,
       editor = {{Festou}, Michel C. and {Keller}, H. Uwe and {Weaver}, Harold A.},
        pages = {317},
       adsurl = {https://ui.adsabs.harvard.edu/abs/2004come.book..317M}
}

@ARTICLE{Strom2020,
       author = {{Str{\o}m}, Paul A. and {Bodewits}, Dennis and {Knight}, Matthew M. and {Kiefer}, Flavien and {Jones}, Geraint H. and {Kral}, Quentin and {Matr{\`a}}, Luca and {Bodman}, Eva and {Capria}, Maria Teresa and {Cleeves}, Ilsedore and {Fitzsimmons}, Alan and {Haghighipour}, Nader and {Harrison}, John H.~D. and {Iglesias}, Daniela and {Kama}, Mihkel and {Linnartz}, Harold and {Majumdar}, Liton and {de Mooij}, Ernst J.~W. and {Milam}, Stefanie N. and {Opitom}, Cyrielle and {Rebollido}, Isabel and {Rogers}, Laura K. and {Snodgrass}, Colin and {Sousa-Silva}, Clara and {Xu}, Siyi and {Lin}, Zhong-Yi and {Zieba}, Sebastian},
        title = "{Exocomets from a Solar System Perspective}",
      journal = {\pasp},
         year = 2020,
        month = oct,
       volume = {132},
       number = {1016},
          eid = {101001},
        pages = {101001},
          doi = {10.1088/1538-3873/aba6a0},
archivePrefix = {arXiv},
       eprint = {2007.09155},
 primaryClass = {astro-ph.EP},
       adsurl = {https://ui.adsabs.harvard.edu/abs/2020PASP..132j1001S}
}

@ARTICLE{aungwerojwit2024,
       author = {{Aungwerojwit}, Amornrat and {G{\"a}nsicke}, Boris T. and {Dhillon}, Vikram S. and {Drake}, Andrew and {Inight}, Keith and {Kaye}, Thomas G. and {Marsh}, T.~R. and {Mullen}, Ed and {Pelisoli}, Ingrid and {Swan}, Andrew},
        title = "{Long-term variability in debris transiting white dwarfs}",
      journal = {\mnras},
         year = 2024,
        month = may,
       volume = {530},
       number = {1},
        pages = {117-128},
          doi = {10.1093/mnras/stae750},
archivePrefix = {arXiv},
       eprint = {2404.04422},
 primaryClass = {astro-ph.EP},
       adsurl = {https://ui.adsabs.harvard.edu/abs/2024MNRAS.530..117A}
}

@ARTICLE{bonsor2024,
       author = {{Bonsor}, Amy},
        title = "{White Dwarf Systems: the Composition of Exoplanets}",
      journal = {arXiv e-prints},
         year = 2024,
        month = sep,
          eid = {arXiv:2409.13294},
        pages = {arXiv:2409.13294},
          doi = {10.48550/arXiv.2409.13294},
archivePrefix = {arXiv},
       eprint = {2409.13294},
 primaryClass = {astro-ph.EP},
       adsurl = {https://ui.adsabs.harvard.edu/abs/2024arXiv240913294B}
}

@ARTICLE{duvvuri2020,
       author = {{Duvvuri}, Girish M. and {Redfield}, Seth and {Veras}, Dimitri},
        title = "{Necroplanetology: Simulating the Tidal Disruption of Differentiated Planetary Material Orbiting WD 1145+017}",
      journal = {\apj},
         year = 2020,
        month = apr,
       volume = {893},
       number = {2},
          eid = {166},
        pages = {166},
          doi = {10.3847/1538-4357/ab7fa0},
archivePrefix = {arXiv},
       eprint = {2003.08410},
 primaryClass = {astro-ph.EP},
       adsurl = {https://ui.adsabs.harvard.edu/abs/2020ApJ...893..166D}
}

@ARTICLE{farihi2013,
       author = {{Farihi}, J. and {G{\"a}nsicke}, B.~T. and {Koester}, D.},
        title = "{Evidence for Water in the Rocky Debris of a Disrupted Extrasolar Minor Planet}",
      journal = {Science},
         year = 2013,
        month = oct,
       volume = {342},
       number = {6155},
        pages = {218-220},
          doi = {10.1126/science.1239447},
archivePrefix = {arXiv},
       eprint = {1310.3269},
 primaryClass = {astro-ph.EP},
       adsurl = {https://ui.adsabs.harvard.edu/abs/2013Sci...342..218F}
}

@ARTICLE{gansicke2016,
       author = {{G{\"a}nsicke}, B.~T. and {Aungwerojwit}, A. and {Marsh}, T.~R. and {Dhillon}, V.~S. and {Sahman}, D.~I. and {Veras}, Dimitri and {Farihi}, J. and {Chote}, P. and {Ashley}, R. and {Arjyotha}, S. and {Rattanasoon}, S. and {Littlefair}, S.~P. and {Pollacco}, D. and {Burleigh}, M.~R.},
        title = "{High-speed Photometry of the Disintegrating Planetesimals at WD1145+017: Evidence for Rapid Dynamical Evolution}",
      journal = {\apjl},
         year = 2016,
        month = feb,
       volume = {818},
       number = {1},
          eid = {L7},
        pages = {L7},
          doi = {10.3847/2041-8205/818/1/L7},
archivePrefix = {arXiv},
       eprint = {1512.09150},
 primaryClass = {astro-ph.EP},
       adsurl = {https://ui.adsabs.harvard.edu/abs/2016ApJ...818L...7G}
}

@ARTICLE{gentilefusillo2017,
       author = {{Gentile Fusillo}, Nicola Pietro and {G{\"a}nsicke}, Boris T. and {Farihi}, Jay and {Koester}, Detlev and {Schreiber}, Matthias R. and {Pala}, Anna F.},
        title = "{Trace hydrogen in helium atmosphere white dwarfs as a possible signature of water accretion}",
      journal = {\mnras},
         year = 2017,
        month = jun,
       volume = {468},
       number = {1},
        pages = {971-980},
          doi = {10.1093/mnras/stx468},
archivePrefix = {arXiv},
       eprint = {1702.06542},
 primaryClass = {astro-ph.SR},
       adsurl = {https://ui.adsabs.harvard.edu/abs/2017MNRAS.468..971G}
}

@ARTICLE{guidry2021,
       author = {{Guidry}, Joseph A. and {Vanderbosch}, Zachary P. and {Hermes}, J.~J. and {Barlow}, Brad N. and {Lopez}, Isaac D. and {Boudreaux}, Thomas M. and {Corcoran}, Kyle A. and {Bell}, Keaton J. and {Montgomery}, M.~H. and {Heintz}, Tyler M. and {Castanheira}, Barbara G. and {Reding}, Joshua S. and {Dunlap}, Bart H. and {Winget}, D.~E. and {Winget}, Karen I. and {Kuehne}, J.~W.},
        title = "{I Spy Transits and Pulsations: Empirical Variability in White Dwarfs Using Gaia and the Zwicky Transient Facility}",
      journal = {\apj},
         year = 2021,
        month = may,
       volume = {912},
       number = {2},
          eid = {125},
        pages = {125},
          doi = {10.3847/1538-4357/abee68},
archivePrefix = {arXiv},
       eprint = {2012.00035},
 primaryClass = {astro-ph.SR},
       adsurl = {https://ui.adsabs.harvard.edu/abs/2021ApJ...912..125G}
}

@ARTICLE{hoskin2020,
       author = {{Hoskin}, Matthew J. and {Toloza}, Odette and {G{\"a}nsicke}, Boris T. and {Raddi}, Roberto and {Koester}, Detlev and {Pala}, Anna F. and {Manser}, Christopher J. and {Farihi}, Jay and {Belmonte}, Maria Teresa and {Hollands}, Mark and {Gentile Fusillo}, Nicola and {Swan}, Andrew},
        title = "{White dwarf pollution by hydrated planetary remnants: hydrogen and metals in WD J204713.76-125908.9}",
      journal = {\mnras},
         year = 2020,
        month = nov,
       volume = {499},
       number = {1},
        pages = {171-182},
          doi = {10.1093/mnras/staa2717},
archivePrefix = {arXiv},
       eprint = {2009.05053},
 primaryClass = {astro-ph.EP},
       adsurl = {https://ui.adsabs.harvard.edu/abs/2020MNRAS.499..171H}
}

@ARTICLE{izquierdo2021,
       author = {{Izquierdo}, Paula and {Toloza}, Odette and {G{\"a}nsicke}, Boris T. and {Rodr{\'\i}guez-Gil}, Pablo and {Farihi}, Jay and {Koester}, Detlev and {Guo}, Jincheng and {Redfield}, Seth},
        title = "{GD 424 - a helium-atmosphere white dwarf with a large amount of trace hydrogen in the process of digesting a rocky planetesimal}",
      journal = {\mnras},
         year = 2021,
        month = mar,
       volume = {501},
       number = {3},
        pages = {4276-4288},
          doi = {10.1093/mnras/staa3987},
archivePrefix = {arXiv},
       eprint = {2012.12957},
 primaryClass = {astro-ph.EP},
       adsurl = {https://ui.adsabs.harvard.edu/abs/2021MNRAS.501.4276I}
}

@ARTICLE{malamud2021,
       author = {{Malamud}, Uri and {Grishin}, Evgeni and {Brouwers}, Marc},
        title = "{Circularization of tidal debris around white dwarfs: implications for gas production and dust variability}",
      journal = {\mnras},
         year = 2021,
        month = mar,
       volume = {501},
       number = {3},
        pages = {3806-3824},
          doi = {10.1093/mnras/staa3940},
archivePrefix = {arXiv},
       eprint = {2012.07854},
 primaryClass = {astro-ph.EP},
       adsurl = {https://ui.adsabs.harvard.edu/abs/2021MNRAS.501.3806M}
}

@ARTICLE{malamud2024,
       author = {{Malamud}, Uri},
        title = "{White dwarf systems: exoplanets and debris disks}",
      journal = {arXiv e-prints},
         year = 2024,
        month = mar,
          eid = {arXiv:2403.07427},
        pages = {arXiv:2403.07427},
          doi = {10.48550/arXiv.2403.07427},
archivePrefix = {arXiv},
       eprint = {2403.07427},
 primaryClass = {astro-ph.EP},
       adsurl = {https://ui.adsabs.harvard.edu/abs/2024arXiv240307427M}
}

@ARTICLE{marshall2023,
       author = {{Marshall}, Jonathan P. and {Ertel}, Steve and {Birtcil}, Eric and {Villaver}, Eva and {Kemper}, Francisca and {Boffin}, Henri and {Scicluna}, Peter and {Kamath}, Devika},
        title = "{Evidence for the Disruption of a Planetary System During the Formation of the Helix Nebula}",
      journal = {\aj},
         year = 2023,
        month = jan,
       volume = {165},
       number = {1},
          eid = {22},
        pages = {22},
          doi = {10.3847/1538-3881/ac9d90},
archivePrefix = {arXiv},
       eprint = {2211.02251},
 primaryClass = {astro-ph.EP},
       adsurl = {https://ui.adsabs.harvard.edu/abs/2023AJ....165...22M}
}

@ARTICLE{oconnor2023,
       author = {{O'Connor}, Christopher E. and {Lai}, Dong and {Seligman}, Darryl Z.},
        title = "{On the pollution of white dwarfs by exo-Oort cloud comets}",
      journal = {\mnras},
         year = 2023,
        month = oct,
       volume = {524},
       number = {4},
        pages = {6181-6197},
          doi = {10.1093/mnras/stad2281},
archivePrefix = {arXiv},
       eprint = {2306.10102},
 primaryClass = {astro-ph.EP},
       adsurl = {https://ui.adsabs.harvard.edu/abs/2023MNRAS.524.6181O}
}

@ARTICLE{okuya2023,
       author = {{Okuya}, Ayaka and {Ida}, Shigeru and {Hyodo}, Ryuki and {Okuzumi}, Satoshi},
        title = "{Modelling the evolution of silicate/volatile accretion discs around white dwarfs}",
      journal = {\mnras},
         year = 2023,
        month = feb,
       volume = {519},
       number = {2},
        pages = {1657-1676},
          doi = {10.1093/mnras/stac3522},
archivePrefix = {arXiv},
       eprint = {2211.16797},
 primaryClass = {astro-ph.EP},
       adsurl = {https://ui.adsabs.harvard.edu/abs/2023MNRAS.519.1657O}
}

@ARTICLE{pham2024,
       author = {{Pham}, Dang and {Rein}, Hanno},
        title = "{Polluting white dwarfs with Oort cloud comets}",
      journal = {\mnras},
         year = 2024,
        month = may,
       volume = {530},
       number = {3},
        pages = {2526-2547},
          doi = {10.1093/mnras/stae986},
archivePrefix = {arXiv},
       eprint = {2404.07160},
 primaryClass = {astro-ph.EP},
       adsurl = {https://ui.adsabs.harvard.edu/abs/2024MNRAS.530.2526P}
}

@ARTICLE{raddi2015,
       author = {{Raddi}, R. and {G{\"a}nsicke}, B.~T. and {Koester}, D. and {Farihi}, J. and {Hermes}, J.~J. and {Scaringi}, S. and {Breedt}, E. and {Girven}, J.},
        title = "{Likely detection of water-rich asteroid debris in a metal-polluted white dwarf}",
      journal = {\mnras},
         year = 2015,
        month = jun,
       volume = {450},
       number = {2},
        pages = {2083-2093},
          doi = {10.1093/mnras/stv701},
archivePrefix = {arXiv},
       eprint = {1503.07864},
 primaryClass = {astro-ph.SR},
       adsurl = {https://ui.adsabs.harvard.edu/abs/2015MNRAS.450.2083R}
}

@ARTICLE{Farihi2017,
       author = {{Farihi}, J. and {von Hippel}, T. and {Pringle}, J.~E.},
        title = "{Magnetospherically-trapped dust and a possible model for the unusual transits at WD 1145+017}",
      journal = {\mnras},
         year = 2017,
        month = oct,
       volume = {471},
       number = {1},
        pages = {L145-L149},
          doi = {10.1093/mnrasl/slx122},
archivePrefix = {arXiv},
       eprint = {1707.09474},
 primaryClass = {astro-ph.SR},
       adsurl = {https://ui.adsabs.harvard.edu/abs/2017MNRAS.471L.145F}
}

@ARTICLE{Crifo1997,
       author = {{Crifo}, F. and {Vidal-Madjar}, A. and {Lallement}, R. and {Ferlet}, R. and {Gerbaldi}, M.},
        title = "{{\ensuremath{\beta}} Pictoris revisited by Hipparcos. Star properties.}",
      journal = {\aap},
         year = 1997,
        month = apr,
       volume = {320},
        pages = {L29-L32},
       adsurl = {https://ui.adsabs.harvard.edu/abs/1997A&A...320L..29C}
}

@ARTICLE{rafikov2011,
       author = {{Rafikov}, Roman R.},
        title = "{Metal Accretion onto White Dwarfs Caused by Poynting-Robertson Drag on their Debris Disks}",
      journal = {\apjl},
         year = 2011,
        month = may,
       volume = {732},
       number = {1},
          eid = {L3},
        pages = {L3},
          doi = {10.1088/2041-8205/732/1/L3},
archivePrefix = {arXiv},
       eprint = {1102.3153},
 primaryClass = {astro-ph.EP},
       adsurl = {https://ui.adsabs.harvard.edu/abs/2011ApJ...732L...3R}
}

@ARTICLE{vanderburg2015,
       author = {{Vanderburg}, Andrew and {Johnson}, John Asher and {Rappaport}, Saul and {Bieryla}, Allyson and {Irwin}, Jonathan and {Lewis}, John Arban and {Kipping}, David and {Brown}, Warren R. and {Dufour}, Patrick and {Ciardi}, David R. and {Angus}, Ruth and {Schaefer}, Laura and {Latham}, David W. and {Charbonneau}, David and {Beichman}, Charles and {Eastman}, Jason and {McCrady}, Nate and {Wittenmyer}, Robert A. and {Wright}, Jason T.},
        title = "{A disintegrating minor planet transiting a white dwarf}",
      journal = {\nat},
         year = 2015,
        month = oct,
       volume = {526},
       number = {7574},
        pages = {546-549},
          doi = {10.1038/nature15527},
archivePrefix = {arXiv},
       eprint = {1510.06387},
 primaryClass = {astro-ph.EP},
       adsurl = {https://ui.adsabs.harvard.edu/abs/2015Natur.526..546V}
}

@ARTICLE{veras2014b,
       author = {{Veras}, Dimitri and {Shannon}, Andrew and {G{\"a}nsicke}, Boris T.},
        title = "{Hydrogen delivery onto white dwarfs from remnant exo-Oort cloud comets}",
      journal = {\mnras},
         year = 2014,
        month = dec,
       volume = {445},
       number = {4},
        pages = {4175-4185},
          doi = {10.1093/mnras/stu2026},
archivePrefix = {arXiv},
       eprint = {1409.7691},
 primaryClass = {astro-ph.SR},
       adsurl = {https://ui.adsabs.harvard.edu/abs/2014MNRAS.445.4175V}
}

@ARTICLE{veras2017,
       author = {{Veras}, Dimitri and {Carter}, Philip J. and {Leinhardt}, Zo{\"e} M. and {G{\"a}nsicke}, Boris T.},
        title = "{Explaining the variability of WD 1145+017 with simulations of asteroid tidal disruption}",
      journal = {\mnras},
         year = 2017,
        month = feb,
       volume = {465},
       number = {1},
        pages = {1008-1022},
          doi = {10.1093/mnras/stw2748},
archivePrefix = {arXiv},
       eprint = {1610.06926},
 primaryClass = {astro-ph.EP},
       adsurl = {https://ui.adsabs.harvard.edu/abs/2017MNRAS.465.1008V}
}

@ARTICLE{veras2022,
       author = {{Veras}, Dimitri and {Birader}, Yusuf and {Zaman}, Uwais},
        title = "{Orbit decay of 2-100 au planetary remnants around white dwarfs with no gravitational assistance from planets}",
      journal = {\mnras},
         year = 2022,
        month = mar,
       volume = {510},
       number = {3},
        pages = {3379-3388},
          doi = {10.1093/mnras/stab3490},
archivePrefix = {arXiv},
       eprint = {2111.13713},
 primaryClass = {astro-ph.EP},
       adsurl = {https://ui.adsabs.harvard.edu/abs/2022MNRAS.510.3379V}
}

@ARTICLE{williams2024,
       author = {{Williams}, Jamie and {Gaensicke}, Boris and {Swan}, Andrew and {O'Brien}, Mairi and {Izquierdo}, Paula and {Cutolo}, Anna-Maria and {Cunningham}, Tim},
        title = "{PEWDD: A database of white dwarfs enriched by exo-planetary material}",
      journal = {arXiv e-prints},
         year = 2024,
        month = sep,
          eid = {arXiv:2409.16046},
        pages = {arXiv:2409.16046},
          doi = {10.48550/arXiv.2409.16046},
archivePrefix = {arXiv},
       eprint = {2409.16046},
 primaryClass = {astro-ph.EP},
       adsurl = {https://ui.adsabs.harvard.edu/abs/2024arXiv240916046W}
}

@ARTICLE{wyatt1950,
       author = {{Wyatt}, S.~P. and {Whipple}, F.~L.},
        title = "{The Poynting-Robertson effect on meteor orbits}",
      journal = {\apj},
         year = 1950,
        month = jan,
       volume = {111},
        pages = {134-141},
          doi = {10.1086/145244},
       adsurl = {https://ui.adsabs.harvard.edu/abs/1950ApJ...111..134W}
}

@ARTICLE{xu2017,
       author = {{Xu}, S. and {Zuckerman}, B. and {Dufour}, P. and {Young}, E.~D. and {Klein}, B. and {Jura}, M.},
        title = "{The Chemical Composition of an Extrasolar Kuiper-Belt-Object}",
      journal = {\apjl},
         year = 2017,
        month = feb,
       volume = {836},
       number = {1},
          eid = {L7},
        pages = {L7},
          doi = {10.3847/2041-8213/836/1/L7},
archivePrefix = {arXiv},
       eprint = {1702.02868},
 primaryClass = {astro-ph.EP},
       adsurl = {https://ui.adsabs.harvard.edu/abs/2017ApJ...836L...7X}
}

@ARTICLE{xu2018,
       author = {{Xu}, S. and {Rappaport}, S. and {van Lieshout}, R. and {Vanderburg}, A. and {Gary}, B. and {Hallakoun}, N. and {Ivanov}, V.~D. and {Wyatt}, M.~C. and {DeVore}, J. and {Bayliss}, D. and {Bento}, J. and {Bieryla}, A. and {Cameron}, A. and {Cann}, J.~M. and {Croll}, B. and {Collins}, K.~A. and {Dalba}, P.~A. and {Debes}, J. and {Doyle}, D. and {Dufour}, P. and {Ely}, J. and {Espinoza}, N. and {Joner}, M.~D. and {Jura}, M. and {Kaye}, T. and {McClain}, J.~L. and {Muirhead}, P. and {Palle}, E. and {Panka}, P.~A. and {Provencal}, J. and {Randall}, S. and {Rodriguez}, J.~E. and {Scarborough}, J. and {Sefako}, R. and {Shporer}, A. and {Strickland}, W. and {Zhou}, G. and {Zuckerman}, B.},
        title = "{A dearth of small particles in the transiting material around the white dwarf WD 1145+017}",
      journal = {\mnras},
         year = 2018,
        month = mar,
       volume = {474},
       number = {4},
        pages = {4795-4809},
          doi = {10.1093/mnras/stx3023},
archivePrefix = {arXiv},
       eprint = {1711.06960},
 primaryClass = {astro-ph.EP},
       adsurl = {https://ui.adsabs.harvard.edu/abs/2018MNRAS.474.4795X}
}

@ARTICLE{xu2024,
       author = {{Xu}, Siyi and {Rogers}, Laura K. and {Blouin}, Simon},
        title = "{The chemistry of extra-solar materials from white dwarf planetary systems}",
      journal = {arXiv e-prints},
         year = 2024,
        month = apr,
          eid = {arXiv:2404.15425},
        pages = {arXiv:2404.15425},
          doi = {10.48550/arXiv.2404.15425},
archivePrefix = {arXiv},
       eprint = {2404.15425},
 primaryClass = {astro-ph.EP},
       adsurl = {https://ui.adsabs.harvard.edu/abs/2024arXiv240415425X}
}

@article{afghan_observations_2023,
	title = {Observations of a dust tail gap in comet {C}/2014 {Q1} ({PanSTARRS})},
	volume = {390},
	issn = {0019-1035},
	url = {https://www.sciencedirect.com/science/article/pii/S0019103522003785},
	doi = {10.1016/j.icarus.2022.115286},
	urldate = {2023-09-25},
	journal = {Icarus},
	author = {Afghan, Qasim and Jones, Geraint H. and Price, Oliver and Coates, Andrew},
	month = jan,
	year = {2023},
	pages = {115286},
}

@article{finson_theory_1968,
	title = {A theory of dust comets. {I}. {Model} and equations},
	volume = {154},
	issn = {0004-637X},
	url = {https://ui.adsabs.harvard.edu/abs/1968ApJ...154..327F},
	doi = {10.1086/149761},
	urldate = {2023-09-29},
	journal = {The Astrophysical Journal},
	author = {Finson, M. J. and Probstein, R. F.},
	month = oct,
	year = {1968},
	note = {ADS Bibcode: 1968ApJ...154..327F},
	pages = {327--352},
}

@article{price_fine-scale_2019,
	title = {Fine-scale structure in cometary dust tails {I}: {Analysis} of striae in {Comet} {C}/2006 {P1} ({McNaught}) through temporal mapping},
	volume = {319},
	issn = {0019-1035},
	shorttitle = {Fine-scale structure in cometary dust tails {I}},
	url = {https://www.sciencedirect.com/science/article/pii/S0019103518301192},
	doi = {10.1016/j.icarus.2018.09.013},
	urldate = {2023-10-02},
	journal = {Icarus},
	author = {Price, Oliver and Jones, Geraint H. and Morrill, Jeff and Owens, Mathew and Battams, Karl and Morgan, Huw and Drückmuller, Miloslav and Deiries, Sebastian},
	month = feb,
	year = {2019},
	pages = {540--557},
}

@article{price_fine-scale_2023,
	title = {Fine-scale structure in cometary dust tails {II}: {Further} evidence for a solar wind influence on cometary dust dynamics from the analysis of striae in comet {C}/2011 {L4} {Pan}-{STARRS}},
	volume = {389},
	issn = {0019-1035},
	shorttitle = {Fine-scale structure in cometary dust tails {II}},
	url = {https://ui.adsabs.harvard.edu/abs/2023Icar..38915218P},
	doi = {10.1016/j.icarus.2022.115218},
	urldate = {2023-10-06},
	journal = {Icarus},
	author = {Price, Oliver and Jones, Geraint H. and Battams, Karl and Owens, Mathew},
	month = jan,
	year = {2023},
	note = {ADS Bibcode: 2023Icar..38915218P},
	pages = {115218},
}

@article{edberg_cme_2016,
	title = {{CME} impact on comet {67P}/{Churyumov}-{Gerasimenko}},
	volume = {462},
	issn = {0035-8711},
	url = {https://ui.adsabs.harvard.edu/abs/2016MNRAS.462S..45E},
	doi = {10.1093/mnras/stw2112},
	urldate = {2023-10-06},
	journal = {Monthly Notices of the Royal Astronomical Society},
	author = {Edberg, Niklas J. T. and Alho, M. and André, M. and Andrews, D. J. and Behar, E. and Burch, J. L. and Carr, C. M. and Cupido, E. and Engelhardt, I. A. D. and Eriksson, A. I. and Glassmeier, K. -H. and Goetz, C. and Goldstein, R. and Henri, P. and Johansson, F. L. and Koenders, C. and Mandt, K. and Möstl, C. and Nilsson, H. and Odelstad, E. and Richter, I. and Simon Wedlund, C. and Stenberg Wieser, G. and Szego, K. and Vigren, E. and Volwerk, M.},
	month = nov,
	year = {2016},
	note = {ADS Bibcode: 2016MNRAS.462S..45E},
	pages = {S45--S56},
}

@article{kimura_electrostatic_2022,
	title = {Electrostatic dust ejection from asteroid (3200) {Phaethon} with the aid of mobile alkali ions at perihelion},
	volume = {382},
	issn = {0019-1035},
	url = {https://www.sciencedirect.com/science/article/pii/S001910352200135X},
	doi = {10.1016/j.icarus.2022.115022},
	urldate = {2023-10-09},
	journal = {Icarus},
	author = {Kimura, Hiroshi and Ohtsuka, Katsuhito and Kikuchi, Shota and Ohtsuki, Keiji and Arai, Tomoko and Yoshida, Fumi and Hirata, Naoyuki and Senshu, Hiroki and Wada, Koji and Hirai, Takayuki and Hong, Peng K. and Kobayashi, Masanori and Ishibashi, Ko and Yamada, Manabu and Okamoto, Takaya},
	month = aug,
	year = {2022},
	pages = {115022},
}

@article{jia_study_2009,
	title = {Study of the 2007 {April} 20 {CME}-{Comet} {Interaction} {Event} with an {MHD} {Model}},
	volume = {696},
	issn = {0004-637X},
	url = {https://ui.adsabs.harvard.edu/abs/2009ApJ...696L..56J},
	doi = {10.1088/0004-637X/696/1/L56},
	urldate = {2023-10-12},
	journal = {The Astrophysical Journal},
	author = {Jia, Y. D. and Russell, C. T. and Jian, L. K. and Manchester, W. B. and Cohen, O. and Vourlidas, A. and Hansen, K. C. and Combi, M. R. and Gombosi, T. I.},
	month = may,
	year = {2009},
	note = {ADS Bibcode: 2009ApJ...696L..56J},
	pages = {L56--L60},
}

@article{kuchar_observations_2008,
	title = {Observations of a comet tail disruption induced by the passage of a {CME}},
	volume = {113},
	issn = {2156-2202},
	url = {https://onlinelibrary.wiley.com/doi/abs/10.1029/2007JA012603},
	doi = {10.1029/2007JA012603},
	language = {en},
	number = {A4},
	urldate = {2024-02-05},
	journal = {Journal of Geophysical Research: Space Physics},
	author = {Kuchar, T. A. and Buffington, A. and Arge, C. N. and Hick, P. P. and Howard, T. A. and Jackson, B. V. and Johnston, J. C. and Mizuno, D. R. and Tappin, S. J. and Webb, D. F.},
	year = {2008},
	note = {\_eprint: https://onlinelibrary.wiley.com/doi/pdf/10.1029/2007JA012603},
}

@article{jian-chun_disconnection_2011,
	title = {A {Disconnection} {Event} of {Comet} {Lulin}†},
	volume = {35},
	issn = {0275-1062},
	url = {https://www.sciencedirect.com/science/article/pii/S0275106211000683},
	doi = {10.1016/j.chinastron.2011.07.007},
	number = {3},
	urldate = {2024-02-06},
	journal = {Chinese Astronomy and Astrophysics},
	author = {Jian-chun, Shi and Chi-sheng, Lin and Zhong-wei, Hu and Hai-bin, Zhao and Yue-hua, Ma},
	month = jul,
	year = {2011},
	pages = {295--303},
}

@article{janson_imaging_2023,
	title = {Imaging of exocomets with infrared interferometry},
	volume = {671},
	issn = {0004-6361},
	url = {https://ui.adsabs.harvard.edu/abs/2023A&A...671A.114J},
	doi = {10.1051/0004-6361/202245402},
	urldate = {2024-03-07},
	journal = {Astronomy and Astrophysics},
	author = {Janson, Markus and Patel, Jayshil and Ringqvist, Simon C. and Lu, Cicero and Rebollido, Isabel and Lichtenberg, Tim and Brandeker, Alexis and Angerhausen, Daniel and Noack, Lena},
	month = mar,
	year = {2023},
	note = {ADS Bibcode: 2023A\&A...671A.114J},
	pages = {A114},
}

@article{donati_magnetic_2023,
	title = {The magnetic field and multiple planets of the young dwarf {AU} {Mic}},
	volume = {525},
	issn = {0035-8711},
	url = {https://doi.org/10.1093/mnras/stad1193},
	doi = {10.1093/mnras/stad1193},
	number = {1},
	urldate = {2024-03-19},
	journal = {Monthly Notices of the Royal Astronomical Society},
	author = {Donati, J-F and Cristofari, P I and Finociety, B and Klein, B and Moutou, C and Gaidos, E and Cadieux, C and Artigau, E and Correia, A C M and Boué, G and Cook, N J and Carmona, A and Lehmann, L T and Bouvier, J and Martioli, E and Morin, J and Fouqué, P and Delfosse, X and Doyon, R and Hébrard, G and Alencar, S H P and Laskar, J and Arnold, L and Petit, P and Kospal, A and Vidotto, A and Folsom, C P and collaboration, the S L S},
	month = oct,
	year = {2023},
	pages = {455--475},
}

@article{zicher_one_2022,
	title = {One year of {AU} {Mic} with {HARPS} - {I}. {Measuring} the masses of the two transiting planets},
	volume = {512},
	issn = {0035-8711},
	url = {https://ui.adsabs.harvard.edu/abs/2022MNRAS.512.3060Z},
	doi = {10.1093/mnras/stac614},
	urldate = {2024-03-20},
	journal = {Monthly Notices of the Royal Astronomical Society},
	author = {Zicher, Norbert and Barragán, Oscar and Klein, Baptiste and Aigrain, Suzanne and Owen, James E. and Gandolfi, Davide and Lagrange, Anne-Marie and Serrano, Luisa Maria and Kaye, Laurel and Nielsen, Louise Dyregaard and Rajpaul, Vinesh Maguire and Grandjean, Antoine and Goffo, Elisa and Nicholson, Belinda},
	month = may,
	year = {2022},
	note = {ADS Bibcode: 2022MNRAS.512.3060Z},
	pages = {3060--3078},
}

@article{sekanina_comet_2012,
	title = {Comet {C}/2011 {W3} ({Lovejoy}): {Orbit} {Determination}, {Outbursts}, {Disintegration} of {Nucleus}, {Dust}-tail {Morphology}, and {Relationship} to {New} {Cluster} of {Bright} {Sungrazers}},
	volume = {757},
	issn = {0004-637X},
	shorttitle = {Comet {C}/2011 {W3} ({Lovejoy})},
	url = {https://ui.adsabs.harvard.edu/abs/2012ApJ...757..127S},
	doi = {10.1088/0004-637X/757/2/127},
	urldate = {2024-05-03},
	journal = {The Astrophysical Journal},
	author = {Sekanina, Zdenek and Chodas, Paul W.},
	month = oct,
	year = {2012},
	note = {Publisher: IOP
ADS Bibcode: 2012ApJ...757..127S},
	pages = {127},
}

@article{klacka_times_2008,
	title = {Times of inspiralling for interplanetary dust grains},
	volume = {390},
	issn = {0035-8711},
	url = {https://ui.adsabs.harvard.edu/abs/2008MNRAS.390.1491K},
	doi = {10.1111/j.1365-2966.2008.13801.x},
	urldate = {2024-07-16},
	journal = {Monthly Notices of the Royal Astronomical Society},
	author = {Klačka, J. and Kocifaj, M.},
	month = nov,
	year = {2008},
	note = {Publisher: OUP
ADS Bibcode: 2008MNRAS.390.1491K},
	pages = {1491--1495},
}

@article{klacka_poyntingrobertson_2014,
	title = {The {Poynting}–{Robertson} effect: {A} critical perspective},
	volume = {232},
	issn = {0019-1035},
	shorttitle = {The {Poynting}–{Robertson} effect},
	url = {https://www.sciencedirect.com/science/article/pii/S0019103512002734},
	doi = {10.1016/j.icarus.2012.06.044},
	urldate = {2024-07-25},
	journal = {Icarus},
	author = {Klačka, J. and Petržala, J. and Pástor, P. and Kómar, L.},
	month = apr,
	year = {2014},
	pages = {249--262},
}

@article{rappaport_drifting_2016,
	title = {Drifting asteroid fragments around {WD} 1145+017},
	volume = {458},
	issn = {0035-8711},
	url = {https://ui.adsabs.harvard.edu/abs/2016MNRAS.458.3904R},
	doi = {10.1093/mnras/stw612},
	urldate = {2024-09-03},
	journal = {Monthly Notices of the Royal Astronomical Society},
	author = {Rappaport, S. and Gary, B. L. and Kaye, T. and Vanderburg, A. and Croll, B. and Benni, P. and Foote, J.},
	month = jun,
	year = {2016},
	note = {Publisher: OUP
ADS Bibcode: 2016MNRAS.458.3904R},
	pages = {3904--3917},
}

@article{nordheim_surface_2015,
	title = {Surface charging and electrostatic dust acceleration at the nucleus of comet {67P} during periods of low activity},
	volume = {119},
	issn = {0032-0633},
	url = {https://www.sciencedirect.com/science/article/pii/S003206331500238X},
	doi = {10.1016/j.pss.2015.08.008},
	urldate = {2024-09-25},
	journal = {Planetary and Space Science},
	author = {Nordheim, T. A. and Jones, G. H. and Halekas, J. S. and Roussos, E. and Coates, A. J.},
	month = dec,
	year = {2015},
	pages = {24--35},
}

@article{hartzell_evaluation_2022,
	title = {An {Evaluation} of {Electrostatic} {Lofting} and {Subsequent} {Particle} {Motion} on {Bennu}},
	volume = {3},
	issn = {2632-3338},
	url = {https://iopscience.iop.org/article/10.3847/PSJ/ac5629/meta},
	doi = {10.3847/PSJ/ac5629},
	language = {en},
	number = {4},
	urldate = {2024-09-26},
	journal = {The Planetary Science Journal},
	author = {Hartzell, Christine and Zimmerman, Michael and Hergenrother, Carl},
	month = apr,
	year = {2022},
	note = {Publisher: IOP Publishing},
	pages = {85},
}

@article{bodman_kic_2016,
	title = {{KIC} 8462852: {Transit} of a {Large} {Comet} {Family}},
	volume = {819},
	issn = {2041-8205},
	shorttitle = {{KIC} 8462852},
	url = {https://dx.doi.org/10.3847/2041-8205/819/2/L34},
	doi = {10.3847/2041-8205/819/2/L34},
	language = {en},
	number = {2},
	urldate = {2024-10-01},
	journal = {The Astrophysical Journal Letters},
	author = {Bodman, Eva H. L. and Quillen, Alice},
	month = mar,
	year = {2016},
	note = {Publisher: The American Astronomical Society},
	pages = {L34},
}

@article{lawson_jwstnircam_2023,
	title = {{JWST}/{NIRCam} {Coronagraphy} of the {Young} {Planet}-hosting {Debris} {Disk} {AU} {Microscopii}},
	volume = {166},
	issn = {0004-6256},
	url = {https://ui.adsabs.harvard.edu/abs/2023AJ....166..150L},
	doi = {10.3847/1538-3881/aced08},
	urldate = {2024-10-02},
	journal = {The Astronomical Journal},
	author = {Lawson, Kellen and Schlieder, Joshua E. and Leisenring, Jarron M. and Bogat, Ell and Beichman, Charles A. and Bryden, Geoffrey and Gáspár, András and Groff, Tyler D. and McElwain, Michael W. and Meyer, Michael R. and Barclay, Thomas and Calissendorff, Per and De Furio, Matthew and Ygouf, Marie and Boccaletti, Anthony and Greene, Thomas P. and Krist, John and Plavchan, Peter and Rieke, Marcia J. and Roellig, Thomas L. and Stansberry, John and Wisniewski, John P. and Young, Erick T.},
	month = oct,
	year = {2023},
	note = {Publisher: IOP
ADS Bibcode: 2023AJ....166..150L},
	pages = {150},
}

@article{arnold_stumbling_2022,
	title = {Stumbling over {Planetary} {Building} {Blocks}: {AU} {Microscopii} as an {Example} of the {Challenge} of {Retrieving} {Debris}-disk {Dust} {Properties}},
	volume = {930},
	issn = {0004-637X},
	shorttitle = {Stumbling over {Planetary} {Building} {Blocks}},
	url = {https://ui.adsabs.harvard.edu/abs/2022ApJ...930..123A},
	doi = {10.3847/1538-4357/ac63a9},
	urldate = {2024-10-02},
	journal = {The Astrophysical Journal},
	author = {Arnold, Jessica A. and Weinberger, Alycia J. and Videen, Gorden and Zubko, Evgenij S.},
	month = may,
	year = {2022},
	note = {Publisher: IOP
ADS Bibcode: 2022ApJ...930..123A},
	pages = {123},
}

@article{alvarado-gomez_simulating_2022,
	title = {Simulating the {Space} {Weather} in the {AU} {Mic} {System}: {Stellar} {Winds} and {Extreme} {Coronal} {Mass} {Ejections}},
	volume = {928},
	issn = {0004-637X},
	shorttitle = {Simulating the {Space} {Weather} in the {AU} {Mic} {System}},
	url = {https://ui.adsabs.harvard.edu/abs/2022ApJ...928..147A},
	doi = {10.3847/1538-4357/ac54b8},
	urldate = {2024-10-02},
	journal = {The Astrophysical Journal},
	author = {Alvarado-Gómez, Julián D. and Cohen, Ofer and Drake, Jeremy J. and Fraschetti, Federico and Poppenhaeger, Katja and Garraffo, Cecilia and Chebly, Judy and Ilin, Ekaterina and Harbach, Laura and Kochukhov, Oleg},
	month = apr,
	year = {2022},
	note = {Publisher: IOP
ADS Bibcode: 2022ApJ...928..147A},
	pages = {147},
}

@article{gilbert_flares_2022,
	title = {Flares, {Rotation}, and {Planets} of the {AU} {Mic} {System} from {TESS} {Observations}},
	volume = {163},
	issn = {1538-3881},
	url = {https://dx.doi.org/10.3847/1538-3881/ac23ca},
	doi = {10.3847/1538-3881/ac23ca},
	language = {en},
	number = {4},
	urldate = {2024-10-02},
	journal = {The Astronomical Journal},
	author = {Gilbert, Emily A. and Barclay, Thomas and Quintana, Elisa V. and Walkowicz, Lucianne M. and Vega, Laura D. and Schlieder, Joshua E. and Monsue, Teresa and Cale, Bryson L. and Collins, Kevin I. and Gaidos, Eric and Mufti, Mohammed El and Reefe, Michael A. and Plavchan, Peter and Tanner, Angelle and Wittenmyer, Robert A. and Wittrock, Justin M. and Jenkins, Jon M. and Latham, David W. and Ricker, George R. and Rose, Mark E. and Seager, S. and Vanderspek, Roland K. and Winn, Joshua N.},
	month = mar,
	year = {2022},
	note = {Publisher: The American Astronomical Society},
	pages = {147},
}

@article{plavchan_planet_2020,
	title = {A planet within the debris disk around the pre-main-sequence star {AU} {Microscopii}},
	volume = {582},
	issn = {0028-0836},
	url = {https://ui.adsabs.harvard.edu/abs/2020Natur.582..497P},
	doi = {10.1038/s41586-020-2400-z},
	urldate = {2024-10-02},
	journal = {Nature},
	author = {Plavchan, Peter and Barclay, Thomas and Gagné, Jonathan and Gao, Peter and Cale, Bryson and Matzko, William and Dragomir, Diana and Quinn, Sam and Feliz, Dax and Stassun, Keivan and Crossfield, Ian J. M. and Berardo, David A. and Latham, David W. and Tieu, Ben and Anglada-Escudé, Guillem and Ricker, George and Vanderspek, Roland and Seager, Sara and Winn, Joshua N. and Jenkins, Jon M. and Rinehart, Stephen and Krishnamurthy, Akshata and Dynes, Scott and Doty, John and Adams, Fred and Afanasev, Dennis A. and Beichman, Chas and Bottom, Mike and Bowler, Brendan P. and Brinkworth, Carolyn and Brown, Carolyn J. and Cancino, Andrew and Ciardi, David R. and Clampin, Mark and Clark, Jake T. and Collins, Karen and Davison, Cassy and Foreman-Mackey, Daniel and Furlan, Elise and Gaidos, Eric J. and Geneser, Claire and Giddens, Frank and Gilbert, Emily and Hall, Ryan and Hellier, Coel and Henry, Todd and Horner, Jonathan and Howard, Andrew W. and Huang, Chelsea and Huber, Joseph and Kane, Stephen R. and Kenworthy, Matthew and Kielkopf, John and Kipping, David and Klenke, Chris and Kruse, Ethan and Latouf, Natasha and Lowrance, Patrick and Mennesson, Bertrand and Mengel, Matthew and Mills, Sean M. and Morton, Tim and Narita, Norio and Newton, Elisabeth and Nishimoto, America and Okumura, Jack and Palle, Enric and Pepper, Joshua and Quintana, Elisa V. and Roberge, Aki and Roccatagliata, Veronica and Schlieder, Joshua E. and Tanner, Angelle and Teske, Johanna and Tinney, C. G. and Vanderburg, Andrew and von Braun, Kaspar and Walp, Bernie and Wang, Jason and Wang, Sharon Xuesong and Weigand, Denise and White, Russel and Wittenmyer, Robert A. and Wright, Duncan J. and Youngblood, Allison and Zhang, Hui and Zilberman, Perri},
	month = jun,
	year = {2020},
	note = {ADS Bibcode: 2020Natur.582..497P},
	pages = {497--500},
}

@article{grady_eroding_2020,
	title = {The {Eroding} {Disk} of {AU} {Mic}},
	volume = {889},
	issn = {0004-637X},
	url = {https://ui.adsabs.harvard.edu/abs/2020ApJ...889L..21G},
	doi = {10.3847/2041-8213/ab65bb},
	urldate = {2024-10-02},
	journal = {The Astrophysical Journal},
	author = {Grady, C. A. and Wisniewski, J. P. and Schneider, G. and Boccaletti, A. and Gaspar, A. and Debes, J. H. and Hines, D. C. and Stark, C. C. and Thalmann, C. and Lagrange, A. -M. and Augereau, J. -C. and Sezestre, E. and Milli, J. and Henning, Th. and Kuchner, M. J.},
	month = jan,
	year = {2020},
	note = {Publisher: IOP
ADS Bibcode: 2020ApJ...889L..21G},
	pages = {L21},
}

@article{boccaletti_observations_2018,
	title = {Observations of fast-moving features in the debris disk of {AU} {Mic} on a three-year timescale: {Confirmation} and new discoveries},
	volume = {614},
	issn = {0004-6361},
	shorttitle = {Observations of fast-moving features in the debris disk of {AU} {Mic} on a three-year timescale},
	url = {https://ui.adsabs.harvard.edu/abs/2018A&A...614A..52B},
	doi = {10.1051/0004-6361/201732462},
	urldate = {2024-10-02},
	journal = {Astronomy and Astrophysics},
	author = {Boccaletti, A. and Sezestre, E. and Lagrange, A. -M. and Thébault, P. and Gratton, R. and Langlois, M. and Thalmann, C. and Janson, M. and Delorme, P. and Augereau, J. -C. and Schneider, G. and Milli, J. and Grady, C. and Debes, J. and Kral, Q. and Olofsson, J. and Carson, J. and Maire, A. L. and Henning, T. and Wisniewski, J. and Schlieder, J. and Dominik, C. and Desidera, S. and Ginski, C. and Hines, D. and Ménard, F. and Mouillet, D. and Pawellek, N. and Vigan, A. and Lagadec, E. and Avenhaus, H. and Beuzit, J. -L. and Biller, B. and Bonavita, M. and Bonnefoy, M. and Brandner, W. and Cantalloube, F. and Chauvin, G. and Cheetham, A. and Cudel, M. and Gry, C. and Daemgen, S. and Feldt, M. and Galicher, R. and Girard, J. and Hagelberg, J. and Janin-Potiron, P. and Kasper, M. and Le Coroller, H. and Mesa, D. and Peretti, S. and Perrot, C. and Samland, M. and Sissa, E. and Wildi, F. and Zurlo, A. and Rochat, S. and Stadler, E. and Gluck, L. and Origné, A. and Llored, M. and Baudoz, P. and Rousset, G. and Martinez, P. and Rigal, F.},
	month = jun,
	year = {2018},
	note = {ADS Bibcode: 2018A\&A...614A..52B},
	pages = {A52},
}

@article{gaia_collaboration_gaia_2018,
	title = {Gaia {Data} {Release} 2. {Summary} of the contents and survey properties},
	volume = {616},
	issn = {0004-6361},
	url = {https://ui.adsabs.harvard.edu/abs/2018A&A...616A...1G},
	doi = {10.1051/0004-6361/201833051},
	urldate = {2024-10-04},
	journal = {Astronomy and Astrophysics},
	author = {{Gaia Collaboration} and Brown, A. G. A. and Vallenari, A. and Prusti, T. and de Bruijne, J. H. J. and Babusiaux, C. and Bailer-Jones, C. A. L. and Biermann, M. and Evans, D. W. and Eyer, L. and Jansen, F. and Jordi, C. and Klioner, S. A. and Lammers, U. and Lindegren, L. and Luri, X. and Mignard, F. and Panem, C. and Pourbaix, D. and Randich, S. and Sartoretti, P. and Siddiqui, H. I. and Soubiran, C. and van Leeuwen, F. and Walton, N. A. and Arenou, F. and Bastian, U. and Cropper, M. and Drimmel, R. and Katz, D. and Lattanzi, M. G. and Bakker, J. and Cacciari, C. and Castañeda, J. and Chaoul, L. and Cheek, N. and De Angeli, F. and Fabricius, C. and Guerra, R. and Holl, B. and Masana, E. and Messineo, R. and Mowlavi, N. and Nienartowicz, K. and Panuzzo, P. and Portell, J. and Riello, M. and Seabroke, G. M. and Tanga, P. and Thévenin, F. and Gracia-Abril, G. and Comoretto, G. and Garcia-Reinaldos, M. and Teyssier, D. and Altmann, M. and Andrae, R. and Audard, M. and Bellas-Velidis, I. and Benson, K. and Berthier, J. and Blomme, R. and Burgess, P. and Busso, G. and Carry, B. and Cellino, A. and Clementini, G. and Clotet, M. and Creevey, O. and Davidson, M. and De Ridder, J. and Delchambre, L. and Dell'Oro, A. and Ducourant, C. and Fernández-Hernández, J. and Fouesneau, M. and Frémat, Y. and Galluccio, L. and García-Torres, M. and González-Núñez, J. and González-Vidal, J. J. and Gosset, E. and Guy, L. P. and Halbwachs, J. -L. and Hambly, N. C. and Harrison, D. L. and Hernández, J. and Hestroffer, D. and Hodgkin, S. T. and Hutton, A. and Jasniewicz, G. and Jean-Antoine-Piccolo, A. and Jordan, S. and Korn, A. J. and Krone-Martins, A. and Lanzafame, A. C. and Lebzelter, T. and Löffler, W. and Manteiga, M. and Marrese, P. M. and Martín-Fleitas, J. M. and Moitinho, A. and Mora, A. and Muinonen, K. and Osinde, J. and Pancino, E. and Pauwels, T. and Petit, J. -M. and Recio-Blanco, A. and Richards, P. J. and Rimoldini, L. and Robin, A. C. and Sarro, L. M. and Siopis, C. and Smith, M. and Sozzetti, A. and Süveges, M. and Torra, J. and van Reeven, W. and Abbas, U. and Abreu Aramburu, A. and Accart, S. and Aerts, C. and Altavilla, G. and Alvarez, M. A. and Alvarez, R. and Alves, J. and Anderson, R. I. and Andrei, A. H. and Anglada Varela, E. and Antiche, E. and Antoja, T. and Arcay, B. and Astraatmadja, T. L. and Bach, N. and Baker, S. G. and Balaguer-Núñez, L. and Balm, P. and Barache, C. and Barata, C. and Barbato, D. and Barblan, F. and Barklem, P. S. and Barrado, D. and Barros, M. and Barstow, M. A. and Bartholomé Muñoz, S. and Bassilana, J. -L. and Becciani, U. and Bellazzini, M. and Berihuete, A. and Bertone, S. and Bianchi, L. and Bienaymé, O. and Blanco-Cuaresma, S. and Boch, T. and Boeche, C. and Bombrun, A. and Borrachero, R. and Bossini, D. and Bouquillon, S. and Bourda, G. and Bragaglia, A. and Bramante, L. and Breddels, M. A. and Bressan, A. and Brouillet, N. and Brüsemeister, T. and Brugaletta, E. and Bucciarelli, B. and Burlacu, A. and Busonero, D. and Butkevich, A. G. and Buzzi, R. and Caffau, E. and Cancelliere, R. and Cannizzaro, G. and Cantat-Gaudin, T. and Carballo, R. and Carlucci, T. and Carrasco, J. M. and Casamiquela, L. and Castellani, M. and Castro-Ginard, A. and Charlot, P. and Chemin, L. and Chiavassa, A. and Cocozza, G. and Costigan, G. and Cowell, S. and Crifo, F. and Crosta, M. and Crowley, C. and Cuypers, J. and Dafonte, C. and Damerdji, Y. and Dapergolas, A. and David, P. and David, M. and de Laverny, P. and De Luise, F. and De March, R. and de Martino, D. and de Souza, R. and de Torres, A. and Debosscher, J. and del Pozo, E. and Delbo, M. and Delgado, A. and Delgado, H. E. and Di Matteo, P. and Diakite, S. and Diener, C. and Distefano, E. and Dolding, C. and Drazinos, P. and Durán, J. and Edvardsson, B. and Enke, H. and Eriksson, K. and Esquej, P. and Eynard Bontemps, G. and Fabre, C. and Fabrizio, M. and Faigler, S. and Falcão, A. J. and Farràs Casas, M. and Federici, L. and Fedorets, G. and Fernique, P. and Figueras, F. and Filippi, F. and Findeisen, K. and Fonti, A. and Fraile, E. and Fraser, M. and Frézouls, B. and Gai, M. and Galleti, S. and Garabato, D. and García-Sedano, F. and Garofalo, A. and Garralda, N. and Gavel, A. and Gavras, P. and Gerssen, J. and Geyer, R. and Giacobbe, P. and Gilmore, G. and Girona, S. and Giuffrida, G. and Glass, F. and Gomes, M. and Granvik, M. and Gueguen, A. and Guerrier, A. and Guiraud, J. and Gutiérrez-Sánchez, R. and Haigron, R. and Hatzidimitriou, D. and Hauser, M. and Haywood, M. and Heiter, U. and Helmi, A. and Heu, J. and Hilger, T. and Hobbs, D. and Hofmann, W. and Holland, G. and Huckle, H. E. and Hypki, A. and Icardi, V. and Janßen, K. and Jevardat de Fombelle, G. and Jonker, P. G. and Juhász, A. L. and Julbe, F. and Karampelas, A. and Kewley, A. and Klar, J. and Kochoska, A. and Kohley, R. and Kolenberg, K. and Kontizas, M. and Kontizas, E. and Koposov, S. E. and Kordopatis, G. and Kostrzewa-Rutkowska, Z. and Koubsky, P. and Lambert, S. and Lanza, A. F. and Lasne, Y. and Lavigne, J. -B. and Le Fustec, Y. and Le Poncin-Lafitte, C. and Lebreton, Y. and Leccia, S. and Leclerc, N. and Lecoeur-Taibi, I. and Lenhardt, H. and Leroux, F. and Liao, S. and Licata, E. and Lindstrøm, H. E. P. and Lister, T. A. and Livanou, E. and Lobel, A. and López, M. and Managau, S. and Mann, R. G. and Mantelet, G. and Marchal, O. and Marchant, J. M. and Marconi, M. and Marinoni, S. and Marschalkó, G. and Marshall, D. J. and Martino, M. and Marton, G. and Mary, N. and Massari, D. and Matijevič, G. and Mazeh, T. and McMillan, P. J. and Messina, S. and Michalik, D. and Millar, N. R. and Molina, D. and Molinaro, R. and Molnár, L. and Montegriffo, P. and Mor, R. and Morbidelli, R. and Morel, T. and Morris, D. and Mulone, A. F. and Muraveva, T. and Musella, I. and Nelemans, G. and Nicastro, L. and Noval, L. and O'Mullane, W. and Ordénovic, C. and Ordóñez-Blanco, D. and Osborne, P. and Pagani, C. and Pagano, I. and Pailler, F. and Palacin, H. and Palaversa, L. and Panahi, A. and Pawlak, M. and Piersimoni, A. M. and Pineau, F. -X. and Plachy, E. and Plum, G. and Poggio, E. and Poujoulet, E. and Prša, A. and Pulone, L. and Racero, E. and Ragaini, S. and Rambaux, N. and Ramos-Lerate, M. and Regibo, S. and Reylé, C. and Riclet, F. and Ripepi, V. and Riva, A. and Rivard, A. and Rixon, G. and Roegiers, T. and Roelens, M. and Romero-Gómez, M. and Rowell, N. and Royer, F. and Ruiz-Dern, L. and Sadowski, G. and Sagristà Sellés, T. and Sahlmann, J. and Salgado, J. and Salguero, E. and Sanna, N. and Santana-Ros, T. and Sarasso, M. and Savietto, H. and Schultheis, M. and Sciacca, E. and Segol, M. and Segovia, J. C. and Ségransan, D. and Shih, I. -C. and Siltala, L. and Silva, A. F. and Smart, R. L. and Smith, K. W. and Solano, E. and Solitro, F. and Sordo, R. and Soria Nieto, S. and Souchay, J. and Spagna, A. and Spoto, F. and Stampa, U. and Steele, I. A. and Steidelmüller, H. and Stephenson, C. A. and Stoev, H. and Suess, F. F. and Surdej, J. and Szabados, L. and Szegedi-Elek, E. and Tapiador, D. and Taris, F. and Tauran, G. and Taylor, M. B. and Teixeira, R. and Terrett, D. and Teyssandier, P. and Thuillot, W. and Titarenko, A. and Torra Clotet, F. and Turon, C. and Ulla, A. and Utrilla, E. and Uzzi, S. and Vaillant, M. and Valentini, G. and Valette, V. and van Elteren, A. and Van Hemelryck, E. and van Leeuwen, M. and Vaschetto, M. and Vecchiato, A. and Veljanoski, J. and Viala, Y. and Vicente, D. and Vogt, S. and von Essen, C. and Voss, H. and Votruba, V. and Voutsinas, S. and Walmsley, G. and Weiler, M. and Wertz, O. and Wevers, T. and Wyrzykowski, L. and Yoldas, A. and Žerjal, M. and Ziaeepour, H. and Zorec, J. and Zschocke, S. and Zucker, S. and Zurbach, C. and Zwitter, T.},
	month = aug,
	year = {2018},
	note = {ADS Bibcode: 2018A\&A...616A...1G},
	pages = {A1},
}

@article{miret-roig_dynamical_2020,
	title = {Dynamical traceback age of the beta {Pictoris} moving group},
	volume = {642},
	copyright = {© N. Miret-Roig et al. 2020},
	issn = {0004-6361, 1432-0746},
	url = {https://www.aanda.org/articles/aa/abs/2020/10/aa38765-20/aa38765-20.html},
	doi = {10.1051/0004-6361/202038765},
	language = {en},
	urldate = {2024-10-04},
	journal = {Astronomy \& Astrophysics},
	author = {Miret-Roig, N. and Galli, P. a. B. and Brandner, W. and Bouy, H. and Barrado, D. and Olivares, J. and Antoja, T. and Romero-Gómez, M. and Figueras, F. and Lillo-Box, J.},
	month = oct,
	year = {2020},
	note = {Publisher: EDP Sciences},
	pages = {A179},
}

@article{klein_one_2022,
	title = {One year of {AU} {Mic} with {HARPS} – {II}. {Stellar} activity and star–planet interaction},
	volume = {512},
	issn = {0035-8711},
	url = {https://doi.org/10.1093/mnras/stac761},
	doi = {10.1093/mnras/stac761},
	number = {4},
	urldate = {2024-10-04},
	journal = {Monthly Notices of the Royal Astronomical Society},
	author = {Klein, Baptiste and Zicher, Norbert and Kavanagh, Robert D and Nielsen, Louise D and Aigrain, Suzanne and Vidotto, Aline A and Barragán, Oscar and Strugarek, Antoine and Nicholson, Belinda and Donati, Jean-François and Bouvier, Jérôme},
	month = jun,
	year = {2022},
	pages = {5067--5084},
}

@article{macgregor_millimeter_2013,
	title = {Millimeter {Emission} {Structure} in the {First} {ALMA} {Image} of the {AU} {Mic} {Debris} {Disk}},
	volume = {762},
	issn = {0004-637X},
	url = {https://ui.adsabs.harvard.edu/abs/2013ApJ...762L..21M},
	doi = {10.1088/2041-8205/762/2/L21},
	urldate = {2024-10-04},
	journal = {The Astrophysical Journal},
	author = {MacGregor, Meredith A. and Wilner, David J. and Rosenfeld, Katherine A. and Andrews, Sean M. and Matthews, Brenda and Hughes, A. Meredith and Booth, Mark and Chiang, Eugene and Graham, James R. and Kalas, Paul and Kennedy, Grant and Sibthorpe, Bruce},
	month = jan,
	year = {2013},
	note = {Publisher: IOP
ADS Bibcode: 2013ApJ...762L..21M},
	pages = {L21},
}

@article{kalas_discovery_2004,
	title = {Discovery of a {Large} {Dust} {Disk} {Around} the {Nearby} {Star} {AU} {Microscopii}},
	volume = {303},
	issn = {0036-8075},
	url = {https://ui.adsabs.harvard.edu/abs/2004Sci...303.1990K},
	doi = {10.1126/science.1093420},
	urldate = {2024-10-04},
	journal = {Science},
	author = {Kalas, Paul and Liu, Michael C. and Matthews, Brenda C.},
	month = mar,
	year = {2004},
	note = {ADS Bibcode: 2004Sci...303.1990K},
	pages = {1990--1992},
}

@article{planes_dust-dust_2024,
	title = {Dust-dust collisions in cometary comas: applications to comet {67P}/{Churyumov}-{Gerasimenko}},
	volume = {531},
	issn = {0035-8711},
	shorttitle = {Dust-dust collisions in cometary comas},
	url = {https://ui.adsabs.harvard.edu/abs/2024MNRAS.531.3168P},
	doi = {10.1093/mnras/stae1078},
	urldate = {2024-10-16},
	journal = {Monthly Notices of the Royal Astronomical Society},
	author = {Planes, María Belén and Parisi, M. Gabriela and Millán, Emmanuel N. and Bringa, Eduardo M. and Cañada-Assandri, Marcela},
	month = jul,
	year = {2024},
	note = {Publisher: OUP
ADS Bibcode: 2024MNRAS.531.3168P},
	pages = {3168--3186},
}

@article{schleicher_narrowband_1998,
	title = {Narrowband {Photometry} of {Comet} {P}/{Halley}: {Variation} with {Heliocentric} {Distance}, {Season}, and {Solar} {Phase} {Angle}},
	volume = {132},
	issn = {0019-1035},
	shorttitle = {Narrowband {Photometry} of {Comet} {P}/{Halley}},
	url = {https://www.sciencedirect.com/science/article/pii/S0019103597959029},
	doi = {10.1006/icar.1997.5902},
	number = {2},
	urldate = {2024-10-31},
	journal = {Icarus},
	author = {Schleicher, David G. and Millis, Robert L. and Birch, Peter V.},
	month = apr,
	year = {1998},
	pages = {397--417},
}

@article{lecavelier_des_etangs_library_1999,
	title = {A library of stellar light variations due to extra-solar comets},
	volume = {140},
	issn = {0365-0138},
	url = {https://ui.adsabs.harvard.edu/abs/1999A&AS..140...15L},
	doi = {10.1051/aas:1999114},
	urldate = {2024-10-31},
	journal = {Astronomy and Astrophysics Supplement Series},
	author = {Lecavelier Des Etangs, A.},
	month = nov,
	year = {1999},
	note = {ADS Bibcode: 1999A\&AS..140...15L},
	pages = {15--20},
}

@article{fink_calculation_2012,
	title = {The calculation of {Afrho} and mass loss rate for comets},
	volume = {221},
	issn = {0019-1035},
	url = {https://ui.adsabs.harvard.edu/abs/2012Icar..221..721F},
	doi = {10.1016/j.icarus.2012.09.001},
	urldate = {2024-10-31},
	journal = {Icarus},
	author = {Fink, Uwe and Rubin, Martin},
	month = nov,
	year = {2012},
	note = {ADS Bibcode: 2012Icar..221..721F},
	pages = {721--734},
}

@article{lukyanyk_numerical_2024,
	title = {Numerical simulations of exocomet transits: {Insights} from beta {Pic} and {KIC} 3542116},
	volume = {688},
	issn = {0004-6361},
	shorttitle = {Numerical simulations of exocomet transits},
	url = {https://ui.adsabs.harvard.edu/abs/2024A&A...688A..65L},
	doi = {10.1051/0004-6361/202348498},
	urldate = {2024-10-31},
	journal = {Astronomy and Astrophysics},
	author = {Luk'yanyk, I. and Kulyk, I. and Shubina, O. and Pavlenko, Ya. and Vasylenko, M. and Dobrycheva, D. and Korsun, P.},
	month = aug,
	year = {2024},
	note = {ADS Bibcode: 2024A\&A...688A..65L},
	pages = {A65},
}

@article{combi_hale-bopp_2002,
	title = {Hale-{Bopp}: {What} {Makes} a {Big} {Comet} {Different}? {Coma} {Dynamics}: {Observations} and {Theory}},
	volume = {89},
	issn = {0167-9295},
	shorttitle = {Hale-{Bopp}},
	url = {https://ui.adsabs.harvard.edu/abs/2002EM&P...89...73C},
	doi = {10.1023/A:1021534117319},
	urldate = {2024-11-01},
	journal = {Earth Moon and Planets},
	author = {Combi, Michael},
	month = oct,
	year = {2002},
	note = {Publisher: Springer
ADS Bibcode: 2002EM\&P...89...73C},
	pages = {73--90},
}

@article{tenishev_numerical_2011,
	title = {{NUMERICAL} {SIMULATION} {OF} {DUST} {IN} {A} {COMETARY} {COMA}: {APPLICATION} {TO} {COMET} {67P}/{CHURYUMOV}-{GERASIMENKO}},
	volume = {732},
	issn = {0004-637X},
	shorttitle = {{NUMERICAL} {SIMULATION} {OF} {DUST} {IN} {A} {COMETARY} {COMA}},
	url = {https://dx.doi.org/10.1088/0004-637X/732/2/104},
	doi = {10.1088/0004-637X/732/2/104},
	language = {en},
	number = {2},
	urldate = {2024-11-01},
	journal = {The Astrophysical Journal},
	author = {Tenishev, Valeriy and Combi, Michael R. and Rubin, Martin},
	month = apr,
	year = {2011},
	note = {Publisher: The American Astronomical Society},
	pages = {104},
}

@article{burns_radiation_1979,
	title = {Radiation forces on small particles in the solar system},
	volume = {40},
	issn = {0019-1035},
	url = {https://www.sciencedirect.com/science/article/pii/0019103579900502},
	doi = {10.1016/0019-1035(79)90050-2},
	number = {1},
	urldate = {2024-11-01},
	journal = {Icarus},
	author = {Burns, Joseph A. and Lamy, Philippe L. and Soter, Steven},
	month = oct,
	year = {1979},
	pages = {1--48},
}

@ARTICLE{budaj2015,
       author = {{Budaj}, J. and {Kocifaj}, M. and {Salmeron}, R. and {Hubeny}, I.},
        title = "{Tables of phase functions, opacities, albedos, equilibrium temperatures, and radiative accelerations of dust grains in exoplanets}",
      journal = {\mnras},
         year = 2015,
        month = nov,
       volume = {454},
       number = {1},
        pages = {2-27},
          doi = {10.1093/mnras/stv1711},
archivePrefix = {arXiv},
       eprint = {1505.08013},
 primaryClass = {astro-ph.EP},
       adsurl = {https://ui.adsabs.harvard.edu/abs/2015MNRAS.454....2B}
}

@ARTICLE{Kimura2002,
       author = {{Kimura}, Hiroshi and {Mann}, Ingrid and {Biesecker}, Douglas A. and {Jessberger}, Elmar K.},
        title = "{Dust Grains in the Comae and Tails of Sungrazing Comets: Modeling of Their Mineralogical and Morphological Properties}",
      journal = {\icarus},
         year = 2002,
        month = oct,
       volume = {159},
       number = {2},
        pages = {529-541},
          doi = {10.1006/icar.2002.6940},
       adsurl = {https://ui.adsabs.harvard.edu/abs/2002Icar..159..529K}
}

@ARTICLE{vanlieshout2014,
       author = {{van Lieshout}, R. and {Min}, M. and {Dominik}, C.},
        title = "{Dusty tails of evaporating exoplanets. I. Constraints on the dust composition}",
      journal = {\aap},
         year = 2014,
        month = dec,
       volume = {572},
          eid = {A76},
        pages = {A76},
          doi = {10.1051/0004-6361/201424876},
archivePrefix = {arXiv},
       eprint = {1410.3494},
 primaryClass = {astro-ph.EP},
       adsurl = {https://ui.adsabs.harvard.edu/abs/2014A&A...572A..76V}
}

@ARTICLE{PHOENIX,
       author = {{Husser}, T. -O. and {Wende-von Berg}, S. and {Dreizler}, S. and {Homeier}, D. and {Reiners}, A. and {Barman}, T. and {Hauschildt}, P.~H.},
        title = "{A new extensive library of PHOENIX stellar atmospheres and synthetic spectra}",
      journal = {\aap},
         year = 2013,
        month = may,
       volume = {553},
          eid = {A6},
        pages = {A6},
          doi = {10.1051/0004-6361/201219058},
archivePrefix = {arXiv},
       eprint = {1303.5632},
 primaryClass = {astro-ph.SR},
       adsurl = {https://ui.adsabs.harvard.edu/abs/2013A&A...553A...6H}
}

@ARTICLE{Phidrates,
       author = {{Huebner}, W.~F. and {Keady}, J.~J. and {Lyon}, S.~P.},
        title = "{Solar Photo Rates for Planetary Atmospheres and Atmospheric Pollutants}",
      journal = {\apss},
         year = 1992,
        month = sep,
       volume = {195},
       number = {1},
        pages = {1-294},
          doi = {10.1007/BF00644558},
       adsurl = {https://ui.adsabs.harvard.edu/abs/1992Ap&SS.195....1H}
}

@ARTICLE{Arnaud1985,
       author = {{Arnaud}, M. and {Rothenflug}, R.},
        title = "{An Updated Evaluation of Recombination and Ionization Rates}",
      journal = {\aaps},
         year = 1985,
        month = jun,
       volume = {60},
        pages = {425},
       adsurl = {https://ui.adsabs.harvard.edu/abs/1985A&AS...60..425A}
}

@ARTICLE{Vrignaud2024b,
       author = {{Vrignaud}, T. and {Lecavelier des Etangs}, A.},
        title = "{Population of excited levels of Fe+, Ni+ and Cr+ in exocomets gaseous tails}",
      journal = {arXiv e-prints},
         year = 2024,
        month = sep,
          eid = {arXiv:2409.15247},
        pages = {arXiv:2409.15247},
          doi = {10.48550/arXiv.2409.15247},
archivePrefix = {arXiv},
       eprint = {2409.15247},
 primaryClass = {astro-ph.EP},
       adsurl = {https://ui.adsabs.harvard.edu/abs/2024arXiv240915247V}
}

@ARTICLE{Shull1982,
       author = {{Shull}, J.~M. and {van Steenberg}, M.},
        title = "{The ionization equilibrium of astrophysically abundant elements.}",
      journal = {\apjs},
         year = 1982,
        month = jan,
       volume = {48},
        pages = {95-107},
          doi = {10.1086/190769},
       adsurl = {https://ui.adsabs.harvard.edu/abs/1982ApJS...48...95S}
}

@ARTICLE{Mazzotta1998,
       author = {{Mazzotta}, P. and {Mazzitelli}, G. and {Colafrancesco}, S. and {Vittorio}, N.},
        title = "{Ionization balance for optically thin plasmas: Rate coefficients for all atoms and ions of the elements H to NI}",
      journal = {\aaps},
         year = 1998,
        month = dec,
       volume = {133},
        pages = {403-409},
          doi = {10.1051/aas:1998330},
archivePrefix = {arXiv},
       eprint = {astro-ph/9806391},
 primaryClass = {astro-ph},
       adsurl = {https://ui.adsabs.harvard.edu/abs/1998A&AS..133..403M}
}

@ARTICLE{Vrignaud2024a,
       author = {{Vrignaud}, T. and {Lecavelier des Etangs}, A. and {Kiefer}, F. and {Lagrange}, A. -M. and {H{\'e}brard}, G. and {Str{\o}m}, P.~A. and {Vidal-Madjar}, A.},
        title = "{Curves of growth for transiting exocomets: Application to Fe II lines in the {\ensuremath{\beta}} Pictoris system}",
      journal = {\aap},
         year = 2024,
        month = apr,
       volume = {684},
          eid = {A210},
        pages = {A210},
          doi = {10.1051/0004-6361/202347588},
archivePrefix = {arXiv},
       eprint = {2402.10169},
 primaryClass = {astro-ph.EP},
       adsurl = {https://ui.adsabs.harvard.edu/abs/2024A&A...684A.210V}
}

@ARTICLE{Beust2024,
       author = {{Beust}, H. and {Milli}, J. and {Morbidelli}, A. and {Lacour}, S. and {Lagrange}, A. -M. and {Chauvin}, G. and {Bonnefoy}, M. and {Wang}, J.},
        title = "{Dynamics of the {\ensuremath{\beta}} Pictoris planetary system and its falling evaporating bodies}",
      journal = {\aap},
         year = 2024,
        month = mar,
       volume = {683},
          eid = {A89},
        pages = {A89},
          doi = {10.1051/0004-6361/202348203},
archivePrefix = {arXiv},
       eprint = {2401.00715},
 primaryClass = {astro-ph.EP},
       adsurl = {https://ui.adsabs.harvard.edu/abs/2024A&A...683A..89B}
}

@PHDTHESIS{Rebollido_PhD,
       author = {{Rebollido}, Isabel},
        title = "{Exocomets: A study of the gaseous environment of A-type main sequence stars}",
       school = {Autonomous University of Madrid, Spain},
         year = 2020,
        month = jan,
       adsurl = {https://ui.adsabs.harvard.edu/abs/2020PhDT.........3R}
}

@ARTICLE{zieba2019,
       author = {{Zieba}, S. and {Zwintz}, K. and {Kenworthy}, M.~A. and {Kennedy}, G.~M.},
        title = "{Transiting exocomets detected in broadband light by TESS in the {\ensuremath{\beta}} Pictoris system}",
      journal = {\aap},
         year = 2019,
        month = may,
       volume = {625},
          eid = {L13},
        pages = {L13},
          doi = {10.1051/0004-6361/201935552},
archivePrefix = {arXiv},
       eprint = {1903.11071},
 primaryClass = {astro-ph.SR},
       adsurl = {https://ui.adsabs.harvard.edu/abs/2019A&A...625L..13Z}
}

@ARTICLE{Rappaport2018,
       author = {{Rappaport}, S. and {Vanderburg}, A. and {Jacobs}, T. and {LaCourse}, D. and {Jenkins}, J. and {Kraus}, A. and {Rizzuto}, A. and {Latham}, D.~W. and {Bieryla}, A. and {Lazarevic}, M. and {Schmitt}, A.},
        title = "{Likely transiting exocomets detected by Kepler}",
      journal = {\mnras},
         year = 2018,
        month = feb,
       volume = {474},
       number = {2},
        pages = {1453-1468},
          doi = {10.1093/mnras/stx2735},
archivePrefix = {arXiv},
       eprint = {1708.06069},
 primaryClass = {astro-ph.EP},
       adsurl = {https://ui.adsabs.harvard.edu/abs/2018MNRAS.474.1453R}
}

@ARTICLE{Gray2006,
       author = {{Gray}, R.~O. and {Corbally}, C.~J. and {Garrison}, R.~F. and {McFadden}, M.~T. and {Bubar}, E.~J. and {McGahee}, C.~E. and {O'Donoghue}, A.~A. and {Knox}, E.~R.},
        title = "{Contributions to the Nearby Stars (NStars) Project: Spectroscopy of Stars Earlier than M0 within 40 pc-The Southern Sample}",
      journal = {\aj},
         year = 2006,
        month = jul,
       volume = {132},
       number = {1},
        pages = {161-170},
          doi = {10.1086/504637},
archivePrefix = {arXiv},
       eprint = {astro-ph/0603770},
 primaryClass = {astro-ph},
       adsurl = {https://ui.adsabs.harvard.edu/abs/2006AJ....132..161G}
}

@ARTICLE{Lagrange1995,
       author = {{Lagrange}, A.~M. and {Vidal-Madjar}, A. and {Deleuil}, M. and {Emerich}, C. and {Beust}, H. and {Ferlet}, R.},
        title = "{The {\ensuremath{\beta}} Pictoris circumstellar disk. XVII. Physical and chemical parameters of the disk.}",
      journal = {\aap},
         year = 1995,
        month = apr,
       volume = {296},
        pages = {499},
       adsurl = {https://ui.adsabs.harvard.edu/abs/1995A&A...296..499L}
}

@ARTICLE{Beust1989,
       author = {{Beust}, H. and {Lagrange-Henri}, A.~M. and {Vidal-Madjar}, A. and {Ferlet}, R.},
        title = "{The beta Pictoris circumstellar disk. IX. Theoretical results on the infall velocities of CA II, AI III and MG II.}",
      journal = {\aap},
         year = 1989,
        month = oct,
       volume = {223},
        pages = {304-312},
       adsurl = {https://ui.adsabs.harvard.edu/abs/1989A&A...223..304B}
}

@BOOK{McDaniel1964,
       author = {{McDaniel}, Earl W.},
        title = "{Collision phenomena in ionized gases}",
         year = 1964,
       adsurl = {https://ui.adsabs.harvard.edu/abs/1964cpig.book.....M}
}

@ARTICLE{Kwok1975,
       author = {{Kwok}, S.},
        title = "{Radiation pressure on grains as a mechanism for mass loss in red giants.}",
      journal = {\apj},
         year = 1975,
        month = jun,
       volume = {198},
        pages = {583-591},
          doi = {10.1086/153637},
       adsurl = {https://ui.adsabs.harvard.edu/abs/1975ApJ...198..583K}
}

@ARTICLE{Lagrange1998,
       author = {{Lagrange}, A. -M. and {Beust}, H. and {Mouillet}, D. and {Deleuil}, M. and {Feldman}, P.~D. and {Ferlet}, R. and {Hobbs}, L. and {Lecavelier Des Etangs}, A. and {Lissauer}, J.~J. and {McGrath}, M.~A. and {McPhate}, J.~B. and {Spyromilio}, J. and {Tobin}, W. and {Vidal-Madjar}, A.},
        title = "{The beta Pictoris circumstellar disk. XXIV. Clues to the origin of the stable gas}",
      journal = {\aap},
         year = 1998,
        month = feb,
       volume = {330},
        pages = {1091-1108},
       adsurl = {https://ui.adsabs.harvard.edu/abs/1998A&A...330.1091L}
}

@ARTICLE{Beust1990,
       author = {{Beust}, H. and {Lagrange-Henri}, A.~M. and {Vidal-Madjar}, A. and {Ferlet}, R.},
        title = "{The beta Pictoris circumstellar disk. X. Numerical simulations of infalling evaporating bodies.}",
      journal = {\aap},
         year = 1990,
        month = sep,
       volume = {236},
        pages = {202},
       adsurl = {https://ui.adsabs.harvard.edu/abs/1990A&A...236..202B}
}

@ARTICLE{Beust2001,
       author = {{Beust}, H. and {Karmann}, C. and {Lagrange}, A. -M.},
        title = "{Falling Evaporating Bodies around Herbig stars. A theoretical study}",
      journal = {\aap},
         year = 2001,
        month = feb,
       volume = {366},
        pages = {945-964},
          doi = {10.1051/0004-6361:20000353},
       adsurl = {https://ui.adsabs.harvard.edu/abs/2001A&A...366..945B}
}

@ARTICLE{Landini1991,
       author = {{Landini}, M. and {Fossi}, B.~C.~M.},
        title = "{Ion equilibrium for minor components in a thin plasma}",
      journal = {\aaps},
         year = 1991,
        month = nov,
       volume = {91},
       number = {1},
        pages = {183-196},
       adsurl = {https://ui.adsabs.harvard.edu/abs/1991A&AS...91..183L}
}

@ARTICLE{Lamy1974,
       author = {{Lamy}, P.~L.},
        title = "{Interaction of interplanetary dust grains with the solar radiation field.}",
      journal = {\aap},
         year = 1974,
        month = oct,
       volume = {35},
       number = {2},
        pages = {197-207},
       adsurl = {https://ui.adsabs.harvard.edu/abs/1974A&A....35..197L}
}

@ARTICLE{Bourdais2024,
       author = {{Le Bourdais}, {\'E}rika and {Dufour}, Patrick and {Xu}, Siyi},
        title = "{Revisiting the Chemical Composition of WD 1145+017: Impact of Circumstellar Disks Contamination on Photospheric Abundances}",
      journal = {arXiv e-prints},
         year = 2024,
        month = oct,
          eid = {arXiv:2410.10948},
        pages = {arXiv:2410.10948},
          doi = {10.48550/arXiv.2410.10948},
archivePrefix = {arXiv},
       eprint = {2410.10948},
 primaryClass = {astro-ph.SR},
       adsurl = {https://ui.adsabs.harvard.edu/abs/2024arXiv241010948L}
}

@ARTICLE{Bruhweiler1991,
       author = {{Bruhweiler}, Frederick C. and {Kondo}, Yoji and {Grady}, C.~A.},
        title = "{Mass Outflow in the Nearby Proto--Planetary System beta Pictoris}",
      journal = {\apjl},
         year = 1991,
        month = apr,
       volume = {371},
        pages = {L27},
          doi = {10.1086/185994},
       adsurl = {https://ui.adsabs.harvard.edu/abs/1991ApJ...371L..27B}
}

@ARTICLE{Grady1997,
       author = {{Grady}, C.~A. and {Sitko}, M.~L. and {Bjorkman}, Karen S. and {P{\'e}rez}, Mario R. and {Lynch}, D.~K. and {Russell}, R.~W. and {Hanner}, M.~S.},
        title = "{The Star-grazing Extrasolar Comets in the HD 100546 System}",
      journal = {\apj},
         year = 1997,
        month = jul,
       volume = {483},
       number = {1},
        pages = {449-456},
          doi = {10.1086/304230},
       adsurl = {https://ui.adsabs.harvard.edu/abs/1997ApJ...483..449G}
}

@ARTICLE{Alvarado2022,
       author = {{Alvarado-G{\'o}mez}, Juli{\'a}n D. and {Cohen}, Ofer and {Drake}, Jeremy J. and {Fraschetti}, Federico and {Poppenhaeger}, Katja and {Garraffo}, Cecilia and {Chebly}, Judy and {Ilin}, Ekaterina and {Harbach}, Laura and {Kochukhov}, Oleg},
        title = "{Simulating the Space Weather in the AU Mic System: Stellar Winds and Extreme Coronal Mass Ejections}",
      journal = {\apj},
         year = 2022,
        month = apr,
       volume = {928},
       number = {2},
          eid = {147},
        pages = {147},
          doi = {10.3847/1538-4357/ac54b8},
archivePrefix = {arXiv},
       eprint = {2202.07949},
 primaryClass = {astro-ph.SR},
       adsurl = {https://ui.adsabs.harvard.edu/abs/2022ApJ...928..147A}
}

@ARTICLE{Beust1993,
       author = {{Beust}, Herve and {Tagger}, Michel},
        title = "{A Hydrodynamical Model for Infalling Evaporating Bodies in the {\ensuremath{\beta}} Pictoris Circumstellar Disk}",
      journal = {\icarus},
         year = 1993,
        month = nov,
       volume = {106},
       number = {1},
        pages = {42-58},
          doi = {10.1006/icar.1993.1157},
       adsurl = {https://ui.adsabs.harvard.edu/abs/1993Icar..106...42B}
}

@ARTICLE{Kiefer2014,
       author = {{Kiefer}, F. and {Lecavelier des Etangs}, A. and {Boissier}, J. and {Vidal-Madjar}, A. and {Beust}, H. and {Lagrange}, A. -M. and {H{\'e}brard}, G. and {Ferlet}, R.},
        title = "{Two families of exocomets in the {\ensuremath{\beta}} Pictoris system}",
      journal = {\nat},
         year = 2014,
        month = oct,
       volume = {514},
       number = {7523},
        pages = {462-464},
          doi = {10.1038/nature13849},
       adsurl = {https://ui.adsabs.harvard.edu/abs/2014Natur.514..462K}
}

@ARTICLE{Ferlet1987,
       author = {{Ferlet}, R. and {Hobbs}, L.~M. and {Vidal-Madjar}, A.},
        title = "{The beta Pictoris circumstellar disk. V. Time variations of the Ca II-K line.}",
      journal = {\aap},
         year = 1987,
        month = oct,
       volume = {185},
        pages = {267-270},
       adsurl = {https://ui.adsabs.harvard.edu/abs/1987A&A...185..267F}
}

@ARTICLE{Beust1998,
       author = {{Beust}, H. and {Lagrange}, A. -M. and {Crawford}, I.~A. and {Goudard}, C. and {Spyromilio}, J. and {Vidal-Madjar}, A.},
        title = "{The beta Pictoris circumstellar disk. XXV. The Ca Ii absorption lines and the Falling Evaporating Bodies model revisited using UHRF observations}",
      journal = {\aap},
         year = 1998,
        month = oct,
       volume = {338},
        pages = {1015-1030},
       adsurl = {https://ui.adsabs.harvard.edu/abs/1998A&A...338.1015B}
}

@MISC{Vrignaud2024_HST,
       author = {{Vrignaud}, Theo and {Lecavelier des Etangs}, Alain and {Hebrard}, Guillaume and {Kiefer}, Flavien and {Strom}, Paul A. and {Vidal-Madjar}, Alfred},
        title = "{The C/Fe and dust-to-ice ratios in Beta Pictoris exocomets}",
 howpublished = {HST Proposal. Cycle 32, ID. \#17790},
         year = 2024,
        month = jul,
        pages = {17790},
       adsurl = {https://ui.adsabs.harvard.edu/abs/2024hst..prop17790V}
}

@ARTICLE{Steckloff2021,
       author = {{Steckloff}, Jordan K. and {Debes}, John and {Steele}, Amy and {Johnson}, Brandon and {Adams}, Elisabeth R. and {Jacobson}, Seth A. and {Springmann}, Alessondra},
        title = "{How Sublimation Delays the Onset of Dusty Debris Disk Formation around White Dwarf Stars}",
      journal = {\apjl},
         year = 2021,
        month = jun,
       volume = {913},
       number = {2},
          eid = {L31},
        pages = {L31},
          doi = {10.3847/2041-8213/abfd39},
archivePrefix = {arXiv},
       eprint = {2104.14035},
 primaryClass = {astro-ph.EP},
       adsurl = {https://ui.adsabs.harvard.edu/abs/2021ApJ...913L..31S}
}

@ARTICLE{Shestakova2022,
       author = {{Shestakova}, Lyubov I. and {Kenzhebekova}, Akmaral I. and {Serebryanskiy}, Aleksander V.},
        title = "{On survival of dust grains in the sublimation zone of cold white dwarfs}",
      journal = {\mnras},
         year = 2022,
        month = jul,
       volume = {514},
       number = {1},
        pages = {997-1005},
          doi = {10.1093/mnras/stac1405},
       adsurl = {https://ui.adsabs.harvard.edu/abs/2022MNRAS.514..997S}
}

@ARTICLE{Lagrange1996,
       author = {{Lagrange}, A. -M. and {Plazy}, F. and {Beust}, H. and {Mouillet}, D. and {Deleuil}, M. and {Ferlet}, R. and {Spyromilio}, J. and {Vidal-Madjar}, A. and {Tobin}, W. and {Hearnshaw}, J.~B. and {Clark}, M. and {Thomas}, K.~W.},
        title = "{The {\ensuremath{\beta}} Pictoris circumstellar disk. XXI. Results from the December 1992 spectroscopic campaign.}",
      journal = {\aap},
         year = 1996,
        month = jun,
       volume = {310},
        pages = {547-563},
       adsurl = {https://ui.adsabs.harvard.edu/abs/1996A&A...310..547L}
}

@ARTICLE{Jewitt2012,
       author = {{Jewitt}, David},
        title = "{The Active Asteroids}",
      journal = {\aj},
         year = 2012,
        month = mar,
       volume = {143},
       number = {3},
          eid = {66},
        pages = {66},
          doi = {10.1088/0004-6256/143/3/66},
archivePrefix = {arXiv},
       eprint = {1112.5220},
 primaryClass = {astro-ph.EP},
       adsurl = {https://ui.adsabs.harvard.edu/abs/2012AJ....143...66J}
}

@ARTICLE{Jewitt2014,
       author = {{Jewitt}, David and {Agarwal}, Jessica and {Li}, Jing and {Weaver}, Harold and {Mutchler}, Max and {Larson}, Stephen},
        title = "{Disintegrating Asteroid P/2013 R3}",
      journal = {\apjl},
         year = 2014,
        month = mar,
       volume = {784},
       number = {1},
          eid = {L8},
        pages = {L8},
          doi = {10.1088/2041-8205/784/1/L8},
archivePrefix = {arXiv},
       eprint = {1403.1237},
 primaryClass = {astro-ph.EP},
       adsurl = {https://ui.adsabs.harvard.edu/abs/2014ApJ...784L...8J}
}

@ARTICLE{Masiero2021,
       author = {{Masiero}, Joseph R. and {Davidsson}, Bj{\"o}rn J.~R. and {Liu}, Yang and {Moore}, Kelsey and {Tuite}, Michael},
        title = "{Volatility of Sodium in Carbonaceous Chondrites at Temperatures Consistent with Low-perihelion Asteroids}",
      journal = {\psj},
         year = 2021,
        month = aug,
       volume = {2},
       number = {4},
          eid = {165},
        pages = {165},
          doi = {10.3847/PSJ/ac0d02},
archivePrefix = {arXiv},
       eprint = {2108.07331},
 primaryClass = {astro-ph.EP},
       adsurl = {https://ui.adsabs.harvard.edu/abs/2021PSJ.....2..165M}
}

@book{probstein1969, 
author = {{Probstein}, R.},
place={Philadelphia - Pa}, 
title={in Problems of Hydrodynamics and Continuum Mechanics},  year={1969}}

@ARTICLE{Absil2021,
       author = {{Absil}, O. and {Marion}, L. and {Ertel}, S. and {Defr{\`e}re}, D. and {Kennedy}, G.~M. and {Romagnolo}, A. and {Le Bouquin}, J. -B. and {Christiaens}, V. and {Milli}, J. and {Bonsor}, A. and {Olofsson}, J. and {Su}, K.~Y.~L. and {Augereau}, J. -C.},
        title = "{A near-infrared interferometric survey of debris-disk stars. VII. The hot-to-warm dust connection}",
      journal = {\aap},
         year = 2021,
        month = jul,
       volume = {651},
          eid = {A45},
        pages = {A45},
          doi = {10.1051/0004-6361/202140561},
archivePrefix = {arXiv},
       eprint = {2104.14216},
 primaryClass = {astro-ph.EP},
       adsurl = {https://ui.adsabs.harvard.edu/abs/2021A&A...651A..45A}
}

@ARTICLE{Defrere2012,
       author = {{Defr{\`e}re}, D. and {Lebreton}, J. and {Le Bouquin}, J. -B. and
         {Lagrange}, A. -M. and {Absil}, O. and {Augereau}, J. -C. and
         {Berger}, J. -P. and {di Folco}, E. and {Ertel}, S. and {Kluska}, J. and
         {Montagnier}, G. and {Millan-Gabet}, R. and {Traub}, W. and {Zins}, G.},
        title = "{Hot circumstellar material resolved around {\ensuremath{\beta}} Pic with VLTI/PIONIER}",
      journal = {\aap},
         year = 2012,
        month = oct,
       volume = {546},
          eid = {L9},
        pages = {L9},
          doi = {10.1051/0004-6361/201220287},
archivePrefix = {arXiv},
       eprint = {1210.1914},
 primaryClass = {astro-ph.SR},
       adsurl = {https://ui.adsabs.harvard.edu/abs/2012A&A...546L...9D}
}

@ARTICLE{Ertel2016,
       author = {{Ertel}, S. and {Defr{\`e}re}, D. and {Absil}, O. and
         {Le Bouquin}, J. -B. and {Augereau}, J. -C. and {Berger}, J. -P. and
         {Blind}, N. and {Bonsor}, A. and {Lagrange}, A. -M. and {Lebreton}, J. and
         {Marion}, L. and {Milli}, J. and {Olofsson}, J.},
        title = "{A near-infrared interferometric survey of debris-disc stars. V. PIONIER search for variability}",
      journal = {\aap},
         year = "2016",
        month = "Oct",
       volume = {595},
          eid = {A44},
        pages = {A44},
          doi = {10.1051/0004-6361/201527721},
archivePrefix = {arXiv},
       eprint = {1608.05731},
 primaryClass = {astro-ph.SR},
       adsurl = {https://ui.adsabs.harvard.edu/abs/2016A&A...595A..44E}
}

@ARTICLE{Kimura2020,
       author = {{Kimura}, Hiroshi and {Kunitomo}, Masanobu and {Suzuki}, Takeru K. and
         {Robrade}, Jan and {Thebault}, Philippe and {Mitsuishi}, Ikuyuki},
        title = "{Hot grain dynamics by electric charging and magnetic trapping in debris disks}",
      journal = {\planss},
         year = 2020,
        month = apr,
       volume = {183},
          eid = {104581},
        pages = {104581},
          doi = {10.1016/j.pss.2018.07.010},
       adsurl = {https://ui.adsabs.harvard.edu/abs/2020P&SS..18304581K}
}

@ARTICLE{Kirchschlager2017,
       author = {{Kirchschlager}, Florian and {Wolf}, Sebastian and {Krivov}, Alexander V. and {Mutschke}, Harald and {Brunngr{\"a}ber}, Robert},
        title = "{Constraints on the structure of hot exozodiacal dust belts}",
      journal = {\mnras},
         year = "2017",
        month = "May",
       volume = {467},
       number = {2},
        pages = {1614-1626},
          doi = {10.1093/mnras/stx202},
archivePrefix = {arXiv},
       eprint = {1701.07271},
 primaryClass = {astro-ph.SR},
       adsurl = {https://ui.adsabs.harvard.edu/abs/2017MNRAS.467.1614K}
}

@ARTICLE{Pearce2020,
       author = {{Pearce}, Tim D. and {Krivov}, Alexander V. and {Booth}, Mark},
        title = "{Gas trapping of hot dust around main-sequence stars}",
      journal = {\mnras},
         year = 2020,
        month = aug,
       volume = {498},
       number = {2},
        pages = {2798-2813},
          doi = {10.1093/mnras/staa2514},
archivePrefix = {arXiv},
       eprint = {2008.07505},
 primaryClass = {astro-ph.EP},
       adsurl = {https://ui.adsabs.harvard.edu/abs/2020MNRAS.498.2798P}
}

@ARTICLE{Pearce2022,
       author = {{Pearce}, Tim D. and others},
        title = "{Hot exozodis: cometary supply without trapping is unlikely to be the mechanism}",
      journal = {\mnras},
         year = 2022,
        month = nov,
       volume = {517},
       number = {1},
        pages = {1436-1451},
          doi = {10.1093/mnras/stac2773},
archivePrefix = {arXiv},
       eprint = {2209.11219},
 primaryClass = {astro-ph.EP},
       adsurl = {https://ui.adsabs.harvard.edu/abs/2022MNRAS.517.1436P}
}

@ARTICLE{Rieke2016,
       author = {{Rieke}, G.~H. and {G{\'a}sp{\'a}r}, Andr{\'a}s and {Ballering}, N.~P.},
        title = "{Magnetic grain trapping and the hot excesses around early-type stars}",
      journal = {\apj},
         year = "2016",
        month = "Jan",
       volume = {816},
       number = {2},
          eid = {50},
        pages = {50},
          doi = {10.3847/0004-637X/816/2/50},
archivePrefix = {arXiv},
       eprint = {1511.04998},
 primaryClass = {astro-ph.EP},
       adsurl = {https://ui.adsabs.harvard.edu/abs/2016ApJ...816...50R}
}

@ARTICLE{Sezestre2019,
       author = {{Sezestre}, {\'E}. and {Augereau}, J. -C. and {Th{\'e}bault}, P.},
        title = "{Hot exozodiacal dust: an exocometary origin?}",
      journal = {\aap},
         year = "2019",
        month = "Jun",
       volume = {626},
          eid = {A2},
        pages = {A2},
          doi = {10.1051/0004-6361/201935250},
archivePrefix = {arXiv},
       eprint = {1903.06130},
 primaryClass = {astro-ph.EP},
       adsurl = {https://ui.adsabs.harvard.edu/abs/2019A&A...626A...2S}
}

@article{zwintz_revisiting_2019,
	title = {Revisiting the pulsational characteristics of the exoplanet host star beta {Pictoris}},
	volume = {627},
	issn = {0004-6361},
	url = {https://ui.adsabs.harvard.edu/abs/2019A&A...627A..28Z},
	doi = {10.1051/0004-6361/201834744},
	urldate = {2025-01-13},
	journal = {Astronomy and Astrophysics},
	author = {Zwintz, K. and Reese, D. R. and Neiner, C. and Pigulski, A. and Kuschnig, R. and Müllner, M. and Zieba, S. and Abe, L. and Guillot, T. and Handler, G. and Kenworthy, M. and Stuik, R. and Moffat, A. F. J. and Popowicz, A. and Rucinski, S. M. and Wade, G. A. and Weiss, W. W. and Bailey, J. I. and Crawford, S. and Ireland, M. and Lomberg, B. and Mamajek, E. E. and Mellon, S. N. and Talens, G. J.},
	month = jul,
	year = {2019},
	note = {ADS Bibcode: 2019A\&A...627A..28Z},
	pages = {A28},
}

@article{kramer_dynamical_2014,
	title = {A dynamical analysis of the dust tail of {Comet} {C}/1995 {O1} ({Hale}–{Bopp}) at high heliocentric distances},
	volume = {236},
	issn = {0019-1035},
	url = {https://www.sciencedirect.com/science/article/pii/S0019103514001596},
	doi = {10.1016/j.icarus.2014.03.033},
	urldate = {2025-01-16},
	journal = {Icarus},
	author = {Kramer, Emily A. and Fernandez, Yanga R. and Lisse, Carey M. and Kelley, Michael S. P. and Woodney, Laura M.},
	month = jul,
	year = {2014},
	pages = {136--145},
}

@ARTICLE{Prsa_2016_IAU,
       author = {{Pr{\v{s}}a}, Andrej and {Harmanec}, Petr and {Torres}, Guillermo and {Mamajek}, Eric and {Asplund}, Martin and {Capitaine}, Nicole and {Christensen-Dalsgaard}, J{\o}rgen and {Depagne}, {\'E}ric and {Haberreiter}, Margit and {Hekker}, Saskia and {Hilton}, James and {Kopp}, Greg and {Kostov}, Veselin and {Kurtz}, Donald W. and {Laskar}, Jacques and {Mason}, Brian D. and {Milone}, Eugene F. and {Montgomery}, Michele and {Richards}, Mercedes and {Schmutz}, Werner and {Schou}, Jesper and {Stewart}, Susan G.},
        title = "{Nominal Values for Selected Solar and Planetary Quantities: IAU 2015 Resolution B3}",
      journal = {\aj},
         year = 2016,
        month = aug,
       volume = {152},
       number = {2},
          eid = {41},
        pages = {41},
          doi = {10.3847/0004-6256/152/2/41},
archivePrefix = {arXiv},
       eprint = {1605.09788},
 primaryClass = {astro-ph.SR},
       adsurl = {https://ui.adsabs.harvard.edu/abs/2016AJ....152...41P}
}

@ARTICLE{Bonanno_2002_Sun,
       author = {{Bonanno}, A. and {Schlattl}, H. and {Patern{\`o}}, L.},
        title = "{The age of the Sun and the relativistic corrections in the EOS}",
      journal = {\aap},
         year = 2002,
        month = aug,
       volume = {390},
        pages = {1115-1118},
          doi = {10.1051/0004-6361:20020749},
archivePrefix = {arXiv},
       eprint = {astro-ph/0204331},
 primaryClass = {astro-ph},
       adsurl = {https://ui.adsabs.harvard.edu/abs/2002A&A...390.1115B}
}

@ARTICLE{Zhang_2023,
       author = {{Zhang}, Qicheng and {Battams}, Karl and {Ye}, Quanzhi and {Knight}, Matthew M. and {Schmidt}, Carl A.},
        title = "{Sodium Brightening of (3200) Phaethon near Perihelion}",
      journal = {\psj},
         year = 2023,
        month = apr,
       volume = {4},
       number = {4},
          eid = {70},
        pages = {70},
          doi = {10.3847/PSJ/acc866},
archivePrefix = {arXiv},
       eprint = {2303.17625},
 primaryClass = {astro-ph.EP},
       adsurl = {https://ui.adsabs.harvard.edu/abs/2023PSJ.....4...70Z}
}

@Misc{NIST_ASD,
author = {A.~Kramida and {Yu.~Ralchenko} and
J.~Reader and {and NIST ASD Team}},
HOWPUBLISHED = {{NIST Atomic Spectra Database
(ver. 5.11), [Online]. Available:
{\tt{https://physics.nist.gov/asd}} [2023, December 15].
National Institute of Standards and Technology,
Gaithersburg, MD.}},
year = {2023},
}

@ARTICLE{Plainaki_2010_europa,
       author = {{Plainaki}, C. and {Milillo}, A. and {Mura}, A. and {Orsini}, S. and {Cassidy}, T.},
        title = "{Neutral particle release from Europa{\textquoteright}s surface}",
      journal = {\icarus},
         year = 2010,
        month = nov,
       volume = {210},
       number = {1},
        pages = {385-395},
          doi = {10.1016/j.icarus.2010.06.041},
archivePrefix = {arXiv},
       eprint = {0911.4602},
 primaryClass = {astro-ph.EP},
       adsurl = {https://ui.adsabs.harvard.edu/abs/2010Icar..210..385P}
}

@ARTICLE{Parkinson2003,
       author = {{Parkinson}, W.~H. and {Yoshino}, K.},
        title = "{Absorption cross-section measurements of water vapor in the wavelength region 181 199 nm}",
      journal = {Chemical Physics},
         year = 2003,
        month = oct,
       volume = {294},
       number = {1},
        pages = {31-35},
          doi = {10.1016/S0301-0104(03)00361-6},
       adsurl = {https://ui.adsabs.harvard.edu/abs/2003CP....294...31P}
}

@ARTICLE{Deleuil1993,
       author = {{Deleuil}, M. and {Gry}, C. and {Lagrange-Henri}, A. -M. and {Vidal-Madjar}, A. and {Beust}, H. and {Ferlet}, R. and {Moos}, H.~W. and {Livengood}, T.~A. and {Ziskin}, D. and {Feldman}, P.~D.},
        title = "{The beta Pictoris circumstellar disk. XV. Highly ionized species near beta Pictoris.}",
      journal = {\aap},
         year = 1993,
        month = jan,
       volume = {267},
        pages = {187-193},
       adsurl = {https://ui.adsabs.harvard.edu/abs/1993A&A...267..187D}
}

@ARTICLE{Huebner1992,
       author = {{Huebner}, W.~F. and {Keady}, J.~J. and {Lyon}, S.~P.},
        title = "{Solar Photo Rates for Planetary Atmospheres and Atmospheric Pollutants}",
      journal = {\apss},
         year = 1992,
        month = sep,
       volume = {195},
       number = {1},
        pages = {1-294},
          doi = {10.1007/BF00644558},
       adsurl = {https://ui.adsabs.harvard.edu/abs/1992Ap&SS.195....1H}
}

@ARTICLE{Fulle2007,
       author = {{Fulle}, M. and {Leblanc}, F. and {Harrison}, R.~A. and {Davis}, C.~J. and {Eyles}, C.~J. and {Halain}, J.~P. and {Howard}, R.~A. and {Bockel{\'e}e-Morvan}, D. and {Cremonese}, G. and {Scarmato}, T.},
        title = "{Discovery of the Atomic Iron Tail of Comet MCNaught Using the Heliospheric Imager on STEREO}",
      journal = {\apjl},
         year = 2007,
        month = may,
       volume = {661},
       number = {1},
        pages = {L93-L96},
          doi = {10.1086/518719},
       adsurl = {https://ui.adsabs.harvard.edu/abs/2007ApJ...661L..93F}
}

@article{kharchuk_model_2009,
	title = {Model analysis of the dust tail of comet {Hale}-{Bopp}},
	volume = {25},
	issn = {1934-8401},
	url = {https://doi.org/10.3103/S0884591309040035},
	doi = {10.3103/S0884591309040035},
	language = {en},
	number = {4},
	urldate = {2025-01-20},
	journal = {Kinematics and Physics of Celestial Bodies},
	author = {Kharchuk, S. V. and Korsun, P. P. and Mikuz, H.},
	month = aug,
	year = {2009},
	pages = {189--193},
}

@ARTICLE{Woods2009,
       author = {{Woods}, Thomas N. and {Chamberlin}, Phillip C. and {Harder}, Jerald W. and {Hock}, Rachel A. and {Snow}, Martin and {Eparvier}, Francis G. and {Fontenla}, Juan and {McClintock}, William E. and {Richard}, Erik C.},
        title = "{Solar Irradiance Reference Spectra (SIRS) for the 2008 Whole Heliosphere Interval (WHI)}",
      journal = {\grl},
         year = 2009,
        month = jan,
       volume = {36},
       number = {1},
          eid = {L01101},
        pages = {L01101},
          doi = {10.1029/2008GL036373},
       adsurl = {https://ui.adsabs.harvard.edu/abs/2009GeoRL..36.1101W}
}

@ARTICLE{Feinstein2022,
       author = {{Feinstein}, Adina D. and {France}, Kevin and {Youngblood}, Allison and {Duvvuri}, Girish M. and {Teal}, D.~J. and {Cauley}, P. Wilson and {Seligman}, Darryl Z. and {Gaidos}, Eric and {Kempton}, Eliza M. -R. and {Bean}, Jacob L. and {Diamond-Lowe}, Hannah and {Newton}, Elisabeth and {Ginzburg}, Sivan and {Plavchan}, Peter and {Gao}, Peter and {Schlichting}, Hilke},
        title = "{AU Microscopii in the Far-UV: Observations in Quiescence, during Flares, and Implications for AU Mic b and c}",
      journal = {\aj},
         year = 2022,
        month = sep,
       volume = {164},
       number = {3},
          eid = {110},
        pages = {110},
          doi = {10.3847/1538-3881/ac8107},
archivePrefix = {arXiv},
       eprint = {2205.09606},
 primaryClass = {astro-ph.SR},
       adsurl = {https://ui.adsabs.harvard.edu/abs/2022AJ....164..110F}
}

@INCOLLECTION{Rodgers2004_cometsII,
       author = {{Rodgers}, S.~D. and {Charnley}, S.~B. and {Huebner}, W.~F. and {Boice}, D.~C.},
        title = "{Physical processes and chemical reactions in cometary comae}",
    booktitle = {Comets II},
         year = 2004,
       editor = {{Festou}, Michel C. and {Keller}, H. Uwe and {Weaver}, Harold A.},
        pages = {505},
       adsurl = {https://ui.adsabs.harvard.edu/abs/2004come.book..505R}
}

@ARTICLE{Protopapa2018,
       author = {{Protopapa}, Silvia and {Kelley}, Michael S.~P. and {Yang}, Bin and {Bauer}, James M. and {Kolokolova}, Ludmilla and {Woodward}, Charles E. and {Keane}, Jacqueline V. and {Sunshine}, Jessica M.},
        title = "{Icy Grains from the Nucleus of Comet C/2013 US$_{10}$ (Catalina)}",
      journal = {\apjl},
         year = 2018,
        month = aug,
       volume = {862},
       number = {2},
          eid = {L16},
        pages = {L16},
          doi = {10.3847/2041-8213/aad33b},
archivePrefix = {arXiv},
       eprint = {1807.08215},
 primaryClass = {astro-ph.EP},
       adsurl = {https://ui.adsabs.harvard.edu/abs/2018ApJ...862L..16P}
}

@ARTICLE{Raymond2022,
       author = {{Raymond}, J.~C. and {Giordano}, S. and {Mancuso}, S. and {Povich}, Matthew S. and {Bemporad}, A.},
        title = "{Ultraviolet Observations of Comet 96/P Machholz at Perihelion}",
      journal = {\apj},
         year = 2022,
        month = feb,
       volume = {926},
       number = {1},
          eid = {93},
        pages = {93},
          doi = {10.3847/1538-4357/ac3cbd},
archivePrefix = {arXiv},
       eprint = {2111.15644},
 primaryClass = {astro-ph.EP},
       adsurl = {https://ui.adsabs.harvard.edu/abs/2022ApJ...926...93R}
}

@ARTICLE{Gloeckler2000,
       author = {{Gloeckler}, G. and {Geiss}, J. and {Schwadron}, N.~A. and {Fisk}, L.~A. and {Zurbuchen}, T.~H. and {Ipavich}, F.~M. and {von Steiger}, R. and {Balsiger}, H. and {Wilken}, B.},
        title = "{Interception of comet Hyakutake's ion tail at a distance of 500 million kilometres}",
      journal = {\nat},
         year = 2000,
        month = apr,
       volume = {404},
       number = {6778},
        pages = {576-578},
          doi = {10.1038/35007015},
       adsurl = {https://ui.adsabs.harvard.edu/abs/2000Natur.404..576G}
}

@ARTICLE{Noonan2018,
       author = {{Noonan}, John W. and {Stern}, S. Alan and {Feldman}, Paul D. and {Broiles}, Thomas and {Simon Wedlund}, Cyril and {Edberg}, Niklas J.~T. and {Schindhelm}, Rebecca and {Parker}, Joel Wm. and {Keeney}, Brian A. and {Vervack}, Jr., Ronald J. and {Steffl}, Andrew J. and {Knight}, Matthew M. and {Weaver}, Harold A. and {Feaga}, Lori M. and {A'Hearn}, Michael and {Bertaux}, Jean-Loup},
        title = "{Ultraviolet Observations of Coronal Mass Ejection Impact on Comet 67P/Churyumov-Gerasimenko by Rosetta Alice}",
      journal = {\aj},
         year = 2018,
        month = jul,
       volume = {156},
       number = {1},
          eid = {16},
        pages = {16},
          doi = {10.3847/1538-3881/aac432},
archivePrefix = {arXiv},
       eprint = {1806.06948},
 primaryClass = {astro-ph.EP},
       adsurl = {https://ui.adsabs.harvard.edu/abs/2018AJ....156...16N}
}

@ARTICLE{Galli2022,
       author = {{Galli}, A. and {Vorburger}, A. and {Carberry Mogan}, S.~R. and {Roussos}, E. and {Stenberg Wieser}, G. and {Wurz}, P. and {F{\"o}hn}, M. and {Krupp}, N. and {Fr{\"a}nz}, M. and {Barabash}, S. and {Futaana}, Y. and {Brandt}, P.~C. and {Kollmann}, P. and {Haggerty}, D.~K. and {Jones}, G.~H. and {Johnson}, R.~E. and {Tucker}, O.~J. and {Simon}, S. and {Tippens}, T. and {Liuzzo}, L.},
        title = "{Callisto's Atmosphere and Its Space Environment: Prospects for the Particle Environment Package on Board JUICE}",
      journal = {Earth and Space Science},
         year = 2022,
        month = may,
       volume = {9},
       number = {5},
          eid = {e02172},
        pages = {e02172},
          doi = {10.1029/2021EA002172},
       adsurl = {https://ui.adsabs.harvard.edu/abs/2022E&SS....902172G}
}

@article{kelley_distribution_2013,
	series = {Stardust/{EPOXI}},
	title = {A distribution of large particles in the coma of {Comet} {103P}/{Hartley} 2},
	volume = {222},
	issn = {0019-1035},
	url = {https://www.sciencedirect.com/science/article/pii/S0019103512004459},
	doi = {10.1016/j.icarus.2012.09.037},
	number = {2},
	urldate = {2025-03-05},
	journal = {Icarus},
	author = {Kelley, Michael S. and Lindler, Don J. and Bodewits, Dennis and A’Hearn, Michael F. and Lisse, Carey M. and Kolokolova, Ludmilla and Kissel, Jochen and Hermalyn, Brendan},
	month = feb,
	year = {2013},
	pages = {634--652},
}

@article{reach_distribution_2009,
	title = {Distribution and properties of fragments and debris from the split {Comet} {73P}/{Schwassmann}-{Wachmann} 3 as revealed by \textit{{Spitzer}} {Space} {Telescope}},
	volume = {203},
	issn = {0019-1035},
	url = {https://www.sciencedirect.com/science/article/pii/S0019103509002097},
	doi = {10.1016/j.icarus.2009.05.027},
	number = {2},
	urldate = {2025-03-05},
	journal = {Icarus},
	author = {Reach, William T. and Vaubaillon, Jeremie and Kelley, Michael S. and Lisse, Carey M. and Sykes, Mark V.},
	month = oct,
	year = {2009},
	pages = {571--588},
}

@article{hanlon_empirical_2025,
	title = {An empirical radiation pressure correction for dust at small stellarcentric distances},
	volume = {543},
	issn = {1745-3925},
	url = {https://doi.org/10.1093/mnrasl/slaf087},
	doi = {10.1093/mnrasl/slaf087},
	number = {1},
	urldate = {2025-08-28},
	journal = {Monthly Notices of the Royal Astronomical Society: Letters},
	author = {Hanlon, Jake and Jones, Geraint H},
	month = oct,
	year = {2025},
	pages = {L14--L19},
}

@ARTICLE{Wittrock_2023_AU_mic_d,
       author = {{Wittrock}, Justin M. and {Plavchan}, Peter P. and {Cale}, Bryson L. and {Barclay}, Thomas and {Ludwig}, Mathis R. and {Schwarz}, Richard P. and {M{\'e}karnia}, Djamel and {Triaud}, Amaury H.~M.~J. and {Abe}, Lyu and {Suarez}, Olga and {Guillot}, Tristan and {Conti}, Dennis M. and {Collins}, Karen A. and {Waite}, Ian A. and {Kielkopf}, John F. and {Collins}, Kevin I. and {Dreizler}, Stefan and {El Mufti}, Mohammed and {Feliz}, Dax L. and {Gaidos}, Eric and {Geneser}, Claire S. and {Horne}, Keith D. and {Kane}, Stephen R. and {Lowrance}, Patrick J. and {Martioli}, Eder and {Radford}, Don J. and {Reefe}, Michael A. and {Roccatagliata}, Veronica and {Shporer}, Avi and {Stassun}, Keivan G. and {Stockdale}, Christopher and {Tan}, Thiam-Guan and {Tanner}, Angelle M. and {Vega}, Laura D.},
        title = "{Validating AU Microscopii d with Transit Timing Variations}",
      journal = {\aj},
         year = 2023,
        month = dec,
       volume = {166},
       number = {6},
          eid = {232},
        pages = {232},
          doi = {10.3847/1538-3881/acfda8},
archivePrefix = {arXiv},
       eprint = {2302.04922},
 primaryClass = {astro-ph.EP},
       adsurl = {https://ui.adsabs.harvard.edu/abs/2023AJ....166..232W}
}

@ARTICLE{Nowak2020,
       author = {{Nowak}, M. and {Lacour}, S. and {Lagrange}, A. -M. and {Rubini}, P. and {Wang}, J. and {Stolker}, T. and {Abuter}, R. and {Amorim}, A. and {Asensio-Torres}, R. and {Baub{\"o}ck}, M. and {Benisty}, M. and {Berger}, J.~P. and {Beust}, H. and {Blunt}, S. and {Boccaletti}, A. and {Bonnefoy}, M. and {Bonnet}, H. and {Brandner}, W. and {Cantalloube}, F. and {Charnay}, B. and {Choquet}, E. and {Christiaens}, V. and {Cl{\'e}net}, Y. and {Coud{\'e} Du Foresto}, V. and {Cridland}, A. and {de Zeeuw}, P.~T. and {Dembet}, R. and {Dexter}, J. and {Drescher}, A. and {Duvert}, G. and {Eckart}, A. and {Eisenhauer}, F. and {Gao}, F. and {Garcia}, P. and {Garcia Lopez}, R. and {Gardner}, T. and {Gendron}, E. and {Genzel}, R. and {Gillessen}, S. and {Girard}, J. and {Grandjean}, A. and {Haubois}, X. and {Hei{\ss}el}, G. and {Henning}, T. and {Hinkley}, S. and {Hippler}, S. and {Horrobin}, M. and {Houll{\'e}}, M. and {Hubert}, Z. and {Jim{\'e}nez-Rosales}, A. and {Jocou}, L. and {Kammerer}, J. and {Kervella}, P. and {Keppler}, M. and {Kreidberg}, L. and {Kulikauskas}, M. and {Lapeyr{\`e}re}, V. and {Le Bouquin}, J. -B. and {L{\'e}na}, P. and {M{\'e}rand}, A. and {Maire}, A. -L. and {Molli{\`e}re}, P. and {Monnier}, J.~D. and {Mouillet}, D. and {M{\"u}ller}, A. and {Nasedkin}, E. and {Ott}, T. and {Otten}, G. and {Paumard}, T. and {Paladini}, C. and {Perraut}, K. and {Perrin}, G. and {Pueyo}, L. and {Pfuhl}, O. and {Rameau}, J. and {Rodet}, L. and {Rodr{\'\i}guez-Coira}, G. and {Rousset}, G. and {Scheithauer}, S. and {Shangguan}, J. and {Stadler}, J. and {Straub}, O. and {Straubmeier}, C. and {Sturm}, E. and {Tacconi}, L.~J. and {van Dishoeck}, E.~F. and {Vigan}, A. and {Vincent}, F. and {von Fellenberg}, S.~D. and {Ward-Duong}, K. and {Widmann}, F. and {Wieprecht}, E. and {Wiezorrek}, E. and {Woillez}, J. and {GRAVITY Collaboration}},
        title = "{Direct confirmation of the radial-velocity planet {\ensuremath{\beta}} Pictoris c}",
      journal = {\aap},
         year = 2020,
        month = oct,
       volume = {642},
          eid = {L2},
        pages = {L2},
          doi = {10.1051/0004-6361/202039039},
archivePrefix = {arXiv},
       eprint = {2010.04442},
 primaryClass = {astro-ph.EP},
       adsurl = {https://ui.adsabs.harvard.edu/abs/2020A&A...642L...2N}
}

@ARTICLE{Feng2022_betapic,
       author = {{Feng}, Fabo and {Butler}, R. Paul and {Vogt}, Steven S. and {Clement}, Matthew S. and {Tinney}, C.~G. and {Cui}, Kaiming and {Aizawa}, Masataka and {Jones}, Hugh R.~A. and {Bailey}, J. and {Burt}, Jennifer and {Carter}, B.~D. and {Crane}, Jeffrey D. and {Flammini Dotti}, Francesco and {Holden}, Bradford and {Ma}, Bo and {Ogihara}, Masahiro and {Oppenheimer}, Rebecca and {O'Toole}, S.~J. and {Shectman}, Stephen A. and {Wittenmyer}, Robert A. and {Wang}, Sharon X. and {Wright}, D.~J. and {Xuan}, Yifan},
        title = "{3D Selection of 167 Substellar Companions to Nearby Stars}",
      journal = {\apjs},
         year = 2022,
        month = sep,
       volume = {262},
       number = {1},
          eid = {21},
        pages = {21},
          doi = {10.3847/1538-4365/ac7e57},
archivePrefix = {arXiv},
       eprint = {2208.12720},
 primaryClass = {astro-ph.EP},
       adsurl = {https://ui.adsabs.harvard.edu/abs/2022ApJS..262...21F}
}

@ARTICLE{Bonnefoy2014_betapic,
       author = {{Bonnefoy}, M. and {Marleau}, G. -D. and {Galicher}, R. and {Beust}, H. and {Lagrange}, A. -M. and {Baudino}, J. -L. and {Chauvin}, G. and {Borgniet}, S. and {Meunier}, N. and {Rameau}, J. and {Boccaletti}, A. and {Cumming}, A. and {Helling}, C. and {Homeier}, D. and {Allard}, F. and {Delorme}, P.},
        title = "{Physical and orbital properties of {\ensuremath{\beta}} Pictoris b}",
      journal = {\aap},
         year = 2014,
        month = jul,
       volume = {567},
          eid = {L9},
        pages = {L9},
          doi = {10.1051/0004-6361/201424041},
archivePrefix = {arXiv},
       eprint = {1407.4001},
 primaryClass = {astro-ph.EP},
       adsurl = {https://ui.adsabs.harvard.edu/abs/2014A&A...567L...9B}
}

@ARTICLE{Wu2025_betapic,
       author = {{Wu}, Yanqin and {Worthen}, Kadin and {Brandeker}, Alexis and {Chen}, Christine},
        title = "{Argon in {\ensuremath{\beta}} Pictoris{\textendash}Entrapment and Release of Volatile in Disks}",
      journal = {\apj},
         year = 2025,
        month = apr,
       volume = {982},
       number = {2},
          eid = {123},
        pages = {123},
          doi = {10.3847/1538-4357/ada287},
archivePrefix = {arXiv},
       eprint = {2407.06035},
 primaryClass = {astro-ph.EP},
       adsurl = {https://ui.adsabs.harvard.edu/abs/2025ApJ...982..123W}
}

@ARTICLE{Ferland2013_cloudy,
       author = {{Ferland}, G.~J. and {Porter}, R.~L. and {van Hoof}, P.~A.~M. and {Williams}, R.~J.~R. and {Abel}, N.~P. and {Lykins}, M.~L. and {Shaw}, G. and {Henney}, W.~J. and {Stancil}, P.~C.},
        title = "{The 2013 Release of Cloudy}",
      journal = {\rmxaa},
         year = 2013,
        month = apr,
       volume = {49},
        pages = {137-163},
          doi = {10.48550/arXiv.1302.4485},
archivePrefix = {arXiv},
       eprint = {1302.4485},
 primaryClass = {astro-ph.GA},
       adsurl = {https://ui.adsabs.harvard.edu/abs/2013RMxAA..49..137F}
}

@ARTICLE{Mayyasi2023_mars_H_escape,
       author = {{Mayyasi}, Majd and {Clarke}, John and {Chaufray}, J. -Y. and {Kass}, D. and {Bougher}, S. and {Bhattacharyya}, D. and {Deighan}, J. and {Jain}, S. and {Schneider}, N. and {Villanueva}, G.~L. and {Montmessin}, F. and {Benna}, M. and {Mahaffy}, P. and {Jakosky}, B.},
        title = "{Solar cycle and seasonal variability of H in the upper atmosphere of Mars}",
      journal = {\icarus},
         year = 2023,
        month = mar,
       volume = {393},
          eid = {115293},
        pages = {115293},
          doi = {10.1016/j.icarus.2022.115293},
       adsurl = {https://ui.adsabs.harvard.edu/abs/2023Icar..39315293M}
}

@ARTICLE{Steinmeyer2023_sublimation,
       author = {{Steinmeyer}, Marie-Luise and {Woitke}, Peter and {Johansen}, Anders},
        title = "{Sublimation of refractory minerals in the gas envelopes of accreting rocky planets}",
      journal = {\aap},
         year = 2023,
        month = sep,
       volume = {677},
          eid = {A181},
        pages = {A181},
          doi = {10.1051/0004-6361/202245636},
archivePrefix = {arXiv},
       eprint = {2308.05647},
 primaryClass = {astro-ph.EP},
       adsurl = {https://ui.adsabs.harvard.edu/abs/2023A&A...677A.181S}
}

@ARTICLE{Roberge2000_betapic_CO,
       author = {{Roberge}, A. and {Feldman}, P.~D. and {Lagrange}, A.~M. and {Vidal-Madjar}, A. and {Ferlet}, R. and {Jolly}, A. and {Lemaire}, J.~L. and {Rostas}, F.},
        title = "{High-Resolution Hubble Space Telescope STIS Spectra of C I and CO in the{\ensuremath{\beta}} Pictoris Circumstellar Disk}",
      journal = {\apj},
         year = 2000,
        month = aug,
       volume = {538},
       number = {2},
        pages = {904-910},
          doi = {10.1086/309157},
archivePrefix = {arXiv},
       eprint = {astro-ph/0003446},
 primaryClass = {astro-ph},
       adsurl = {https://ui.adsabs.harvard.edu/abs/2000ApJ...538..904R}
}

@ARTICLE{Dent_2014_betapic_CO,
       author = {{Dent}, W.~R.~F. and {Wyatt}, M.~C. and {Roberge}, A. and {Augereau}, J. -C. and {Casassus}, S. and {Corder}, S. and {Greaves}, J.~S. and {de Gregorio-Monsalvo}, I. and {Hales}, A. and {Jackson}, A.~P. and {Hughes}, A. Meredith and {Lagrange}, A. -M. and {Matthews}, B. and {Wilner}, D.},
        title = "{Molecular Gas Clumps from the Destruction of Icy Bodies in the {\ensuremath{\beta}} Pictoris Debris Disk}",
      journal = {Science},
         year = 2014,
        month = mar,
       volume = {343},
       number = {6178},
        pages = {1490-1492},
          doi = {10.1126/science.1248726},
archivePrefix = {arXiv},
       eprint = {1404.1380},
 primaryClass = {astro-ph.SR},
       adsurl = {https://ui.adsabs.harvard.edu/abs/2014Sci...343.1490D}
}

@ARTICLE{Vrignaud2025,
       author = {{Vrignaud}, T. and {Lecavelier des Etangs}, A. and {Str{\o}m}, P.~A. and {Kiefer}, F.},
        title = "{Abundances of refractory ions in Beta Pictoris exocomets}",
      journal = {\aap},
         year = 2025,
        month = may,
       volume = {697},
          eid = {A21},
        pages = {A21},
          doi = {10.1051/0004-6361/202453568},
archivePrefix = {arXiv},
       eprint = {2503.17346},
 primaryClass = {astro-ph.EP},
       adsurl = {https://ui.adsabs.harvard.edu/abs/2025A&A...697A..21V}
}

@ARTICLE{Furusho_2005_Na1_Swings,
       author = {{Furusho}, Reiko and {Kawakita}, Hideyo and {Fujii}, Mitsugu and {Watanabe}, Jun-ichi},
        title = "{Heliocentric Dependence of Sodium Emission of Comet Hale-Bopp (C/1995O1)}",
      journal = {\apj},
         year = 2005,
        month = jan,
       volume = {618},
       number = {1},
        pages = {543-546},
          doi = {10.1086/425953},
       adsurl = {https://ui.adsabs.harvard.edu/abs/2005ApJ...618..543F}
}

@ARTICLE{McKellar1944_swings,
       author = {{McKellar}, Andrew},
        title = "{Comparison of the {\ensuremath{\lambda}} 3883 CN Band in the Spectra of Comets 1940c and 1942g.}",
      journal = {\apj},
         year = 1944,
        month = jul,
       volume = {100},
        pages = {69},
          doi = {10.1086/144638},
       adsurl = {https://ui.adsabs.harvard.edu/abs/1944ApJ...100...69M}
}

@ARTICLE{Swings1941,
       author = {{Swings}, P. and {Elvey}, C.~T. and {Babcock}, H.~W.},
        title = "{The Spectrum of Comet Cunningham, 1940C.}",
      journal = {\apj},
         year = 1941,
        month = sep,
       volume = {94},
        pages = {320},
          doi = {10.1086/144336},
       adsurl = {https://ui.adsabs.harvard.edu/abs/1941ApJ....94..320S}
}

@ARTICLE{Ferrario2015_white_dwarfs_magnetic_fields,
       author = {{Ferrario}, Lilia and {de Martino}, Domitilla and {G{\"a}nsicke}, Boris T.},
        title = "{Magnetic White Dwarfs}",
      journal = {\ssr},
         year = 2015,
        month = oct,
       volume = {191},
       number = {1-4},
        pages = {111-169},
          doi = {10.1007/s11214-015-0152-0},
archivePrefix = {arXiv},
       eprint = {1504.08072},
 primaryClass = {astro-ph.SR},
       adsurl = {https://ui.adsabs.harvard.edu/abs/2015SSRv..191..111F}
}

@ARTICLE{Hogg2021_WD_magnetic_fields,
       author = {{Hogg}, M.~A. and {Cutter}, R. and {Wynn}, G.~A.},
        title = "{The effect of a magnetic field on the dynamics of debris discs around white dwarfs}",
      journal = {\mnras},
         year = 2021,
        month = jan,
       volume = {500},
       number = {3},
        pages = {2986-3001},
          doi = {10.1093/mnras/staa3316},
archivePrefix = {arXiv},
       eprint = {2009.03444},
 primaryClass = {astro-ph.SR},
       adsurl = {https://ui.adsabs.harvard.edu/abs/2021MNRAS.500.2986H}
}

@ARTICLE{Iseli2002_sungrazing,
       author = {{Iseli}, M. and {K{\"u}ppers}, M. and {Benz}, W. and {Bochsler}, P.},
        title = "{Sungrazing Comets: Properties of Nuclei and in Situ Detectability of Cometary Ions at 1 AU}",
      journal = {\icarus},
         year = 2002,
        month = feb,
       volume = {155},
       number = {2},
        pages = {350-364},
          doi = {10.1006/icar.2001.6722},
archivePrefix = {arXiv},
       eprint = {astro-ph/0110091},
 primaryClass = {astro-ph},
       adsurl = {https://ui.adsabs.harvard.edu/abs/2002Icar..155..350I}
}

@ARTICLE{Preston1967,
       author = {{Preston}, G.~W.},
        title = "{The spectrum of Ikeya-Seki (1965f)}",
      journal = {\apj},
         year = 1967,
        month = feb,
       volume = {147},
        pages = {718-742},
          doi = {10.1086/149049},
       adsurl = {https://ui.adsabs.harvard.edu/abs/1967ApJ...147..718P}
}

@INCOLLECTION{Rickman2010_cometary_dynamics,
       author = {{Rickman}, H.},
        title = "{Cometary Dynamics}",
    booktitle = {Lecture Notes in Physics, Berlin Springer Verlag},
         year = 2010,
       editor = {{Souchay}, J. and {Dvorak}, R.},
       volume = {790},
        pages = {341-399},
          doi = {10.1007/978-3-642-04458-8_7},
       adsurl = {https://ui.adsabs.harvard.edu/abs/2010LNP...790..341R}
}

@ARTICLE{Cochran2015_composition_comets,
       author = {{Cochran}, Anita L. and {Levasseur-Regourd}, Anny-Chantal and {Cordiner}, Martin and {Hadamcik}, Edith and {Lasue}, J{\'e}r{\'e}mie and {Gicquel}, Adeline and {Schleicher}, David G. and {Charnley}, Steven B. and {Mumma}, Michael J. and {Paganini}, Lucas and {Bockel{\'e}e-Morvan}, Dominique and {Biver}, Nicolas and {Kuan}, Yi-Jehng},
        title = "{The Composition of Comets}",
      journal = {\ssr},
         year = 2015,
        month = dec,
       volume = {197},
       number = {1-4},
        pages = {9-46},
          doi = {10.1007/s11214-015-0183-6},
archivePrefix = {arXiv},
       eprint = {1507.00761},
 primaryClass = {astro-ph.EP},
       adsurl = {https://ui.adsabs.harvard.edu/abs/2015SSRv..197....9C}
}

@ARTICLE{Bockel2017_cometary_ices,
       author = {{Bockelée-Morvan}, D. and {Biver}, N.},
        title = "{The composition of cometary ices}",
      journal = {Philosophical Transactions of the Royal Society of London Series A},
         year = 2017,
        month = may,
       volume = {375},
       number = {2097},
          eid = {20160252},
        pages = {20160252},
          doi = {10.1098/rsta.2016.0252},
       adsurl = {https://ui.adsabs.harvard.edu/abs/2017RSPTA.37560252B}
}

@ARTICLE{Brasser2015_comet_fading,
       author = {{Brasser}, R. and {Wang}, J. -H.},
        title = "{An updated estimate of the number of Jupiter-family comets using a simple fading law}",
      journal = {\aap},
         year = 2015,
        month = jan,
       volume = {573},
          eid = {A102},
        pages = {A102},
          doi = {10.1051/0004-6361/201423687},
archivePrefix = {arXiv},
       eprint = {1412.1198},
 primaryClass = {astro-ph.EP},
       adsurl = {https://ui.adsabs.harvard.edu/abs/2015A&A...573A.102B}
}

@ARTICLE{Seligman2022,
       author = {{Seligman}, Darryl Z. and {Rogers}, Leslie A. and {Cabot}, Samuel H.~C. and {Noonan}, John W. and {Kareta}, Theodore and {Mandt}, Kathleen E. and {Ciesla}, Fred and {McKay}, Adam and {Feinstein}, Adina D. and {Levine}, W. Garrett and {Bean}, Jacob L. and {Nordlander}, Thomas and {Krumholz}, Mark R. and {Mansfield}, Megan and {Hoover}, Devin J. and {Van Clepper}, Eric},
        title = "{The Volatile Carbon-to-oxygen Ratio as a Tracer for the Formation Locations of Interstellar Comets}",
      journal = {\psj},
         year = 2022,
        month = jul,
       volume = {3},
       number = {7},
          eid = {150},
        pages = {150},
          doi = {10.3847/PSJ/ac75b5},
archivePrefix = {arXiv},
       eprint = {2204.13211},
 primaryClass = {astro-ph.EP},
       adsurl = {https://ui.adsabs.harvard.edu/abs/2022PSJ.....3..150S}
}

@ARTICLE{Farnham2021_C_2014_distant_activity,
       author = {{Farnham}, Tony L. and {Kelley}, Michael S.~P. and {Bauer}, James M.},
        title = "{Early Activity in Comet C/2014 UN271 Bernardinelli-Bernstein as Observed by TESS}",
      journal = {\psj},
         year = 2021,
        month = dec,
       volume = {2},
       number = {6},
          eid = {236},
        pages = {236},
          doi = {10.3847/PSJ/ac323d},
       adsurl = {https://ui.adsabs.harvard.edu/abs/2021PSJ.....2..236F}
}

@ARTICLE{Hui2024_distant_activity_C_2019,
       author = {{Hui}, Man-To and {Weryk}, Robert and {Micheli}, Marco and {Huang}, Zhong and {Wainscoat}, Richard},
        title = "{Serendipitous Archival Observations of a New Ultradistant Comet C/2019 E3 (ATLAS)}",
      journal = {\aj},
         year = 2024,
        month = mar,
       volume = {167},
       number = {3},
          eid = {140},
        pages = {140},
          doi = {10.3847/1538-3881/ad2500},
       adsurl = {https://ui.adsabs.harvard.edu/abs/2024AJ....167..140H}
}

@ARTICLE{Welsh2016_fe1_betapic,
       author = {{Welsh}, Barry Y. and {Montgomery}, Sharon},
        title = "{Exocomet Circumstellar Fe I Absorption in the Beta Pictoris Gas Disk}",
      journal = {\pasp},
         year = 2016,
        month = jun,
       volume = {128},
       number = {964},
        pages = {064201},
          doi = {10.1088/1538-3873/128/964/064201},
       adsurl = {https://ui.adsabs.harvard.edu/abs/2016PASP..128f4201W}
}

@ARTICLE{Hoeijmakers2025_betapic,
       author = {{Hoeijmakers}, H.~J. and {Jaworska}, K.~P. and {Prinoth}, B.},
        title = "{Exocomets of {\ensuremath{\beta}} Pictoris: I. Exocomet destruction, sodium absorption and disk line variability in 17 years of HARPS observations}",
      journal = {\aap},
         year = 2025,
        month = aug,
       volume = {700},
          eid = {A239},
        pages = {A239},
          doi = {10.1051/0004-6361/202451594},
archivePrefix = {arXiv},
       eprint = {2505.21625},
 primaryClass = {astro-ph.EP},
       adsurl = {https://ui.adsabs.harvard.edu/abs/2025A&A...700A.239H}
}

@MISC{Lagrange1997hst,
       author = {{Lagrange}, Anne-Marie},
        title = "{STIS observations of Beta Pictoris}",
 howpublished = {HST Proposal ID 7512. Cycle 7},
         year = 1997,
        month = jul,
        pages = {7512},
       adsurl = {https://ui.adsabs.harvard.edu/abs/1997hst..prop.7512L}
}

@ARTICLE{Brandeker2011_betapic_braking,
       author = {{Brandeker}, Alexis},
        title = "{Exposing the Gas Braking Mechanism of the {\ensuremath{\beta}} Pictoris Disk}",
      journal = {\apj},
         year = 2011,
        month = mar,
       volume = {729},
       number = {2},
          eid = {122},
        pages = {122},
          doi = {10.1088/0004-637X/729/2/122},
archivePrefix = {arXiv},
       eprint = {1101.4032},
 primaryClass = {astro-ph.SR},
       adsurl = {https://ui.adsabs.harvard.edu/abs/2011ApJ...729..122B}
}

@ARTICLE{Spinrad1968_sodium_velocity,
       author = {{Spinrad}, H. and {Miner}, E.~D.},
        title = "{Sodium velocity fields in Comet 1965f}",
      journal = {\apj},
         year = 1968,
        month = aug,
       volume = {153},
        pages = {355-366},
          doi = {10.1086/149672},
       adsurl = {https://ui.adsabs.harvard.edu/abs/1968ApJ...153..355S}
}

\end{document}